\documentclass[12pt,twoside,letterpaper]{book}

\usepackage{amsfonts,amsmath,amssymb}
\usepackage{bm,bbm}
\usepackage{multirow}
\usepackage[spanish]{babel}
\usepackage[dvips]{graphicx}
\usepackage{url}

\newtheorem{theorem}{Teorema}[chapter]
\newtheorem{definition}[theorem]{Definici\'{o}n}
\newenvironment{proof}[1][Demostraci\'{o}n]{\noindent\textbf{#1.} }{\ \rule{0.5em}{0.5em}}

\begin{document}

\begin{titlepage}
\begin{center}
{\large UNIVERSIDAD DE CONCEPCI\'{O}N} \\
{\large FACULTAD DE CIENCIAS F\'{I}SICAS Y MATEM\'{A}TICAS} \\
{\large DEPARTAMENTO DE F\'{I}SICA}

\vspace{\stretch{15}}
\includegraphics[width=.14\textwidth]{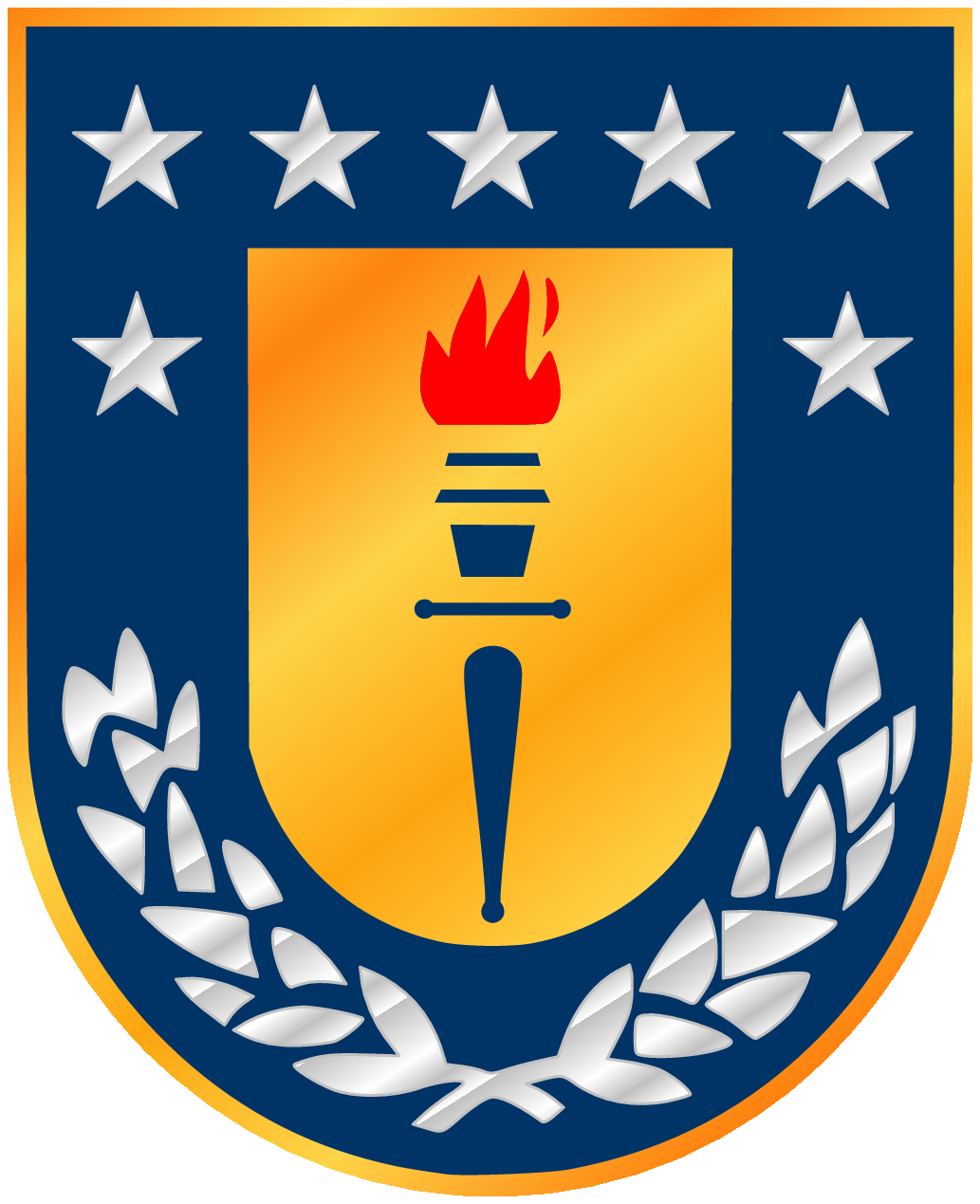}

\vspace{\stretch{20}}
{\huge \textbf{Formas de Transgresi\'{o}n}} \\
{\huge \textbf{y Semigrupos Abelianos}} \\
{\huge \textbf{en Supergravedad}}

\vspace{\stretch{20}}
Tesis en cumplimiento parcial de los requisitos

para optar al grado acad\'{e}mico de

Doctor en Ciencias F\'{\i}sicas

\vspace{\stretch{10}}
por

\vspace{\stretch{10}}
{\Large Eduardo Antonio Rodr\'{\i}guez Salgado}

\vspace{\stretch{20}}
\begin{tabular}[c]{lll}
Director de Tesis & : & Dr. Patricio Salgado \\
& & \\
Comisi\'{o}n & : & Dr. Jos\'{e} A. de Azc\'{a}rraga\\
& & Dr. Sergio del Campo\\
& & Dr. Jorge Zanelli
\end{tabular}

\vspace{\stretch{20}}
Concepci\'{o}n, Chile

Octubre 2006
\end{center}
\end{titlepage}

\ \\
\newpage


\vspace*{.2\textheight}

\begin{flushright}
{\Large \textit{Para Daniela,}} \\
\smallskip
{\Large \textit{por el amor y la paciencia.}}
\end{flushright}

\newpage


\ \\
\newpage


\tableofcontents

\chapter*{Agradecimientos}
\addcontentsline{toc}{chapter}{Agradecimientos}

\pagestyle{plain}

He adquirido una deuda de gratitud con mucha gente en los a\~n{o}s que ha tardado preparar esta Tesis. De la primera \'{e}poca en Concepci\'{o}n, quisiera agradecer a Marcelo Alid, Patricio Mella, Julio Oliva, Jos\'{e} Luis Romero, C\'{e}sar S\'{a}nchez y Ruth Sandoval, por el compa\~{n}erismo y el buen humor.

Tambi\'{e}n quisiera reconocer el trabajo de Sophie von Werder y Maria Hartmann, del DAAD, quienes siempre tuvieron la mejor disposici\'{o}n para ayudarme, desde los primeros d\'{\i}as de postulaci\'{o}n a la beca hasta el momento en que hubo que comprar los pasajes de vuelta.

Gracias tambi\'{e}n a Dolores K\"{u}bler, Sylke Strehle, Ulrike Trebesius y las dem\'{a}s profesoras del Instituto Herder en Leipzig; ich bedanke mich sehr bei Ihnen f\"{u}r die Geduld! Un abrazo asimismo para quienes compartieron conmigo esos meses de aprendizaje duro; ojal\'{a} que les vaya bien donde est\'{e}n.

El Prof. Dieter L\"{u}st fue lo suficientemente amable como para aceptarnos en su grupo sin conocernos; agradezco la cordial hospitalidad y el constante apoyo que nos brind\'{o} tanto en la Humboldt-Universit\"{a}t zu Berlin como en el Arnold Sommerfeld Center for Theoretical Physics de la Ludwig-Maximilians-Universit\"{a}t M\"{u}nchen. Gracias tambi\'{e}n al Prof.~Jorge Zanelli, quien contribuy\'{o} crucialmente para hacer posible la estad\'{\i}a.

Un abrazo para todos aquellos que contribuyeron a mantener una agradable atm\'{o}sfera de trabajo y camarader\'{\i}a, tanto en Berl\'{\i}n como en M\'{u}nich: Natalia Borodatchenkova, Danilo D\'{\i}az, Marija Dimitrijevi\'{c}, Florian Gmeiner, Viviane Gra\ss, Florian Koch, Daniel Krefl, Frank Meyer, Johannes Oberreuter, Dan Oprisa, Jan Perz, Susanne Reffert, Waldemar Schulgin, Maren Stein y Prasanta Tripathy. Un agradecimiento especial a mis compa\~{n}eros de oficina, Murad Alim, Rachid Benhamid, Matteo Cardella, Enrico Pajer y Mario Salizzoni.

El trabajo reportado en esta Tesis fue realizado a cuatro manos con mi amigo Fernando Izaurieta. Es un placer reconocer aqu\'{\i} esta colaboraci\'{o}n y agradecerle por todo lo que hemos aprendido juntos en estos a\~{n}os.

Mi m\'{a}s profunda gratitud es tambi\'{e}n para el Prof. Patricio Salgado, quien crey\'{o} en nosotros desde el primer momento y quien nos ha apoyado incondicionalmente todo este tiempo, incluso a la distancia.

El m\'{a}s cordial agradecimiento al Prof. Jos\'{e} A. de Azc\'{a}rraga por su amable hospitalidad en la Universitat de Val\`{e}ncia durante los meses de noviembre y diciembre de 2005.

Agradezco tambi\'{e}n el trabajo arduo de las secretarias del Departamento de F\'{\i}sica de la Universidad de Concepci\'{o}n, Marta Astudillo, Patricia Luarte y Marcela Sanhueza, as\'{\i} como tambi\'{e}n de los auxiliares, Heraldo Manr\'{\i}quez y V\'{\i}ctor Mora.

Mi familia ha sido una fuente constante de apoyo; gracias por comprender y por confiar.

Finalmente, el m\'{a}s grande de los abrazos va para mi amada Daniela, compa\~{n}era incansable de estas lides. Gracias por el amor, la comprensi\'{o}n, el apoyo, la paciencia, y por haber dejado todo para irse conmigo a un pa\'{\i}s fr\'{\i}o al otro lado del mundo.

Mi dedicaci\'{o}n exclusiva al Programa de Doctorado ha sido posible gracias al apoyo econ\'{o}mico entregado a trav\'{e}s de becas por la Universidad de Concepci\'{o}n (2001, 2006), la Comisi\'{o}n Nacional de Investigaci\'{o}n Cient\'{\i}fica y Tecnol\'{o}gica CONICYT (2002--2003) y el Servicio Alem\'{a}n de Intercambio Acad\'{e}mico DAAD (2003--2006). Agradezco tambi\'{e}n el apoyo circunstancial prestado por el Ministerio de Educaci\'{o}n a trav\'{e}s del Proyecto MECESUP UCO 0209, la Ludwig-Maximilians-Universit\"{a}t M\"{u}nchen por medio del Arnold Sommerfeld Center for Theoretical Physics y la Universitat de Val\`{e}ncia a trav\'{e}s del Departament de F\'{\i}sica Te\`{o}rica.

\chapter*{Resumen}
\addcontentsline{toc}{chapter}{Resumen}

Dos temas principales recorren las p\'{a}ginas de esta Tesis: las formas de transgresi\'{o}n como lagrangeanos para teor\'{\i}as de gauge y la expansi\'{o}n en semigrupos abelianos de \'{a}lgebras de Lie.

Una forma de transgresi\'{o}n es una funci\'{o}n de dos conexiones de gauge cuya propiedad principal es su completa invariancia bajo transformaciones de gauge. A partir de esta forma se construye un lagrangeano, se derivan ecuaciones de movimiento, condiciones de borde y cargas de Noether asociadas. Se propone un m\'{e}todo de separaci\'{o}n en subespacios, basado en la f\'{o}rmula extendida de homotop\'{\i}a de Cartan, que permite (i) separar el lagrangeano en contribuciones de `volumen' y de `borde', y (ii) dividir el t\'{e}rmino de volumen en sublagrangeanos correspondientes a los subespacios del \'{a}lgebra de gauge. A modo de ejemplo se reconstruye una acci\'{o}n transgresora para Gravedad en dimensiones impares.

Se hace uso de semigrupos abelianos para desarrollar un m\'{e}todo de expansi\'{o}n para (super)\'{a}lgebras de Lie, basado en el trabajo de de~Azc\'{a}rraga, Izquierdo, Pic\'{o}n y Varela. La idea central consiste en considerar el producto directo entre un semigrupo abeliano $S$ y una (super)\'{a}lgebra de Lie $\mathfrak{g}$. Se proporcionan condiciones generales bajo las cuales \'{a}lgebras m\'{a}s peque\~{n}as (sub\'{a}lgebras y las llamadas `\'{a}lgebras forzadas') pueden ser extra\'{\i}das de $S \otimes \mathfrak{g}$. Se muestra como recuperar los casos conocidos de expansiones en este nuevo contexto. Algunas super\'{a}lgebras en $d=11$ son obtenidas como ejemplos de aplicaci\'{o}n del m\'{e}todo. Se formulan teoremas generales que permiten encontrar un tensor invariante para el \'{a}lgebra expandida a partir de un tensor invariante para el \'{a}lgebra original.

Finalmente se considera una teor\'{\i}a de gauge en $d=11$ para el \'{a}lgebra~M utilizando las ideas desarrolladas en la Tesis. Las propiedades din\'{a}micas de esta teor\'{\i}a son brevemente analizadas.

\chapter*{Abstract}
\addcontentsline{toc}{chapter}{Abstract}

Two main themes populate this Thesis's pages: transgression forms as Lagrangians for gauge theories and the Abelian semigroup expansion of Lie algebras.

A transgression form is a function of two gauge connections whose main property is its full invariance under gauge transformations. From this form a Lagrangian is built, and equations of motion, boundary conditions and associated Noether currents are derived. A subspace separation method, based on the extended Cartan homotopy formula, is proposed, which allows to (i) split the Lagrangian in `bulk' and `boundary' contributions and (ii) separate the bulk term in sublagrangians corresponding to the subspaces of the gauge algebra. As an example, a transgression action for odd-dimensional Gravity is reconstructed.

Use is made of Abelian semigroups to develop an expansion method for Lie (super)algebras, based on the work by de~Azc\'{a}rraga, Izquierdo, Pic\'{o}n and Varela. The main idea consists in considering the direct product between an Abelian semigroup $S$ and a Lie (super)algebra $\mathfrak{g}$. General conditions under which smaller algebras (subalgebras and the so-called `forced algebras') can be extracted from $S \otimes \mathfrak{g}$ are given. It is shown how to recover the known expansion cases in this new context. Several $d=11$ superalgebras are obtained as examples of the application of the method. General theorems that allow to find an invariant tensor for the expanded algebra from an invariant tensor for the original algebra are formulated.

Finally, a $d=11$ gauge theory for the M~Algebra is considered by using the ideas developed in the Thesis. The dynamical properties of this theory are briefly analyzed.

\chapter{\label{ch:Intro}Introducci\'{o}n}

\pagestyle{headings}

\begin{quote}
\textit{I want to know God's thoughts; the rest are details.}

A.~Einstein (1879--1955).
\end{quote}

\section{Motivaci\'{o}n General}

Hace unos a\~{n}os, en una conversaci\'{o}n sincera con uno de mis profesores, me encontr\'{e} de pronto en la situaci\'{o}n de explicar cu\'{a}les eran mis intenciones de largo plazo en cuanto al estudio profesional de la f\'{\i}sica. No es una pregunta f\'{a}cil de responder, de modo que recurr\'{\i} a un viejo truco: citar una frase para el bronce de un f\'{\i}sico famoso. La frase escogida, una de las m\'{a}s frecuentemente citadas del cient\'{\i}fico con m\'{a}s presencia en el mundo actual, dice as\'{\i} (la traducci\'{o}n es m\'{\i}a): ``Quiero conocer los pensamientos de Dios. El resto son detalles''. No es completamente seguro que Albert Einstein haya jam\'{a}s dicho cosa parecida, pues se le atribuyen un sinn\'{u}mero de frases, pero ciertamente el esp\'{\i}ritu de estas palabras se ajusta a la percepci\'{o}n que hoy tenemos de \'{e}l. Es una frase muy repetida y, sin embargo, sigue siendo la mejor explicaci\'{o}n que puedo dar acerca de por qu\'{e} estoy interesado en el estudio de la f\'{\i}sica.

\textquestiondown Qu\'{e} significa ``conocer los pensamientos de Dios''? El sentido de esta pregunta, al menos para m\'{\i}, es el de entender el mundo; entender por qu\'{e} las cosas son como son, por qu\'{e} ocurren de la manera en que lo hacen. Por qu\'{e}, por ejemplo, podemos recordar el pasado pero no el futuro. Por qu\'{e} el espacio en el que nos movemos tiene tres dimensiones. Por qu\'{e} hay galaxias, luz, seres humanos. Cu\'{a}les son los principios fundamentales, las leyes b\'{a}sicas que rigen el universo y que permiten que todo exista. \textquestiondown Son estas leyes arbitrarias? \textquestiondown Pudieron haber sido quiz\'{a}s distintas?

Entender el c\'{o}mo funcionan las cosas ya es un desaf\'{\i}o formidable. Es claro que ``conocer los pensamientos de Dios'' es una tarea que va m\'{a}s all\'{a} de las capacidades individuales de cualquier persona. Sin embargo, y en este punto podemos recurrir a una frase famosa de otro f\'{\i}sico c\'{e}lebre, si en los \'{u}ltimos cien a\~{n}os hemos sido capaces de ver m\'{a}s all\'{a} de lo que hab\'{\i}a sido comprendido antes, es porque est\'{a}bamos de pie sobre los hombros de gigantes\footnote{La frase es en realidad anterior a Newton, habiendo sido atribuida a Bernard de Chartres, un fil\'{o}sofo franc\'{e}s del siglo XII, por su disc\'{\i}pulo John of Salisbury. El original en lat\'{\i}n es \textit{Pigmaei gigantum humeris impositi plusquam ipsi gigantes vident}.}. La naturaleza acumulativa del conocimiento, y el trabajo de generaciones de pensadores a trav\'{e}s de los siglos, permiten que hoy nos sean de f\'{a}cil acceso nociones que fueron inalcanzables, por ejemplo, para los antiguos griegos.

El punto en el que nos encontramos hoy tiene a los f\'{\i}sicos siguiendo el camino de la Unificaci\'{o}n, cuyos primeros pasos fueron dados por Maxwell. El paradigma de la unificaci\'{o}n sostiene que la naturaleza es una sola, siendo las diferentes ramas de su estudio nada m\'{a}s que construcciones humanas realizadas en el intento de comprenderla mejor, y que deber\'{\i}a existir por lo tanto un modo \textit{unificado} para describirla.

Esta Tesis se enmarca dentro de la b\'{u}squeda de una teor\'{\i}a unificada de las interacciones f\'{\i}sicas. El tema escogido es la construcci\'{o}n de una teor\'{\i}a de Supergravedad haciendo uso de m\'{e}todos matem\'{a}ticos especializados. El concepto fundamental en el que se basa la construcci\'{o}n es el de Simetr\'{\i}a. La primera parte de la Tesis trata de una clase de teor\'{\i}as escogidas por su alto grado de simetr\'{\i}a. Resulta por decir lo menos sorprendente el que la b\'{u}squeda de simetr\'{\i}a en las teor\'{\i}as a menudo conduzca a resultados f\'{\i}sicamente relevantes, es decir, teor\'{\i}as cuyas predicciones pueden ser comprobadas a trav\'{e}s de la experimentaci\'{o}n~\cite{Wig60}. La segunda parte de la Tesis introduce un m\'{e}todo para expandir \textit{\'{a}lgebras de Lie}, que son el instrumento matem\'{a}tico utilizado para describir la simetr\'{\i}a. Este procedimiento de expansi\'{o}n permite comprender la simetr\'{\i}a desde una nueva perspectiva, que resulta provechosa para la f\'{\i}sica. En el cap\'{\i}tulo final de la Tesis se propone una aplicaci\'{o}n de este m\'{e}todo a una teor\'{\i}a concreta, de la clase estudiada en la primera parte. Si bien el resultado obtenido est\'{a} con seguridad a\'{u}n lejos de ser capaz de dar respuesta a todas nuestras preguntas, constituye un paso en la direcci\'{o}n que nos interesa, y una base sobre la cual continuar trabajando.

\section{Los temas tratados en la Tesis}

Los primeros tres cap\'{\i}tulos de esta Tesis describen una clase de teor\'{\i}as de gauge en espacio-tiempos de dimensionalidad impar cuyo funcional de acci\'{o}n es la integral de una forma de transgresi\'{o}n. Las formas de transgresi\'{o}n son funciones de dos potenciales de gauge (uno-formas conexiones) independientes, cada uno con su correspondiente intensidad de campo (dos-forma curvatura). Su principal caracter\'{\i}stica es su invariancia bajo transformaciones de gauge para un grupo o supergrupo de Lie arbitrario. Los lagrangeanos transgresores pueden ser considerados como una generalizaci\'{o}n de los lagrangeanos de Chern--Simons (CS). Por un lado, la forma de CS corresponde al caso particular de una transgresi\'{o}n en que una de las conexiones es puesta igual a cero; por otro, una forma de transgresi\'{o}n puede, en general, ser escrita como la diferencia entre dos formas de CS, una para cada conexi\'{o}n, m\'{a}s una forma exacta.

Escogiendo apropiadamente el grupo de gauge, los lagrangeanos transgresores y de CS permiten realizar, si bien s\'{o}lo en dimensiones impares, el viejo anhelo de interpretar la gravitaci\'{o}n como una teor\'{\i}a de gauge~\cite{Cha89,Kib61,MacDo77,Mil54,Uti56,Wit88}. Debe destacarse que esta realizaci\'{o}n es lograda no en el contexto de una teor\'{\i}a de Yang--Mills, sino para una acci\'{o}n carente de una m\'{e}trica de \textit{background}, en el esp\'{\i}ritu de Relatividad General. Las extensiones supersim\'{e}tricas de estos grupos permiten formular teor\'{\i}as de Supergravedad~\cite{Ach86,Banh96,Cha90,Cre78,Tro96,Tro97,Tro98} donde las transformaciones de supersimetr\'{\i}a se cierran \textit{off-shell} sin que sea necesario introducir campos auxiliares~\cite{Ede06b,Zan05}. El sector gravitacional de estas teor\'{\i}as generaliza la acci\'{o}n de Einstein--Hilbert para incluir potencias m\'{a}s altas de la curvatura~\cite{Banh93,Cha89,Lan38,Lov71,Mar91,Reg86,Zum85,Zwi85}.

En el cap\'{\i}tulo~\ref{ch:CS} se revisan brevemente los aspectos m\'{a}s importantes relacionados con las teor\'{\i}as de CS, enfatizando aquellos que son generalizados al introducir las formas de transgresi\'{o}n.

La acci\'{o}n transgresora~\cite{Aro99c,Aro06,Bor03,Bor05,Iza05,Iza06a,Iza06c,Mor04a,Mor05,Mor06a,Mor06b,Ole05} es estudiada en t\'{e}rminos generales en el cap\'{\i}tulo~\ref{ch:trans}. Las ecuaciones de movimiento, las condiciones de borde inducidas y las cargas de Noether correspondientes son calculadas para un grupo de gauge arbitrario. Tambi\'{e}n se clarifica la relaci\'{o}n entre la forma de transgresi\'{o}n y la de CS, enfatizando los problemas relacionados con la invariancia de gauge~\cite{Iza05,Iza06a}.

En el cap\'{\i}tulo~\ref{ch:metsepsub} se introduce un m\'{e}todo de separaci\'{o}n en subespacios basado en la f\'{o}rmula extendida de la homotop\'{\i}a de Cartan~\cite{Man85}. Este m\'{e}todo permite escribir expl\'{\i}citamente el lagrangeano transgresor usando informaci\'{o}n acerca de la estructura algebraica del grupo de gauge utilizado. El m\'{e}todo prueba ser una herramienta invaluable para el estudio de lagrangeanos transgresores para teor\'{\i}as de Gravedad y Supergravedad, aunque su \'{a}mbito de aplicaci\'{o}n no est\'{a} restringido a ellas.

La acci\'{o}n finita para gravitaci\'{o}n en dimensiones impares introducida en~\cite{Mor04a} es recuperada en el cap\'{\i}tulo~\ref{ch:TGFTGrav} como un ejemplo de acci\'{o}n transgresora para el \'{a}lgebra de anti-de Sitter.

El tema de las formas de transgresi\'{o}n es abandonado temporalmente en el cap\'{\i}tulo~\ref{ch:expansion} para pasar al estudio de un m\'{e}todo de expansi\'{o}n de \'{a}lgebras y super\'{a}lgebras de Lie. El m\'{e}todo propuesto, que llamaremos `$S$-Expansi\'{o}n'~\cite{Iza06b}, est\'{a} inspirado en un m\'{e}todo de expansi\'{o}n introducido por de~Azc\'{a}rraga, Izquierdo, Pic\'{o}n y Varela en~\cite{deAz02}. Las diferencias metodol\'{o}gicas, no obstante, son significativas. Por ejemplo, mientras el m\'{e}todo original de~\cite{deAz02} recurre a la formulaci\'{o}n de Maurer--Cartan para las \'{a}lgebras de Lie, el m\'{e}todo que aqu\'{\i} se propone se basa en operaciones realizadas directamente sobre los generadores del \'{a}lgebra. Un rol fundamental del nuevo m\'{e}todo es jugado por semigrupos abelianos discretos. El uso de semigrupos permite generalizar la expansi\'{o}n en serie realizada en~\cite{deAz02}, la cual puede ser visualizada como una elecci\'{o}n de un semigrupo particular. Como resultado, todos los casos de expansi\'{o}n estudiados en~\cite{deAz02} pueden ser recuperados en este enfoque, mientras que nuevas \'{a}lgebras expandidas son obtenidas al escoger semigrupos distintos.

El m\'{e}todo es ilustrado a trav\'{e}s de diagramas y ejemplos. Los diagramas constituyen una herramienta poderosa para comprender el funcionamiento de la $S$-Expansi\'{o}n. Los ejemplos han sido escogidos (a excepci\'{o}n del primero, cuyo inter\'{e}s es puramente pedag\'{o}gico) de acuerdo a su relevancia para el objetivo de largo plazo de comprender la formulaci\'{o}n geom\'{e}trica de las teor\'{\i}as de supergravedad. Se estudian diferentes posibilidades de expansi\'{o}n de la super\'{a}lgebra $\mathfrak{osp} \left( 32|1 \right)$, las cuales conducen al \'{a}lgebra~M, a una super\'{a}lgebra extendida similar a las estudiadas por D'Auria y Fr\'{e}, y a una nueva super\'{a}lgebra con $\mathcal{N}=2$ que mezcla aspectos de todas ellas. Algunos casos de expansi\'{o}n no directamente relacionados con la aplicaci\'{o}n en Supergravedad son revisados en~\cite{deAz02,Iza06b,Iza06t}.

Una caracter\'{\i}stica relevante del m\'{e}todo de $S$-Expansi\'{o}n es que permite construir tensores invariantes para las \'{a}lgebras expandidas a partir de un tensor invariante para el \'{a}lgebra original. Esta propiedad permite hacer contacto con las teor\'{\i}as de gauge transgresoras estudiadas en los cap\'{\i}tulos anteriores, donde el conocimiento de un tensor invariante para el \'{a}lgebra de gauge constituye un elemento fundamental.

En el cap\'{\i}tulo~\ref{ch:TGFTMAlg} se considera una teor\'{\i}a de gauge en $d=11$ para el \'{a}lgebra~M, construida haciendo uso de lagrangeanos transgresores y del m\'{e}todo de expansi\'{o}n en semigrupos abelianos~\cite{Iza06c}. La acci\'{o}n considerada corresponde a la integral de una forma de transgresi\'{o}n, como las estudiadas en el cap\'{\i}tulo~\ref{ch:trans}. El tensor invariante escogido para el \'{a}lgebra~M es obtenido a partir de la $S$-expansi\'{o}n de la super\'{a}lgebra $\mathfrak{osp} \left( 32|1 \right)$ con un semigrupo apropiado (ver cap\'{\i}tulo~\ref{ch:expansion}). Una forma expl\'{\i}cita para el lagrangeano es obtenida a trav\'{e}s del m\'{e}todo de separaci\'{o}n en subespacios introducido en el cap\'{\i}tulo~\ref{ch:metsepsub}. La teor\'{\i}a presentada en este cap\'{\i}tulo debe ser considerada m\'{a}s una ilustraci\'{o}n de los conceptos mencionados que un intento de producir un modelo realista.

La Tesis concluye en el cap\'{\i}tulo~\ref{ch:final} con un resumen de los resultados y una visi\'{o}n de las perspectivas abiertas por la presente investigaci\'{o}n.

Los Ap\'{e}ndices contienen detalles sobre distintos aspectos de los c\'{a}lculos involucrados en la Tesis as\'{\i} como un resumen de la notaci\'{o}n y las convenciones utilizadas.

\chapter{\label{ch:CS}Teor\'{\i}as de Chern--Simons}

Las teor\'{\i}as de gauge constituyen el fundamento de la f\'{\i}sica de part\'{\i}culas moderna (para una introducci\'{o}n hist\'{o}rica, ver, e.g.,~\cite{Rai00} y las referencias all\'{\i} citadas). En este cap\'{\i}tulo describimos brevemente una clase de teor\'{\i}as de gauge caracterizada por una acci\'{o}n independiente de la m\'{e}trica y por la pseudo-invariancia de gauge: las teor\'{\i}as de Chern--Simons (CS).

\section{\label{sec:CSgral}Formalismo General}

\subsection{\label{sec:1nc}Definiciones}

Sea $\mathfrak{g}$ una (s\'{u}per)algebra de Lie y sea $\bm{P}$ una
$p$-forma valuada en el \'{a}lgebra\footnote{Usamos caracteres en negrita para denotar objetos valuados en una (s\'{u}per)algebra (ver Ap\'{e}ndice~\ref{Ap:NC} para la notaci\'{o}n y las convenciones).}, i.e.
\begin{equation}
\bm{P}=P^{A}\bm{G}_{A},
\end{equation}
donde $\left\{  \bm{G}_{A},A=1,\dotsc,\dim\left(  \mathfrak{g}\right)
\right\}  $ es una base para $\mathfrak{g}$ y los coeficientes $P^{A}$ son
$p$-formas. Las componentes $P_{\mu_{1}\cdots\mu_{p}}^{A}$ de la $p$-forma
$P^{A}$ pueden ser n\'{u}meros complejos ordinarios (conmutantes) o de
Grassmann (anticonmutantes). De acuerdo a la Estad\'{\i}stica est\'{a}ndar, los n\'{u}meros conmutantes se utilizan en la descripci\'{o}n de campos bos\'{o}nicos y los anticonmutantes en la descripci\'{o}n de campos fermi\'{o}nicos~\cite{Ram01}.

El \emph{conmutador} entre $\bm{P}$ y una $q$-forma $\bm{Q}$
est\'{a} definido por\footnote{El producto entre elementos de $\mathfrak{g}$
ha de entenderse en t\'{e}rminos del \emph{\'{a}lgebra envolvente universal},
la cual existe para cualquier \'{a}lgebra de Lie \cite{deAz95}.}
\begin{equation}
\left[  \bm{P},\bm{Q}\right]  =\bm{PQ}-\left(
-1\right)  ^{pq}\bm{QP}. \label{defconm}%
\end{equation}
Esta definici\'{o}n reproduce las nociones usuales de conmutador y
anticonmutador para 0-formas. En efecto, cuando $p=q=0$ tenemos
\begin{align}
\left[  \bm{P},\bm{Q}\right]   &  =P^{A}Q^{B}\left[
\bm{G}_{A},\bm{G}_{B}\right]  ,\quad\text{si }P^{A}\text{
\'{o} }Q^{B}\text{ son n\'{u}meros ordinarios,}\\
\left[  \bm{P},\bm{Q}\right]   &  =P^{A}Q^{B}\left\{
\bm{G}_{A},\bm{G}_{B}\right\}  ,\quad\text{si }P^{A}\text{ y
}Q^{B}\text{ son n\'{u}meros de Grassmann,}%
\end{align}
con
\begin{align}
\left[  \bm{G}_{A},\bm{G}_{B}\right]   &  =\bm{G}%
_{A}\bm{G}_{B}-\bm{G}_{B}\bm{G}_{A},\\
\left\{  \bm{G}_{A},\bm{G}_{B}\right\}   &  =\bm{G}%
_{A}\bm{G}_{B}+\bm{G}_{B}\bm{G}_{A}.
\end{align}

Para formas de grado m\'{a}s alto, la def.~(\ref{defconm}) asegura que el
conmutador entre formas est\'{e} siempre valuado en la (s\'{u}per)algebra
$\mathfrak{g}$.

Una primera aplicaci\'{o}n del conmutador entre formas diferenciales puede apreciarse en la siguiente definici\'{o}n.
\begin{definition}
La \emph{derivada covariante} $\mathrm{D}\bm{Z}$ de una $p$-forma $\bm{Z}$ valuada en $\mathfrak{g}$ est\'{a} definida por
\begin{equation}
\mathrm{D}\bm{Z}=\mathrm{d}\bm{Z}+\left[  \bm{A},\bm{Z}\right]  , \label{DefDerCov}
\end{equation}
donde $\bm{A}$ es una 1-forma. Esta derivada tiene la propiedad que,
si $\bm{Z}$ transforma como un tensor y $\bm{A}$ como una
\emph{conexi\'{o}n} bajo $\mathfrak{g}$, entonces $\mathrm{D}\bm{Z}$
tambi\'{e}n transforma como tensor. La ley de transformaci\'{o}n para una
1-forma conexi\'{o}n $\bm{A}$ es
\begin{equation}
\bm{A}\rightarrow\bm{A}^{\prime}=g\left(  \bm{A}-g^{-1}\mathrm{d}g\right)  g^{-1}, \label{TransConex}
\end{equation}
donde $g=\exp\left(  \lambda^{A}\bm{G}_{A}\right)  $ es un elemento
del \emph{grupo} de gauge cuya \'{a}lgebra es $\mathfrak{g}$.
\end{definition}
La versi\'{o}n
infinitesimal de~(\ref{TransConex}) puede tambi\'{e}n ser utilizada para
motivar la definici\'{o}n de derivada covariante, pues tiene la forma
\begin{equation}
\delta\bm{A}=-\mathrm{D}\bm{\lambda}, \label{deltaAinf}%
\end{equation}
con $\bm{\lambda}=\lambda^{A}\bm{G}_{A}$.

\subsection{\label{sec:TenInvCS}El tensor invariante}

Un elemento esencial en la definici\'{o}n de una forma de CS es un tensor
sim\'{e}trico invariante\footnote{La presentaci\'{o}n dada en esta secci\'{o}n constituye s\'{o}lo un esbozo. Recomendamos al lector interesado consultar las Refs.~\cite{deAz97,Mar91}.} bajo el \'{a}lgebra $\mathfrak{g}$. Un tensor invariante de grado $r$ es
una funci\'{o}n multilineal de la forma
\begin{equation}
\left\langle \cdots\right\rangle _{r}:\;\underset{r}{\underbrace
{\mathfrak{g}\times\dotsb\times\mathfrak{g}}}\rightarrow\mathbb{C}.
\end{equation}

El requerimiento de \emph{simetr\'{\i}a} se traduce en la condici\'{o}n
\begin{equation}
\left\langle \cdots\bm{PQ}\cdots\right\rangle _{r}=\left(  -1\right)
^{pq}\left\langle \cdots\bm{QP}\cdots\right\rangle _{r},
\end{equation}
donde $\bm{P}$ es una $p$-forma y $\bm{Q}$ es una $q$-forma.
Es importante notar que esta condici\'{o}n es v\'{a}lida tanto para
generadores bos\'{o}nicos como fermi\'{o}nicos, dado que ambos van siempre
multiplicados por formas conmutantes o de Grassmann, seg\'{u}n corresponda.

El requerimiento de \emph{invariancia} para $\left\langle \cdots\right\rangle
_{r}$ puede ser escrito de varias maneras equivalentes. Posiblemente la
m\'{a}s fundamental de ellas es
\begin{equation}
\left\langle \left(  g\bm{Z}_{1}g^{-1}\right)  \cdots\left(
g\bm{Z}_{r}g^{-1}\right)  \right\rangle _{r}=\left\langle
\bm{Z}_{1}\cdots\bm{Z}_{r}\right\rangle _{r},
\label{DefTenInv}%
\end{equation}
donde $g=\exp\left(  \lambda^{A}\bm{G}_{A}\right)  $. En t\'{e}rminos de
$\bm{\lambda}=\lambda^{A}\bm{G}_{A}\in\mathfrak{g}$, la
ec.~(\ref{DefTenInv}) toma la forma
\begin{equation}
\left\langle \left[  \bm{\lambda},\bm{Z}_{1}\right]
\bm{Z}_{2}\cdots\bm{Z}_{r}\right\rangle _{r}+\dotsb
+\left\langle \bm{Z}_{1}\cdots\bm{Z}_{r-1}\left[
\bm{\lambda},\bm{Z}_{r}\right]  \right\rangle _{r}=0.
\end{equation}
Admitiendo la posibilidad de que $\bm{\lambda}$ sea una 1-forma (como
la conexi\'{o}n $\bm{A}$) est\'{a} condici\'{o}n se transforma en
\begin{eqnarray}
\left\langle \left[  \bm{A},\bm{Z}_{1}\right]
\bm{Z}_{2}\cdots\bm{Z}_{r}\right\rangle _{r}+\left(
-1\right)  ^{p_{1}}\left\langle \bm{Z}_{1}\left[  \bm{A},\bm{Z}_{2}\right]  \bm{Z}_{3}\cdots\bm{Z}_{r}\right\rangle _{r}+ & & \nonumber\\
+\dotsb+\left(  -1\right)  ^{p_{1}+\dotsb+p_{r-1}}\left\langle
\bm{Z}_{1}\cdots\bm{Z}_{r-1}\left[  \bm{A},\bm{Z}_{r}\right]  \right\rangle _{r} & = & 0, \label{invconcasi}
\end{eqnarray}
donde $p_{k}$ denota el grado de $\bm{Z}_{k}$. Recordando la
definici\'{o}n de derivada covariante, ec.~(\ref{DefDerCov}), vemos que la
condici\'{o}n de invariancia~(\ref{invconcasi}) puede ser escrita tambi\'{e}n
en el modo alternativo
\begin{equation}
\left\langle \mathrm{D}\left(  \bm{Z}_{1}\cdots\bm{Z}%
_{r}\right)  \right\rangle _{r}=\mathrm{d}\left\langle \bm{Z}%
_{1}\cdots\bm{Z}_{r}\right\rangle _{r}. \label{ReqInv}%
\end{equation}
Conviene notar que el lado izquierdo de~(\ref{ReqInv}) no es m\'{a}s que una
manera abreviada y no particularmente rigurosa de escribir
\begin{align}
\left\langle \mathrm{D}\left(  \bm{Z}_{1}\cdots\bm{Z}%
_{r}\right)  \right\rangle _{r} = & \left\langle \left(  \mathrm{D}%
\bm{Z}_{1}\right)  \bm{Z}_{2}\cdots\bm{Z}%
_{r}\right\rangle _{r}+\left(  -1\right)  ^{p_{1}}\left\langle \bm{Z}%
_{1}\left(  \mathrm{D}\bm{Z}_{2}\right)  \bm{Z}_{3}%
\cdots\bm{Z}_{r}\right\rangle _{r}+ \nonumber \\
&  +\dotsb+\left(  -1\right)  ^{p_{1}+\dotsb+p_{r-1}}\left\langle
\bm{Z}_{1}\cdots\bm{Z}_{r-1}\left(  \mathrm{D}\bm{Z}%
_{r}\right)  \right\rangle _{r}.
\end{align}
Todas estas formas equivalentes del requerimiento de invariancia son
utilizadas a lo largo de la Tesis, dependiendo del contexto. La forma~(\ref{DefTenInv}%
), por ejemplo, es \'{u}til para calcular la variaci\'{o}n de gauge de cargas
conservadas (ver secci\'{o}n~\ref{sec:NoCh}, p\'{a}g.~\pageref{sec:NoCh}). La
versi\'{o}n~(\ref{ReqInv}), en cambio, permite demostrar el teorema de
Chern--Weil (secci\'{o}n~\ref{sec:cwt}) y calcular ecuaciones de movimiento y
condiciones de borde (secci\'{o}n~\ref{sec:eomTr}) a partir de una acci\'{o}n invariante de gauge.

A modo de ejemplo, consideremos el \'{a}lgebra de Poincar\'{e} en $d=3$,
$\mathfrak{g}=\mathfrak{iso}\left(  2,1\right)  $. Una base para esta
\'{a}lgebra es dada por $\bm{J}_{ab}$ y $\bm{P}_{a}$, los
cuales satisfacen las relaciones de conmutaci\'{o}n
\begin{align}
\left[  \bm{P}_{a},\bm{P}_{b}\right]   &  =0,\label{PP}\\
\left[  \bm{J}_{ab},\bm{P}_{c}\right]   &  =\eta
_{cb}\bm{P}_{a}-\eta_{ca}\bm{P}_{b},\label{JP}\\
\left[  \bm{J}_{ab},\bm{J}_{cd}\right]   &  =\eta
_{cb}\bm{J}_{ad}-\eta_{ca}\bm{J}_{bd}+\eta_{db}\bm{J}%
_{ca}-\eta_{da}\bm{J}_{cb}, \label{JJ}%
\end{align}
donde $a,b,\dotsc=1,2,3$ y $\eta_{ab}=\operatorname{diag}\left(
-1,1,1\right)  $. Un tensor invariante de rango 2 para $\mathfrak{g}$ es
encontrado en el tensor de Levi-Civita,
\begin{equation}
\left\langle \bm{J}_{ab}\bm{P}_{c}\right\rangle =\varepsilon
_{abc}, \label{levi}%
\end{equation}
con todas las dem\'{a}s componentes iguales a cero. La condici\'{o}n de
invariancia~(\ref{invconcasi}) sobre~(\ref{levi}) toma la forma
\begin{equation}
\left\langle \left[  \bm{A},\bm{J}_{ab}\right]  \bm{P}%
_{c}\right\rangle +\left\langle \bm{J}_{ab}\left[  \bm{A}%
,\bm{P}_{c}\right]  \right\rangle =0, \label{InvPoinc}%
\end{equation}
donde $\bm{A}$ es una 1-forma valuada en $\mathfrak{g}$,
\begin{equation}
\bm{A}=e^{a}\bm{P}_{a}+\frac{1}{2}\omega^{ab}\bm{J}%
_{ab}. \label{ConexPoinc}%
\end{equation}
Reemplazando~(\ref{ConexPoinc}) en~(\ref{InvPoinc}) y utilizando el
\'{a}lgebra~(\ref{PP})--(\ref{JJ}) es f\'{a}cil ver que la condici\'{o}n de
invariancia se reduce a
\begin{equation}
\omega_{\phantom{e}a}^{e}\varepsilon_{ebc}+\omega_{\phantom{e}b}%
^{e}\varepsilon_{aec}+\omega_{\phantom{e}c}^{e}\varepsilon_{abe}=0.
\end{equation}
Esta condici\'{o}n, por supuesto, no es otra cosa que la conocida identidad
\begin{equation}
\mathrm{D}_{\omega}\varepsilon_{abc}=0.
\end{equation}

Para demostrar esta identidad resulta conveniente partir de la delta de
Kronecker generalizada\footnote{Las propiedades m\'{a}s importantes de la delta de Kronecker generalizada est\'{a}n resumidas en el Ap\'{e}ndice~\ref{Ap:Delta}.} $\delta_{abcd}^{efgh}$, la cual se anula id\'{e}nticamente en $d=3$:
\begin{equation}
\delta_{abcd}^{efgh}=0.
\end{equation}
Multiplicando este cero por $\omega_{\phantom{d}e}^{d}$ hallamos
\begin{align*}
0  &  =\omega_{\phantom{d}e}^{d}\delta_{abcd}^{efgh}\\
&  =\omega_{\phantom{d}e}^{d}\left(  \delta_{a}^{e}\delta_{bcd}^{fgh}%
-\delta_{b}^{e}\delta_{acd}^{fgh}+\delta_{c}^{e}\delta_{abd}^{fgh}-\delta
_{d}^{e}\delta_{abc}^{fgh}\right) \\
&  =\omega_{\phantom{d}a}^{d}\delta_{bcd}^{fgh}-\omega_{\phantom{d}b}%
^{d}\delta_{acd}^{fgh}+\omega_{\phantom{d}c}^{d}\delta_{abd}^{fgh}\\
&  =\omega_{\phantom{e}a}^{e}\delta_{ebc}^{fgh}+\omega_{\phantom{e}b}%
^{e}\delta_{aec}^{fgh}+\omega_{\phantom{e}c}^{e}\delta_{abe}^{fgh}.
\end{align*}
Poniendo ahora $f=1$, $g=2$, $h=3$ obtenemos el resultado deseado:
\begin{equation}
\omega_{\phantom{e}a}^{e}\varepsilon_{ebc}+\omega_{\phantom{e}b}%
^{e}\varepsilon_{aec}+\omega_{\phantom{e}c}^{e}\varepsilon_{abe}=0.
\end{equation}
La generalizaci\'{o}n a dimensiones m\'{a}s altas no presenta dificultad
alguna. El tensor de Levi-Civita tambi\'{e}n proporciona un
tensor invariante para los grupos de de Sitter y anti-de Sitter, cuya
contracci\'{o}n de \.{I}n\"{o}n\"{u}--Wigner corresponde a Poincar\'{e}.

\subsection{\label{sec:CSCS}La forma de Chern--Simons}

La forma de CS es una funci\'{o}n polinomial local de una 1-forma
$\bm{A}$ valuada en una (s\'{u}per)algebra de Lie $\mathfrak{g}$.
Expl\'{\i}citamente,
\begin{equation}
\mathcal{Q}_{\mathrm{CS}}^{\left(  2n+1\right)  }\equiv\left(  n+1\right)
\int_{0}^{1}dt\left\langle \bm{A}\left(  t\mathrm{d}\bm{A}%
+t^{2}\bm{A}^{2}\right)  ^{n}\right\rangle , \label{DefCS}%
\end{equation}
donde $\left\langle \cdots\right\rangle $ denota un tensor sim\'{e}trico
invariante bajo $\mathfrak{g}$ de rango $n+1$. Los casos m\'{a}s simples
corresponden a $n=1,2$:
\begin{align}
\mathcal{Q}_{\mathrm{CS}}^{\left(  3\right)  }  &  =\left\langle
\bm{A}\mathrm{d}\bm{A}+\frac{2}{3}\bm{A}%
^{3}\right\rangle ,\\
\mathcal{Q}_{\mathrm{CS}}^{\left(  5\right)  }  &  =\left\langle
\bm{A}\left(  \mathrm{d}\bm{A}\right)  ^{2}+\frac{3}%
{2}\bm{A}^{3}\mathrm{d}\bm{A}+\frac{3}{5}\bm{A}%
^{5}\right\rangle .
\end{align}

Como se aprecia en estos ejemplos, el rol de la integraci\'{o}n sobre $t$
en~(\ref{DefCS}) es el de fijar los coeficientes del polinomio.

La propiedad m\'{a}s importante de la forma de CS tiene que ver con su comportamiento bajo transformaciones de gauge.

Denotando por $\bm{F}=\mathrm{d}\bm{A}+\bm{A}^{2}$ a la curvatura correspondiente a la conexi\'{o}n $\bm{A}$, puede demostrarse que la forma de CS satisface la identidad
\begin{equation}
\mathrm{d} \mathcal{Q}_{\mathrm{CS}}^{\left( 2n+1 \right)} = \left\langle \bm{F}^{n+1} \right\rangle.
\label{dQ=Fn}
\end{equation}
Efectuando una transformaci\'{o}n de gauge a ambos lados de~(\ref{dQ=Fn}) encontramos
\begin{equation}
\mathrm{d} \delta_{\mathrm{gauge}} \mathcal{Q}_{\mathrm{CS}}^{\left( 2n+1 \right)} = 0.
\end{equation}
Esta identidad muestra que la variaci\'{o}n de la forma de CS bajo transformaciones de gauge es una forma exacta.

Realizando expl\'{\i}citamente la variaci\'{o}n, uno encuentra que la forma de CS cambia a
\begin{equation}
\mathcal{Q}_{\mathrm{CS}}^{\left(  2n+1\right)} \left( \bm{A}^{\prime} \right) =\mathcal{Q}_{\mathrm{CS}}^{\left(  2n+1\right)} \left( \bm{A} \right)+\left(
-1\right)  ^{n+1}\frac{n!\left(  n+1\right)  !}{\left(  2n+1\right)
!}\left\langle \left(  g^{-1}\mathrm{d}g\right)  ^{2n+1}\right\rangle
+\mathrm{d}\Omega^{\left( 2n \right)}_{\mathrm{fin}}, \label{finiteCSgt}
\end{equation}
donde $\Omega^{\left( 2n \right)}_{\mathrm{fin}}$ es una $2n$-forma. El segundo t\'{e}rmino en el lado derecho de~(\ref{finiteCSgt}) es una forma exacta que no depende de $\bm{A}$.

Las consideraciones anteriores implican que \emph{localmente} podemos escribir
\begin{equation}
\delta_{\mathrm{gauge}}\mathcal{Q}_{\mathrm{CS}}^{\left(  2n+1\right)
}=\mathrm{d}\Omega^{\left(  2n\right)  },\label{deltaCSinf}
\end{equation}
donde $\Omega^{\left(  2n\right)  }$ es una $2n$-forma que depende de
$\bm{A}$ y su derivada exterior $\mathrm{d}\bm{A}$.

La invariancia m\'{o}dulo derivadas totales de la forma de CS (por lo menos en
el caso de transformaciones infinitesimales) hace que sea interesante
considerarla como lagrangeano para una teor\'{\i}a de gauge basada en
$\mathfrak{g}$. Esta teor\'{\i}a de gauge difiere de la teor\'{\i}a usual de
Yang--Mills (YM) en diversos aspectos, comenzando por el lagrangeano. Para
facilitar la comparaci\'{o}n, partamos escribiendo ambos\footnote{Aqu\'{\i} $k$ es una constante arbitraria adimensional (cuando $\hbar=1$).}:
\begin{align}
L_{\mathrm{YM}}  &  =-\frac{1}{4}\left\langle \bm{F}\wedge\left.
\star\bm{F}\right.  \right\rangle ,\\
L_{\mathrm{CS}}^{\left(  2n+1\right)  }  &  =\left(  n+1\right) k \int_{0}%
^{1}dt\left\langle \bm{A}\left(  t\mathrm{d}\bm{A}%
+t^{2}\bm{A}^{2}\right)  ^{n}\right\rangle . \label{LagCS2n+1}
\end{align}

A continuaci\'{o}n listamos una serie de aspectos relevantes a ambos
lagrangeanos y comentamos sobre las semejanzas y diferencias entre ellos.

\begin{itemize}
\item \textit{Dimensionalidad}: Mientras el lagrangeano de YM puede ser
escrito en cualquier n\'{u}mero de dimensiones, el de CS s\'{o}lo existe en
dimensiones impares.

\item \textit{Tensor invariante}: Ambos lagrangeanos hacen uso de un tensor
sim\'{e}trico $\left\langle \cdots\right\rangle $ invariante bajo el
\'{a}lgebra. En el caso de YM, el tensor es de rango 2 y es usualmente
identificado con la m\'{e}trica de Killing.
Para CS, el tensor es de rango $n+1$ y s\'{o}lo puede vincularse con la m\'{e}trica de Killing en $d=3$.

\item \textit{La m\'{e}trica}: El lagrangeano de CS es independiente de
cualquier m\'{e}trica que pueda existir (o no) en la variedad espacio-temporal
$M$. El lagrangeano de YM, en cambio, necesita de la existencia de una
m\'{e}trica en $M$ para poder ser escrito, debido a la presencia del dual
$\star$ de Hodge\footnote{El dual $\star$ de Hodge relaciona una $p$-forma $\alpha = \left( 1/p! \right) \alpha_{\mu_{1} \cdots \mu_{p}} \mathrm{d}x^{\mu_{1}} \wedge \cdots \wedge \mathrm{d}x^{\mu_{p}}$ en una variedad espacio-temporal $d$-dimensional con la $\left( d-p \right)$-forma $\star \alpha$ definida por $\star \alpha = \left( 1/p! \left( d-p \right)! \right) \sqrt{-g} \varepsilon_{\mu_{1} \cdots \mu_{d}} \alpha^{\mu_{1} \cdots \mu_{p}} \mathrm{d}x^{\mu_{p+1}} \wedge \cdots \wedge \mathrm{d}x^{\mu_{d}}$. Como se desprende de su definici\'{o}n, el dual de Hodge requiere la presencia de una m\'{e}trica invertible en la variedad~\cite{Egu80,Nak03}.}.

\item \textit{Campos independientes}: El \'{u}nico campo independiente en
ambos lagrangeanos es la 1-forma conexi\'{o}n $\bm{A}$. La m\'{e}trica
presente en el lagrangeano de YM es usualmente considerada como un campo de
\textit{background}. Esta m\'{e}trica puede hacerse din\'{a}mica recurriendo a un lagrangeano adicional (el de Einstein--Hilbert, por ejemplo), pero no puede interpretarse como un campo de gauge en el mismo sentido que $\bm{A}$.

\item \textit{Curvatura}: El lagrangeano de YM es cuadr\'{a}tico en la
curvatura $\bm{F}$. El lagrangeano de CS, en cambio, contiene potencias m\'{a}s altas de la curvatura (para $d\geq5$).

\item \textit{Din\'{a}mica}: Las ecuaciones de movimiento\footnote{Las
ecuaciones de movimiento para YM que se obtienen mediante variaci\'{o}n
directa del lagrangeano son $\left\langle \left(  \mathrm{D}\left.
\star\bm{F}\right.  \right)  \bm{G}_{A}\right\rangle =0$.
Cuando la `m\'{e}trica de Killing' $K_{AB}\equiv\left\langle \bm{G}%
_{A}\bm{G}_{B}\right\rangle $ es \emph{invertible}, entonces \'{e}stas
pueden escribirse m\'{a}s sencillamente como $\mathrm{D}\left.  \star
\bm{F}\right.  =0$.} para YM y CS son $\mathrm{D}\left.  \star
F^{A}\right.  =0$ y $\left\langle \bm{F}^{n}\bm{G}%
_{A}\right\rangle =0$, respectivamente. A pesar de la presencia de potencias
de la curvatura mayores que dos en el lagrangeano, las ecuaciones de
movimiento para CS son de primer orden en $\bm{A}$.
\end{itemize}

En esta Tesis nos dedicamos en forma exclusiva al estudio de lagrangeanos de CS y de su generalizaci\'{o}n inmediata, las formas de transgresi\'{o}n (ver cap\'{\i}tulo~\ref{ch:trans}). Esta elecci\'{o}n est\'{a} motivada por el deseo de escribir una teor\'{\i}a unificada, que incluya gravitaci\'{o}n, en donde todos los campos independientes puedan ser interpretados como componentes de una conexi\'{o}n para un grupo de gauge.

\section{\label{sec:CSGrav}Chern--Simons y Gravitaci\'{o}n}

En esta secci\'{o}n consideramos una teor\'{\i}a de CS en $d=2n+1$ dimensiones para el
\'{a}lgebra de anti-de Sitter, $\mathfrak{so}\left(  2n,2\right)  $. La base
can\'{o}nica para esta \'{a}lgebra est\'{a} dada por los generadores
$\bm{P}_{a}$ y $\bm{J}_{ab}$, con las relaciones de
conmutaci\'{o}n
\begin{align}
\left[  \bm{P}_{a},\bm{P}_{b}\right]   &  =\bm{J}%
_{ab},\\
\left[  \bm{J}_{ab},\bm{P}_{c}\right]   &  =\eta
_{cb}\bm{P}_{a}-\eta_{ca}\bm{P}_{b},\\
\left[  \bm{J}_{ab},\bm{J}_{cd}\right]   &  =\eta
_{cb}\bm{J}_{ad}-\eta_{ca}\bm{J}_{bd}+\eta_{db}\bm{J}%
_{ca}-\eta_{da}\bm{J}_{cb},
\end{align}
donde $\eta_{ab}=\operatorname{diag}\left(  -,+,\dotsc,+\right)  $ es la
m\'{e}trica de Minkowski. La estructura de esta \'{a}lgebra permite
intrepretar los campos de gauge asociados a $\bm{P}_{a}$ y
$\bm{J}_{ab}$ como el \emph{vielbein} y la \emph{conexi\'{o}n de spin}
en una formulaci\'{o}n de primer orden de gravedad. Para comprobar esto,
consideremos la 1-forma conexi\'{o}n\footnote{La introducci\'{o}n de la escala de longitud $\ell$ es necesaria debido a que la interpretaci\'{o}n de $e^{a}$ como vielbein requiere que \'{e}ste tenga dimensiones de longitud.}
\begin{equation}
\bm{A}=\frac{1}{\ell}e^{a}\bm{P}_{a}+\frac{1}{2}\omega
^{ab}\bm{J}_{ab} \label{ConexAdS}%
\end{equation}
y un elemento arbitrario $\bm{\lambda}\in\mathfrak{so}\left(
2n,2\right)  $,
\begin{equation}
\bm{\lambda}=\frac{1}{\ell}\lambda^{a}\bm{P}_{a}+\frac{1}%
{2}\lambda^{ab}\bm{J}_{ab}.
\end{equation}
La transformaci\'{o}n de gauge infinitesimal $\delta\bm{A}%
=-\mathrm{D}\bm{\lambda}$ puede ser descompuesta en la forma
\begin{align}
\delta e^{a}  &  =\lambda_{\phantom{a}b}^{a}e^{b}-\mathrm{D}_{\omega}%
\lambda^{a},\label{delta-e}\\
\delta\omega^{ab}  &  =-\mathrm{D}_{\omega}\lambda^{ab}+\frac{1}{\ell^{2}%
}\left(  \lambda^{a}e^{b}-\lambda^{b}e^{a}\right)  . \label{delta-omega}%
\end{align}
Cuando $\bm{\lambda}$ corresponde a una transformaci\'{o}n de Lorentz,
i.e. cuando $\lambda^{a}=0$, las ecs.~(\ref{delta-e})--(\ref{delta-omega})
reproducen las familiares leyes de transformaci\'{o}n para el vielbein y la
conexi\'{o}n de spin. El primero resulta ser un vector de Lorentz y la segunda
una conexi\'{o}n.

La curvatura asociada a la conexi\'{o}n~(\ref{ConexAdS}) es
\[
\bm{F}=\frac{1}{\ell}T^{a}\bm{P}_{a}+\frac{1}{2}\left(
R^{ab}+\frac{1}{\ell^{2}}e^{a}e^{b}\right)  \bm{J}_{ab},
\]
donde
\begin{align}
T^{a}  &  =\mathrm{D}_{\omega}e^{a},\\
R^{ab}  &  =\mathrm{d}\omega^{ab}+\omega_{\phantom{a}c}^{a}\omega^{cb},
\end{align}
corresponden a las definiciones usuales de torsi\'{o}n y curvatura de Lorentz.

Para poder escribir un lagrangeano de CS para esta \'{a}lgebra hace falta un
tensor sim\'{e}trico invariante de rango $n+1$. Al igual que para el
\'{a}lgebra de Poincar\'{e} en $d=3$ (ver secci\'{o}n~\ref{sec:TenInvCS},
p\'{a}g.~\pageref{levi}), podemos recurrir a
\begin{equation}
\left\langle \bm{J}_{a_{1}a_{2}}\cdots\bm{J}_{a_{2n-1}a_{2n}%
}\bm{P}_{a_{2n+1}}\right\rangle =\frac{2^{n}}{n+1}\varepsilon
_{a_{1}\cdots a_{2n+1}}, \label{TenInvEps2n+1}%
\end{equation}
con todas las dem\'{a}s componentes iguales a cero.

En este punto resulta conveniente introducir formas diferenciales valuadas en
el \'{a}lgebra, como
\begin{align}
\bm{e}  &  =\frac{1}{\ell}e^{a}\bm{P}_{a},\\
\bm{\omega}  &  =\frac{1}{2}\omega^{ab}\bm{J}_{ab}.
\end{align}
En t\'{e}rminos de ellas, la curvatura de Lorentz y la torsi\'{o}n adquieren
las expresiones sencillas
\begin{align}
\bm{T}  &  =\mathrm{d}\bm{e}+\left[  \bm{\omega
},\bm{e}\right]  ,\\
\bm{R}  &  =\mathrm{d}\bm{\omega}+\bm{\omega}^{2}.
\end{align}
Las componentes $R^{ab}$ y $T^{a}$ pueden ser recuperadas por medio de
\begin{align}
\bm{T}  &  =\frac{1}{\ell}T^{a}\bm{P}_{a},\\
\bm{R}  &  =\frac{1}{2}R^{ab}\bm{J}_{ab}.
\end{align}
Estas definiciones son extremadamente \'{u}tiles para manipular expresiones
complejas, ya que minimizan la escritura sin comprometer la rigurosidad o la
claridad de la exposici\'{o}n.

Usando esta notaci\'{o}n, el lagrangeano de CS [cf.~ec.~(\ref{LagCS2n+1})] toma la forma\footnote{El lagrangeano~(\ref{LagCSGrav1}) pertenece a una familia de lagrangeanos para gravitaci\'{o}n en dimensiones mayores a cuatro originalmente introducida por D.~Lovelock en~\cite{Lov71} (ver tambi\'{e}n~\cite{Lan38}).}
\begin{equation}
L_{\mathrm{CS}}^{\left(  2n+1\right)  }=\left(  n+1\right)  k\int_{0}%
^{1}dt\left\langle \left(  \bm{R}+t^{2}\bm{e}^{2}\right)
^{n}\bm{e}\right\rangle +\mathrm{d}B_{\mathrm{CS}}^{\left(  2n\right)
}. \label{LagCSGrav1}%
\end{equation}
Si bien en~(\ref{LagCSGrav1}) no hemos reemplazado expl\'{\i}citamente el
tensor invariante~(\ref{TenInvEps2n+1}), s\'{\i} hemos tenido en cuenta
cu\'{a}les de sus componentes son distintas de cero. Una consecuencia directa
de la estructura particular de este tensor invariante es la ausencia de
torsi\'{o}n en el lagrangeano. Otros tensores invariantes producir\'{a}n, en general, lagrangeanos con torsi\'{o}n~\cite{Mar91}.

A modo de comparaci\'{o}n citamos tambi\'{e}n la forma expl\'{\i}cita del
lagrangeano~(\ref{LagCSGrav1}) obtenida al reemplazar el tensor
invariante~(\ref{TenInvEps2n+1}):
\begin{align}
L_{\mathrm{CS}}^{\left(  2n+1\right)  } = & \frac{k}{\ell}\varepsilon
_{a_{1}\cdots a_{2n+1}}\int_{0}^{1}dt\left(  R^{a_{1}a_{2}}+\frac{t^{2}}%
{\ell^{2}}e^{a_{1}}e^{a_{2}}\right)  \times\dotsb\times\nonumber\\
&  \times\left(  R^{a_{2n-1}a_{2n}}+\frac{t^{2}}{\ell^{2}}e^{a_{2n-1}%
}e^{a_{2n}}\right)  e^{a_{2n+1}}+\mathrm{d}B_{\mathrm{CS}}^{\left(  2n\right)
}. \label{LagCSGrav2}%
\end{align}

Hay algo enga\~{n}oso en la aparente sencillez de las
expresiones~(\ref{LagCSGrav1})--(\ref{LagCSGrav2}). Como el lector diligente
habr\'{a} notado, no es inmediato obtenerlas a partir de la definici\'{o}n de
la forma de CS [cf.~ec.~(\ref{DefCS})]. En general, conseguir una
expresi\'{o}n manifiestamente invariante de Lorentz para un lagrangeano de CS
gravitacional a partir de la definici\'{o}n~(\ref{DefCS}) es una tarea no
trivial. Esto se debe a la naturaleza de la forma de CS, la cual depende
de $\mathrm{d}\bm{A}$ y $\bm{A}^{2}$ en forma separada y no a
trav\'{e}s de la combinaci\'{o}n tensorial $\bm{F}=\mathrm{d}%
\bm{A}+\bm{A}^{2}$ (en otras palabras, el lagrangeano de CS no
puede ser escrito s\'{o}lo como funci\'{o}n de $\bm{F}$, siendo
imprescindible incluir tambi\'{e}n $\bm{A}$).

El modo de obtener las expresiones~(\ref{LagCSGrav1}) y~(\ref{LagCSGrav2})
consiste en visualizar la forma de CS como un caso particular de objetos
m\'{a}s generales conocidos como \emph{formas de transgresi\'{o}n}. Las formas
de transgresi\'{o}n son uno de los temas recurrentes de esta Tesis, y como tales son
analizadas en detalle en el cap\'{\i}tulo~\ref{ch:trans}. El formalismo de las transgresiones facilita
enormemente la obtenci\'{o}n de expresiones invariantes de Lorentz para
lagrangeanos gravitacionales de CS, y permite encontrar una forma
expl\'{\i}cita para el t\'{e}rmino de borde $B_{\mathrm{CS}}^{\left(
2n\right)  }$ que aparece en~(\ref{LagCSGrav1}) y~(\ref{LagCSGrav2}).

\chapter{\label{ch:trans}Formas de Transgresi\'{o}n como Lagrangeanos para Teor\'{\i}as de Gauge}
\chaptermark{Formas de Transgresi\'{o}n}

En este cap\'{\i}tulo introducimos uno de los conceptos fundamentales de esta
Tesis: las \emph{formas de transgresi\'{o}n}. Referencias relevantes para el
fundamento matem\'{a}tico de las formas de transgresi\'{o}n son el libro de
J.~A.~de~Azc\'{a}rraga y J.~M.~Izquierdo~\cite{deAz95} y el libro de M.
Nakahara~\cite{Nak03}. La idea de usar las formas de transgresi\'{o}n como
lagrangeanos para teor\'{\i}as de gauge es relativamente nueva,
remont\'{a}ndose a apenas un pu\~{n}ado de art\'{\i}culos
recientes~\cite{Aro99c,Aro06,Bor03,Bor05,Iza05,Iza06a,Iza06c,Mor04a,Mor05,Mor06a,Mor06b,Ole05}. La tesis de doctorado de P.~Mora~\cite{Mor05}
tiene como tema principal el de las formas de transgresi\'{o}n en teor\'{\i}a
de campos.

Las formas de transgresi\'{o}n constituyen la matriz de donde surgen las
formas de CS. En las secciones siguientes revisamos brevemente los aspectos
m\'{a}s importantes relacionados con las formas de transgresi\'{o}n, dejando para el Cap\'{\i}tulo~\ref{ch:metsepsub} la presentaci\'{o}n de un m\'{e}todo de separaci\'{o}n en subespacios que ayuda a interpretar f\'{\i}sicamente un lagrangeano transgresor.

\section{\label{sec:cwt}El Teorema de Chern--Weil}

El teorema de Chern--Weil liga la derivada exterior de la forma de
transgresi\'{o}n con la diferencia entre dos densidades topol\'{o}gicas
asociadas a dos nociones distintas de curvatura.

Sean $\bm{A}$ y $\bar{\bm{A}}$ dos conexiones para una
(s\'{u}per)\'{a}lgebra de Lie $\mathfrak{g}$. Las curvaturas correspondientes
ser\'{a}n denotadas por $\bm{F}$\ y $\bar{\bm{F}}$. Sea
$\left\langle \cdots\right\rangle $ un tensor sim\'{e}trico de rango $r=n+1$
invariante bajo $\mathfrak{g}$ (ver secci\'{o}n~\ref{sec:TenInvCS},
p\'{a}g.~\pageref{sec:TenInvCS}). La \emph{forma de transgresi\'{o}n}
$\mathcal{Q}_{\bm{A}\leftarrow\bar{\bm{A}}}^{\left(
2n+1\right)  }$ es definida como
\begin{equation}
\mathcal{Q}_{\bm{A}\leftarrow\bar{\bm{A}}}^{\left(
2n+1\right)  }\equiv\left(  n+1\right)  \int_{0}^{1}dt\left\langle
\bm{\Theta F}_{t}^{n}\right\rangle , \label{DefTra}%
\end{equation}
donde
\begin{align}
\bm{\Theta}  &  \equiv\bm{A}-\bar{\bm{A}}, \label{Theta} \\
\bm{A}_{t}  &  \equiv\bar{\bm{A}}+t\bm{\Theta}, \label{At} \\
\bm{F}_{t}  &  \equiv\mathrm{d}\bm{A}_{t}+\bm{A}_{t}^{2}. \label{Ft}
\end{align}
La conexi\'{o}n $\bm{A}_{t}$ interpola entre $\bar{\bm{A}}$ y
$\bm{A}$ a medida que $t$ barre el intervalo $0\leq t\leq1$ (esto
explica la notaci\'{o}n escogida para denotar la transgresi\'{o}n).

La propiedad m\'{a}s importante de las formas de transgresi\'{o}n es su
invariancia bajo transformaciones de gauge. Esta invariancia queda establecida
f\'{a}cilmente por medio de la siguiente cadena de argumentos:

\begin{enumerate}
\item La diferencia entre dos conexiones es un tensor: $\bm{\Theta}$
transforma como tensor.

\item La suma de una conexi\'{o}n con un tensor produce otra conexi\'{o}n:
$\bm{A}_{t}$ transforma como conexi\'{o}n.

\item La curvatura asociada a una conexi\'{o}n es un tensor: $\bm{F}%
_{t}$ transforma como tensor.

\item Dado que todos sus argumentos son tensores, la propiedad de invariancia [ver
ec.~(\ref{DefTenInv})] del tensor sim\'{e}trico $\left\langle \cdots
\right\rangle $ garantiza que la forma de transgresi\'{o}n $\mathcal{Q}%
_{\bm{A}\leftarrow\bar{\bm{A}}}^{\left(  2n+1\right)  }$
permanezca invariante bajo transformaciones de gauge.
\end{enumerate}

Consideremos ahora el car\'{a}cter de Chern asociado a cada una de las
conexiones $\bm{A}$ y $\bar{\bm{A}}$, $\left\langle \bm{F}^{n+1} \right\rangle$ y $\left\langle \bar{\bm{F}}^{n+1} \right\rangle$. La \emph{diferencia} entre estos caracteres puede ser expresada en la forma
\begin{equation}
\left\langle \bm{F}^{n+1}\right\rangle -\left\langle \bar
{\bm{F}}^{n+1}\right\rangle =\int_{0}^{1}dt\frac{d}{dt}\left\langle
\bm{F}_{t}^{n+1}\right\rangle . \label{cw1}%
\end{equation}
Aplicando la regla de Leibniz para $d/dt$ y recurriendo a la simetr\'{\i}a de
$\left\langle \cdots\right\rangle $ podemos escribir
\begin{equation}
\left\langle \bm{F}^{n+1}\right\rangle -\left\langle \bar
{\bm{F}}^{n+1}\right\rangle =\left(  n+1\right)  \int_{0}%
^{1}dt\left\langle \bm{F}_{t}^{n}\frac{d}{dt}\bm{F}%
_{t}\right\rangle . \label{cw2}%
\end{equation}

De la definici\'{o}n de $\bm{F}_{t}$ [cf.~ecs.~(\ref{Theta})--(\ref{Ft})] es directo establecer la
identidad
\begin{equation}
\frac{d}{dt}\bm{F}_{t}=\mathrm{D}_{t}\bm{\Theta}, \label{cw3}%
\end{equation}
donde $\mathrm{D}_{t}$ denota la derivada covariante en la conexi\'{o}n
$\bm{A}_{t}$.

Reemplazando~(\ref{cw3}) en~(\ref{cw2}) obtenemos
\begin{equation}
\left\langle \bm{F}^{n+1}\right\rangle -\left\langle \bar
{\bm{F}}^{n+1}\right\rangle =\left(  n+1\right)  \int_{0}%
^{1}dt\left\langle \bm{F}_{t}^{n}\mathrm{D}_{t}\bm{\Theta
}\right\rangle .
\end{equation}
La identidad de Bianchi $\mathrm{D}_{t}\bm{F}_{t}=0$ nos permite ahora
escribir
\begin{equation}
\left\langle \bm{F}^{n+1}\right\rangle -\left\langle \bar
{\bm{F}}^{n+1}\right\rangle =\left(  n+1\right)  \int_{0}%
^{1}dt\left\langle \mathrm{D}_{t}\left(  \bm{F}_{t}^{n}%
\bm{\Theta}\right)  \right\rangle .
\end{equation}
Usando la propiedad de invariancia del tensor sim\'{e}trico $\left\langle
\cdots\right\rangle $ [cf.~ec.~(\ref{ReqInv})] encontramos finalmente
\begin{equation}
\left\langle \bm{F}^{n+1}\right\rangle -\left\langle \bar
{\bm{F}}^{n+1}\right\rangle =\left(  n+1\right)  \mathrm{d}\int
_{0}^{1}dt\left\langle \bm{F}_{t}^{n}\bm{\Theta}\right\rangle
. \label{cw6}%
\end{equation}
En virtud de la simetr\'{\i}a de $\left\langle \cdots\right\rangle $
reconocemos la forma de transgresi\'{o}n en el lado derecho de~(\ref{cw6}). Luego, podemos escribir
\begin{equation}
\left\langle \bm{F}^{n+1}\right\rangle -\left\langle \bar
{\bm{F}}^{n+1}\right\rangle =\mathrm{d}\mathcal{Q}_{\bm{A}%
\leftarrow\bar{\bm{A}}}^{\left(  2n+1\right)  }. \label{cw7}%
\end{equation}

Los pasos anteriores constituyen un esbozo simple de la demostraci\'{o}n del
Teorema de Chern--Weil. Integrando~(\ref{cw7}) sobre una variedad $\left(
2n+2\right)  $-dimensional con borde, el teorema de Chern--Weil expresa la
diferencia entre los caracteres de Chern asociados a cada una de las
conexiones con la integral de la $\left(  2n+1\right)  $-forma
transgresi\'{o}n sobre el borde.

El teorema de Chern--Weil tambi\'{e}n provee de una justificaci\'{o}n para la
pseudo-invariancia de la forma de CS. En efecto, es claro que la forma de CS
puede ser considerada como el caso particular $\bar{\bm{A}}=0$ de una
forma de transgresi\'{o}n. En este caso el teorema de Chern--Weil toma la
forma
\begin{equation}
\left\langle \bm{F}^{n+1}\right\rangle =\mathrm{d}\mathcal{Q}%
_{\bm{A}\leftarrow0}^{\left(  2n+1\right)  }. \label{cwtCS}%
\end{equation}
Realizando una transformaci\'{o}n de gauge a ambos lados de~(\ref{cwtCS})
encontramos $\mathrm{d}\delta\mathcal{Q}_{\bm{A}\leftarrow0}^{\left(
2n+1\right)  }=0$, lo cual implica que, al menos localmente, podemos escribir
$\delta\mathcal{Q}_{\bm{A}\leftarrow0}^{\left(  2n+1\right)
}=\mathrm{d}\Omega$ para alguna $2n$-forma $\Omega$. La p\'{e}rdida de la invariancia
de gauge total al pasar de la forma de transgresi\'{o}n a la forma de CS puede ser entendida notando
que la igualdad $\bar{\bm{A}}=0$ no es preservada bajo
transformaciones de gauge. Una vez que se ha fijado $\bar{\bm{A}}=0$
para obtener la forma de CS, el resultado depende de una \'{u}nica
conexi\'{o}n $\bm{A}$, sin que vuelva a haber mencion alguna de
$\bar{\bm{A}}$. Luego, el t\'{e}rmino no homog\'{e}neo $\mathrm{d}%
gg^{-1}$ que deber\'{\i}a aparecer en la variaci\'{o}n de $\bar{\bm{A}%
}=0$ est\'{a} ausente de la variaci\'{o}n de la forma de CS $\mathcal{Q}_{\bm{A}%
\leftarrow0}^{\left(  2n+1\right)  }$, provocando el quiebre en la invariancia
de gauge.

Esta diferencia crucial --- la invariancia de gauge --- entre la forma de
transgresi\'{o}n y la forma de CS es la motivaci\'{o}n principal para
considerar las formas de transgresi\'{o}n como posibles lagrangeanos para una
teor\'{\i}a de gauge, posibilidad que es explorada en las secciones siguientes
de este cap\'{\i}tulo.

\section{\label{sec:AcTra}La Acci\'{o}n Transgresora}

En esta secci\'{o}n estudiamos en t\'{e}rminos generales (i.e. sin
especificar el \'{a}lgebra de gauge $\mathfrak{g}$ ni la dimensi\'{o}n
del espacio-tiempo) la teor\'{\i}a de gauge definida en una
variedad $\left(  2n+1\right)  $-dimensional $M$ por la acci\'{o}n
\begin{equation}
S_{\mathrm{T}}\left[  \bm{A},\bar{\bm{A}}\right]  =k\int
_{M}\mathcal{Q}_{\bm{A}\leftarrow\bar{\bm{A}}}^{\left(
2n+1\right)  }, \label{AcTra}%
\end{equation}
donde $k$ es una constante adimensional arbitraria y $\mathcal{Q}%
_{\bm{A}\leftarrow\bar{\bm{A}}}^{\left(  2n+1\right)  }$ es la
forma de transgresi\'{o}n definida en~(\ref{DefTra}). La
acci\'{o}n~(\ref{AcTra}) es un funcional de dos campos independientes, las 1-formas conexiones $\bm{A}$ y $\bar{\bm{A}}$.

\subsection{\label{sec:symtra}Simetr\'{\i}as}

La \emph{acci\'{o}n transgresora}~(\ref{AcTra}) es invariante bajo dos grupos de simetr\'{\i}as
independientes; difeomorfismos y transformaciones de gauge.

La invariancia bajo difeomorfismos es la m\'{a}s directa de las dos. Ella
est\'{a} garantizada por el hecho que el lagrangeano en~(\ref{AcTra}) es una
forma diferencial. Todas las formas diferenciales, por construcci\'{o}n, son
invariantes bajo el grupo de difeomorfismos. Esta invariancia puede expresarse
diciendo que la \emph{variaci\'{o}n funcional} infinitesimal de una $p$-forma
$\alpha$ es igual a $\delta\alpha=-\pounds _{\xi}\alpha$, donde $\xi$ es el
campo vectorial (infinitesimal) que genera el difeomorfismo, $\delta x^{\mu
}=\xi^{\mu}\left(  x\right)  $, y $\pounds _{\xi}$ denota la derivada de
Lie\footnote{La \emph{derivada de Lie} $\pounds _{\xi}$ actuando sobre formas
diferenciales puede escribirse en t\'{e}rminos del operador de contracci\'{o}n
$\mathrm{I}_{\xi}$ y la derivada exterior\ $\mathrm{d}$ como $\pounds _{\xi
}=\mathrm{dI}_{\xi}+\mathrm{I}_{\xi}\mathrm{d}$. El \emph{operador de
contracci\'{o}n} $\mathrm{I}_{\xi}$ toma una $p$-forma $\alpha=\left(
1/p!\right)  \alpha_{\mu_{1}\cdots\mu_{p}}\mathrm{d}x^{\mu_{1}}\cdots
\mathrm{d}x^{\mu_{p}}$ y produce la $\left(  p-1\right)  $-forma
$\mathrm{I}_{\xi}\alpha=\left(  1/\left(  p-1\right)  !\right)  \xi^{\mu_{1}%
}\alpha_{\mu_{1}\cdots\mu_{p}}\mathrm{d}x^{\mu_{2}}\cdots\mathrm{d}x^{\mu_{p}%
}$. Este operador es a veces llamado producto interno y denotado por $\xi \rfloor$.}. A modo de referencia citamos aqu\'{\i} las variaciones funcionales de
$\bm{A}$ y $\bar{\bm{A}}$:
\begin{align}
\delta_{\mathrm{dif}}\bm{A}  &  =-\pounds _{\xi}\bm{A}%
,\label{ddA}\\
\delta_{\mathrm{dif}}\bar{\bm{A}}  &  =-\pounds _{\xi}\bar
{\bm{A}}. \label{ddAb}%
\end{align}

La invariancia bajo transformaciones de gauge de la acci\'{o}n~(\ref{AcTra})
est\'{a} garantizada por las propiedades especiales de la forma de transgresi\'{o}n, y es en este sentido
menos evidente que la invariancia bajo difeomorfismos. Las leyes de
transformaci\'{o}n infinitesimales para $\bm{A}$ y $\bar
{\bm{A}}$ son en este caso
\begin{align}
\delta_{\mathrm{gauge}}\bm{A}  &  =-\mathrm{D}\bm{\lambda
},\label{dgA}\\
\delta_{\mathrm{gauge}}\bar{\bm{A}}  &  =-\mathrm{\bar{D}%
}\bm{\lambda}, \label{dgAb}%
\end{align}
donde $\bm{\lambda}\in\mathfrak{g}$ es el elemento del \'{a}lgebra que
define la transformaci\'{o}n.

La invariancia de gauge de la acci\'{o}n transgresora contraste fuertemente
con lo que sucede en el caso de CS, donde la acci\'{o}n cambia por un
t\'{e}rmino de borde bajo transformaciones de gauge.

Como principio de simetr\'{\i}a, la invariancia de gauge prohibe la
adici\'{o}n de un t\'{e}rmino de borde arbitrario a la acci\'{o}n
transgresora, dado que esto en general destruir\'{\i}a la invariancia. Esto
significa, en particular, que las condiciones de borde y las cargas de Noether
asociadas a la acci\'{o}n transgresora tienen significado intr\'{\i}nseco, a
diferencia del caso de CS, donde pueden ser modificadas por la adici\'{o}n de
un t\'{e}rmino de borde arbitrario a la acci\'{o}n.

M\'{a}s all\'{a} de las transformaciones de gauge y difeomorfismos, la
acci\'{o}n transgresora tiene otra propiedad de simetr\'{\i}a importante,
discreta esta vez. Bajo el intercambio $\bm{A}\leftrightarrow
\bar{\bm{A}}$, la acci\'{o}n~(\ref{AcTra}) cambia de signo:
\begin{equation}
S_{\mathrm{T}}^{\left(  2n+1\right)  }\left[  \bar{\bm{A}%
},\bm{A}\right]  =-S_{\mathrm{T}}^{\left(  2n+1\right)  }\left[
\bm{A},\bar{\bm{A}}\right]  . \label{reflex}%
\end{equation}
Para demostrar esta \emph{simetr\'{\i}a de intercambio} basta con notar que,
al realizar la transformaci\'{o}n $\bm{A}\leftrightarrow\bar{\bm{A}}$, las variables
definidas en~(\ref{Theta})--(\ref{Ft}) cambian seg\'{u}n
\begin{align}
\bm{\Theta}  &  \rightarrow-\bm{\Theta},\\
\bm{A}_{t}  &  \rightarrow\bm{A}_{1-t},\\
\bm{F}_{t}  &  \rightarrow\bm{F}_{1-t}.
\end{align}
La ec.~(\ref{reflex}) se sigue de la identidad
\begin{equation}
\int_{0}^{1}f\left(  t\right)  dt=\int_{0}^{1}f\left(  1-t\right)  dt,
\end{equation}
la cual es v\'{a}lida para una funci\'{o}n $f$ cualquiera.

Las consecuencias de~(\ref{reflex}) a\'{u}n son materia de exploraci\'{o}n, si
bien es posible que se relacionen con la eventual invariancia de~(\ref{AcTra})
bajo la transformaci\'{o}n discreta combinada CPT (conjugaci\'{o}n de carga,
paridad e inversi\'{o}n temporal; ver secci\'{o}n~\ref{sec:NoCh}).

\subsection{\label{sec:eomTr}Ecuaciones de movimiento y condiciones de borde}

Bajo una variaci\'{o}n infinitesimal arbitraria $\bm{A}\rightarrow
\bm{A}^{\prime}=\bm{A}+\delta\bm{A}$, $\bar
{\bm{A}}\rightarrow\bar{\bm{A}}^{\prime}=\bar{\bm{A}%
}+\delta\bar{\bm{A}}$, la acci\'{o}n transgresora~(\ref{AcTra}) cambia
en
\begin{equation}
\delta S_{\mathrm{T}}^{\left(  2n+1\right)  }=\left(  n+1\right)  k\int
_{M}\left(  \left\langle \delta\bm{AF}^{n}\right\rangle -\left\langle
\delta\bar{\bm{A}}\bar{\bm{F}}^{n}\right\rangle \right)
+\int_{\partial M}\Xi, \label{var}%
\end{equation}
donde el t\'{e}rmino de borde $\Xi$\ tiene la forma
\begin{equation}
\Xi\equiv n\left(  n+1\right)  k\int_{0}^{1}dt\left\langle \delta
\bm{A}_{t}\bm{\Theta F}_{t}^{n-1}\right\rangle . \label{th}%
\end{equation}

La deducci\'{o}n de~(\ref{var}) a partir de~(\ref{AcTra}) no presenta mayores
dificultades, y es presentada a continuaci\'{o}n como ejemplo arquet\'{\i}pico
de c\'{a}lculo en el formalismo de las formas de transgresi\'{o}n.

Partimos de la definici\'{o}n de forma de transgresi\'{o}n
[cf.~ec.~(\ref{DefTra})],
\begin{equation}
\mathcal{Q}_{\bm{A}\leftarrow\bar{\bm{A}}}^{\left(
2n+1\right)  }=\left(  n+1\right)  \int_{0}^{1}dt\left\langle
\bm{\Theta F}_{t}^{n}\right\rangle . \label{Tra1}%
\end{equation}
Bajo variaciones infinitesimales arbitrarias en $\bm{A}$ y $\bar{\bm{A}}$, las variables $\bm{\Theta}$, $\bm{A}_{t}$ y $\bm{F}_{t}$ definidas en las ecs.~(\ref{Theta})--(\ref{Ft}) cambian de acuerdo a
\begin{align}
\delta\bm{\Theta}  &  =\delta\bm{A}-\delta\bar{\bm{A}%
},\\
\delta\bm{A}_{t}  &  =\delta\bar{\bm{A}}+t\delta
\bm{\Theta},\label{deltaAt}\\
\delta\bm{F}_{t}  &  =\mathrm{D}_{t}\delta\bm{A}_{t},
\label{dFt}%
\end{align}
donde $\mathrm{D}_{t}$ denota la derivada covariante en la conexi\'{o}n
$\bm{A}_{t}$.

Efectuando la variaci\'{o}n en~(\ref{Tra1}) encontramos
\begin{equation}
\delta\mathcal{Q}_{\bm{A}\leftarrow\bar{\bm{A}}}^{\left(
2n+1\right)  }=\left(  n+1\right)  \int_{0}^{1}dt\left\langle \delta
\bm{\Theta F}_{t}^{n}\right\rangle +n\left(  n+1\right)  \int_{0}%
^{1}dt\left\langle \bm{\Theta}\mathrm{D}_{t}\delta\bm{A}%
_{t}\bm{F}_{t}^{n-1}\right\rangle , \label{varQ}%
\end{equation}
donde ya hemos reemplazado expl\'{\i}citamente $\delta\bm{F}_{t}$ de
acuerdo a~(\ref{dFt}).

La regla de Leibniz para $\mathrm{D}_{t}$ y la propiedad de invariancia del
tensor sim\'{e}trico $\left\langle \cdots\right\rangle $ nos permiten escribir
el \'{u}ltimo t\'{e}rmino en~(\ref{varQ}) en la forma
\begin{equation}
\left\langle \bm{\Theta}\mathrm{D}_{t}\delta\bm{A}%
_{t}\bm{F}_{t}^{n-1}\right\rangle =\left\langle \mathrm{D}%
_{t}\bm{\Theta}\delta\bm{A}_{t}\bm{F}_{t}%
^{n-1}\right\rangle +\mathrm{d}\left\langle \delta\bm{A}%
_{t}\bm{\Theta F}_{t}^{n-1}\right\rangle . \label{auxeom1}%
\end{equation}
Aqu\'{\i} hemos usado tambi\'{e}n la identidad de Bianchi $\mathrm{D}_{t}\bm{F}_{t}=0$.

De las identidades [cf.~ec.~(\ref{cw3})]
\begin{align}
\frac{d}{dt}\bm{F}_{t}  &  =\mathrm{D}_{t}\bm{\Theta},\\
\frac{d}{dt}\delta\bm{A}_{t}  &  =\delta\bm{\Theta},
\end{align}
y la regla de Leibniz para $d/dt$ se sigue que
\begin{equation}
n\left\langle \mathrm{D}_{t}\bm{\Theta}\delta\bm{A}%
_{t}\bm{F}_{t}^{n-1}\right\rangle =\frac{d}{dt}\left\langle
\delta\bm{A}_{t}\bm{F}_{t}^{n}\right\rangle -\left\langle
\delta\bm{\Theta F}_{t}^{n}\right\rangle . \label{auxeom2}%
\end{equation}

Reemplazando~(\ref{auxeom2}) en~(\ref{auxeom1}) obtenemos la identidad
\begin{equation}
n\left\langle \bm{\Theta}\mathrm{D}_{t}\delta\bm{A}%
_{t}\bm{F}_{t}^{n-1}\right\rangle =\frac{d}{dt}\left\langle
\delta\bm{A}_{t}\bm{F}_{t}^{n}\right\rangle -\left\langle
\delta\bm{\Theta F}_{t}^{n}\right\rangle +n\mathrm{d}\left\langle
\delta\bm{A}_{t}\bm{\Theta F}_{t}^{n-1}\right\rangle .
\end{equation}

Ocupando esta igualdad en~(\ref{varQ}) encontramos
\begin{equation}
\delta\mathcal{Q}_{\bm{A}\leftarrow\bar{\bm{A}}}^{\left(
2n+1\right)  }=\left(  n+1\right)  \int_{0}^{1}dt\frac{d}{dt}\left\langle
\delta\bm{A}_{t}\bm{F}_{t}^{n}\right\rangle +n\left(
n+1\right)  \mathrm{d}\int_{0}^{1}dt\left\langle \delta\bm{A}%
_{t}\bm{\Theta F}_{t}^{n-1}\right\rangle ,
\end{equation}
lo cual nos conduce directamente al resultado final:
\begin{equation}
\delta\mathcal{Q}_{\bm{A} \leftarrow\bar{\bm{A}}}^{\left(
2n+1 \right)  } = \left(  n+1 \right)  \left(  \left\langle \delta
\bm{AF}^{n} \right\rangle - \left\langle \delta\bar{\bm{A}}
\bar{\bm{F}}^{n} \right\rangle \right)  + n \left(  n+1 \right)
\mathrm{d} \int_{0}^{1} dt \left\langle \delta\bm{A}_{t}
\bm{\Theta F}_{t}^{n-1} \right\rangle .
\end{equation}

Las ecuaciones de movimiento y las condiciones de borde pueden leerse
directamente de esta variaci\'{o}n. Las primeras pueden escribirse como
\begin{align}
\left\langle \bm{F}^{n}\bm{G}_{A}\right\rangle  &
=0,\label{FnGa1}\\
\left\langle \bar{\bm{F}}^{n}\bm{G}_{A}\right\rangle  &  =0,
\label{FnGa2}%
\end{align}
en tanto que las segundas son
\begin{equation}
\left.  \int_{0}^{1}dt\left\langle \delta\bm{A}_{t}\bm{\Theta
F}_{t}^{n-1}\right\rangle \right\vert _{\partial M}=0. \label{bctrans}%
\end{equation}

La din\'{a}mica producida por la acci\'{o}n~(\ref{AcTra}) determina que
$\bm{A}$ y $\bar{\bm{A}}$ sean campos independientes en $M$,
obedeciendo cada uno a ecuaciones de movimiento de CS desacopladas. Las
condiciones de borde, en cambio, ligan ambas conexiones en $\partial M$,
produciendo la primera diferencia significativa, a nivel din\'{a}mico, entre
la acci\'{o}n de CS y la acci\'{o}n transgresora~(\ref{AcTra}).

\subsection{\label{sec:NoCh}Cargas de Noether}

Como con cualquier sistema de gauge, el Teorema de Noether proporciona un
medio de extraer corrientes conservadas a partir de la acci\'{o}n
transgresora. Estas corrientes son $\left(  d-1\right)  $-formas (donde $d$ es
la dimensi\'{o}n de la variedad espacio-temporal $M$) cuya derivada exterior
se anula cuando las ecuaciones del movimiento son satisfechas (i.e.
\textit{on-shell}). Es usual, si bien en modo alguno necesario, interpretar
estas corrientes como el dual $\star$ de Hodge de una 1-forma $J$, de modo que
la `ecuaci\'{o}n de continuidad' puede ser escrita en la forma
\begin{equation}
\mathrm{d}\left.  \star J\right.  =0. \label{dJ=0}%
\end{equation}
En el caso que nos ocupa, el lagrangeano es invariante bajo dos conjuntos
independientes de simetr\'{\i}as; a saber, gauge y difeomorfismos. El teorema
de Noether proporciona consecuentemente dos corrientes conservadas
independientes para cada uno de ellos.

En el Ap\'{e}ndice~\ref{Ap:Noether} se revisa brevemente el Teorema de Noether
en el lenguaje de formas diferenciales, con el objetivo principal de fijar la
notaci\'{o}n y las convenciones utilizadas. Una deducci\'{o}n detallada de las corrientes
conservadas para la acci\'{o}n~(\ref{AcTra}) es presentada en el
Ap\'{e}ndice~\ref{Ap:Cargas}. El resultado es
\begin{align}
\left.  \star J_{\mathrm{gauge}}\right.   &  =n\left(  n+1\right)
k\mathrm{d}\int_{0}^{1}dt\left\langle \bm{\lambda\Theta F}_{t}%
^{n-1}\right\rangle ,\label{Jgauge}\\
\left.  \star J_{\mathrm{dif}}\right.   &  =n\left(  n+1\right)
k\mathrm{d}\int_{0}^{1}dt\left\langle \mathrm{I}_{\xi}\bm{A}%
_{t}\bm{\Theta F}_{t}^{n-1}\right\rangle , \label{Jdif}%
\end{align}
donde $d=2n+1$ es la dimensi\'{o}n de la variedad espacio-temporal $M$,
$\bm{\lambda}\in\mathfrak{g}$ es un elemento del \'{a}lgebra y $\xi$
es el campo vectorial infinitesimal que genera el difeomorfismo. Para deducir
estas expresiones ha sido necesario omitir t\'{e}rminos proporcionales a las
ecuaciones del movimiento, de manera que las corrientes s\'{o}lo est\'{a}n
definidas \textit{on-shell}. Una consecuencia de este procedimiento es que
tanto~(\ref{Jgauge}) como~(\ref{Jdif}) pueden ser escritas como formas
exactas, haciendo que la verificaci\'{o}n de la ley de
conservaci\'{o}n~(\ref{dJ=0}) sea trivial.

Asumiendo que $M$ tiene la topolog\'{\i}a $M=\mathbb{R}\times\Sigma$ es
posible integrar~(\ref{Jgauge})--(\ref{Jdif}) sobre la `secci\'{o}n espacial'
$\Sigma$ para obtener las cargas
\begin{align}
Q_{\mathrm{gauge}}\left(  \bm{\lambda}\right)   &  =n\left(
n+1\right)  k\int_{\partial\Sigma}\int_{0}^{1}dt\left\langle
\bm{\lambda\Theta F}_{t}^{n-1}\right\rangle ,\label{Qgauge}\\
Q_{\mathrm{dif}}\left(  \xi\right)   &  =n\left(  n+1\right)  k\int
_{\partial\Sigma}\int_{0}^{1}dt\left\langle \mathrm{I}_{\xi}\bm{A}%
_{t}\bm{\Theta F}_{t}^{n-1}\right\rangle . \label{Qdif}%
\end{align}
Es interesante notar que el teorema de Stokes nos permite escribir estas
cargas como integrales sobre el \emph{borde} de la secci\'{o}n espacial
$\Sigma$. En el espacio-tiempo tetradimensional usual,~(\ref{Qgauge})
y~(\ref{Qdif}) corresponder\'{\i}an a integrales de superficie.

Las cargas de gauge y difeomorfismos~(\ref{Qgauge}) y~(\ref{Qdif}) son trivialmente invariantes bajo
difeomorfismos, ya que est\'{a}n construidas a partir de formas diferenciales.
En cambio, bajo la transformaci\'{o}n de gauge infinitesimal
$g=1+\bm{\lambda}$, ellas cambian de acuerdo a
\begin{align}
\delta_{\bm{\lambda}}Q_{\mathrm{gauge}}\left(  \bm{\eta
}\right)   &  =-Q_{\mathrm{gauge}}\left(  \left[  \bm{\lambda
},\bm{\eta}\right]  \right)  ,\label{deltaQg}\\
\delta_{\bm{\lambda}}Q_{\mathrm{dif}}\left(  \xi\right)   &
=-Q_{\mathrm{gauge}}\left(  \pounds _{\xi}\bm{\lambda}\right)  .
\label{deltaQd}%
\end{align}

Como se muestra en~\cite{Mor06a}, las cargas de gauge~(\ref{Qgauge})
reproducen el \'{a}lgebra $\mathfrak{g}$ en el sentido que, definiendo el
\textit{bracket} entre cargas a partir de $\delta_{\bm{\lambda}%
}Q_{\bm{\eta}}\equiv\left\{  Q_{\bm{\eta}}%
,Q_{\bm{\lambda}}\right\}  $, se cumple que
\begin{equation}
\left\{ Q_{\bm{\eta}}, Q_{\bm{\lambda}} \right\} = Q_{\left[ \bm{\eta}, \bm{\lambda} \right]}
\label{AlgCar}
\end{equation}
La ausencia de t\'{e}rminos centrales en~(\ref{AlgCar}) es una consecuencia directa de la completa invariancia de gauge de la acci\'{o}n. Para la carga de
difeomorfismos, la ec.~(\ref{deltaQd}) implica que $Q_{\mathrm{dif}}$ es
invariante bajo aquellas transformaciones de gauge restringidas por la
condici\'{o}n $\pounds _{\xi}\bm{\lambda}=0$ en $\partial M$.

Una \'{u}ltima observaci\'{o}n acerca de las cargas~(\ref{Qgauge}) y
(\ref{Qdif}) es que ambas cambian de signo bajo el intercambio $\bm{A}%
\leftrightarrow\bar{\bm{A}}$; esto podr\'{\i}a se\~{n}alar que esta
operaci\'{o}n debiera ser considerada como `conjugaci\'{o}n de carga' en la
acci\'{o}n transgresora. Si aceptamos esta interpretaci\'{o}n, entonces la
acci\'{o}n~(\ref{AcTra}) resulta ser invariante bajo la transformaci\'{o}n
combinada CPT; en efecto, dado que
\begin{align}
\mathrm{C}\left(  S_{\mathrm{T}}^{\left(  2n+1\right)  }\right)   &
=-S_{\mathrm{T}}^{\left(  2n+1\right)  },\\
\mathrm{PT}\left(  S_{\mathrm{T}}^{\left(  2n+1\right)  }\right)   &
=-S_{\mathrm{T}}^{\left(  2n+1\right)  },
\end{align}
uno encuentra inmediatamente
\begin{equation}
\mathrm{CPT} \left( S_{\mathrm{T}}^{\left( 2n+1 \right)} \right) = S_{\mathrm{T}}^{\left( 2n+1 \right)}.
\end{equation}

La relevancia f\'{\i}sica de las cargas~(\ref{Qgauge}) y~(\ref{Qdif}) es
esencialmente un problema abierto. Un paso adelante en este sentido ser\'{\i}a
poder demostrar que ellas son finitas para una configuraci\'{o}n cl\'{a}sica
arbitraria. Aunque esta demostraci\'{o}n general a\'{u}n no est\'{a}
disponible, s\'{\i} se ha mostrado, para el caso particular en que
$\mathfrak{g}$ corresponde al \'{a}lgebra de anti-de Sitter, que las
cargas~(\ref{Qgauge}) y~(\ref{Qdif}), evaluadas para una soluci\'{o}n exacta
de la teor\'{\i}a, resultan ser finitas y f\'{\i}sicamente relevantes, es
decir, que reproducen lo que se obtiene por m\'{e}todos alternativos~\cite{Mor05,Mor06a}.

\subsection{\label{sec:noeoff}Conservaci\'{o}n \textit{off-shell} de las
corrientes de Noether}

La discusi\'{o}n sobre cargas conservadas dada en la
secci\'{o}n~\ref{sec:NoCh} est\'{a} basada en el uso del teorema de Noether
(ver Ap\'{e}ndice~\ref{Ap:Noether}), el cual es v\'{a}lido para cualquier
lagrangeano con simetr\'{\i}as de gauge o difeomorfismos. En esta secci\'{o}n
presentamos una deducci\'{o}n, particular a la acci\'{o}n
transgresora~(\ref{AcTra}), que muestra que las corrientes de
Noether~(\ref{Jgauge}) y~(\ref{Jdif}) no s\'{o}lo se conservan
\textit{on-shell}, sino tambi\'{e}n \emph{off-shell} (i.e. sin hacer uso de
las ecuaciones de movimiento).

Para demostrar la invariancia \textit{off-shell} de las
corrientes~(\ref{Jgauge}) y~(\ref{Jdif}), partimos de la variaci\'{o}n general
del lagrangeano transgresor [cf.~ecs.~(\ref{var})--(\ref{th})],
\begin{align}
\delta L_{\mathrm{T}}^{\left(  2n+1\right)  }  &  =\left(  n+1\right)
k\left(  \left\langle \delta\bm{AF}^{n}\right\rangle -\left\langle
\delta\bar{\bm{A}}\bar{\bm{F}}^{n}\right\rangle \right)
+\mathrm{d}\Xi,\label{varLt}\\
\Xi &  =n\left(  n+1\right)  k\int_{0}^{1}dt\left\langle \delta\bm{A}%
_{t}\bm{\Theta F}_{t}^{n-1}\right\rangle . \label{BTvarLt}%
\end{align}

\subsubsection{Conservaci\'{o}n \textit{off-shell} de la corriente de gauge}

La variaci\'{o}n de las conexiones $\bm{A}$ y $\bar{\bm{A}}$
bajo una transformaci\'{o}n de gauge infinitesimal generada por
$\bm{\lambda}\in\mathfrak{g}$ est\'{a} dada por [cf.~ecs.~(\ref{dgA}%
)--(\ref{dgAb})]
\begin{align}
\delta_{\mathrm{gauge}}\bm{A}  &  =-\mathrm{D}\bm{\lambda
},\label{dgA2}\\
\delta_{\mathrm{gauge}}\bar{\bm{A}}  &  =-\mathrm{\bar{D}%
}\bm{\lambda}. \label{dgAb2}%
\end{align}
Reemplazando~(\ref{dgA2})--(\ref{dgAb2}) en~(\ref{varLt})--(\ref{BTvarLt}) y
usando la identidad de Bianchi, podemos escribir la variaci\'{o}n del lagrangeano en la forma
\begin{equation}
\delta_{\mathrm{gauge}}L_{\mathrm{T}}^{\left(  2n+1\right)  }=-\left(
n+1\right)  k\mathrm{d}\left(  \left\langle \bm{\lambda F}%
^{n}\right\rangle -\left\langle \bm{\lambda}\bar{\bm{F}}%
^{n}\right\rangle \right)  +\mathrm{d}\Xi_{\mathrm{gauge}},
\end{equation}
donde la notaci\'{o}n $\Xi_{\mathrm{gauge}}$ indica que debemos reemplazar en $\Xi$ la transformaci\'{o}n de gauge co\-rrespondiente. Como $\delta_{\mathrm{gauge}}L_{\mathrm{T}}^{\left(  2n+1\right)  }=0$, hallamos que la corriente
\begin{equation}
\left.  \star J_{\mathrm{gauge}}^{\prime}\right.  \equiv\left(  n+1\right)
k\left(  \left\langle \bm{\lambda F}^{n}\right\rangle -\left\langle
\bm{\lambda}\bar{\bm{F}}^{n}\right\rangle \right)
-\Xi_{\mathrm{gauge}} \label{Jpg1}%
\end{equation}
es conservada \textit{off-shell}, i.e. sin hacer uso de las ecuaciones de
movimiento. Reemplazando expl\'{\i}citamente $\Xi_{\mathrm{gauge}}$ [cf.~ec.~(\ref{Xi-gauge})]
\begin{equation}
\Xi_{\mathrm{gauge}}=-n\left(  n+1\right)  k\mathrm{d}\int_{0}^{1}%
dt\left\langle \bm{\lambda\Theta F}_{t}^{n-1}\right\rangle +\left(
n+1\right)  k\left(  \left\langle \bm{\lambda F}^{n}\right\rangle
-\left\langle \bm{\lambda}\bar{\bm{F}}^{n}\right\rangle
\right)  \label{Xi-gauge2}
\end{equation}
en~(\ref{Jpg1}), encontramos
\begin{equation}
\left.  \star J_{\mathrm{gauge}}^{\prime}\right.  =n\left(  n+1\right)
k\mathrm{d}\int_{0}^{1}dt\left\langle \bm{\lambda\Theta F}_{t}%
^{n-1}\right\rangle .
\end{equation}
La corriente conservada \textit{off-shell} $\left.  \star J_{\mathrm{gauge}%
}^{\prime}\right.  $ coincide exactamente con la corriente de Noether $\left.
\star J_{\mathrm{gauge}}\right.  $, debido a la cancelaci\'{o}n ocurrida entre
el primer t\'{e}rmino en~(\ref{Jpg1}) y el \'{u}ltimo t\'{e}rmino
en~(\ref{Xi-gauge2}).

\subsubsection{Conservaci\'{o}n \textit{off-shell} de la corriente de
difeomorfismos}

La variaci\'{o}n infinitesimal de las conexiones $\bm{A}$ y
$\bar{\bm{A}}$ bajo un difeomorfismo $\delta x^{\mu}=\xi^{\mu}\left(
x\right)  $ est\'{a} dada por [cf.~ecs.~(\ref{ddA})--(\ref{ddAb})]
\begin{align}
\delta_{\mathrm{dif}}\bm{A}  &  =-\pounds _{\xi}\bm{A}%
,\label{ddA2}\\
\delta_{\mathrm{dif}}\bar{\bm{A}}  &  =-\pounds _{\xi}\bar
{\bm{A}}. \label{ddAb2}%
\end{align}
Reemplazando~(\ref{ddA2})--(\ref{ddAb2}) en~(\ref{varLt})--(\ref{BTvarLt})
hallamos que la variaci\'{o}n funcional del lagrangeano puede ser escrita en la forma
\begin{equation}
\delta_{\mathrm{dif}}L_{\mathrm{T}}^{\left(  2n+1\right)  }=-\left(
n+1\right)  k\left(  \left\langle \pounds _{\xi}\bm{AF}^{n}%
\right\rangle -\left\langle \pounds _{\xi}\bar{\bm{A}}\bar
{\bm{F}}^{n}\right\rangle \right)  +\mathrm{d}\Xi_{\mathrm{dif}},
\label{aux27}%
\end{equation}
donde la notaci\'{o}n $\Xi_{\mathrm{dif}}$ indica que debemos reemplazar en $\Xi$ la variaci\'{o}n de las conexiones bajo difeomorfismos. Usando la identidad\footnote{Esta igualdad no es preservada bajo transformaciones de gauge.}
\begin{equation}
\pounds _{\xi}\bm{A}=\mathrm{I}_{\xi}\bm{F}+\mathrm{DI}_{\xi}\bm{A}
\end{equation}
es posible escribir
\begin{equation}
\left\langle \pounds _{\xi}\bm{AF}^{n}\right\rangle =\left\langle
\mathrm{I}_{\xi}\bm{FF}^{n}\right\rangle +\left\langle \mathrm{DI}%
_{\xi}\bm{AF}^{n}\right\rangle .
\end{equation}
Ocupando ahora la identidad de Bianchi y la regla de Leibniz para
$\mathrm{I}_{\xi}$ (y observando tambi\'{e}n que $\left\langle \bm{F}%
^{n+1}\right\rangle =0$ para $d=2n+1$) encontramos finalmente
\begin{equation}
\left\langle \pounds _{\xi}\bm{AF}^{n}\right\rangle =\mathrm{d}%
\left\langle \mathrm{I}_{\xi}\bm{AF}^{n}\right\rangle . \label{aux38}%
\end{equation}
Reemplazando~(\ref{aux38}) en~(\ref{aux27}) obtenemos
\begin{equation}
\delta_{\mathrm{dif}}L_{\mathrm{T}}^{\left(  2n+1\right)  }=-\left(
n+1\right)  k\mathrm{d}\left(  \left\langle \mathrm{I}_{\xi}\bm{AF}%
^{n}\right\rangle -\left\langle \mathrm{I}_{\xi}\bar{\bm{A}}%
\bar{\bm{F}}^{n}\right\rangle \right)  +\mathrm{d}\Xi_{\mathrm{dif}}.
\end{equation}
Recordando que
\begin{align*}
\delta_{\mathrm{dif}}L_{\mathrm{T}}^{\left(  2n+1\right)  }  &
=-\pounds _{\xi}L_{\mathrm{T}}^{\left(  2n+1\right)  }\\
&  =-\mathrm{dI}_{\xi}L_{\mathrm{T}}^{\left(  2n+1\right)  },
\end{align*}
podemos afirmar que la corriente
\begin{equation}
\left.  \star J_{\mathrm{dif}}^{\prime}\right.  \equiv\left(  n+1\right)
k\left(  \left\langle \mathrm{I}_{\xi}\bm{AF}^{n}\right\rangle
-\left\langle \mathrm{I}_{\xi}\bar{\bm{A}}\bar{\bm{F}}%
^{n}\right\rangle \right)  -\Xi_{\mathrm{dif}}-\mathrm{I}_{\xi}L_{\mathrm{T}%
}^{\left(  2n+1\right)  } \label{aux99}%
\end{equation}
es conservada \textit{off-shell}, i.e. sin hacer uso de las ecuaciones de movimiento.

Reemplazando [cf.~ec.~(\ref{Xi-dif})]
\begin{align}
\Xi_{\mathrm{dif}}+\mathrm{I}_{\xi}L_{\mathrm{T}}^{\left(  2n+1\right)  }
= & -n\left(  n+1\right)  k\mathrm{d}\int_{0}^{1}dt\left\langle \mathrm{I}_{\xi
}\bm{A}_{t}\bm{\Theta F}_{t}^{n-1}\right\rangle +\nonumber\\
&  +\left(  n+1\right)  k\left(  \left\langle \mathrm{I}_{\xi}\bm{AF}%
^{n}\right\rangle -\left\langle \mathrm{I}_{\xi}\bar{\bm{A}}%
\bar{\bm{F}}^{n}\right\rangle \right)  \label{Xi-dif2}%
\end{align}
en~(\ref{aux99}), encontramos
\begin{equation}
\left.  \star J_{\mathrm{dif}}^{\prime}\right.  =n\left(  n+1\right)
k\mathrm{d}\int_{0}^{1}dt\left\langle \mathrm{I}_{\xi}\bm{A}%
_{t}\bm{\Theta F}_{t}^{n-1}\right\rangle .
\end{equation}
La corriente conservada \textit{off-shell} $\left.  \star J_{\mathrm{dif}%
}^{\prime}\right.  $ coincide exactamente con la corriente de Noether $\left.
\star J_{\mathrm{dif}}\right.  $, debido a la cancelaci\'{o}n ocurrida entre
el primer t\'{e}rmino en~(\ref{aux99}) y el \'{u}ltimo t\'{e}rmino
en~(\ref{Xi-dif2}).

Desde un punto de vista estrictamente algebraico, la raz\'{o}n para la
existencia de una corriente conservada \textit{off-shell} yace en el
hecho que $\left\langle \delta\bm{AF}^{n}\right\rangle $ puede
escribirse como una derivada total, para cada una de las conexiones, cuando la
variaci\'{o}n corresponde a una transformaci\'{o}n de gauge o a un difeomorfismo.

Por otro lado, independientemente del procedimiento usado para obtenerlas, el hecho que las corrientes de Noether~(\ref{Jgauge})--(\ref{Jdif}) puedan escribirse como formas exactas apunta a que su conservaci\'{o}n es m\'{a}s bien una identidad algebraica que una consecuencia de la din\'{a}mica. Desde este punto de vista, la deducci\'{o}n alternativa presentada en esta secci\'{o}n, la cual no hace uso de las ecuaciones de movimiento, no hace m\'{a}s que confirmar estas sospechas.

\subsection{\label{sec:comtra}Acerca de qu\'{e} puede hacerse con dos conexiones}

La acci\'{o}n transgresora~(\ref{AcTra}) puede ser escrita como la
\emph{diferencia} de dos acciones de CS, una para $\bm{A}$ y otra para
$\bar{\bm{A}}$, m\'{a}s un t\'{e}rmino de borde,
\begin{equation}
S_{\mathrm{T}}^{\left(  2n+1\right)  }\left[  \bm{A},\bar
{\bm{A}}\right]  =S_{\mathrm{CS}}^{\left(  2n+1\right)  }\left[
\bm{A}\right]  -S_{\mathrm{CS}}^{\left(  2n+1\right)  }\left[
\bar{\bm{A}}\right]  +\int_{\partial M}\mathcal{B}^{\left(  2n\right)
}. \label{t=c+s}%
\end{equation}
Una estructura de este tipo puede ser inferida a partir de las ecuaciones de
movimiento, las cuales corresponden a dos sistemas de CS desacoplados. La
demostraci\'{o}n de~(\ref{t=c+s}) es dada en la secci\'{o}n~\ref{sec:identri}
como un corolario trivial de la identidad triangular~(\ref{q34}).

El signo relativo entre ambas acciones de CS en~(\ref{t=c+s}) es importante,
pues indica un t\'{e}rmino cin\'{e}tico con el signo `equivocado' para
$\bar{\bm{A}}$. Un campo con estas caracter\'{\i}sticas es denominado
\emph{fantasma} (ghost, phantom), pues su teor\'{\i}a cu\'{a}ntica no est\'{a}
bien definida (ver, e.g., \cite{Cli03}).

Una alternativa para evitar este problema es considerar dos variedades
cobordantes $M$ y $\bar{M}$, con cada conexi\'{o}n definida s\'{o}lo en una de
ellas. Este punto de vista ha sido propuesto en las Refs.~\cite{Mor04a,Mor05}.
Otra interpretaci\'{o}n para la diferencia de signo consiste en asociar cada
conexi\'{o}n a una \emph{orientaci\'{o}n} distinta de una \'{u}nica variedad
$M$, como se sugiere en~\cite{Iza06c}. En este cuadro, el signo negativo en el
segundo t\'{e}rmino de~(\ref{t=c+s}) es entendido como una consecuencia de
integrar sobre $M$ con la orientaci\'{o}n opuesta,
\begin{equation}
-S_{\mathrm{CS}}^{\left(  2n+1\right)  }\left[  \bar{\bm{A}}\right]
=-\int_{M}L_{\mathrm{CS}}^{\left(  2n+1\right)  }\left(  \bar{\bm{A}%
}\right)  =\int_{-M}L_{\mathrm{CS}}^{\left(  2n+1\right)  }\left(
\bar{\bm{A}}\right)  .
\end{equation}

M\'{a}s adelante veremos ejemplos de acciones transgresoras donde,
restringiendo $\bar{\bm{A}}$ a estar valuada s\'{o}lo en una
sub\'{a}lgebra de $\mathfrak{g}$, se consigue anular la segunda acci\'{o}n de
CS en~(\ref{t=c+s}), confinando toda la dependencia de $S_{\mathrm{T}%
}^{\left(  2n+1\right)  }\left[  \bm{A},\bar{\bm{A}}\right]  $
en $\bar{\bm{A}}$ a un t\'{e}rmino de borde.

\chapter{\label{ch:metsepsub}El M\'{e}todo de Separaci\'{o}n en Subespacios}

Este cap\'{\i}tulo est\'{a} basado en resultados obtenidos en las Refs.~\cite{Iza05,Iza06a}.

\section{\label{sec:motif}Motivaci\'{o}n}

En el cap\'{\i}tulo~\ref{ch:trans} hemos presentado un an\'{a}lisis general,
muy somero, de la acci\'{o}n transgresora~(\ref{AcTra}). En principio, el
lagrangeano transgresor en su forma m\'{a}s general [cf.~ec.~(\ref{AcTra})]
\begin{equation}
L_{\mathrm{T}}^{\left(  2n+1\right)  }\left(  \bm{A},\bar
{\bm{A}}\right)  =\left(  n+1\right)  k\int_{0}^{1}dt\left\langle
\bm{\Theta F}_{t}^{n}\right\rangle , \label{LaTra}%
\end{equation}
contiene toda la informaci\'{o}n que uno pueda desear acerca de la
teor\'{\i}a. A partir de \'{e}l es posible deducir ecuaciones de movimiento,
condiciones de borde y cargas conservadas, sin que sea necesario especificar
un \'{a}lgebra de gauge $\mathfrak{g}$. En este sentido, la acci\'{o}n
transgresora, al igual que las acciones de CS o YM, proporciona m\'{a}s un
marco general para una clase de teor\'{\i}as de gauge que un modelo concreto. En efecto, buena parte de la
f\'{\i}sica generada por esta acci\'{o}n proviene de la elecci\'{o}n del
\'{a}lgebra $\mathfrak{g}$. Esta elecci\'{o}n determina cu\'{a}les ser\'{a}n
los campos independientes de la teor\'{\i}a y, junto con el tensor invariante
$\left\langle \cdots\right\rangle $, fija la forma de las distintas
interacciones que ocurrir\'{a}n entre ellos.

En este cap\'{\i}tulo analizamos una herramienta matem\'{a}tica, la f\'{o}rmula
extendida de la homotop\'{\i}a de Cartan, sobre la cual puede construirse un
m\'{e}todo que permite extraer informaci\'{o}n adicional del
lagrangeano~(\ref{LaTra}). En cierto sentido, esta informaci\'{o}n es
superflua; como mencionamos en el p\'{a}rrafo anterior, las propiedades
din\'{a}micas m\'{a}s importantes de la teor\'{\i}a pueden ser deducidas en
completa generalidad, para cualquier \'{a}lgebra de gauge. Sin embargo, la ec.~(\ref{LaTra}) para el lagrangeano comunica escasa informaci\'{o}n a quien desee hacerse una idea del contenido f\'{\i}sico del modelo, debido precisamente a su generalidad y a su lenguaje extremadamente compacto. Una versi\'{o}n m\'{a}s expl\'{\i}cita del lagrangeano, que tome en cuenta las particularidades del \'{a}lgebra de gauge en cuesti\'{o}n, es esencial para dar una interpretaci\'{o}n en t\'{e}rminos m\'{a}s f\'{\i}sicos a la teor\'{\i}a.

Por otro lado, las \'{a}lgebras de gauge que son usadas en la
pr\'{a}ctica contienen a menudo distintos subespacios (cuando no
necesariamente sub\'{a}lgebras) con significado f\'{\i}sico propio. Un claro
ejemplo de esto es proporcionado por una super\'{a}lgebra, donde uno puede
distinguir inequ\'{\i}vocamente generadores bos\'{o}nicos de fermi\'{o}nicos.
Resulta interesante entonces escribir el lagrangeano de una manera tal que
refleje la estructura de subespacios del \'{a}lgebra.

\'{E}l m\'{e}todo de separaci\'{o}n en subespacios descrito en esta
secci\'{o}n permite separar el lagrangeano transgresor en dos niveles
distintos: primero, la acci\'{o}n es descompuesta en contribuciones de volumen
y de borde\footnote{Tal separaci\'{o}n no es, por supuesto, un\'{\i}voca, y distintas maneras de aplicar el m\'{e}todo conducir\'{a}n, en general, a descomposiciones diferentes.}. Posteriormente, el lagrangeano de volumen es separado en trozos
que se corresponden con los distintos subespacios del \'{a}lgebra de gauge.

La observaci\'{o}n clave detr\'{a}s del m\'{e}todo proviene del teorema de
Chern--Weil (ver secci\'{o}n~\ref{sec:cwt}, p\'{a}g.~\pageref{sec:cwt}). En
efecto, si consideramos \emph{tres} conexiones de gauge $\bm{A}$,
$\bar{\bm{A}}$ y $\tilde{\bm{A}}$, entonces es directo
demostrar que la siguiente combinaci\'{o}n de derivadas exteriores de formas
de transgresi\'{o}n se anula id\'{e}nticamente:
\begin{equation}
\mathrm{d}\mathcal{Q}_{\bm{A}\leftarrow\bar{\bm{A}}}^{\left(
2n+1\right)  }+\mathrm{d}\mathcal{Q}_{\bar{\bm{A}}\leftarrow
\tilde{\bm{A}}}^{\left(  2n+1\right)  }+\mathrm{d}\mathcal{Q}%
_{\tilde{\bm{A}}\leftarrow\bm{A}}^{\left(  2n+1\right)  }=0.
\end{equation}
En efecto, el teorema de Chern--Weil implica que cada una de estas derivadas es igual a la
diferencia de los caracteres de Chern respectivos, de modo que
\begin{eqnarray*}
\mathrm{d}\mathcal{Q}_{\bm{A}\leftarrow\bar{\bm{A}}}^{\left(
2n+1\right)  }+\mathrm{d}\mathcal{Q}_{\bar{\bm{A}}\leftarrow
\tilde{\bm{A}}}^{\left(  2n+1\right)  }+\mathrm{d}\mathcal{Q}%
_{\tilde{\bm{A}}\leftarrow\bm{A}}^{\left(  2n+1\right)  } & = & \left\langle \bm{F}^{n+1}\right\rangle -\left\langle \bar
{\bm{F}}^{n+1}\right\rangle +\left\langle \bar{\bm{F}}%
^{n+1}\right\rangle +\\
& & -\left\langle \tilde{\bm{F}}^{n+1}\right\rangle +\left\langle
\tilde{\bm{F}}^{n+1}\right\rangle -\left\langle \bm{F}%
^{n+1}\right\rangle \\
& = & 0.
\end{eqnarray*}

Esta propiedad de las formas de transgresi\'{o}n hace que sea posible escribir, al menos
localmente,
\begin{equation}
\mathcal{Q}_{\bm{A}\leftarrow\bar{\bm{A}}}^{\left(
2n+1\right)  }+\mathcal{Q}_{\bar{\bm{A}}\leftarrow\tilde
{\bm{A}}}^{\left(  2n+1\right)  }+\mathcal{Q}_{\tilde{\bm{A}%
}\leftarrow\bm{A}}^{\left(  2n+1\right)  }=\mathrm{d}\mathcal{Q}%
_{\bm{A}\leftarrow\tilde{\bm{A}}\leftarrow\bar{\bm{A}%
}}^{\left(  2n\right)  }, \label{treq1}%
\end{equation}
donde $\mathcal{Q}_{\bm{A}\leftarrow\tilde{\bm{A}}%
\leftarrow\bar{\bm{A}}}^{\left(  2n\right)  }$ es una $2n$-forma hasta
ahora no especificada. Ocupando la simetr\'{\i}a de intercambio de las formas
de transgresi\'{o}n (ver secci\'{o}n~\ref{sec:symtra},
p\'{a}g.~\pageref{sec:symtra}), podemos reescribir~(\ref{treq1}) en la forma
\begin{equation}
\mathcal{Q}_{\bm{A}\leftarrow\bar{\bm{A}}}^{\left(
2n+1\right)  }=\mathcal{Q}_{\bm{A}\leftarrow\tilde{\bm{A}}%
}^{\left(  2n+1\right)  }+\mathcal{Q}_{\tilde{\bm{A}}\leftarrow
\bar{\bm{A}}}^{\left(  2n+1\right)  }+\mathrm{d}\mathcal{Q}%
_{\bm{A}\leftarrow\tilde{\bm{A}}\leftarrow\bar{\bm{A}%
}}^{\left(  2n\right)  }. \label{treq2}%
\end{equation}
La identidad~(\ref{treq2}) tiene una interpretaci\'{o}n muy interesante: la forma de
transgresi\'{o}n $\mathcal{Q}_{\bm{A}\leftarrow\bar{\bm{A}}%
}^{\left(  2n+1\right)  }$, que `interpola' entre las conexiones
$\bar{\bm{A}}$ y $\bm{A}$, puede descomponerse en la suma de
dos transgresiones; una, $\mathcal{Q}_{\tilde{\bm{A}}\leftarrow
\bar{\bm{A}}}^{\left(  2n+1\right)  }$, que `interpola' entre
$\bar{\bm{A}}$ y $\tilde{\bm{A}}$, y otra, $\mathcal{Q}%
_{\bm{A}\leftarrow\tilde{\bm{A}}}^{\left(  2n+1\right)  }$,
que `interpola' entre $\tilde{\bm{A}}$ y $\bm{A}$, m\'{a}s una
forma exacta que depende de $\bm{A}$, $\tilde{\bm{A}}$ y
$\bar{\bm{A}}$. La \emph{identidad triangular}~(\ref{treq2}) permite
dividir un lagrangeano transgresor usando una conexi\'{o}n intermedia
arbitraria $\tilde{\bm{A}}$. El uso iterativo de esta identidad
permite llevar a cabo la separaci\'{o}n en subespacios mencionada m\'{a}s
arriba. Sin embargo, para completar esta separaci\'{o}n necesitamos conocer la
forma expl\'{\i}cita del \'{u}ltimo t\'{e}mino en~(\ref{treq2}),
$\mathcal{Q}_{\bm{A}\leftarrow\tilde{\bm{A}}\leftarrow
\bar{\bm{A}}}^{\left(  2n\right)  }$. La f\'{o}rmula extendida de la
homotop\'{\i}a de Cartan, o m\'{a}s bien un caso particular de ella, permite
responder esta pregunta.

\section[La f\'{o}rmula extendida de la homotop\'{\i}a de Cartan]%
{\label{sec:fehc}La f\'{o}rmula extendida de la homotop\'{\i}a de Cartan
\sectionmark{F\'{o}rmula extendida de homotop\'{\i}a de Cartan}}

\sectionmark{F\'{o}rmula extendida de homotop\'{\i}a de Cartan}

Esta secci\'{o}n introduce la f\'{o}rmula extendida de la homotop\'{\i}a de
Cartan (FEHC). La referencia principal es~\cite{Man85}, aunque la
demostraci\'{o}n que incluimos aqu\'{\i} es diferente a la dada por aquellos autores.

Consideremos un conjunto $\left\{  \bm{A}_{i},i=0,\dotsc,r+1\right\}
$ de 1-formas conexiones en un \textit{fibre bundle} sobre una variedad $d$-dimensional $M$ y un simplex
orientable $\left(  r+1\right)  $-dimensional $T_{r+1}$ parametrizado por
$\left\{  t^{i},i=0,\dotsc,r+1\right\}  $. Estos par\'{a}metros deben
satisfacer las relaciones
\begin{align}
t^{i}  &  \geq0,\qquad i=0,\dotsc,r+1,\label{tgt0}\\
\sum_{i=0}^{r+1}t^{i}  &  =1. \label{St=1}%
\end{align}
La ec.~(\ref{St=1}), en particular, implica que la combinaci\'{o}n lineal
\begin{equation}
\bm{A}_{t}=\sum_{i=0}^{r+1}t^{i}\bm{A}_{i}%
\end{equation}
tambi\'{e}n transforma como una conexi\'{o}n, al igual que cada
$\bm{A}_{i}$. Por lo tanto, tiene sentido definir la curvatura
\begin{equation}
\bm{F}_{t}=\mathrm{d}\bm{A}_{t}+\bm{A}_{t}^{2}.
\end{equation}

\begin{figure}
\begin{center}
\includegraphics[width=.8\textwidth]{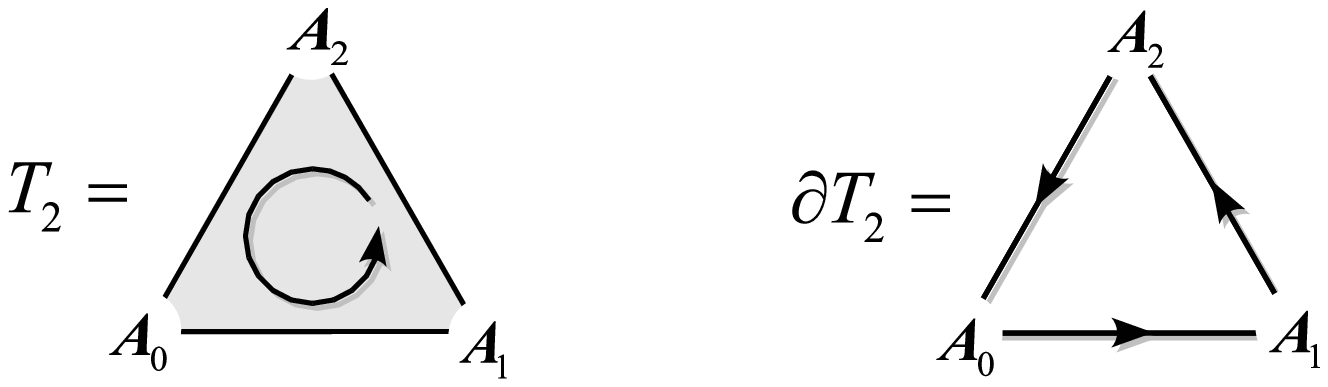}
\end{center}
\caption{\label{T2andDT2}Un simplex $2$-dimensional $T_{2} = \left(  \bm{A}_{0}
\bm{A}_{1} \bm{A}_{2} \right)  $ y su borde, $\partial T_{2} =
\left(  \bm{A}_{1} \bm{A}_{2} \right)  - \left(
\bm{A}_{0} \bm{A}_{2} \right)  + \left(  \bm{A}_{0}
\bm{A}_{1} \right)  $.}
\end{figure}

Como se sugiere en la fig.~\ref{T2andDT2} para el caso $r=1$, es posible asociar cada una de las conexiones $\bm{A}_{i}$ con un v\'{e}rtice del simplex $T_{r+1}$, el cual representamos consistentemente
como
\begin{equation}
T_{r+1}=\left(  \bm{A}_{0}\bm{A}_{1}\cdots\bm{A}_{r+1}\right)  .
\end{equation}

Las derivadas exteriores en $M$ y en $T_{r+1}$ ser\'{a}n denotadas por
$\mathrm{d}$ y $\mathrm{d}_{t}$, respectivamente. Estas derivadas son mapeos
de la forma
\begin{align}
\mathrm{d}  &  :\Omega^{a}\left(  M\right)  \times\Omega^{b}\left(
T_{r+1}\right)  \rightarrow\Omega^{a+1}\left(  M\right)  \times\Omega
^{b}\left(  T_{r+1}\right)  ,\\
\mathrm{d}_{t}  &  :\Omega^{a}\left(  M\right)  \times\Omega^{b}\left(
T_{r+1}\right)  \rightarrow\Omega^{a}\left(  M\right)  \times\Omega
^{b+1}\left(  T_{r+1}\right)  ,
\end{align}
donde $\Omega^{a}\left(  M\right)  $ denota el espacio de las $a$-formas en
$M$ y $\Omega^{b}\left(  T_{r+1}\right)  $ el espacio de las $b$-formas en
$T_{r+1}$. Es tambi\'{e}n posible definir un operador $l_{t}$, llamado
\emph{derivaci\'{o}n homot\'{o}pica}~\cite{Man85},
\begin{equation}
l_{t}:\Omega^{a}\left(  M\right)  \times\Omega^{b}\left(  T_{r+1}\right)
\rightarrow\Omega^{a-1}\left(  M\right)  \times\Omega^{b+1}\left(
T_{r+1}\right)  ,
\end{equation}
el cual disminuye en una unidad el grado de una forma diferencial en $M$,
mientras que aumenta en una unidad el grado de una forma diferencial en
$T_{r+1}$. La \'{u}nica manera consistente de definir su acci\'{o}n sobre
$\bm{A}_{t}$ y $\bm{F}_{t}$ es
\begin{align}
l_{t}\bm{A}_{t}  &  =0,\label{ltAt}\\
l_{t}\bm{F}_{t}  &  =\mathrm{d}_{t}\bm{A}_{t}. \label{ltFt}%
\end{align}
Los tres operadores $\mathrm{d}$, $\mathrm{d}_{t}$ y $l_{t}$ cumplen con la
regla de Leibniz y satisfacen adem\'{a}s la siguiente \'{a}lgebra gradada:
\begin{eqnarray}
\mathrm{d}^{2} & = & 0, \label{ga1} \\
\mathrm{d}_{t}^{2} & = & 0, \\
\left[ l_{t},\mathrm{d} \right] & = & \mathrm{d}_{t}, \label{ltd=dt} \\
\left[ l_{t},\mathrm{d}_{t} \right] & = & 0, \\
\left\{ \mathrm{d},\mathrm{d}_{t} \right\} & = & 0. \label{ga5}
\end{eqnarray}

Con la notaci\'{o}n anterior, la FEHC puede escribirse en la forma
\begin{equation}
\int_{\partial T_{r+1}}\frac{l_{t}^{p}}{p!}\pi=\int_{T_{r+1}}\frac{l_{t}%
^{p+1}}{\left(  p+1\right)  !}\mathrm{d}\pi+\left(  -1\right)  ^{p+q}%
\mathrm{d}\int_{T_{r+1}}\frac{l_{t}^{p+1}}{\left(  p+1\right)  !}\pi,
\label{fehc1}%
\end{equation}
donde $\pi$ es

\begin{itemize}
\item un polinomio en las formas $\left\{  \bm{A}_{t},\bm{F}%
_{t},\mathrm{d}_{t}\bm{A}_{t},\mathrm{d}_{t}\bm{F}%
_{t}\right\}  $,

\item una $m$-forma en $M$,

\item una $q$-forma en $T_{r+1}$,
\end{itemize}

con $m\geq p$ y $p+q=r$. Un poco m\'{a}s expl\'{\i}citamente,
\begin{equation}
\pi=\sum_{p}\alpha_{p}\left\langle \bm{A}_{t}^{a_{p}}\bm{F}%
_{t}^{b_{p}}\left(  \mathrm{d}_{t}\bm{A}_{t}\right)  ^{c_{p}}\left(
\mathrm{d}_{t}\bm{F}_{t}\right)  ^{d_{p}}\right\rangle ,
\end{equation}
donde $\alpha_{p}$ son constantes arbitrarias, $\left\langle \cdots
\right\rangle $ es una forma multilineal en el \'{a}lgebra y los exponentes
$a_{p},b_{b},c_{p},d_{p}$ satisfacen las relaciones
\begin{align}
a_{p}+2b_{p}+c_{p}+2d_{p}  &  =m,\\
c_{p}+d_{p}  &  =q.
\end{align}

La FEHC puede ser considerada como la versi\'{o}n integrada de la identidad
diferencial
\begin{equation}
\left(  p+1\right)  \mathrm{d}_{t}l_{t}^{p}\pi=l_{t}^{p+1}\mathrm{d}%
\pi-\mathrm{d}l_{t}^{p+1}\pi, \label{fehc2}%
\end{equation}
la cual es claramente equivalente a
\begin{equation}
\left[  l_{t}^{p+1},\mathrm{d}\right]  =\left(  p+1\right)  \mathrm{d}%
_{t}l_{t}^{p}.
\end{equation}
Esta identidad puede ser demostrada usando el m\'{e}todo de inducci\'{o}n
matem\'{a}\-tica y el \'{a}lgebra~(\ref{ga1})--(\ref{ga5}). Para $p=1$, tenemos
\begin{align*}
\left[  l_{t}^{2},\mathrm{d}\right]   &  =l_{t}\left[  l_{t},\mathrm{d}%
\right]  +\left[  l_{t},\mathrm{d}\right]  l_{t}\\
&  =l_{t}\mathrm{d}_{t}+\mathrm{d}_{t}l_{t}\\
&  =2\mathrm{d}_{t}l_{t}.
\end{align*}
Asumiendo su validez para $p=k$, el caso $p=k+1$ se demuestra f\'{a}cilmente:
\begin{align*}
\left[  l_{t}^{k+2},\mathrm{d}\right]   &  =l_{t}\left[  l_{t}^{k+1}%
,\mathrm{d}\right]  +\left[  l_{t},\mathrm{d}\right]  l_{t}^{k+1}\\
&  =\left(  k+1\right)  l_{t}\mathrm{d}_{t}l_{t}^{k}+\mathrm{d}_{t}l_{t}%
^{k+1}\\
&  =\left(  k+2\right)  \mathrm{d}_{t}l_{t}^{k+1}.
\end{align*}

Integrando~(\ref{fehc2}) sobre el simplex $T_{r+1}$ encontramos
\begin{equation}
\left(  p+1\right)  \int_{\partial T_{r+1}}l_{t}^{p}\pi=\int_{T_{r+1}}%
l_{t}^{p+1}\mathrm{d}\pi-\int_{T_{r+1}}\mathrm{d}l_{t}^{p+1}\pi,
\end{equation}
donde el teorema de Stokes para $T_{r+1}$ ha sido usado en el lado izquierdo.
Ocupando ahora la regla de integraci\'{o}n\footnote{Esta regla de
integraci\'{o}n est\'{a} basada en la convenci\'{o}n de escribir primero los
diferenciales en $T$ y luego los diferenciales en $M$; el factor de signo $\left( -1 \right)^{s}$ surge por el intercambio de la 1-forma (en $M$) $\mathrm{d}$ con la $s$-forma (en $T$)
$\alpha$.}~\cite{Man85}
\begin{equation}
\mathrm{d}\int_{T_{s}}\alpha=\left(  -1\right)  ^{s}\int_{T_{s}}%
\mathrm{d}\alpha,
\end{equation}
hallamos finalmente
\begin{equation}
\left(  p+1\right)  \int_{\partial T_{r+1}}l_{t}^{p}\pi=\int_{T_{r+1}}%
l_{t}^{p+1}\mathrm{d}\pi+\left(  -1\right)  ^{p+q}\mathrm{d}\int_{T_{r+1}%
}l_{t}^{p+1}\pi. \label{fehc3}%
\end{equation}
Esta identidad es trivialmente equivalente a la FEHC, ec.~(\ref{fehc1}).

En lo sucesivo estaremos interesados en un caso particular de la FEHC, el cual
corresponde a una elecci\'{o}n concreta para $\pi$,
\begin{equation}
\pi=\left\langle \bm{F}_{t}^{n+1}\right\rangle ,
\end{equation}
donde $\left\langle \cdots\right\rangle $ es un tensor sim\'{e}trico
invariante de rango $n+1$. Esta elecci\'{o}n tiene las siguientes propiedades:

\begin{itemize}
\item $\pi$ es cerrado\footnote{Esta propiedad se deduce f\'{a}cilmente de la
invariancia de $\pi$ (ver secci\'{o}n~\ref{sec:TenInvCS},
p\'{a}g.~\pageref{sec:TenInvCS}) y de la identidad de Bianchi $\mathrm{D}%
_{t}\bm{F}_{t}=0$.} en $M$, i.e. $\mathrm{d}\pi=0$,

\item $\pi$ es una 0-forma en $T_{r+1}$, i.e. $q=0$,

\item $\pi$ es una $\left(  2n+2\right)  $-forma en $M$, i.e. $m=2n+2$.
\end{itemize}

Los valores permitidos para $p$ son $p=0,\dotsc,2n+2$. La FEHC se reduce en
este caso a
\begin{equation}
\int_{\partial T_{p+1}}\frac{l_{t}^{p}}{p!}\left\langle \bm{F}%
_{t}^{n+1}\right\rangle =\left(  -1\right)  ^{p}\mathrm{d}\int_{T_{p+1}}%
\frac{l_{t}^{p+1}}{\left(  p+1\right)  !}\left\langle \bm{F}_{t}%
^{n+1}\right\rangle . \label{fehc4}%
\end{equation}
Nos referiremos a la ec.~(\ref{fehc4}) como la versi\'{o}n `restringida' de la FEHC.

\subsection{\label{sec:cwt2}El Teorema de Chern--Weil, otra vez}

El teorema de Chern--Weil, introducido en la sec.~\ref{sec:cwt}, puede ser
considerado como el caso particular $p=0$ de la FEHC, versi\'{o}n restringida
[cf.~ec.~(\ref{fehc4})]. Esta interpretaci\'{o}n provee de una
demostraci\'{o}n alternativa a la dada en la sec.~\ref{sec:cwt}. Poniendo
$p=0$ en~(\ref{fehc4}), hallamos
\begin{equation}
\int_{\partial T_{1}}\left\langle \bm{F}_{t}^{n+1}\right\rangle
=\mathrm{d}\int_{T_{1}}l_{t}\left\langle \bm{F}_{t}^{n+1}\right\rangle
, \label{fehc.p=0}%
\end{equation}
donde conviene recordar que $\bm{F}_{t}$ es la curvatura asociada a la
conexi\'{o}n
\begin{equation}
\bm{A}_{t}=t^{0}\bm{A}_{0}+t^{1}\bm{A}_{1},
\end{equation}
con $t^{0}+t^{1}=1$. El borde del simplex unidimensional $T_{1}=\left(
\bm{A}_{0}\bm{A}_{1}\right)  $ es simplemente
\begin{equation}
\partial T_{1}=\left(  \bm{A}_{1}\right)  -\left(  \bm{A}%
_{0}\right)  ,
\end{equation}
de modo que la integraci\'{o}n en el lado izquierdo de~(\ref{fehc.p=0}) es
trivial:
\begin{equation}
\int_{\partial T_{1}}\left\langle \bm{F}_{t}^{n+1}\right\rangle
=\left\langle \bm{F}_{1}^{n+1}\right\rangle -\left\langle
\bm{F}_{0}^{n+1}\right\rangle .
\end{equation}

Por otro lado, la regla de Leibniz para $l_{t}$ y la naturaleza sim\'{e}trica
del tensor invariante $\left\langle \cdots\right\rangle $ implican que podemos
escribir
\begin{equation}
l_{t}\left\langle \bm{F}_{t}^{n+1}\right\rangle =\left(  n+1\right)
\left\langle \left(  l_{t}\bm{F}_{t}\right)  \bm{F}_{t}%
^{n}\right\rangle .
\end{equation}
De la definici\'{o}n del operador de derivaci\'{o}n homot\'{o}pica $l_{t}$ [cf.~ec.~(\ref{ltFt})] tenemos que
\begin{eqnarray*}
l_{t}\bm{F}_{t} & = & \mathrm{d}_{t}\bm{A}_{t}\\
& = & \mathrm{d}t^{0}\bm{A}_{0}+\mathrm{d}t^{1}\bm{A}_{1}\\
& = & \mathrm{d}t^{1}\left(  \bm{A}_{1}-\bm{A}_{0}\right)  .
\end{eqnarray*}
Reemplazando en~(\ref{fehc.p=0}) encontramos
\begin{equation}
\left\langle \bm{F}_{1}^{n+1}\right\rangle -\left\langle
\bm{F}_{0}^{n+1}\right\rangle =\left(  n+1\right)  \mathrm{d}%
\int_{T_{1}}\mathrm{d}t^{1}\left\langle \left(  \bm{A}_{1}%
-\bm{A}_{0}\right)  \bm{F}_{t}^{n}\right\rangle .
\end{equation}

Dado que integrar sobre el simplex $T_{1}$ es equivalente a integrar con
$t^{1}$ desde $t^{1}=0$ hasta $t^{1}=1$, podemos escribir simplemente
\begin{equation}
\left\langle \bm{F}_{1}^{n+1}\right\rangle -\left\langle
\bm{F}_{0}^{n+1}\right\rangle =\mathrm{d}\mathcal{Q}_{\bm{A}%
_{1}\leftarrow\bm{A}_{0}}^{\left(  2n+1\right)  },
\end{equation}
donde identificamos la forma de transgresi\'{o}n $\mathcal{Q}_{\bm{A}%
_{1}\leftarrow\bm{A}_{0}}^{\left(  2n+1\right)  }$ con
\begin{align}
\mathcal{Q}_{\bm{A}_{1}\leftarrow\bm{A}_{0}}^{\left(
2n+1\right)  }  &  =\int_{\left(  \bm{A}_{0}\bm{A}_{1}\right)
}l_{t}\left\langle \bm{F}_{t}^{n+1}\right\rangle \label{Qdef-alt}\\
&  =\left(  n+1\right)  \int_{0}^{1}\mathrm{d}t^{1}\left\langle \left(
\bm{A}_{1}-\bm{A}_{0}\right)  \bm{F}_{t}%
^{n}\right\rangle .
\end{align}

Esto concluye nuestra derivaci\'{o}n del teorema de Chern--Weil como un
corolario de la FEHC.

\subsection{\label{sec:identri}La Identidad Triangular}

En esta secci\'{o}n estudiamos el caso $p=1$ de la versi\'{o}n restringida de
la FEHC, ec.~(\ref{fehc4}). Poniendo $p=1$ en~(\ref{fehc4}) obtenemos
inmediatamente
\begin{equation}
\int_{\partial T_{2}}l_{t}\left\langle \bm{F}_{t}^{n+1}\right\rangle
=-\mathrm{d}\int_{T_{2}}\frac{l_{t}^{2}}{2}\left\langle \bm{F}%
_{t}^{n+1}\right\rangle , \label{fehc.p=1}%
\end{equation}
donde $\bm{F}_{t}$ es la curvatura asociada a la conexi\'{o}n
\begin{equation}
\bm{A}_{t}=t^{0}\bm{A}_{0}+t^{1}\bm{A}_{1}%
+t^{2}\bm{A}_{2},
\end{equation}
con $t^{0}+t^{1}+t^{2}=1$. El borde del simplex 2-dimensional $T_{2}=\left(
\bm{A}_{0}\bm{A}_{1}\bm{A}_{2}\right)  $ puede ser
escrito como la suma (ver fig.~\ref{T2andDT2})
\begin{equation}
\partial T_{2}=\left(  \bm{A}_{1}\bm{A}_{2}\right)  -\left(
\bm{A}_{0}\bm{A}_{2}\right)  +\left(  \bm{A}%
_{0}\bm{A}_{1}\right)  ,
\end{equation}
de modo que la integraci\'{o}n en el lado izquierdo de~(\ref{fehc.p=1}) puede
descomponerse en la forma
\begin{equation}
\int_{\partial T_{2}}l_{t}\left\langle \bm{F}_{t}^{n+1}\right\rangle
=\int_{\left(  \bm{A}_{1}\bm{A}_{2}\right)  }l_{t}\left\langle
\bm{F}_{t}^{n+1}\right\rangle -\int_{\left(  \bm{A}%
_{0}\bm{A}_{2}\right)  }l_{t}\left\langle \bm{F}_{t}%
^{n+1}\right\rangle +\int_{\left(  \bm{A}_{0}\bm{A}%
_{1}\right)  }l_{t}\left\langle \bm{F}_{t}^{n+1}\right\rangle .
\end{equation}
Recordando la definici\'{o}n de forma de transgresi\'{o}n dada
en~(\ref{Qdef-alt}), podemos escribir
\begin{equation}
\int_{\partial T_{2}}l_{t}\left\langle \bm{F}_{t}^{n+1}\right\rangle
=\mathcal{Q}_{\bm{A}_{2}\leftarrow\bm{A}_{1}}^{\left(
2n+1\right)  }-\mathcal{Q}_{\bm{A}_{2}\leftarrow\bm{A}_{0}%
}^{\left(  2n+1\right)  }+\mathcal{Q}_{\bm{A}_{1}\leftarrow
\bm{A}_{0}}^{\left(  2n+1\right)  }.
\end{equation}

Por otro lado, la regla de Leibniz para $l_{t}$ y la propiedad sim\'{e}trica
del tensor invariante $\left\langle \cdots\right\rangle $ implican que
\begin{equation}
l_{t}^{2}\left\langle \bm{F}_{t}^{n+1}\right\rangle =n\left(
n+1\right)  \left\langle \left(  \mathrm{d}_{t}\bm{A}_{t}\right)
^{2}\bm{F}_{t}^{n-1}\right\rangle .
\end{equation}

Integrando sobre el simplex $T_{2}$ obtenemos
\begin{equation}
\int_{T_{2}}\frac{l_{t}^{2}}{2}\left\langle \bm{F}_{t}^{n+1}%
\right\rangle =\mathcal{Q}_{\bm{A}_{2}\leftarrow\bm{A}%
_{1}\leftarrow\bm{A}_{0}}^{\left(  2n\right)  },
\end{equation}
donde $\mathcal{Q}_{\bm{A}_{2}\leftarrow\bm{A}_{1}%
\leftarrow\bm{A}_{0}}^{\left(  2n\right)  }$ est\'{a} dada por
\begin{equation}
\mathcal{Q}_{\bm{A}_{2}\leftarrow\bm{A}_{1}\leftarrow
\bm{A}_{0}}^{\left(  2n\right)  }\equiv n\left(  n+1\right)  \int
_{0}^{1}dt\int_{0}^{t}ds\left\langle \left(  \bm{A}_{2}-\bm{A}%
_{1}\right)  \left(  \bm{A}_{1}-\bm{A}_{0}\right)
\bm{F}_{t}^{n-1}\right\rangle . \label{q3def}%
\end{equation}
En~(\ref{q3def}) hemos introducido las variables auxiliares $t=1-t^{0}$,
$s=t^{2}$, en t\'{e}rminos de las cuales $\bm{A}_{t}$ toma la forma
\begin{equation}
\bm{A}_{t}=\bm{A}_{0}+s\left(  \bm{A}_{2}%
-\bm{A}_{1}\right)  +t\left(  \bm{A}_{1}-\bm{A}%
_{0}\right)  . \label{A2t}%
\end{equation}

Resumiendo, encontramos la identidad triangular
\begin{equation}
\mathcal{Q}_{\bm{A}_{2}\leftarrow\bm{A}_{1}}^{\left(
2n+1\right)  }-\mathcal{Q}_{\bm{A}_{2}\leftarrow\bm{A}_{0}%
}^{\left(  2n+1\right)  }+\mathcal{Q}_{\bm{A}_{1}\leftarrow
\bm{A}_{0}}^{\left(  2n+1\right)  }=-\mathrm{d}\mathcal{Q}%
_{\bm{A}_{2}\leftarrow\bm{A}_{1}\leftarrow\bm{A}_{0}%
}^{\left(  2n\right)  },
\end{equation}
o alternativamente,
\begin{equation}
\mathcal{Q}_{\bm{A}_{2}\leftarrow\bm{A}_{0}}^{\left(
2n+1\right)  }=\mathcal{Q}_{\bm{A}_{2}\leftarrow\bm{A}_{1}%
}^{\left(  2n+1\right)  }+\mathcal{Q}_{\bm{A}_{1}\leftarrow
\bm{A}_{0}}^{\left(  2n+1\right)  }+\mathrm{d}\mathcal{Q}%
_{\bm{A}_{2}\leftarrow\bm{A}_{1}\leftarrow\bm{A}_{0}%
}^{\left(  2n\right)  }. \label{q34}%
\end{equation}
En este punto enfatizamos que el uso de la FEHC ha sido crucial para obtener
la forma expl\'{\i}cita del \'{u}timo t\'{e}rmino en~(\ref{q34}), cuya
existencia hab\'{\i}a sido deducida a partir del teorema de Chern--Weil (ver
sec.~\ref{sec:motif}).

\section{\label{sec:formmet}Formulaci\'{o}n del M\'{e}todo}

El m\'{e}todo de separaci\'{o}n en subespacios para el lagrangeano transgresor
[cf.~ec.~(\ref{AcTra})]
\begin{equation}
L_{\mathrm{T}}^{\left(  2n+1\right)  }\left(  \bm{A},\bar
{\bm{A}}\right)  =k\mathcal{Q}_{\bm{A}\leftarrow
\bar{\bm{A}}}^{\left(  2n+1\right)  }%
\end{equation}
consiste en una aplicaci\'{o}n iterativa de la identidad triangular
[cf.~ec.~(\ref{q34})]%
\begin{equation}
\mathcal{Q}_{\bm{A}\leftarrow\bar{\bm{A}}}^{\left(
2n+1\right)  }=\mathcal{Q}_{\bm{A}\leftarrow\tilde{\bm{A}}%
}^{\left(  2n+1\right)  }+\mathcal{Q}_{\tilde{\bm{A}}\leftarrow
\bar{\bm{A}}}^{\left(  2n+1\right)  }+\mathrm{d}\mathcal{Q}%
_{\bm{A}\leftarrow\tilde{\bm{A}}\leftarrow\bar{\bm{A}%
}}^{\left(  2n\right)  }. \label{treqdef}%
\end{equation}
Aqu\'{\i} $\mathcal{Q}_{\bm{A}\leftarrow\tilde{\bm{A}%
}\leftarrow\bar{\bm{A}}}^{\left(  2n\right)  }$ est\'{a} dada por
[cf.~ec.~(\ref{q3def})]
\begin{equation}
\mathcal{Q}_{\bm{A}\leftarrow\tilde{\bm{A}}\leftarrow
\bar{\bm{A}}}^{\left(  2n\right)  }\equiv n\left(  n+1\right)
\int_{0}^{1}dt\int_{0}^{t}ds\left\langle \left(  \bm{A}-\tilde
{\bm{A}}\right)  \left(  \tilde{\bm{A}}-\bar{\bm{A}%
}\right)  \bm{F}_{st}^{n-1}\right\rangle , \label{borde}%
\end{equation}
donde $\bm{F}_{st}$ es la curvatura asociada a la conexi\'{o}n
[cf.~ec.~(\ref{A2t})]
\begin{equation}
\bm{A}_{st}=\bar{\bm{A}}+s\left(  \bm{A}%
-\tilde{\bm{A}}\right)  +t\left(  \tilde{\bm{A}}%
-\bar{\bm{A}}\right)  .
\end{equation}
El m\'{e}todo puede ser descrito esquem\'{a}ticamente por medio de la
siguiente serie de pasos:

\begin{enumerate}
\item Identificar los subespacios relevantes presentes en el \'{a}lgebra de
gauge $\mathfrak{g}$, i.e. escribir $\mathfrak{g}=V_{0}\oplus\cdots\oplus
V_{p}$.

\item Escribir las conexiones $\bm{A}$ y $\bar{\bm{A}}$ como
sumas de trozos valuados cada uno en un subespacio distinto de $\mathfrak{g}$,
i.e. $\bm{A}=\bm{a}_{0}+\dotsb+\bm{a}_{p}$,
$\bar{\bm{A}}=\bar{\bm{a}}_{0}+\dotsb+\bar{\bm{a}}%
_{p}$, con $\bm{a}_{k},\bar{\bm{a}}_{k} \in V_{k}$.

\item Usar la ec.~(\ref{treqdef}) con $\tilde{\bm{A}}=\bm{a}%
_{0}+\dotsb+\bm{a}_{p-1}$ para obtener un lagrangeano parcial para
$\bm{a}_{p}$.

\item Iterar el paso 3 para cada uno de los trozos $\bm{a}_{k}$.
\end{enumerate}

El caso m\'{a}s simple corresponde a cuando existen s\'{o}lo dos subespacios,
$\mathfrak{g}=V_{0}\oplus V_{1}$. Siguiendo los pasos descritos m\'{a}s
arriba, escribimos las conexiones $\bm{A}$ y $\bar{\bm{A}}$
como
\begin{align}
\bm{A}  &  =\bm{a}_{0}+\bm{a}_{1},\\
\bar{\bm{A}}  &  =\bar{\bm{a}}_{0}+\bar{\bm{a}}_{1}.
\end{align}
Ahora usamos la ec.~(\ref{treqdef}) con $\tilde{\bm{A}}=\bm{a}%
_{0}$ para encontrar
\begin{equation}
\mathcal{Q}_{\bm{a}_{0}+\bm{a}_{1}\leftarrow\bar
{\bm{a}}_{0}+\bar{\bm{a}}_{1}}^{\left(  2n+1\right)
}=\mathcal{Q}_{\bm{a}_{0}+\bm{a}_{1}\leftarrow\bm{a}%
_{0}}^{\left(  2n+1\right)  }+\mathcal{Q}_{\bm{a}_{0}\leftarrow
\bar{\bm{a}}_{0}+\bar{\bm{a}}_{1}}^{\left(  2n+1\right)
}+\mathrm{d}\mathcal{Q}_{\bm{a}_{0}+\bm{a}_{1}\leftarrow
\bm{a}_{0}\leftarrow\bar{\bm{a}}_{0}+\bar{\bm{a}}_{1}%
}^{\left(  2n\right)  }. \label{trex1}%
\end{equation}
Iterando el m\'{e}todo, la transgresi\'{o}n $\mathcal{Q}_{\bm{a}%
_{0}\leftarrow\bar{\bm{a}}_{0}+\bar{\bm{a}}_{1}}^{\left(
2n+1\right)  }$ puede a su vez ser descompuesta usando $\tilde{\bm{A}%
}=\bar{\bm{a}}_{0}$:
\begin{equation}
\mathcal{Q}_{\bm{a}_{0}\leftarrow\bar{\bm{a}}_{0}%
+\bar{\bm{a}}_{1}}^{\left(  2n+1\right)  }=\mathcal{Q}_{\bm{a}%
_{0}\leftarrow\bar{\bm{a}}_{0}}^{\left(  2n+1\right)  }+\mathcal{Q}%
_{\bar{\bm{a}}_{0}\leftarrow\bar{\bm{a}}_{0}+\bar
{\bm{a}}_{1}}^{\left(  2n+1\right)  }+\mathrm{d}\mathcal{Q}%
_{\bm{a}_{0}\leftarrow\bar{\bm{a}}_{0}\leftarrow
\bar{\bm{a}}_{0}+\bar{\bm{a}}_{1}}^{\left(  2n\right)  }.
\label{trex2}%
\end{equation}
Introduciendo~(\ref{trex2}) en~(\ref{trex1}) hallamos finalmente
\begin{eqnarray}
\mathcal{Q}_{\bm{a}_{0}+\bm{a}_{1}\leftarrow\bar
{\bm{a}}_{0}+\bar{\bm{a}}_{1}}^{\left(  2n+1\right)  }  &
= & \mathcal{Q}_{\bm{a}_{0}+\bm{a}_{1}\leftarrow\bm{a}%
_{0}}^{\left(  2n+1\right)  }+\mathcal{Q}_{\bar{\bm{a}}_{0}%
\leftarrow\bar{\bm{a}}_{0}+\bar{\bm{a}}_{1}}^{\left(
2n+1\right)  }+\mathcal{Q}_{\bm{a}_{0}\leftarrow\bar{\bm{a}%
}_{0}}^{\left(  2n+1\right)  }+\nonumber\\
& & +\mathrm{d}\mathcal{Q}_{\bm{a}_{0}\leftarrow\bar{\bm{a}%
}_{0}\leftarrow\bar{\bm{a}}_{0}+\bar{\bm{a}}_{1}}^{\left(
2n\right)  }+\mathrm{d}\mathcal{Q}_{\bm{a}_{0}+\bm{a}%
_{1}\leftarrow\bm{a}_{0}\leftarrow\bar{\bm{a}}_{0}%
+\bar{\bm{a}}_{1}}^{\left(  2n\right)  }. \label{trex3}%
\end{eqnarray}

Esta descomposici\'{o}n presenta varios aspectos interesantes. De la
simetr\'{\i}a de intercambio para las formas de transgresi\'{o}n (ver
secci\'{o}n~\ref{sec:symtra}, p\'{a}g.~\pageref{sec:symtra}) tenemos que los
dos primeros t\'{e}rminos en el lado derecho de~(\ref{trex3}) son reflejos
especulares el uno del otro, en el sentido que son id\'{e}nticos bajo el
intercambio $\bm{A}\leftrightarrow\bar{\bm{A}}$:
\begin{eqnarray}
\mathcal{Q}_{\bm{a}_{0}+\bm{a}_{1}\leftarrow\bar
{\bm{a}}_{0}+\bar{\bm{a}}_{1}}^{\left(  2n+1\right)  }  &
= & \mathcal{Q}_{\bm{a}_{0}+\bm{a}_{1}\leftarrow\bm{a}%
_{0}}^{\left(  2n+1\right)  }-\mathcal{Q}_{\bar{\bm{a}}_{0}%
+\bar{\bm{a}}_{1}\leftarrow\bar{\bm{a}}_{0}}^{\left(
2n+1\right)  }+\mathcal{Q}_{\bm{a}_{0}\leftarrow\bar{\bm{a}%
}_{0}}^{\left(  2n+1\right)  }+\nonumber\\
& & +\mathrm{d}\mathcal{Q}_{\bm{a}_{0}\leftarrow\bar{\bm{a}%
}_{0}\leftarrow\bar{\bm{a}}_{0}+\bar{\bm{a}}_{1}}^{\left(
2n\right)  }+\mathrm{d}\mathcal{Q}_{\bm{a}_{0}+\bm{a}%
_{1}\leftarrow\bm{a}_{0}\leftarrow\bar{\bm{a}}_{0}%
+\bar{\bm{a}}_{1}}^{\left(  2n\right)  }.
\end{eqnarray}
El tercer t\'{e}rmino es una suerte de `espejo', pues, aparte de los
t\'{e}rminos de borde, es el \'{u}nico que liga componentes de $\bm{A}%
$ con componentes de $\bar{\bm{A}}$. Interesantemente, siempre es
posible descomponer esta clase de t\'{e}rminos en una suma de dos
transgresiones `desacopladas' m\'{a}s un t\'{e}rmino de borde, usando la
identidad triangular~(\ref{treqdef}) con $\tilde{\bm{A}}=0$ (ver
secci\'{o}n~\ref{sec:tr-cs} m\'{a}s adelante).

El m\'{e}todo de separaci\'{o}n en subespacios encuentra aplicaciones en los
cap\'{\i}tulos~\ref{ch:TGFTGrav} y~\ref{ch:TGFTMAlg}, donde ser\'{a} usado para encontrar versiones expl\'{\i}citas de lagrangeanos transgresores en el contexto de teor\'{\i}as de gauge para gravedad en dimensiones impares y Supergravedad en $d=11$.

\section{\label{sec:tr-cs}La importancia de la invariancia}

La identidad triangular [cf.~ec.~(\ref{treqdef})]
\begin{equation}
\mathcal{Q}_{\bm{A}\leftarrow\bar{\bm{A}}}^{\left(
2n+1\right)  }=\mathcal{Q}_{\bm{A}\leftarrow\tilde{\bm{A}}%
}^{\left(  2n+1\right)  }+\mathcal{Q}_{\tilde{\bm{A}}\leftarrow
\bar{\bm{A}}}^{\left(  2n+1\right)  }+\mathrm{d}\mathcal{Q}%
_{\bm{A}\leftarrow\tilde{\bm{A}}\leftarrow\bar{\bm{A}%
}}^{\left(  2n\right)  } \label{treq5}
\end{equation}
arroja nueva luz sobre la acci\'{o}n transgresora discutida en el cap\'{\i}tulo~\ref{ch:trans}.

Es importante destacar que, si bien cada t\'{e}rmino en el lado derecho de~(\ref{treq5}) depende de la conexi\'{o}n intermedia $\tilde{\bm{A}}$, lo hacen de modo tal que su suma es independiente de $\tilde{\bm{A}}$. Este hecho cobra especial relevancia al analizar las propiedades de invariancia de la forma de transgresi\'{o}n. En efecto, \'{e}sta es invariante bajo transformaciones arbitrarias sobre $\tilde{\bm{A}}$, y no s\'{o}lo bajo transformaciones de gauge. Esta simple observaci\'{o}n ayuda a clarificar algunos aspectos aparentemente enigm\'{a}ticos de la acci\'{o}n transgresora para gravedad descrita en el cap\'{\i}tulo~\ref{ch:TGFTGrav}.

Como se menciona en la secci\'{o}n~\ref{sec:cwt}, la forma de CS corresponde
al caso particular $\bar{\bm{A}}=0$ de la forma de transgresi\'{o}n:
\begin{equation}
\mathcal{Q}_{\mathrm{CS}}^{\left(  2n+1\right)  }\left(  \bm{A}%
\right)  =\mathcal{Q}_{\bm{A}\leftarrow0}^{\left(  2n+1\right)  }.
\end{equation}
La no preservaci\'{o}n de la igualdad $\bar{\bm{A}}=0$ bajo
transformaciones de gauge es la ra\'{\i}z de la pseudo-invariancia del
lagrangeano de CS.

La forma de transgresi\'{o}n puede escribirse como la diferencia de dos formas
de CS m\'{a}s un t\'{e}rmino de borde. En efecto, usando la identidad
triangular~(\ref{treq5}) con $\tilde{\bm{A}}=0$ hallamos
\begin{equation}
\mathcal{Q}_{\bm{A}\leftarrow\bar{\bm{A}}}^{\left(
2n+1\right)  }=\mathcal{Q}_{\mathrm{CS}}^{\left(  2n+1\right)  }\left(
\bm{A}\right)  -\mathcal{Q}_{\mathrm{CS}}^{\left(  2n+1\right)
}\left(  \bar{\bm{A}}\right)  +\mathrm{d} \mathcal{B}^{\left( 2n \right)},
\end{equation}
con $\mathcal{B}^{\left( 2n \right)} = \mathcal{Q}_{\bm{A} \leftarrow 0 \leftarrow \bar{\bm{A}}}^{\left(  2n \right)}$. La forma expl\'{\i}cita de este t\'{e}rmino de borde es
\begin{equation}
\mathcal{B}^{\left( 2n \right)} = -n\left(  n+1\right)  \int_{0}^{1}dt\int_{0}^{t}ds\left\langle
\bm{A}\bar{\bm{A}}\bm{F}_{st}^{n-1}\right\rangle ,
\end{equation}
donde $\bm{F}_{st}$ es la curvatura\footnote{La manera m\'{a}s
eficiente de calcular esta curvatura consiste en usar una versi\'{o}n
generalizada de las ecuaciones de Gauss--Codazzi. Sean $\bm{A}$
y\ $\bar{\bm{A}}$ dos conexiones, y sea $\bm{\Delta}%
\equiv\bm{A}-\bar{\bm{A}}$. Las curvaturas asociadas a estas
conexiones est\'{a}n relacionadas mediante la identidad $\bm{F}%
=\bar{\bm{F}}+\bar{\mathrm{D}}\bm{\Delta}+\bm{\Delta
}^{2}$, donde $\bar{\mathrm{D}}$ es la derivada covariante en la conexi\'{o}n
$\bar{\bm{A}}$.} asociada a la conexi\'{o}n $\bm{A}_{st}$, con
\begin{align}
\bm{A}_{st}  &  =s\bm{A}+\left(  1-t\right)  \bar
{\bm{A}},\\
\bm{F}_{st}  &  =\bar{\bm{F}}+\bar{\mathrm{D}}\left(
s\bm{A}-t\bar{\bm{A}}\right)  +\left(  s\bm{A}%
-t\bar{\bm{A}}\right)  ^{2}.
\end{align}

Esta manera de escribir la transgresi\'{o}n clarifica la forma desacoplada de
las ecuaciones de movimiento, mostrando que en efecto la acci\'{o}n puede
escribirse como la suma de dos acciones de CS m\'{a}s un t\'{e}rmino de borde.
La presencia de este t\'{e}rmino de borde es crucial para asegurar la
invariancia de la forma de transgresi\'{o}n, raz\'{o}n por la cual no puede
ser ignorado.

\chapter{\label{ch:TGFTGrav}Gravedad Transgresora en $\lowercase{d}=2\lowercase{n}+1$}

En este cap\'{\i}tulo exploramos una teor\'{\i}a de gauge transgresora para el \'{a}lgebra de anti-de Sitter en dimensiones impares.

El lagrangeano es una forma de transgresi\'{o}n donde una de las conexiones involucradas, $\bar{\bm{A}} = \bar{\bm{\omega}}$, est\'{a} valuada s\'{o}lo en la sub\'{a}lgebra de Lorentz del \'{a}lgebra de AdS. Esta configuraci\'{o}n de campos, junto con la elecci\'{o}n de tensor invariante, permite relegar toda la dependencia de la acci\'{o}n en $\bar{\bm{\omega}}$ a un t\'{e}rmino de borde, evitando as\'{\i} los potenciales problemas de interpretaci\'{o}n surgidos de tener dos conexiones independientes en la variedad.

La acci\'{o}n que encontramos fue discutida inicialmente en la Ref.~\cite{Mor04a}, donde se encontr\'{o} que la adici\'{o}n a la acci\'{o}n de CS para gravedad del t\'{e}rmino de borde proveniente de la transgresi\'{o}n permite describir correctamente la termodin\'{a}mica de agujeros negros y calcular cargas conservadas independientes del \textit{background} como una aplicaci\'{o}n directa del Teorema de Noether (ver tambi\'{e}n~\cite{All03,Aro99a,Aro99b,Kof06,Mor04b,Obu06}).

\section{\label{sec:tgac}La Acci\'{o}n}

El \'{a}lgebra y el tensor invariante que ocupamos para escribir el
lagrangeano transgresor son los mismos de la secci\'{o}n~\ref{sec:CSGrav};
vale decir, el \'{a}lgebra de AdS, generada por $\bm{J}_{ab}$ y
$\bm{P}_{a}$, con las relaciones de conmutaci\'{o}n
\begin{align}
\left[  \bm{P}_{a},\bm{P}_{b}\right]   &  =\bm{J}%
_{ab},\\
\left[  \bm{J}_{ab},\bm{P}_{c}\right]   &  =\eta
_{cb}\bm{P}_{a}-\eta_{ca}\bm{P}_{b},\\
\left[  \bm{J}_{ab},\bm{J}_{cd}\right]   &  =\eta
_{cb}\bm{J}_{ad}-\eta_{ca}\bm{J}_{bd}+\eta_{db}\bm{J}%
_{ca}-\eta_{da}\bm{J}_{cb},
\end{align}
y el tensor de Levi-Civita $\varepsilon_{a_{1}\cdots a_{2n+1}}$ como tensor
sim\'{e}trico invariante,
\begin{equation}
\left\langle \bm{J}_{a_{1}a_{2}}\cdots\bm{J}_{a_{2n-1}a_{2n}%
}\bm{P}_{a_{2n+1}}\right\rangle =\frac{2^{n}}{n+1}\varepsilon
_{a_{1}\cdots a_{2n+1}}, \label{tsieps}%
\end{equation}
con todas las dem\'{a}s componentes iguales a cero.

Como lagrangeano tomamos
\begin{equation}
L_{\mathrm{G}}^{\left( 2n+1 \right)} = k \mathcal{Q}_{\bm{A} \leftarrow \bar{\bm{A}}}^{\left( 2n+1 \right)}, \label{Lg2n+1}
\end{equation}
donde $k$ es una constante adimensional arbitraria y $\mathcal{Q}_{\bm{A} \leftarrow \bar{\bm{A}}}^{\left( 2n+1 \right)}$ es la forma de transgresi\'{o}n que interpola entre las conexiones
\begin{align}
\bar{\bm{A}} & = \bar{\bm{\omega}}, \\
\bm{A}       & = \bm{e} + \bm{\omega}.
\end{align}
Las curvaturas asociadas a estas conexiones son
\begin{align}
\bar{\bm{F}} & =\bar{\bm{R}}, \label{F0} \\
\bm{F}       & =\bm{R} + \bm{e}^{2} + \bm{T}, \label{F2}
\end{align}
donde
\begin{align}
\bm{R}  &  =\mathrm{d}\bm{\omega}+\bm{\omega}%
^{2},\label{R7}\\
\bm{T}  &  =\mathrm{d}\bm{e}+\left[  \bm{\omega
},\bm{e}\right]  ,
\end{align}
son la curvatura y la torsi\'{o}n de Lorentz, respectivamente [una
expresi\'{o}n completamente an\'{a}loga a~(\ref{R7}) es v\'{a}lida para
$\bar{\bm{R}}$].

El lagrangeano~(\ref{Lg2n+1}) puede ser reescrito usando el m\'{e}todo de
separaci\'{o}n en subespacios presentado en el cap\'{\i}tulo~\ref{ch:metsepsub}%
. Para ello introducimos la conexi\'{o}n intermedia
\begin{equation}
\tilde{\bm{A}} = \bm{\omega}
\end{equation}
y ocupamos la identidad triangular~(\ref{treqdef}) en la forma
\begin{equation}
\mathcal{Q}_{\bm{e}+\bm{\omega} \leftarrow \bar{\bm{\omega}}}^{\left( 2n+1 \right)} =
\mathcal{Q}_{\bm{e}+\bm{\omega} \leftarrow \bm{\omega}}^{\left( 2n+1 \right)} +
\mathcal{Q}_{\bm{\omega} \leftarrow \bar{\bm{\omega}}}^{\left( 2n+1 \right)} +
\mathrm{d} \mathcal{Q}_{\bm{e}+\bm{\omega} \leftarrow \bm{\omega} \leftarrow \bar{\bm{\omega}}}^{\left( 2n \right)}.
\label{trgrav}
\end{equation}

El primer t\'{e}rmino en~(\ref{trgrav}) es
\begin{equation}
\mathcal{Q}_{\bm{e}+\bm{\omega}\leftarrow\bm{\omega}%
}^{\left(  2n+1\right)  }=\left(  n+1\right)  \int_{0}^{1}dt\left\langle
\bm{eF}_{t}^{n}\right\rangle ,
\end{equation}
donde $\bm{F}_{t}$ es la curvatura asociada a la conexi\'{o}n
interpolante $\bm{A}_{t}$,
\begin{align}
\bm{A}_{t}  &  =\bm{\omega}+t\bm{e},\\
\bm{F}_{t}  &  =\bm{R}+t^{2}\bm{e}^{2}+t\bm{T}%
.
\end{align}
Teniendo en cuenta la forma particular del tensor sim\'{e}trico
invariante~(\ref{tsieps}) podemos escribir
\begin{equation}
\mathcal{Q}_{\bm{e}+\bm{\omega}\leftarrow\bm{\omega}%
}^{\left(  2n+1\right)  }=\left(  n+1\right)  \int_{0}^{1}dt\left\langle
\bm{e}\left(  \bm{R}+t^{2}\bm{e}^{2}\right)
^{n}\right\rangle . \label{qeww}%
\end{equation}
Como vemos, la elecci\'{o}n~(\ref{tsieps}) implica que este trozo del
lagrangeano sea independiente de la torsi\'{o}n. Es tambi\'{e}n id\'{e}ntico
al t\'{e}rmino de volumen del lagrangeano de CS para gravedad en $d=2n+1$,
ec.~(\ref{LagCSGrav1}).

El segundo t\'{e}rmino en~(\ref{trgrav}) se anula id\'{e}nticamente:
\begin{equation}
\mathcal{Q}_{\bm{\omega}\leftarrow\bar{\bm{\omega}}}^{\left(
2n+1\right)  }=0. \label{Qww=0}%
\end{equation}
Algebraicamente, esto es una consecuencia del hecho que el tensor
invariante~(\ref{tsieps}) no tiene componentes no nulas de la forma
$\left\langle \bm{J}\cdots\bm{J}\right\rangle $, que
habr\'{\i}a sido la \'{u}nica en contribuir a~(\ref{Qww=0}). La anulaci\'{o}n
de este trozo del lagrangeano tiene consecuencias interesantes, pues relega
toda la dependencia en $\bar{\bm{\omega}}$ a un t\'{e}rmino de borde.
Esto elimina en gran medida el principal problema de interpretaci\'{o}n de los
lagrangeanos transgresores, a saber, la presencia de dos conexiones
independientes en la variedad $M$. A\'{u}n es necesario justificar la
presencia de $\bar{\bm{\omega}}$ en el borde de $M$; para ello
recurriremos (m\'{a}s adelante) a las condiciones de borde generadas por el
lagrangeano~(\ref{Lg2n+1}).

Volviendo a la ec.~(\ref{borde}), encontramos que el \'{u}ltimo t\'{e}rmino
en~(\ref{trgrav}) puede ser escrito en la forma
\begin{equation}
\mathcal{Q}_{\bm{e}+\bm{\omega}\leftarrow\bm{\omega
}\leftarrow\bar{\bm{\omega}}}^{\left(  2n\right)  }=n\left(
n+1\right)  \int_{0}^{1}dt\int_{0}^{t}ds\left\langle \bm{e\theta
F}_{st}^{n-1}\right\rangle ,
\end{equation}
donde
\begin{equation}
\bm{\theta}\equiv\bm{\omega}-\bar{\bm{\omega}}%
\end{equation}
y $\bm{F}_{st}$ es la curvatura asociada a la conexi\'{o}n $\bm{A}_{st}$, con
\begin{align}
\bm{A}_{st} & = \bar{\bm{\omega}} + s \bm{e} + t \bm{\theta}, \\
\bm{F}_{st} & = \bar{\bm{R}} + \mathrm{D}_{\bar{\bm{\omega}}} \left( s \bm{e} + t \bm{\theta} \right) + s^{2} \bm{e}^{2} + st \left[ \bm{e}, \bm{\theta} \right] + t^{2} \bm{\theta}^{2}.
\end{align}
Teniendo en consideraci\'{o}n la forma del tensor invariante~(\ref{tsieps}) podemos
escribir $\mathcal{Q}_{\bm{e}+\bm{\omega}\leftarrow
\bm{\omega}\leftarrow\bar{\bm{\omega}}}^{\left(  2n\right)  }$
como
\begin{equation}
\mathcal{Q}_{\bm{e}+\bm{\omega}\leftarrow\bm{\omega
}\leftarrow\bar{\bm{\omega}}}^{\left(  2n\right)  }=n\left(
n+1\right)  \int_{0}^{1}dt\int_{0}^{t}ds\left\langle \bm{e\theta
}\left(  \bar{\bm{R}}+t\mathrm{D}_{\bar{\bm{\omega}}%
}\bm{\theta}+s^{2}\bm{e}^{2}+t^{2}\bm{\theta}%
^{2}\right)  ^{n-1}\right\rangle .
\end{equation}

Reuniendo los trozos, el lagrangeano~(\ref{trgrav}) toma la forma
\begin{align}
L_{\mathrm{G}}^{\left(  2n+1\right)  } = & \left(  n+1\right)  k\int_{0}%
^{1}dt\left\langle \bm{e}\left(  \bm{R}+t^{2}\bm{e}%
^{2}\right)  ^{n}\right\rangle +\nonumber\\
&  +n\left(  n+1\right)  k\mathrm{d}\int_{0}^{1}dt\int_{0}^{t}ds\left\langle
\bm{e\theta}\left(  \bar{\bm{R}}+t\mathrm{D}_{\bar
{\bm{\omega}}}\bm{\theta}+s^{2}\bm{e}^{2}%
+t^{2}\bm{\theta}^{2}\right)  ^{n-1}\right\rangle . \label{Lg2n+1b}%
\end{align}
Este lagrangeano corresponde a un lagrangeano de CS para $\left(
\bm{e},\bm{\omega}\right)  $ m\'{a}s un t\'{e}rmino de borde
que incluye un campo extra $\bar{\bm{\omega}}$.

\section[Ecuaciones de Movimiento y Condiciones de Borde]%
{\label{sec:tgeom}Ecuaciones de Movimiento y Condiciones de Borde
\sectionmark{Ecs. de Movimiento y Condiciones de Borde}}
\sectionmark{Ecs. de Movimiento y Condiciones de Borde}

Las ecuaciones de movimiento para el lagrangeano~(\ref{trgrav}) est\'{a}n
dadas por
\begin{align}
\left\langle \bm{J}_{ab}\left(  \bm{R}+\bm{e}%
^{2}\right)  ^{n-1}\bm{T}\right\rangle  &  =0,\\
\left\langle \bm{P}_{a}\left(  \bm{R}+\bm{e}%
^{2}\right)  ^{n}\right\rangle  &  =0.
\end{align}
Estas pueden obtenerse por variaci\'{o}n directa del lagrangeano~(\ref{Lg2n+1b}) o reemplazando~(\ref{F0})--(\ref{F2}) en las f\'{o}rmulas generales~(\ref{FnGa1})--(\ref{FnGa2}).
Definiendo
\begin{equation}
\mathcal{R}_{abc} \equiv \left\langle \bm{F}^{n-1} \bm{J}_{ab} \bm{P}_{c} \right\rangle ,
\end{equation}
ellas adoptan la forma
\begin{align}
\mathcal{R}_{abc} T^{c} & = 0, \\
\mathcal{R}_{abc} \left( R^{ab} + \frac{1}{\ell^{2}} e^{a} e^{b} \right) & = 0.
\end{align}

Como $\bar{\bm{\omega}}$ entra en el lagrangeano s\'{o}lo a trav\'{e}s
de un t\'{e}rmino de borde, la ecuaci\'{o}n de movimiento asociada a su
variaci\'{o}n es vac\'{\i}a.

Una versi\'{o}n expl\'{\i}cita para $\mathcal{R}_{abc}$\ es encontrada
haciendo uso del tensor invariante~(\ref{tsieps}):%
\begin{eqnarray}
\mathcal{R}_{abc} & = & \frac{2}{n+1} \varepsilon_{abc a_{1} \cdots a_{2n-2}} \left( R^{a_{1} a_{2}} + \frac{1}{\ell^{2}} e^{a_{1} a_{2}} \right) \cdots \nonumber \\
& & \cdots \left( R^{a_{2n-3} a_{2n-2}} + \frac{1}{\ell^{2}} e^{a_{2n-3} a_{2n-2}} \right).
\end{eqnarray}

Claramente, ambas ecuaciones de movimiento son satisfechas cuando se cumple
$\mathcal{R}_{abc}=0$. Otra soluci\'{o}n obtenida por simple inspecci\'{o}n es
un espacio de curvatura negativa constante, $R^{ab}=-\left(  1/\ell
^{2}\right)  e^{a}e^{b}$. En ambos casos la torsi\'{o}n queda indeterminada
por las ecuaciones de movimiento.

Al igual que para las ecuaciones de movimiento, tambi\'{e}n hay dos rutas
posibles para obtener condiciones de borde: por variaci\'{o}n directa del
lagrangeano~(\ref{trgrav}) o por sustituci\'{o}n de las cantidades relevantes
en la f\'{o}rmula general~(\ref{bctrans}). De cualquiera de estas dos maneras
obtenemos
\begin{equation}
\left.  \int_{0}^{1}dt\left\langle \left(  \delta\bar{\bm{\omega}%
}+t\delta\bm{\theta}+t\delta\bm{e}\right)  \left(
\bm{\theta}+\bm{e}\right)  \bm{F}_{t}^{n-1}%
\right\rangle \right\vert _{\partial M}=0, \label{bcgravtrans}%
\end{equation}
donde en este caso la conexi\'{o}n $\bm{A}_{t}$ y la curvatura
asociada $\bm{F}_{t}$ est\'{a}n dadas por
\begin{align}
\bm{A}_{t}  &  =\bar{\bm{\omega}}+t\left(  \bm{e}%
+\bm{\theta}\right)  ,\\
\bm{F}_{t}  &  =\bar{\bm{R}}+t\mathrm{D}_{\bar
{\bm{\omega}}}\left(  \bm{e}+\bm{\theta}\right)
+t^{2}\left(  \bm{e}^{2}+\left[  \bm{e},\bm{\theta
}\right]  +\bm{\theta}^{2}\right)  .
\end{align}

Hay muchos modos alternativos de satisfacer las condiciones de
borde~(\ref{bcgravtrans}). En~\cite{Mor04a} se proponen argumentos
f\'{\i}sicos\footnote{El requerimiento f\'{\i}sico principal es la
equivalencia del concepto de transporte paralelo inducido por
$\bm{\omega}$ y $\bar{\bm{\omega}}$ en el borde de la variedad
espacio-temporal $M$.} que permiten fijar parcialmente estas condiciones;
probablemente el m\'{a}s significativo de ellos es exigir que $\bar
{\bm{\omega}}$ tenga un valor fijo en $\partial M$, es decir,
\begin{equation}
\left.  \delta\bar{\bm{\omega}}\right\vert _{\partial M}=0.
\end{equation}
Las condiciones de borde restantes pueden ser escritas en la forma
\begin{equation}
\left.  \int_{0}^{1}dt \left\langle t \left(  \delta\bm{\theta
e}-\bm{\theta}\delta\bm{e}\right)  \left(  \bar{\bm{R}%
}+t^{2}\bm{e}^{2}+t^{2}\bm{\theta}^{2}\right)  ^{n-1}%
\right\rangle \right\vert _{\partial M}=0.
\end{equation}
Estas pueden ser satisfechas requiriendo
\begin{equation}
\delta\theta^{\lbrack ab}e^{c]}=\theta^{\lbrack ab}\delta e^{c]}.
\end{equation}

Quisi\'{e}ramos resaltar aqu\'{\i} que la presencia de $\bar
{\bm{\omega}}$ en el lagrangeano, si bien no afecta las ecuaciones de
movimiento, s\'{\i} cambia las condiciones de borde y permite que tanto la
acci\'{o}n como las cargas conservadas que se derivan de ella tengan un valor
finito al ser evaluadas en una soluci\'{o}n~\cite{Mor04a}.

\section{\label{sec:tgdis}Discusi\'{o}n}

La forma de transgresi\'{o}n usada como lagrageano en la secci\'{o}n~\ref{sec:tgac} tiene la particularidad de que una de las transgresiones involucradas, $\bm{A}_{0}=\bar{\bm{\omega}}$,
est\'{a} valuada s\'{o}lo en la sub\'{a}lgebra de Lorentz y no en toda el \'{a}lgebra de AdS. En esta secci\'{o}n consideramos algunas implicaciones de este hecho.

El problema con la afirmaci\'{o}n `$\bm{A}_{0}=\bar{\bm{\omega}}$ est\'{a} valuada s\'{o}lo en la sub\'{a}lgebra de Lorentz' es que ella no es preservada bajo transformaciones de gauge. En efecto, bajo un
\textit{boost} infinitesimal de AdS, $g=1+\lambda^{a}\bm{P}_{a}$, $\bar{\bm{\omega}}$ cambia a
\begin{equation}
\bar{\bm{\omega}} \rightarrow \bar{\bm{\omega}} + \bar{\bm{e}}_{\mathrm{g}},
\end{equation}
donde $\bar{\bm{e}}_{\mathrm{g}} = - \mathrm{D}_{\bar{\bm{\omega}}} \bm{\lambda}$ es un vielbein que corresponde a puro gauge. En esta situaci\'{o}n, si bien $\mathcal{Q}_{\bm{\omega} \leftarrow \bar{\bm{\omega}}}^{\left( 2n+1 \right)}$ es cero, su variaci\'{o}n bajo el \textit{boost} no se anula, i.e.
\begin{equation}
\mathcal{Q}_{\bm{\omega} \leftarrow \bar{\bm{\omega}}}^{\left( 2n+1 \right)} = 0 \rightarrow \mathcal{Q}_{\bm{\omega} + \bm{e}_{\mathrm{g}} \leftarrow \bar{\bm{\omega}} + \bar{\bm{e}}_{\mathrm{g}}}^{\left( 2n+1 \right)} \neq 0.
\end{equation}

Por supuesto, si escribimos el lagrangeano $L_{\mathrm{G}}^{\left(
2n+1\right)  }$ como en~(\ref{Lg2n+1b}), omitiendo $\mathcal{Q}%
_{\bm{\omega}\leftarrow\bar{\bm{\omega}}}^{\left(
2n+1\right)  }$, entonces el resultado no es m\'{a}s invariante de gauge
(aunque, gracias a las propiedades especiales del tensor de Levi-Civita, el
cambio producido es a lo m\'{a}s una forma cerrada).

En el cap\'{\i}tulo~\ref{ch:TGFTMAlg} se considera una teor\'{\i}a de gauge transgresora para dos conexiones completamente valudas en el \'{a}lgebra correspondiente, obviando as\'{\i} esta clase de sutilezas.

\chapter{\label{ch:expansion}Expansi\'{o}n de \'{A}lgebras de Lie con Semigrupos Abelianos}
\chaptermark{$S$-Expansi\'{o}n de \'{A}lgebras de Lie}

\begin{quote}
\textit{Presiento que por lo emp\'{\i}rico \\
se ha enloquecido la br\'{u}jula.}

E.~Llona y J.~Seves, \textit{C\'{a}ndidos}, Inti-Illimani (1986).
\end{quote}

En este cap\'{\i}tulo presentamos un m\'{e}todo mediante el cual nuevas
\'{a}lgebras de Lie\footnote{El m\'{e}todo funciona tambi\'{e}n para
super\'{a}lgebras, como se muestra en los ejemplos de la
secci\'{o}n~\ref{sec:sexpex}. Para abreviar, seguiremos refiri\'{e}ndonos
al\ `\'{a}lgebra $\mathfrak{g}$', entendiendo que puede ser tambi\'{e}n una
super\'{a}lgebra.} pueden ser obtenidas a partir de un \'{a}lgebra dada
$\mathfrak{g}$. Este m\'{e}todo es una reinterpretaci\'{o}n y una
extensi\'{o}n del m\'{e}todo de \emph{expansi\'{o}n} de \'{a}lgebras de Lie
introducido por de~Azc\'{a}rraga, Izquierdo, Pic\'{o}n y Varela
en~\cite{deAz02} (ver tambi\'{e}n~\cite{deAz04,Hat01}).

La dimensi\'{o}n del \'{a}lgebra resultante es, en general,
mayor o igual que la del \'{a}lgebra original. Por ejemplo, el
\'{a}lgebra~M~\cite{Tow95}, con 583 generadores bos\'{o}nicos, puede ser
considerada como una expansi\'{o}n del \'{a}lgebra ortosimpl\'{e}ctica
$\mathfrak{osp}\left(  32|1\right)  $, la cual posee s\'{o}lo 528 (ambas
\'{a}lgebras tienen el mismo n\'{u}mero de generadores fermi\'{o}nicos). Este punto de vista puede ser ventajoso para ayudar a comprender los fundamentos geom\'{e}tricos de la teor\'{\i}a de Supergravedad en 11 dimensiones. Algunas aplicaciones f\'{\i}sicas del m\'{e}todo de expansi\'{o}n pueden ser encontradas en las Refs.~\cite{And05,Ban03,Ban04a,Ban04b,Dal05,DAu05,deAz05,Hat03,Hat04,Iza06c,Mee03,Sak06}.

El enfoque que presentamos aqu\'{\i} est\'{a} basado \'{\i}ntegramente en
operaciones llevadas a cabo directamente sobre los generadores del
\'{a}lgebra, y como tal difiere desde el comienzo con el presentado
en~\cite{deAz02}, donde se utiliza la formulaci\'{o}n dual de Maurer--Cartan
(MC). Una consecuencia de este cambio de punto de vista es que la
expansi\'{o}n de \'{a}lgebras diferenciales libres~\cite{DAu82,Fre84,Sul77}
queda fuera del alcance de nuestro m\'{e}todo.

Una caracter\'{\i}stica fundamental del presente m\'{e}todo es la
utilizaci\'{o}n de semigrupos abelianos\footnote{M\'{a}s espec\'{\i}ficamente,
utilizamos s\'{o}lo semigrupos abelianos \emph{finitos}. La
generalizaci\'{o}n al caso de semigrupos con un n\'{u}mero infinito contable de elementos parece factible.}. El \'{a}lgebra expandida que se obtiene a partir de $\mathfrak{g}$
queda determinada en gran medida por la elecci\'{o}n de semigrupo $S$. Todos
los casos de expansi\'{o}n\footnote{En este cap\'{\i}tulo hemos omitido el an\'{a}lisis de algunos casos de expansi\'{o}n no directamente relevantes para su aplicaci\'{o}n en Supergravedad. El lector interesado puede encontrarlos en las Refs.~\cite{deAz02,deAz04,Iza06b,Iza06t}.} en formas de MC presentados en~\cite{deAz02} pueden
ser recuperados en este contexto mediante una elecci\'{o}n particular de $S$.
Elecciones distintas conducen, en general, a \'{a}lgebras expandidas que no
pueden ser obtenidas por los m\'{e}todos de~\cite{deAz02}. A menudo nos
referiremos al proceso de expansi\'{o}n en semigrupos abelianos como `$S$-expansi\'{o}n'.

Una de las ventajas del m\'{e}todo de $S$-expansi\'{o}n radica en la posibilidad de extraer tensores
invariantes no triviales para las \'{a}lgebras expandidas, los cuales son
ingredientes esenciales, como se ha visto en los cap\'{\i}tulos~\ref{ch:CS}
y~\ref{ch:trans}, para la formulaci\'{o}n de teor\'{\i}as de CS y de
transgresi\'{o}n. Una aplicaci\'{o}n en este sentido es presentada en el
cap\'{\i}tulo~\ref{ch:TGFTMAlg}.

Este cap\'{\i}tulo est\'{a} basado en la Ref.~\cite{Iza06b}.

\section{\label{sec:pre}Preliminares}

Antes de analizar el procedimiento de $S$-expansi\'{o}n mismo, resulta
conveniente introducir algunas definiciones y notaci\'{o}n b\'{a}sica.

\subsection{\label{sec:semigru}Semigrupos}

Sea $S$ un semigrupo\footnote{No existe aparentemente un consenso acerca de la
definici\'{o}n de semigrupo. Nosotros adoptaremos aquella seg\'{u}n la cual un
semigrupo es un conjunto dotado de una multiplicaci\'{o}n cerrada y
asociativa. No requerimos que contenga inverso ni identidad.} abeliano
finito, de elementos $S=\left\{  \lambda_{\alpha}\right\}  $. El
producto de $n$ elementos de $S$ puede ser escrito como
\begin{equation}
\lambda_{\alpha_{1}}\cdots\lambda_{\alpha_{n}}=\lambda_{\gamma\left(
\alpha_{1},\dotsc,\alpha_{n}\right)  }, \label{tms}%
\end{equation}
donde $\gamma\left(  \alpha_{1},\dotsc,\alpha_{n}\right)  $ es una funci\'{o}n
que codifica la informaci\'{o}n de la tabla de multiplicaci\'{o}n de $S$. Una
manera alternativa, m\'{a}s pr\'{a}ctica para nuestros prop\'{o}sitos, es dada
en la siguiente definici\'{o}n.

\begin{definition}
El $n$-selector $K_{\alpha_{1} \cdots \alpha_{n}}^{\phantom{\alpha_{1} \cdots \alpha_{n}}\rho}$ es un s\'{\i}mbolo asociado al
semigrupo $S$, definido como
\begin{equation}
K_{\alpha_{1}\cdots\alpha_{n}}^{\phantom{\alpha_{1} \cdots \alpha_{n}}\rho
}=\left\{
\begin{array}
[c]{cl}%
1, & \text{cuando }\rho=\gamma\left(  \alpha_{1},\dotsc,\alpha_{n}\right) \\
0, & \text{en cualquier otro caso}%
\end{array}
\right.  . \label{n-sel}%
\end{equation}
Dado que $S$ es asociativo, el $n$-selector satisface la identidad
\begin{equation}
K_{\alpha_{1}\cdots\alpha_{n}}^{\phantom{\alpha_{1}\cdots\alpha_{n}}\rho
}=K_{\alpha_{1}\cdots\alpha_{n-1}}%
^{\phantom{\alpha_{1}\cdots\alpha_{n-1}}\sigma}K_{\sigma\alpha_{n}%
}^{\phantom{\sigma\alpha_{n}}\rho}=K_{\alpha_{1}\sigma}%
^{\phantom{\alpha_{1}\sigma}\rho}K_{\alpha_{2}\cdots\alpha_{n}}%
^{\phantom{\alpha_{2}\cdots\alpha_{n}}\sigma}. \label{idsel}%
\end{equation}
\end{definition}

Usando la identidad~(\ref{idsel}) es siempre posible escribir el $n$-selector
en t\'{e}rminos de $2$-selectores, los cuales codifican la informaci\'{o}n
b\'{a}sica de la tabla de multiplicaci\'{o}n de $S$.

Otra forma de decir lo mismo es que los $2$-selectores proporcionan una
representaci\'{o}n matricial para $S$; en efecto, escribiendo
\begin{equation}
\left(  \lambda_{\alpha}\right)  _{\mu}^{\phantom{\mu}\nu}=K_{\mu\alpha
}^{\phantom{\mu \alpha}\nu}, \label{lambda=K}%
\end{equation}
encontramos
\begin{equation}
\left(  \lambda_{\alpha}\right)  _{\mu}^{\phantom{\mu}\sigma}\left(
\lambda_{\beta}\right)  _{\sigma}^{\phantom{\sigma}\nu}=K_{\alpha\beta
}^{\phantom{\alpha \beta}\sigma}\left(  \lambda_{\sigma}\right)  _{\mu
}^{\phantom{\mu}\nu}=\left(  \lambda_{\gamma\left(  \alpha,\beta\right)
}\right)  _{\mu}^{\phantom{\mu}\nu}.
\end{equation}

En lo sucesivo restringiremos nuestra atenci\'{o}n a semigrupos
\emph{abelianos}, lo cual implica que los $n$-selectores ser\'{a}n
completamente sim\'{e}tricos en sus \'{\i}ndices inferiores.

La definici\'{o}n siguiente introduce un producto entre subconjuntos de un
semigrupo abeliano que ser\'{a} usado ampliamente m\'{a}s adelante.

\begin{definition}
\label{def:Sprod}Sean $S_{p}$ y $S_{q}$ dos subconjuntos de un semigrupo $S$.
El producto $S_{p}\times S_{q}$ es definido como
\begin{equation}
S_{p} \times S_{q} = \left\{ \lambda_{\gamma} \middle| \lambda_{\gamma} = \lambda_{\alpha_{p}} \lambda_{\alpha_{q}}, \text{ con } \lambda_{\alpha_{p}} \in S_{p}, \lambda_{\alpha_{q}} \in S_{q} \right\} .
\end{equation}
En palabras, $S_{p}\times S_{q}\subset S$ es el conjunto que resulta de tomar
todos los productos entre todos los elementos de $S_{p}$ con todos los
elementos de $S_{q}$. Como $S$ es abeliano, tenemos que $S_{q}\times
S_{p}=S_{p}\times S_{q}$.
\end{definition}

Vale la pena enfatizar que, en general, $S_{p}$, $S_{q}$ y $S_{p}\times S_{q}$
no tienen por qu\'{e} ser semigrupos ellos mismos.

Notemos tambi\'{e}n que un semigrupo $S$ puede estar dotado de un elemento
nulo, el cual denotamos por $0_{S}$. El elemento $0_{S}$ es definido como
aquel que satisface
\begin{equation}
0_{S}\lambda_{\alpha}=\lambda_{\alpha}0_{S}=0_{S},
\end{equation}
para cualquier $\lambda_{\alpha}\in S$.

\subsection{\label{sec:algfor}\'{A}lgebras Forzadas}

La siguiente definici\'{o}n introduce el concepto de \'{a}lgebras forzadas.

\begin{definition}
Consideremos un \'{a}lgebra de Lie $\mathfrak{g}$ de la forma $\mathfrak{g}%
=V_{0}\oplus V_{1}$, siendo $\left\{  \bm{T}_{a_{0}}\right\}  $ una
base para $V_{0}$ y $\left\{  \bm{T}_{a_{1}}\right\}  $ una base para
$V_{1}$. Cuando $\left[  V_{0},V_{1}\right]  \subset V_{1}$, i.e. cuando las
relaciones de conmutaci\'{o}n tienen la forma general
\begin{align}
\left[  \bm{T}_{a_{0}},\bm{T}_{b_{0}}\right]   &
=C_{a_{0}b_{0}}^{\phantom{a_{0} b_{0}}c_{0}}\bm{T}_{c_{0}}%
+C_{a_{0}b_{0}}^{\phantom{a_{0} b_{0}}c_{1}}\bm{T}_{c_{1}%
},\label{FA00}\\
\left[  \bm{T}_{a_{0}},\bm{T}_{b_{1}}\right]   &
=C_{a_{0}b_{1}}^{\phantom{a_{0} b_{1}}c_{1}}\bm{T}_{c_{1}%
},\label{FA01}\\
\left[  \bm{T}_{a_{1}},\bm{T}_{b_{1}}\right]   &
=C_{a_{1}b_{1}}^{\phantom{a_{1} b_{1}}c_{0}}\bm{T}_{c_{0}}%
+C_{a_{1}b_{1}}^{\phantom{a_{1} b_{1}}c_{1}}\bm{T}_{c_{1}},
\label{FA11}%
\end{align}
entonces es posible demostrar que las constantes de estructura $C_{a_{0}b_{0}}^{\phantom{a_{0} b_{0}}c_{0}}$ satisfacen la identidad de Jacobi por s\'{\i} mismas, de manera que $\left[  \bm{T}_{a_{0}},\bm{T}_{b_{0}%
}\right]  =C_{a_{0}b_{0}}^{\phantom{a_{0} b_{0}}c_{0}}\bm{T}_{c_{0}}$
son las relaciones de conmutaci\'{o}n de un \'{a}lgebra de Lie diferente a
$\mathfrak{g}$. Esta \'{a}lgebra, cuyas constantes de estructura son
$C_{a_{0}b_{0}}^{\phantom{a_{0} b_{0}}c_{0}}$, es llamada un \emph{\'{a}lgebra
forzada} de $\mathfrak{g}$ y simbolizada por $\left\vert V_{0}\right\vert $.
\end{definition}

Un \'{a}lgebra forzada puede ser considerada en cierto modo como la `inversa' de un \'{a}lgebra \emph{extendida}, si bien no es necesario que $V_{1}$ sea un ideal. Es importante notar que un \'{a}lgebra forzada \emph{no corresponde}, en general, a una sub\'{a}lgebra.

\section{\label{sec:sexppro}El Procedimiento de $S$-Expansi\'{o}n}

\subsection{\label{sec:sexparb}$S$-Expansi\'{o}n para un semigrupo arbitrario
$S$}

El siguiente teorema contiene el primer resultado central del m\'{e}todo de
$S$-expansiones.

\begin{theorem}
\label{th:Sxg}Sea $S=\left\{  \lambda_{\alpha}\right\}  $ un semigrupo
abeliano con $2$-selector $K_{\alpha\beta}^{\phantom{\alpha \beta}\gamma}$ y
sea $\mathfrak{g}$ un \'{a}lgebra de Lie con base $\left\{  \bm{T}%
_{A}\right\}  $ y constantes de estructura $C_{AB}^{\phantom{AB}C}$. Denotemos
un elemento del producto directo $S\otimes\mathfrak{g}$ por $\bm{T}%
_{\left(  A,\alpha\right)  }=\lambda_{\alpha}\bm{T}_{A}$ y
consideremos el conmutador inducido $\left[  \bm{T}_{\left(
A,\alpha\right)  },\bm{T}_{\left(  B,\beta\right)  }\right]
\equiv\lambda_{\alpha}\lambda_{\beta}\left[  \bm{T}_{A},\bm{T}%
_{B}\right]  $. Con estas definiciones, $S\otimes\mathfrak{g}$ es tambi\'{e}n
un \'{a}lgebra de Lie con constantes de estructura
\begin{equation}
C_{\left(  A,\alpha\right)  \left(  B,\beta\right)  }%
^{\phantom{\left( A, \alpha \right) \left( B, \beta \right)}\left(
C,\gamma\right)  }=K_{\alpha\beta}^{\phantom{\alpha \beta}\gamma}%
C_{AB}^{\phantom{AB}C}. \label{C=KC}%
\end{equation}

\end{theorem}

\begin{proof}
De la definici\'{o}n del conmutador inducido, y usando la regla de
multiplicaci\'{o}n~(\ref{tms}), tenemos
\begin{align*}
\left[  \bm{T}_{\left(  A,\alpha\right)  },\bm{T}_{\left(
B,\beta\right)  }\right]   &  \equiv\lambda_{\alpha}\lambda_{\beta}\left[
\bm{T}_{A},\bm{T}_{B}\right] \\
&  =\lambda_{\gamma\left(  \alpha,\beta\right)  }C_{AB}^{\phantom{AB}C}%
\bm{T}_{C}\\
&  =C_{AB}^{\phantom{AB}C}\bm{T}_{\left(  C,\gamma\left(  \alpha
,\beta\right)  \right)  }.
\end{align*}
La definici\'{o}n del $2$-selector $K_{\alpha\beta}%
^{\phantom{\alpha \beta}\rho}$ [cf.~ec.~(\ref{n-sel})],
\begin{equation}
K_{\alpha\beta}^{\phantom{\alpha \beta}\rho}=\left\{
\begin{array}
[c]{cl}%
1, & \text{cuando }\rho=\gamma\left(  \alpha,\beta\right) \\
0, & \text{en cualquier otro caso}%
\end{array}
\right.  ,
\end{equation}
nos permite escribir
\begin{equation}
\left[  \bm{T}_{\left(  A,\alpha\right)  },\bm{T}_{\left(
B,\beta\right)  }\right]  =K_{\alpha\beta}^{\phantom{\alpha \beta}\rho}%
C_{AB}^{\phantom{AB}C}\bm{T}_{\left(  C,\rho\right)  }.
\end{equation}
Por lo tanto, el \'{a}lgebra generada por $\left\{  \bm{T}_{\left(
A,\alpha\right)  }\right\}  $ se cierra, y las constantes de estructura son
\begin{equation}
C_{\left(  A,\alpha\right)  \left(  B,\beta\right)  }%
^{\phantom{\left( A, \alpha \right) \left( B, \beta \right)}\left(
C,\gamma\right)  }=K_{\alpha\beta}^{\phantom{\alpha \beta}\gamma}%
C_{AB}^{\phantom{AB}C}. \label{cesex}%
\end{equation}
Dado que $S$ es abeliano, las constantes de estructura $C_{\left(
A,\alpha\right)  \left(  B,\beta\right)  }%
^{\phantom{\left( A, \alpha \right) \left( B, \beta \right)}\left(
C,\gamma\right)  }$ tienen las mismas simetr\'{\i}as que $C_{AB}%
^{\phantom{AB}C}$, a saber,
\begin{equation}
C_{\left(  A,\alpha\right)  \left(  B,\beta\right)  }%
^{\phantom{\left( A, \alpha \right) \left( B, \beta \right)}\left(
C,\gamma\right)  }=-\left(  -1\right)  ^{\mathfrak{q}\left(  A\right)
\mathfrak{q}\left(  B\right)  }C_{\left(  B,\beta\right)  \left(
A,\alpha\right)  }%
^{\phantom{\left( A, \alpha \right) \left( B, \beta \right)}\left(
C,\gamma\right)  },
\end{equation}
donde $\mathfrak{q}\left(  A\right)  $ denota el \emph{grado} de
$\bm{T}_{A}$ (1 para Fermi y 0 para Bose). Notemos que el
procedimiento de $S$-expansi\'{o}n no cambia la naturaleza fermi\'{o}nica o
bos\'{o}nica de los generadores, i.e. $\mathfrak{q} \left( A, \alpha \right) = \mathfrak{q} \left( A \right)$.

Para demostrar que las constantes de estructura~(\ref{cesex}) satisfacen la
identidad de Jacobi basta con utilizar las propiedades de los selectores
[cf.~ec.~(\ref{idsel})] y el hecho que, por hip\'{o}tesis, las constantes de
estructura $C_{AB}^{\phantom{AB}C}$ satisfacen la identidad de Jacobi. Esto
concluye la demostraci\'{o}n.
\end{proof}

La siguiente definici\'{o}n es una consecuencia natural del
teorema~\ref{th:Sxg}.

\begin{definition}
Sea $S$ un semigrupo abeliano finito y sea $\mathfrak{g}$ un
\'{a}lgebra de Lie. El \'{a}lgebra de Lie $\mathfrak{G}$ definida por
$\mathfrak{G}=S\otimes\mathfrak{g}$ es llamada \emph{\'{A}lgebra }%
$S$\emph{-Expandida} de $\mathfrak{g}$.
\end{definition}

El \'{a}lgebra $S$-expandida $\mathfrak{G}=S\otimes\mathfrak{g}$, interesante
en s\'{\i} misma, constituye tambi\'{e}n el punto de partida para obtener
otras \'{a}lgebras, ya sean sub\'{a}lgebras o \'{a}lgebras forzadas de
$\mathfrak{G}$.

En la secci\'{o}n siguiente se analiza un caso importante de $S$-expansi\'{o}n que ocurre cada vez que el semigrupo $S$ contiene un elemento cero.

\subsection{\label{sec:0F}$0_{S}$-Forzamiento de un \'{A}lgebra $S$-Expandida}

Consideremos el caso en que el semigrupo $S$ est\'{a} provisto de un elemento $0_{S}\in S$ que satisface $0_{S} \lambda_{\alpha} = \lambda_{\alpha} 0_{S} = 0_{S}$ para todo $\lambda_{\alpha} \in S$. Separemos los elementos de $S$ en elementos no nulos $\lambda_{i}$, $i=0,\dotsc,N$ y $\lambda_{N+1}=0_{S}$. El $2$-selector de $S$ satisface
\begin{align}
K_{i,N+1}^{\phantom{i,N+1}j}  &  =K_{N+1,i}^{\phantom{N+1,i}j}=0, \label{KK1} \\
K_{i,N+1}^{\phantom{i,N+1}N+1}  &  =K_{N+1,i}^{\phantom{N+1,i}N+1}=1,\\
K_{N+1,N+1}^{\phantom{N+1,N+1}j}  &  =0,\\
K_{N+1,N+1}^{\phantom{N+1,N+1}N+1}  &  =1. \label{KK4}
\end{align}

Las ecs.~(\ref{KK1})--(\ref{KK4}) implican que el \'{a}lgebra $S$-expandida $\mathfrak{G} = S \otimes \mathfrak{g}$ puede ser separada en la forma
\begin{align}
\left[  \bm{T}_{\left(  A,i\right)  },\bm{T}_{\left(
B,j\right)  }\right]   &  =K_{ij}^{\phantom{ij}k}C_{AB}^{\phantom{AB}C}%
\bm{T}_{\left(  C,k\right)  }+K_{ij}^{\phantom{ij}N+1}C_{AB}%
^{\phantom{AB}C}\bm{T}_{\left(  C,N+1\right)  },\label{TTij}\\
\left[  \bm{T}_{\left(  A,N+1\right)  },\bm{T}_{\left(
B,j\right)  }\right]   &  =C_{AB}^{\phantom{AB}C}\bm{T}_{\left(
C,N+1\right)  },\label{TTNj}\\
\left[  \bm{T}_{\left(  A,N+1\right)  },\bm{T}_{\left(
B,N+1\right)  }\right]   &  =C_{AB}^{\phantom{AB}C}\bm{T}_{\left(
C,N+1\right)  }. \label{TTNN}%
\end{align}

Hay dos observaciones importantes con respecto a las ecs.~(\ref{TTij})--(\ref{TTNN}). En primer lugar, de~(\ref{TTNN}) notamos que los generadores
$\bm{T}_{\left(  A,N+1\right)  }$ forman una sub\'{a}lgebra del
\'{a}lgebra $S$-expandida $\mathfrak{G}=S\otimes\mathfrak{g}$, la cual es
isomorfa a $\mathfrak{g}$. En segundo lugar, comparando con~(\ref{FA00}%
)--(\ref{FA11}) vemos que $\mathfrak{G}$ tiene la forma $\mathfrak{G}%
=V_{0}\oplus V_{1}$, con $V_{0}=\left\{  \bm{T}_{\left(  A,i\right)
}\right\}  $, $V_{1}=\left\{  \bm{T}_{\left(  A,N+1\right)  }\right\}
$ y $\left[  V_{0},V_{1}\right]  \subset V_{1}$. Por lo tanto, hemos
encontrado que la presencia de $0_{S}\in S$ implica que es posible extraer un
\'{a}lgebra forzada de $\mathfrak{G}=S\otimes\mathfrak{g}$, la cual est\'{a}
dada por [ver secci\'{o}n~\ref{sec:algfor}]
\begin{equation}
\left[ \bm{T}_{\left( A,i \right)}, \bm{T}_{\left( B,j \right)} \right] = C_{\left( A,i \right) \left( B,j \right)}^{\phantom{\left( A,i \right) \left( B,j \right)} \left( C,k \right)} \bm{T}_{\left( C,k \right)},
\end{equation}
con las constantes de estructura
\begin{equation}
C_{\left( A,i \right) \left( B,j \right)}^{\phantom{\left( A,i \right) \left( B,j \right)} \left( C,k \right)} = K_{ij}^{\phantom{ij}k} C_{AB}^{\phantom{AB}C}. \label{C=KC0F}
\end{equation}

El forzamiento es en este caso equivalente a imponer la condici\'{o}n
\begin{equation}
\bm{T}_{\left(  A,N+1\right)  }=0_{S}\bm{T}_{A}=\bm{0}%
.
\end{equation}
Si bien esta condici\'{o}n puede parecer muy natural (sobre todo si uno
identifica el cero del semigrupo con el cero del campo sobre el cual est\'{a}
definida el \'{a}lgebra), su justificaci\'{o}n est\'{a} basada en la
interpretaci\'{o}n como forzamiento, dado que es s\'{o}lo aquella la que
garantiza que los generadores restantes sigan formando un \'{a}lgebra de Lie.

Notemos que este forzamiento abelianiza algunos sectores del \'{a}lgebra; en efecto, cuando $\lambda_{i} \lambda_{j} = 0_{S}$, entonces tenemos $\left[ \bm{T}_{\left( A,i \right)}, \bm{T}_{\left( B,j \right)} \right] = \bm{0}$.

La siguiente definici\'{o}n resume la discusi\'{o}n anterior.

\begin{definition}
Sea $S$ un semigrupo abeliano con un elemento cero $0_{S}\in S$ y sea
$\mathfrak{G}=S\otimes\mathfrak{g}$ un \'{a}lgebra $S$-expandida. El
\'{a}lgebra de Lie obtenida imponiendo la condici\'{o}n $0_{S}\bm{T}%
_{A}=\bm{0}$ sobre $\mathfrak{G}$ (o sobre una sub\'{a}lgebra de
$\mathfrak{G}$) es llamada \emph{\'{a}lgebra }$0_{S}$\emph{-forzada} de
$\mathfrak{G}$ (o de la sub\'{a}lgebra).
\end{definition}

El $0_{S}$-forzamiento juega un rol esencial a la hora de reproducir los
resultados de la expansi\'{o}n de MC en el contexto de la $S$-expansi\'{o}n,
como ser\'{a} visto en la secci\'{o}n siguiente.

\subsection{\label{sec:MCexp}Expansi\'{o}n de MC como una $S$-Expansi\'{o}n}

En esta secci\'{o}n explicamos c\'{o}mo reproducir la expansi\'{o}n de MC~\cite{deAz02} en el contexto de la $S$-expansi\'{o}n.

En pocas palabras, la idea de la expansi\'{o}n de MC es considerar el \'{a}lgebra $\mathfrak{g}$ seg\'{u}n la descripci\'{o}n dual dada por las formas de MC en la variedad del grupo y, luego de reescalar algunos de los par\'{a}metros del grupo en un factor $\lambda$, expandir las formas de MC como una serie de potencias en $\lambda$. Esta serie es finalmente truncada de modo tal de asegurar la clausura del \'{a}lgebra.

El Teorema~1 de la Ref.~\cite{deAz02} muestra que, en el caso m\'{a}s general, el \'{a}lgebra expandida tiene las constantes de estructura
\begin{equation}
C_{\left(  A,i\right)  \left(  B,j\right)  }%
^{\phantom{\left( A, i \right) \left( B, j \right)}\left(  C,k\right)
}=\left\{
\begin{array}
[c]{cl}%
0, & \text{cuando }i+j\neq k\\
C_{AB}^{\phantom{AB}C}, & \text{cuando }i+j=k
\end{array}
\right.  , \label{scazc}%
\end{equation}
donde los par\'{a}metros $i,j,k=0,\dotsc,N$ corresponden al orden de la
expansi\'{o}n, siendo $N$ el orden de truncamiento.

Estas constantes de estructura pueden ser obtenidas en el contexto de la
$S$-expansi\'{o}n mediante el $0_{S}$-forzamiento de un \'{a}lgebra
$S_{\mathrm{E}}^{\left(  N\right)  }$-expandida, donde $S_{\mathrm{E}%
}^{\left(  N\right)  }$ es el semigrupo introducido en la siguiente definici\'{o}n.

\begin{definition}
\label{def:SEN}Sea $S_{\mathrm{E}}^{\left(  N\right)  }$ el semigrupo de
elementos
\begin{equation}
S_{\mathrm{E}}^{\left(  N\right)  }=\left\{  \lambda_{\alpha},\alpha
=0,\dotsc,N,N+1\right\}  ,
\end{equation}
provisto de la regla de multiplicaci\'{o}n
\begin{equation}
\lambda_{\alpha}\lambda_{\beta}=\left\{
\begin{array}
[c]{cl}%
\lambda_{\alpha+\beta}, & \text{cuando }\alpha+\beta\leq N\\
\lambda_{N+1}, & \text{cuando }\alpha+\beta\geq N+1
\end{array}
\right.  . \label{SENdef}%
\end{equation}
Los $2$-selectores para $S_{\mathrm{E}}^{\left(  N\right)  }$ tienen la forma
\begin{equation}
K_{\alpha\beta}^{\phantom{\alpha \beta}\gamma}=\left\{
\begin{array}
[c]{cl}%
\delta_{\alpha+\beta}^{\gamma}, & \text{cuando }\alpha+\beta\leq N\\
\delta_{N+1}^{\gamma}, & \text{cuando }\alpha+\beta\geq N+1
\end{array}
\right.  , \label{2-selSEN}%
\end{equation}
donde $\delta_{\sigma}^{\rho}$ es la delta de Kronecker. De~(\ref{SENdef})
vemos que $\lambda_{N+1}$ puede ser identificado con el elemento cero de
$S_{\mathrm{E}}^{\left(  N\right)  }$, i.e. $\lambda_{N+1}=0_{S}$.
\end{definition}

Para obtener las constantes de estructura correspondientes a la $S_{\mathrm{E}%
}^{\left(  N\right)  }$-expansi\'{o}n de un \'{a}lgebra de Lie $\mathfrak{g}$,
basta con particularizar la f\'{o}rmula general~(\ref{C=KC}) al caso
$S=S_{\mathrm{E}}^{\left(  N\right)  }$. Esto corresponde a reemplazar la
forma del $2$-selector~(\ref{2-selSEN}) en~(\ref{C=KC}). Llevando a cabo este
reemplazo obtenemos
\begin{equation}
C_{\left(  A,\alpha\right)  \left(  B,\beta\right)  }%
^{\phantom{\left( A, \alpha \right) \left( B, \beta \right)}\left(
C,\gamma\right)  }=\left\{
\begin{array}
[c]{cl}%
\delta_{\alpha+\beta}^{\gamma}C_{AB}^{\phantom{AB}C}, & \text{cuando }\alpha
+\beta\leq N\\
\delta_{N+1}^{\gamma}C_{AB}^{\phantom{AB}C}, & \text{cuando }\alpha+\beta\geq
N+1
\end{array}
\right.  , \label{C=dC1}%
\end{equation}
con $\alpha,\beta,\gamma=0,\dotsc,N,N+1$.

El paso siguiente consiste en realizar el $0_{S}$-forzamiento del \'{a}lgebra
$S_{\mathrm{E}}^{\left(  N\right)  }$-expandida. Imponiendo la condici\'{o}n
$\lambda_{N+1}\bm{T}_{A}=\bm{0}$ (ver secci\'{o}n~\ref{sec:0F}%
, p\'{a}g.~\pageref{sec:0F}), la ec.~(\ref{C=dC1}) se reduce a
\begin{equation}
C_{\left(  A,i\right)  \left(  B,j\right)  }%
^{\phantom{\left( A, i \right) \left( B, j \right)}\left(  C,k\right)
}=\delta_{i+j}^{k}C_{AB}^{\phantom{AB}C}, \label{C=dC2}%
\end{equation}
con $i,j,k=0,\dotsc,N$. Estas constantes de estructura se corresponden
exactamente con las de la expansi\'{o}n de MC, ec.~(\ref{scazc}).

Esta coincidencia no es tal, por supuesto. Volvamos por un momento al esquema
propuesto en~\cite{deAz02} y consideremos las potencias del par\'{a}metro de
reescalamiento $\lambda$. El producto de dos de estas potencias satisface
\begin{equation}
\lambda^{\alpha}\lambda^{\beta}=\lambda^{\alpha+\beta}, \label{lambdapro}%
\end{equation}
en tanto que el truncamiento a orden $N$ puede ser impuesto simb\'{o}licamente
como
\begin{equation}
\lambda^{\alpha}=0\qquad\text{cuando }\alpha>N. \label{lambdanull}%
\end{equation}
Es claro que las ecs.~(\ref{lambdapro}) y~(\ref{lambdanull}) reproducen
exactamente la ley de multiplicaci\'{o}n del semigrupo $S_{\mathrm{E}%
}^{\left(  N\right)  }$. Esta \emph{estructura algebraica} es todo cuanto se requiere
para llevar a cabo la expansi\'{o}n. Tambi\'{e}n resulta claro al hacer esta
comparaci\'{o}n que es necesario imponer la condici\'{o}n de $0_{S}%
$-forzamiento $\lambda_{N+1}\bm{T}_{A}=\bm{0}$ para obtener
las constantes de estructura~(\ref{C=dC2}).

El procedimiento de $S$-expansi\'{o}n es v\'{a}lido para cualquier \'{a}lgebra
de Lie $\mathfrak{g}$, y es en este sentido muy general. Sin embargo, cuando
se conoce algo de la estructura algebraica de $\mathfrak{g}$, es posible ir
m\'{a}s all\'{a}. Por ejemplo, en el contexto de la expansi\'{o}n de MC, el
reescalamiento y el truncamiento pueden ser ejecutados de varias maneras
dependiendo de la estructura de $\mathfrak{g}$, conduciendo en cada caso a
diferentes tipos de \'{a}lgebras expandidas. La contracci\'{o}n de
\.{I}n\"{o}n\"{u}--Wigner generalizada, o la obtenci\'{o}n del \'{a}lgebra~M a
partir de $\mathfrak{osp} \left( 32|1 \right)$ son ejemplos importantes de estos casos.
En el contexto de la $S$-expansi\'{o}n, la informaci\'{o}n sobre la estructura
algebraica de $\mathfrak{g}$ puede ser usada para extraer sub\'{a}lgebras y
\'{a}lgebras forzadas del \'{a}lgebra $S$-expandida. Utilizando esta
informaci\'{o}n es posible reproducir todos los casos de expansi\'{o}n
encontrados en~\cite{deAz02}. Nuevos tipos de \'{a}lgebras expandidas pueden
tambi\'{e}n ser obtenidos ocupando semigrupos distintos a $S_{\mathrm{E}}$.

\section{\label{sec:SubAlgRes}Sub\'{a}lgebras Resonantes}

Un \'{a}lgebra $S$-expandida tiene una estructura bastante simple,
reproduciendo el \'{a}lgebra original $\mathfrak{g}$ en una serie de `niveles'
que corresponden a los elementos del semigrupo. En esta secci\'{o}n estudiamos
un modo de extraer un \'{a}lgebra m\'{a}s peque\~{n}a a partir de
$S\otimes\mathfrak{g}$; el m\'{e}todo de las sub\'{a}lgebras resonantes.

\subsection{\label{sec:SubAlgResArb}Sub\'{a}lgebras Resonantes para un
semigrupo arbitrario $S$}

Para poder extraer sub\'{a}lgebras resonantes a partir de un \'{a}lgebra
$S$-expandida $S\otimes\mathfrak{g}$, es indispensable contar con alguna
informaci\'{o}n acerca de la estructura algebraica de $\mathfrak{g}$. Esta
informaci\'{o}n es codificada de la siguiente manera.

Sea $\mathfrak{g}=\bigoplus_{p\in I}V_{p}$ una descomposici\'{o}n de
$\mathfrak{g}$ en un cierto n\'{u}mero de \emph{subespacios} $V_{p}$, con $I$
un conjunto de \'{\i}ndices. Para cada $p,q\in I$ es siempre posible definir
un subconjunto de $I$, $i_{\left(  p,q\right)  }\subset I$, tal que
\begin{equation}
\left[  V_{p},V_{q}\right]  \subset\bigoplus_{r\in i_{\left(  p,q\right)  }%
}V_{r}. \label{VpVqSVr}%
\end{equation}
De este modo, los subconjuntos $i_{\left(  p,q\right)  }$ almacenan la
informaci\'{o}n sobre la estructura de subespacios del \'{a}lgebra. Notemos
que los subespacios $V_{p}$ no corresponden, en general, a sub\'{a}lgebras\footnote{La condici\'{o}n para que $V_{p}$ corresponda a una sub\'{a}lgebra de $\mathfrak{g}$\ es $i_{\left( p,p \right)} = \left\{ p \right\}$.} de $\mathfrak{g}$.

De forma an\'{a}loga a la partici\'{o}n del \'{a}lgebra en subespacios, es
siempre posible descomponer un semigrupo arbitrario $S$ en \emph{subconjuntos}
$S_{p}\subset S$ en la forma $S=\bigcup_{p\in I}S_{p}$, donde $I$ es el mismo
conjunto de \'{\i}ndices del p\'{a}rrafo anterior. Notemos que los
subconjuntos $S_{p}$ no tiene por qu\'{e} ser semigrupos ellos mismos.

En principio, la descomposici\'{o}n de $S$ en subconjuntos es completamente
arbitraria; sin embargo, recurriendo al producto entre subconjuntos dado en la
def.~\ref{def:Sprod}, es a veces posible escoger una descomposici\'{o}n muy
particular, como se detalla en la siguiente definici\'{o}n.

\begin{definition}
\label{def:ParRes}Sea $\mathfrak{g}=\bigoplus_{p\in I}V_{p}$ una
descomposici\'{o}n de $\mathfrak{g}$ en subespacios $V_{p}$, con una
estructura descrita por los subconjuntos $i_{\left(  p,q\right)  }\subset I$,
como se muestra en la ec.~(\ref{VpVqSVr}). Sea $S=\bigcup_{p\in I}S_{p}$ una
descomposici\'{o}n en subconjuntos del semigrupo abeliano $S$ tal que
\begin{equation}
S_{p}\times S_{q}\subset\bigcap_{r\in i_{\left(  p,q\right)  }}S_{r},
\label{SpSqISr}%
\end{equation}
donde el producto $\times$ entre subconjuntos es el de la def.~\ref{def:Sprod}%
. Cuando una descomposici\'{o}n $S=\bigcup_{p\in I}S_{p}$ de $S$ que cumple
con~(\ref{SpSqISr}) existe, decimos que esta descomposici\'{o}n est\'{a} en
\emph{resonancia} con la descomposici\'{o}n de $\mathfrak{g}$ en subespacios,
$\mathfrak{g}=\bigoplus_{p\in I}V_{p}$.
\end{definition}

La existencia de una `partici\'{o}n resonante' de $S$ es una
hip\'{o}tesis esencial en el teorema siguiente, el cual permite extraer
sistem\'{a}ticamente sub\'{a}lgebras a partir del \'{a}lgebra $S$-expandida
$S\otimes\mathfrak{g}$.

\begin{theorem}
\label{th:SubAlgRes}Sea $\mathfrak{g}=\bigoplus_{p\in I}V_{p}$ una
descomposici\'{o}n de $\mathfrak{g}$ en subespacios, con una estructura dada
por la ec.~(\ref{VpVqSVr}), y sea $S=\bigcup_{p\in I}S_{p}$ una partici\'{o}n
resonante del semigrupo abeliano $S$, con la estructura dada en la
ec.~(\ref{SpSqISr}). Para cada $p\in I$, se definen los subespacios $W_{p}$
del \'{a}lgebra $S$-expandida $\mathfrak{G}=S\otimes\mathfrak{g}$ como
\begin{equation}
W_{p}\equiv S_{p}\otimes V_{p}.
\end{equation}
Entonces,
\begin{equation}
\mathfrak{G}_{\mathrm{R}}\equiv\bigoplus_{p\in I}W_{p}%
\end{equation}
es una sub\'{a}lgebra de $\mathfrak{G}$.
\end{theorem}

\begin{proof}
Usando las ecs.~(\ref{VpVqSVr})--(\ref{SpSqISr}), tenemos
\begin{align}
\left[  W_{p},W_{q}\right]   &  \subset\left(  S_{p}\times S_{q}\right)
\otimes\left[  V_{p},V_{q}\right] \nonumber\\
&  \subset\bigcap_{s\in i_{\left(  p,q\right)  }}S_{s}\otimes\bigoplus_{r\in
i_{\left(  p,q\right)  }}V_{r}\nonumber\\
&  \subset\bigoplus_{r\in i_{\left(  p,q\right)  }}\left(  \bigcap_{s\in
i_{\left(  p,q\right)  }}S_{s}\right)  \otimes V_{r}.
\end{align}
Es claro que para cada $r\in i_{\left(  p,q\right)  }$ uno puede escribir
\begin{equation}
\bigcap_{s\in i_{\left(  p,q\right)  }}S_{s}\subset S_{r}.
\end{equation}
Luego,
\begin{equation}
\left[  W_{p},W_{q}\right]  \subset\bigoplus_{r\in i_{\left(  p,q\right)  }%
}S_{r}\otimes V_{r}%
\end{equation}
y encontramos
\begin{equation}
\left[  W_{p},W_{q}\right]  \subset\bigoplus_{r\in i_{\left(  p,q\right)  }%
}W_{r}. \label{WpWqSWr}%
\end{equation}
Por lo tanto, el \'{a}lgebra se cierra y $\mathfrak{G}_{\mathrm{R}}%
=\bigoplus_{p\in I}W_{p}$ corresponde a una sub\'{a}lgebra de $\mathfrak{G}$.
\end{proof}

\begin{definition}
El \'{a}lgebra $\mathfrak{G}_{\mathrm{R}}=\bigoplus_{p\in I}W_{p}$ hallada en
el teorema~\ref{th:SubAlgRes} es llamada una \emph{Sub\'{a}lgebra Resonante}
del \'{a}lgebra $S$-expandida $\mathfrak{G}=S\otimes\mathfrak{g}$.
\end{definition}

La adopci\'{o}n del nombre `resonancia' se debe a la similitud formal entre
las ecs.~(\ref{VpVqSVr}) y~(\ref{SpSqISr}); la ec.~(\ref{SpSqISr}) ser\'{a} denominada tambi\'{e}n `\emph{condici\'{o}n de resonancia}'.

El teorema~\ref{th:SubAlgRes} convierte el problema de encontrar
sub\'{a}lgebras de un \'{a}lgebra $S$-expandida $\mathfrak{G}=S\otimes
\mathfrak{g}$ en el problema de encontrar una partici\'{o}n resonante para el
semigrupo $S$. Como se muestra m\'{a}s adelante por medio de ejemplos, resolver la condici\'{o}n de resonancia~(\ref{SpSqISr}) resulta ser un problema f\'{a}cilmente abordable. Por lo tanto, el teorema~\ref{th:SubAlgRes} puede ser considerado como una \'{u}til herramienta para extraer sub\'{a}lgebras de un \'{a}lgebra $S$-expandida.

Usando la ec.~(\ref{C=KC}) y la partici\'{o}n resonante de $S$ es posible
encontrar una expresi\'{o}n expl\'{\i}cita para las constantes de estructura
de la sub\'{a}lgebra resonante $\mathfrak{G}_{\mathrm{R}}$. Denotando por
$\left\{  \bm{T}_{a_{p}}\right\}  $ la base para el subespacio
$V_{p}$, podemos escribir
\begin{equation}
C_{\left(  a_{p},\alpha_{p}\right)  \left(  b_{q},\beta_{q}\right)
}%
^{\phantom{\left( a_{p}, \alpha_{p} \right) \left( b_{q}, \beta_{q} \right)}\left(
c_{r},\gamma_{r}\right)  }=K_{\alpha_{p}\beta_{q}}%
^{\phantom{\alpha_{p} \beta_{q}}\gamma_{r}}C_{a_{p}b_{q}}%
^{\phantom{a_{p} b_{q}}c_{r}}, \label{C=KCRes}%
\end{equation}
con $\alpha_{p},\beta_{q},\gamma_{r}$ tales que $\lambda_{\alpha_{p}}\in
S_{p}$, $\lambda_{\beta_{q}}\in S_{q}$ y $\lambda_{\gamma_{r}}\in S_{r}$.

Una observaci\'{o}n interesante es que la estructura de subespacios de la
sub\'{a}lgebra resonante $\mathfrak{G}_{\mathrm{R}}$ es la misma que la del
\'{a}lgebra original $\mathfrak{g}$, como puede observarse de la
ec.~(\ref{WpWqSWr}).

Las sub\'{a}lgebras resonantes juegan un rol central en este cap\'{\i}tulo y a
lo largo de la Tesis. Vale la pena notar que la mayor parte de los casos
particulares considerados en~\cite{deAz02} pueden ser recuperados utilizando
el teorema~\ref{th:SubAlgRes} para $S=S_{\mathrm{E}}^{\left(  N\right)  }$,
como veremos en la secci\'{o}n siguiente. Todos los casos restantes pueden ser
obtenidos como un forzamiento de una sub\'{a}lgebra resonante.

\subsection{\label{sec:SubAlgResSEN}Sub\'{a}lgebras Resonantes con
$S=S_{\mathrm{E}}^{\left(  N\right)  }$}

En esta secci\'{o}n rederivamos algunos resultados de expansiones de
\'{a}lgebras presentados en~\cite{deAz02} en el marco de las $S$-expansiones.
Para obtener estas \'{a}lgebras es necesario proceder de acuerdo a los tres
pasos siguientes:

\begin{enumerate}
\item Realizar una $S$-expansi\'{o}n usando el semigrupo $S=S_{\mathrm{E}%
}^{\left(  N\right)  }$,

\item Encontrar una partici\'{o}n resonante para $S_{\mathrm{E}}^{\left(
N\right)  }$ y construir la sub\'{a}lgebra resonante $\mathfrak{G}%
_{\mathrm{R}}$,

\item Aplicar un $0_{S}$-forzamiento a la sub\'{a}lgebra resonante.
\end{enumerate}

Escogiendo un semigrupo distinto u omitiendo el forzamiento uno encuentra
\'{a}lgebras que no est\'{a}n contenidas dentro de la expansi\'{o}n de MC. Ejemplos de esta clase son dados en la secciones~\ref{sec:so43} y~\ref{sec:newsuper}.

\subsubsection{Caso cuando $\mathfrak{g}=V_{0}\oplus V_{1}$, siendo $V_{0}$
una sub\'{a}lgebra y $V_{1}$ un coseto sim\'{e}trico}

Sea $\mathfrak{g}=V_{0}\oplus V_{1}$ una descomposici\'{o}n de $\mathfrak{g}$
en subespacios tal que
\begin{align}
\left[  V_{0},V_{0}\right]   &  \subset V_{0},\label{V0V0SV0}\\
\left[  V_{0},V_{1}\right]   &  \subset V_{1},\label{V0V1SV1}\\
\left[  V_{1},V_{1}\right]   &  \subset V_{0}. \label{V1V1SV0}%
\end{align}
Estas relaciones de conmutaci\'{o}n corresponden al caso en que $V_{0}$ es una
sub\'{a}lgebra y $V_{1}$ un coseto sim\'{e}trico.

Sea $S_{\mathrm{E}}^{\left(  N\right)  }=S_{0}\cup S_{1}$, con $N$ arbitrario,
una partici\'{o}n de $S_{\mathrm{E}}^{\left(  N\right)  }$ en subconjuntos
dada por\footnote{Aqu\'{\i} $\left[  x\right]  $ denota la parte entera de
$x$.}
\begin{align}
S_{0}  &  =\left\{  \lambda_{2m},\text{ con }m=0,\dotsc,\left[  \frac{N}%
{2}\right]  \right\}  \cup\left\{  \lambda_{N+1}\right\}  ,\label{S0}\\
S_{1}  &  =\left\{  \lambda_{2m+1},\text{ con }m=0,\dotsc,\left[  \frac
{N-1}{2}\right]  \right\}  \cup\left\{  \lambda_{N+1}\right\}  . \label{S1}%
\end{align}

Esta partici\'{o}n de $S_{\mathrm{E}}^{\left(  N\right)  }$ es resonante con
respecto a la estructura del \'{a}lgebra [ecs.~(\ref{V0V0SV0})--(\ref{V1V1SV0}%
)], ya que satisface [cf.~ec.~(\ref{SpSqISr})]%
\begin{align}
S_{0}\times S_{0}  &  \subset S_{0},\\
S_{0}\times S_{1}  &  \subset S_{1},\\
S_{1}\times S_{1}  &  \subset S_{0}.
\end{align}
Como se mostr\'{o} en la secci\'{o}n~\ref{sec:SubAlgResArb}, la resonancia de
la partici\'{o}n~(\ref{S0})--(\ref{S1}) permite extraer una sub\'{a}lgebra
resonante a partir del \'{a}lgebra $S_{\mathrm{E}}^{\left(  N\right)  }%
$-expandida $\mathfrak{G}=S_{\mathrm{E}}^{\left(  N\right)  }\otimes
\mathfrak{g}$. En efecto, de acuerdo al teorema~\ref{th:SubAlgRes}, tenemos
que
\begin{equation}
\mathfrak{G}_{\mathrm{R}}=W_{0}\oplus W_{1},
\end{equation}
con
\begin{align}
W_{0}  &  =S_{0}\otimes V_{0},\\
W_{1}  &  =S_{1}\otimes V_{1},
\end{align}
es una sub\'{a}lgebra resonante de $\mathfrak{G}$.

Notemos que es la estructura de los subconjuntos $S_{0}$ y $S_{1}$ la que
determina la forma de la sub\'{a}lgebra resonante $\mathfrak{G}_{\mathrm{R}}$.

Las constantes de estructura para $\mathfrak{G}_{\mathrm{R}}$ son obtenidas
directamente de la f\'{o}rmula general para sub\'{a}lgebras resonantes,
ec.~(\ref{C=KCRes}). Ellas pueden ser escritas en la forma
\begin{equation}
C_{\left(  a_{p},\alpha_{p}\right)  \left(  b_{q},\beta_{q}\right)
}%
^{\phantom{\left( a_{p}, \alpha_{p} \right) \left( b_{q}, \beta_{q} \right)}\left(
c_{r},\gamma_{r}\right)  }=\left\{
\begin{array}
[c]{cl}%
\delta_{\alpha_{p}+\beta_{q}}^{\gamma_{r}}C_{a_{p}b_{q}}%
^{\phantom{a_{p} b_{q}}c_{r}}, & \text{cuando }\alpha_{p}+\beta_{q}\leq N\\
\delta_{N+1}^{\gamma_{r}}C_{a_{p}b_{q}}^{\phantom{a_{p} b_{q}}c_{r}}, &
\text{cuando }\alpha_{p}+\beta_{q}\geq N+1
\end{array}
\right.  ,
\label{CE-cossym0}
\end{equation}
con $p,q=0,1$, $\alpha_{p},\beta_{p},\gamma_{p}=2m+p$, $m=0,\dotsc,\left[
\frac{N-p}{2}\right]  ,\frac{N+1-p}{2}$.

Una vez obtenida la sub\'{a}lgebra resonante, el \'{u}ltimo paso para
reproducir el resultado correspondiente de~\cite{deAz02} en este contexto es
realizar un $0_{S}$-forzamiento de $\mathfrak{G}_{\mathrm{R}}$, con
$\lambda_{N+1}=0_{S}$. Dado que este forzamiento es equivalente a imponer la
condici\'{o}n $\lambda_{N+1}\bm{T}_{A}=\bm{0}$, no es
dif\'{\i}cil ver que el \'{u}nico efecto sobre las constantes de estructura~(\ref{CE-cossym0}) concierne el rango de los \'{\i}ndices $\alpha_{p},\beta_{p},\gamma_{p}$; en
efecto, las nuevas constantes de estructura son
\begin{equation}
C_{\left(  a_{p},\alpha_{p}\right)  \left(  b_{q},\beta_{q}\right)
}%
^{\phantom{\left( a_{p}, \alpha_{p} \right) \left( b_{q}, \beta_{q} \right)}\left(
c_{r},\gamma_{r}\right)  }=\left\{
\begin{array}
[c]{cl}%
\delta_{\alpha_{p}+\beta_{q}}^{\gamma_{r}}C_{a_{p}b_{q}}%
^{\phantom{a_{p} b_{q}}c_{r}}, & \text{cuando }\alpha_{p}+\beta_{q}\leq N\\
\delta_{N+1}^{\gamma_{r}}C_{a_{p}b_{q}}^{\phantom{a_{p} b_{q}}c_{r}}, &
\text{cuando }\alpha_{p}+\beta_{q}\geq N+1
\end{array}
\right.  , \label{CE-cossym}%
\end{equation}
con $p,q=0,1$, $\alpha_{p},\beta_{p},\gamma_{p}=2m+p$, $m=0,\dotsc,\left[
\frac{N-p}{2}\right]  $. La restricci\'{o}n en el rango de $\alpha_{p}%
,\beta_{p},\gamma_{p}$ impide que tomen el valor $N+1$, el
cual est\'{a} ahora excluido del \'{a}lgebra.

Con la notaci\'{o}n de~\cite{deAz02}, el $0_{S}$-forzamiento del \'{a}lgebra
$S_{\mathrm{E}}^{\left(  N\right)  }$-expandida corresponde a $\mathcal{G}%
\left(  N_{0},N_{1}\right)  $ para el caso del coseto sim\'{e}trico, con
\begin{align}
N_{0}  &  =2\left[  \frac{N}{2}\right]  ,\\
N_{1}  &  =2\left[  \frac{N-1}{2}\right]  +1.
\end{align}
Las constantes de estructura~(\ref{CE-cossym}) corresponden a las de la
ec.~(3.31) de la Ref.~\cite{deAz02}.

\begin{figure}
\begin{center}
\includegraphics[width=\textwidth]{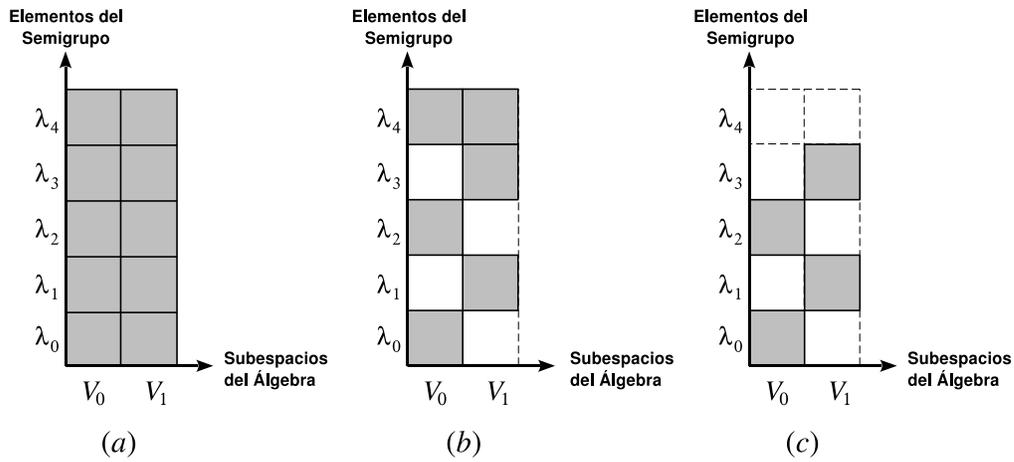}
\end{center}
\caption{\label{fig:ResonantSubAlg}Procedimiento de $S$-expansi\'{o}n de un \'{a}lgebra de Lie $\mathfrak{g}=V_{0} \oplus V_{1}$, con $S=S_{\mathrm{E}}^{\left( 3 \right)}$. (\textit{a}) El \'{a}lgebra $S_{\mathrm{E}}^{\left( 3 \right)}$-expandida $\mathfrak{G}=S_{\mathrm{E}}^{\left( 3 \right)} \otimes \mathfrak{g}$ contiene todos los subespacios de la forma $\lambda_{\alpha} \otimes V_{p}$. (\textit{b}) La sub\'{a}lgebra resonante $\mathfrak{G}_{\mathrm{R}}=W_{0} \oplus W_{1}$ contiene s\'{o}lo aquellos subespacios de la forma $\lambda_{\alpha_{p}} \otimes V_{p}$, con $\lambda_{\alpha_{p}} \in S_{p}$ [ver ecs.~(\ref{S0.01})--(\ref{S1.01})]. (\textit{c}) El $\lambda_{4}$-forzamiento (con $\lambda_{4}=0_{S}$) de $\mathfrak{G}_{\mathrm{R}}$ elimina todos los subespacios de la forma $\lambda_{4} \otimes V_{p}$.}
\end{figure}

Una idea m\'{a}s intuitiva del proceso de $S$-expansi\'{o}n, extracci\'{o}n de sub\'{a}lgebra resonante y $0_{S}$-forzamiento puede ser obtenida a partir de un diagrama. En la fig.~\ref{fig:ResonantSubAlg} se muestra el caso cuando $\mathfrak{g}=V_{0} \oplus V_{1}$, con $V_{0}$ una sub\'{a}lgebra y $V_{1}$ un coseto sim\'{e}trico, y $S=S_{\mathrm{E}}^{\left(  3\right)  }$.

Los subespacios ($V_{p}$) de $\mathfrak{g}$ son dispuestos en el eje horizontal, en tanto que los elementos del semigrupo ($\lambda_{\alpha}$) ocupan el eje vertical. El \'{a}lgebra $S_{\mathrm{E}}^{\left(  3\right)}$-expandida $\mathfrak{G}=S_{\mathrm{E}}^{\left(  3\right)  }\otimes \mathfrak{g}$ corresponde a la regi\'{o}n sombreada en la parte~(\textit{a}) de la fig.~\ref{fig:ResonantSubAlg}. Esta \'{a}lgebra incluye todos los subespacios de la forma $\lambda_{\alpha} \otimes V_{p}$. En la parte~(\textit{b}) de la fig.~\ref{fig:ResonantSubAlg}, el \'{a}rea sombreada corresponde a la sub\'{a}lgebra resonante $\mathfrak{G}_{\mathrm{R}}=W_{0}\oplus W_{1}$, con [cf.~ecs.~(\ref{S0})--(\ref{S1})]
\begin{align}
S_{0}  &  =\left\{  \lambda_{0},\lambda_{2},\lambda_{4}\right\} \label{S0.01} ,\\
S_{1}  &  =\left\{  \lambda_{1},\lambda_{3},\lambda_{4}\right\} \label{S1.01} .
\end{align}
Las columnas en esta parte de la figura corresponden a los subespacios $W_{0}$ y $W_{1}$ de la sub\'{a}lgebra resonante $\mathfrak{G}_{\mathrm{R}}$; notemos que la elecci\'{o}n de la partici\'{o}n resonante determina cu\'{a}les subespacios de $\mathfrak{G}$ estar\'{a}n presentes en $\mathfrak{G}_{\mathrm{R}}$. Finalmente, la parte~(\textit{c}) de la fig.~\ref{fig:ResonantSubAlg} indica el $0_{S}$-forzamiento de la sub\'{a}lgebra resonante $\mathfrak{G}_{\mathrm{R}}$, con $\lambda_{4}=0_{S}$, exluyendo en efecto todos aquellos subespacios de la forma $\lambda_{4} \otimes \mathfrak{g}$. El \'{a}lgebra $\mathcal{G}\left( N_{0}, N_{1} \right)$ de la Ref.~\cite{deAz02} corresponde a esta parte de la figura.

El $0_{S}$-forzamiento de la sub\'{a}lgebra resonante $\mathfrak{G}%
_{\mathrm{R}}$ con $N=1$ reproduce la contracci\'{o}n de
\.{I}n\"{o}n\"{u}--Wigner para $\mathfrak{g}=V_{0}\oplus V_{1}$, la cual
abelianiza los generadores del coseto sim\'{e}trico $V_{1}$.

\subsubsection{\label{sec:ExpSuper}Caso cuando $\mathfrak{g}=V_{0}\oplus
V_{1}\oplus V_{2}$ es una Super\'{a}lgebra}

Una super\'{a}lgebra $\mathfrak{g}$ viene naturalmente dividida en tres
subespacios $V_{0}$, $V_{1}$ y $V_{2}$, donde $V_{1}$ corresponde al sector
fermi\'{o}nico y $V_{0}\oplus V_{2}$ al bos\'{o}nico, siendo $V_{0}$ una
sub\'{a}lgebra. Esta estructura algebraica puede ser escrita en la forma
\begin{align}
\left[  V_{0},V_{0}\right]   &  \subset V_{0},\label{SuperV0V0=V0}\\
\left[  V_{0},V_{1}\right]   &  \subset V_{1},\label{SuperV0V1=V1}\\
\left[  V_{0},V_{2}\right]   &  \subset V_{2},\label{SuperV0V2=V2}\\
\left[  V_{1},V_{1}\right]   &  \subset V_{0}\oplus V_{2}%
,\label{SuperV1V1=V0+V2}\\
\left[  V_{1},V_{2}\right]   &  \subset V_{1},\label{SuperV1V2=V1}\\
\left[  V_{2},V_{2}\right]   &  \subset V_{0}\oplus V_{2}.
\label{SuperV2V2=V0+V2}%
\end{align}

Para extraer una sub\'{a}lgebra resonante del \'{a}lgebra $S_{\mathrm{E}%
}^{\left(  N\right)  }$-expandida $\mathfrak{G}=S_{\mathrm{E}}^{\left(
N\right)  }\otimes\mathfrak{g}$ es necesario encontrar una partici\'{o}n
resonante del semigrupo $S_{\mathrm{E}}^{\left(  N\right)  }$. Esta
partici\'{o}n est\'{a} dada por $S_{\mathrm{E}}^{\left(  N\right)  }=S_{0}\cup
S_{1}\cup S_{2}$, con
\begin{equation}
S_{p}=\left\{  \lambda_{2m+p},\text{ con }m=0,\dotsc,\left[  \frac{N-p}%
{2}\right]  \right\}  \cup\left\{  \lambda_{N+1}\right\}  ,\qquad p=0,1,2.
\label{ParResSuper}%
\end{equation}

Esta partici\'{o}n es resonante en el sentido de la def.~\ref{def:ParRes},
pues satisface
\begin{align}
S_{0}\times S_{0}  &  \subset S_{0},\label{SuperS0S0=S0}\\
S_{0}\times S_{1}  &  \subset S_{1},\label{SuperS0S1=S1}\\
S_{0}\times S_{2}  &  \subset S_{2},\label{SuperS0S2=S2}\\
S_{1}\times S_{1}  &  \subset S_{0}\cap S_{2},\label{SuperS1S1=S0nS2}\\
S_{1}\times S_{2}  &  \subset S_{1},\label{SuperS1S2=S1}\\
S_{2}\times S_{2}  &  \subset S_{0}\cap S_{2}. \label{SuperS2S2=S0nS2}%
\end{align}

Dada la partici\'{o}n resonante~(\ref{SuperS0S0=S0})--(\ref{SuperS2S2=S0nS2}),
el teorema~\ref{th:SubAlgRes} nos asegura que $\mathfrak{G}_{\mathrm{R}}%
=W_{0}\oplus W_{1}\oplus W_{2}$, con $W_{p}=S_{p}\otimes V_{p}$, $p=0,1,2$, es
una sub\'{a}lgebra resonante. Dado que esta sub\'{a}lgebra puede ser de
inter\'{e}s por s\'{\i} misma, citamos aqu\'{\i} sus constantes de estructura,
las cuales son obtenidas particularizando la ec. general~(\ref{C=KCRes}) al
caso actual:
\begin{equation}
C_{\left(  a_{p},\alpha_{p}\right)  \left(  b_{q},\beta_{q}\right)
}%
^{\phantom{\left( a_{p}, \alpha_{p} \right) \left( b_{q}, \beta_{q} \right)}\left(
c_{r},\gamma_{r}\right)  }=\left\{
\begin{array}
[c]{cl}%
\delta_{\alpha_{p}+\beta_{q}}^{\gamma_{r}}C_{a_{p}b_{q}}%
^{\phantom{a_{p} b_{q}}c_{r}}, & \text{cuando }\alpha_{p}+\beta_{q}\leq N\\
\delta_{N+1}^{\gamma_{r}}C_{a_{p}b_{q}}^{\phantom{a_{p} b_{q}}c_{r}}, &
\text{cuando }\alpha_{p}+\beta_{q}\geq N+1
\end{array}
\right.  ,
\end{equation}
con $p,q,r=0,1,2$, $\alpha_{p},\beta_{p},\gamma_{p}=2m+p$, $m=0,\dotsc,\left[
\frac{N-p}{2}\right]  ,\frac{N+1-p}{2}$.

Alternativamente, uno puede dar un paso adicional y considerar el $0_{S}%
$-forzamiento de esta sub\'{a}lgebra resonante. Al hacerlo, todos los
generadores de la forma $\lambda_{N+1}\bm{T}_{A}$ son eliminados del
\'{a}lgebra, de modo que las constantes de estructura se transforman en%
\begin{equation}
C_{\left(  a_{p},\alpha_{p}\right)  \left(  b_{q},\beta_{q}\right)
}%
^{\phantom{\left( a_{p}, \alpha_{p} \right) \left( b_{q}, \beta_{q} \right)}\left(
c_{r},\gamma_{r}\right)  }=\left\{
\begin{array}
[c]{cl}%
\delta_{\alpha_{p}+\beta_{q}}^{\gamma_{r}}C_{a_{p}b_{q}}%
^{\phantom{a_{p} b_{q}}c_{r}}, & \text{cuando }\alpha_{p}+\beta_{q}\leq N\\
\delta_{N+1}^{\gamma_{r}}C_{a_{p}b_{q}}^{\phantom{a_{p} b_{q}}c_{r}}, &
\text{cuando }\alpha_{p}+\beta_{q}\geq N+1
\end{array}
\right.  , \label{CE-sa0}%
\end{equation}
con $p,q,r=0,1,2$, $\alpha_{p},\beta_{p},\gamma_{p}=2m+p$, $m=0,\dotsc,\left[
\frac{N-p}{2}\right]  $.

Esta \'{a}lgebra $0_{S}$-forzada corresponde al \'{a}lgebra $\mathcal{G}%
\left(  N_{0},N_{1},N_{2}\right)  $ (de acuerdo a la notaci\'{o}n de la Ref.~\cite{deAz02}), con
\begin{equation}
N_{p}=2\left[  \frac{N-p}{2}\right]  +p,\qquad p=0,1,2.
\end{equation}
Las constantes de estructura~(\ref{CE-sa0}) corresponden a las de la ec.~(5.6)
de esta referencia.

\begin{figure}
\begin{center}
\includegraphics[width=\textwidth]{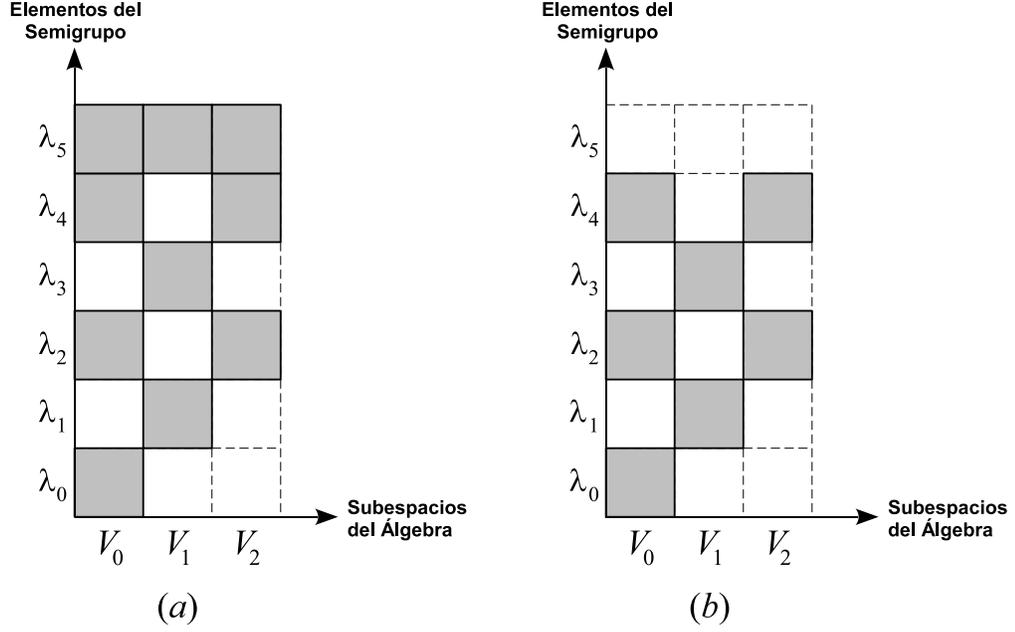}
\end{center}
\caption{\label{fig:SubAlgResSuper}(\textit{a}) Sub\'{a}lgebra resonante obtenida a partir del
\'{a}lgebra $S_{\mathrm{E}}^{(4)}$-expandida $\mathfrak{G} = S_{\mathrm{E}%
}^{(4)} \otimes\mathfrak{g}$ en el caso cuando $\mathfrak{g} = V_{0} \oplus
V_{1} \oplus V_{2}$ es una super\'{a}lgebra. (\textit{b}) El sector
$\lambda_{5} \otimes\mathfrak{g}$ es eliminado de la sub\'{a}lgebra resonante
$\mathfrak{G}_{\mathrm{R}}$ a trav\'{e}s del proceso de $0_{S}$-forzamiento.
El \'{a}lgebra resultante corresponde a $\mathcal{G}(4,3,4)$ en el contexto de
la Ref.~\cite{deAz02}.}
\end{figure}

La fig.~\ref{fig:SubAlgResSuper} ilustra gr\'{a}ficamente el proceso que
acabamos de describir para el caso $N=4$. En la parte~(\textit{a}) se muestra la
sub\'{a}lgebra resonante $\mathfrak{G}_{\mathrm{R}}$ y en la parte~(\textit{b}) su
$0_{S}$-forzamiento. Notemos como las columnas del gr\'{a}fico corresponden a
los subconjuntos $S_{p}$ de la partici\'{o}n resonante, y como el $0_{S}%
$-forzamiento consiste en eleminar todos los subespacios de la forma
$\lambda_{5}\otimes\mathfrak{g}$ (con $\lambda_{5}=0_{S}$).

\section{\label{sec:sexpex}Ejemplos de \'{A}lgebras $S$-Expandidas}

En esta secci\'{o}n presentamos algunos ejemplos del proceso de $S$-expansi\'{o}n. A excepci\'{o}n del primero de ellos, cuyo inter\'{e}s es puramente pedag\'{o}gico, su elecci\'{o}n est\'{a} motivada por nuestro inter\'{e}s en una formulaci\'{o}n geom\'{e}\-trica de la teor\'{\i}a de Supergravedad\footnote{Para un an\'{a}lisis en profundidad de distintas \'{a}lgebras de Supersimetr\'{\i}a, ver~\cite{deAz89,vanHo82} y las referencias all\'{\i} citadas.} en $d=11$.

Brevemente, los ejemplos son:
\begin{itemize}
\item El \'{a}lgebra $\mathfrak{so}\left(  4\right)  $ como una $S$%
-expansi\'{o}n de $\mathfrak{so}\left(  3\right)  $, con $S=\mathbb{Z}_{2}$.
Posiblemente el ejemplo m\'{a}s simple, pues parte de un \'{a}lgebra con
s\'{o}lo tres generadores y un semigrupo de apenas dos elementos.

\item El \'{a}lgebra~M como una $S$-expansi\'{o}n de $\mathfrak{osp}\left(
32|1\right)  $, con $S=S_{\mathrm{E}}^{\left(  2\right)  }$. Un ejemplo
importante con consecuencias f\'{\i}sicas interesantes, como se ver\'{a} en el cap\'{\i}tulo~\ref{ch:TGFTMAlg}.

\item Una super\'{a}lgebra extendida ($\mathcal{N}=2$), similar a las introducidas por D'Auria y Fr\'{e} en~\cite{DAu82}, obtenida como una $S$-expansi\'{o}n de
$\mathfrak{osp}\left(  32|1\right)  $ con $S=S_{\mathrm{E}}^{\left(
3\right)  }$. El paso de $S_{\mathrm{E}}^{\left(  2\right)  }$ a
$S_{\mathrm{E}}^{\left(  3\right)  }$ produce un generador fermi\'{o}nico
extra y cambia algunas relaciones de conmutaci\'{o}n.

\item Una nueva super\'{a}lgebra extendida ($\mathcal{N}=2$) con generadores
bos\'{o}nicos extra es obtenida a trav\'{e}s de una $S$-expansi\'{o}n de
$\mathfrak{osp}\left(  32|1\right)  $ con un semigrupo distinto a
$S_{\mathrm{E}}^{\left(  N\right)  }$ ($S=\mathbb{Z}_{4}$).
\end{itemize}

\subsection{\label{sec:so43}$\mathfrak{so}\left(  4\right)  $ como una
$S$-expansi\'{o}n de $\mathfrak{so}\left(  3\right)  $}

En esta secci\'{o}n mostramos c\'{o}mo el \'{a}lgebra $\mathfrak{so}\left(
4\right)  $ puede ser considerada como una $S$-expansi\'{o}n de $\mathfrak{so}%
\left(  3\right)  $. El ejemplo es particularmente simple, no s\'{o}lo por la
sencillez de las \'{a}lgebras en juego, sino tambi\'{e}n porque no involucra
ni forzamiento ni la extracci\'{o}n de una sub\'{a}lgebra resonante.

Cualquier \'{a}lgebra de la familia $\mathfrak{so}\left(  n\right)  $ puede
ser descrita mediante las relaciones de conmutaci\'{o}n
\begin{equation}
\left[  \bm{S}_{\kappa\lambda},\bm{S}_{\mu\nu}\right]
=-i\left(  \delta_{\mu\lambda}\bm{S}_{\kappa\nu}-\delta_{\mu\kappa
}\bm{S}_{\lambda\nu}+\delta_{\nu\lambda}\bm{S}_{\mu\kappa
}-\delta_{\nu\kappa}\bm{S}_{\mu\lambda}\right)  , \label{son}%
\end{equation}
con $\kappa,\lambda,\mu,\nu=1,\dotsc,n$ y $\delta_{\mu\nu}=\operatorname{diag}%
\left(  +1,\dotsc,+1\right)  $. En el caso particular de $\mathfrak{so}\left(
3\right)  $, uno puede definir $\bm{G}_{1}=\bm{S}_{23}$,
$\bm{G}_{2}=\bm{S}_{31}$, $\bm{G}_{3}=\bm{S}%
_{12}$, de manera de escribir~(\ref{son}) en la forma
\[
\left[  \bm{G}_{i},\bm{G}_{j}\right]  =i\varepsilon
_{ijk}\bm{G}_{k},
\]
con $i,j,k=1,2,3$.

\begin{table}
\begin{center}
\begin{tabular}[c]{c|cc}
& $a$ & $b$\\
\hline
$a$ & $a$ & $b$\\
$b$ & $b$ & $a$
\end{tabular}
\end{center}
\caption{\label{tab:Z2mt}Tabla de multiplicaci\'{o}n del semigrupo abeliano $\mathbb{Z}_{2} = \left\{ a,b \right\}$. El \'{a}lgebra $\mathfrak{so} \left( 4 \right)$ puede ser considerada como una $\mathbb{Z}_{2}$-expansi\'{o}n de $\mathfrak{so} \left( 3 \right)$.}
\end{table}

Consideremos ahora el semigrupo\footnote{Un grupo es tambi\'{e}n un semigrupo, despu\'{e}s de todo.} $\mathbb{Z}_{2}=\left\{  a,b\right\}  $ definido por la tabla de multiplicaci\'{o}n dada en el cuadro~\ref{tab:Z2mt}. Llamando
\begin{align}
\bm{J}_{i}  &  =a\bm{G}_{i},\\
\bm{K}_{i}  &  =b\bm{G}_{i},
\end{align}
resulta directo verificar que las relaciones de conmutaci\'{o}n del \'{a}lgebra $\mathbb{Z}_{2}$-expandida $\mathfrak{G} = \mathbb{Z}_{2} \otimes \mathfrak{so} \left( 3 \right)$ tienen la forma
\begin{align}
\left[  \bm{J}_{i},\bm{J}_{j}\right]   &  =i\varepsilon
_{ijk}\bm{J}_{k},\label{JJ3}\\
\left[  \bm{J}_{i},\bm{K}_{j}\right]   &  =i\varepsilon
_{ijk}\bm{K}_{k},\label{JK3}\\
\left[  \bm{K}_{i},\bm{K}_{j}\right]   &  =i\varepsilon
_{ijk}\bm{J}_{k}. \label{KK3}%
\end{align}

Reunificando ahora $\bm{J}_{i}$ y $\bm{K}_{i}$ en generadores
$\bm{M}_{\kappa\lambda}$, $\kappa,\lambda=1,2,3,4$, de acuerdo a la
regla
\begin{align}
\bm{M}_{ij}  &  =\varepsilon_{ijk}\bm{J}_{k},\\
\bm{M}_{i4}  &  =\bm{K}_{i},
\end{align}
uno puede reescribir las relaciones de conmutaci\'{o}n~(\ref{JJ3}%
)--(\ref{KK3}) en la forma
\begin{equation}
\left[  \bm{M}_{\kappa\lambda},\bm{M}_{\mu\nu}\right]
=-i\left(  \delta_{\mu\lambda}\bm{M}_{\kappa\nu}-\delta_{\mu\kappa
}\bm{M}_{\lambda\nu}+\delta_{\nu\lambda}\bm{M}_{\mu\kappa
}-\delta_{\nu\kappa}\bm{M}_{\mu\lambda}\right)  ,
\end{equation}
demostrando as\'{\i} [por comparaci\'{o}n con~(\ref{son})] que ellas
corresponden a las del \'{a}lgebra $\mathfrak{so}\left(  4\right)  $.

\subsection{\label{sec:MAlgSExp}El \'{A}lgebra~M como una $S$-expansi\'{o}n de $\mathfrak{osp} \left( 32|1 \right)$}

En esta secci\'{o}n presentamos la derivaci\'{o}n del
\'{a}lgebra~M~\cite{Tow95} como una $S_{\mathrm{E}}^{\left(  2\right)  }%
$-expansi\'{o}n del \'{a}lgebra ortosimpl\'{e}ctica $\mathfrak{osp}\left(
32|1\right)  $.

La inclusi\'{o}n del \'{a}lgebra~M dentro de los ejemplos es motivada por dos
hechos distintos pero relacionados. En primer lugar, el paradigma de la
$S$-expansi\'{o}n permite considerar el \'{a}lgebra~M como un miembro de una
familia de \'{a}lgebras, todas generadas a partir de $\mathfrak{osp}\left(
32|1\right)  $ a trav\'{e}s de diferentes elecciones de semigrupo y diferentes
alternativas de forzamiento. Esto puede resultar relevante desde un punto de
vista f\'{\i}sico, dado que todas ellas comparten caracter\'{\i}sticas
importantes. En segundo lugar, este enfoque permite escribir un tensor
invariante para el \'{a}lgebra~M a partir de uno para $\mathfrak{osp}\left(
32|1\right)  $ (ver secci\'{o}n~\ref{sec:TenInvSExp}), haciendo contacto con
las acciones transgresoras y de CS discutidas en los
cap\'{\i}tulos~\ref{ch:CS} y~\ref{ch:trans}.

\begin{table}
\begin{align}
\left[ \bm{P}_{a}, \bm{P}_{b} \right] = & \bm{J}_{ab}, \\
\left[ \bm{J}^{ab}, \bm{P}_{c} \right] = & \delta_{ec}^{ab} \bm{P}^{e}, \\
\left[ \bm{P}_{a}, \bm{Z}_{b_{1} \cdots b_{5}} \right] = & - \frac{1}{5!} \varepsilon_{a b_{1} \cdots b_{5} c_{1} \cdots c_{5}} \bm{Z}^{c_{1} \cdots c_{5}}, \\
\left[ \bm{J}^{ab}, \bm{J}_{cd} \right] = & \delta_{ecd}^{abf} \bm{J}_{\phantom{e}f}^{e}, \\
\left[ \bm{J}^{ab}, \bm{Z}_{c_{1} \cdots c_{5}} \right] = & \frac{1}{4!} \delta_{dc_{1} \cdots c_{5}}^{ab e_{1} \cdots e_{4}} \bm{Z}_{\phantom{d}e_{1} \cdots e_{4}}^{d}, \\
\left[ \bm{Z}^{a_{1} \cdots a_{5}}, \bm{Z}_{b_{1} \cdots b_{5}} \right] = & \eta^{\left[ a_{1} \cdots a_{5} \right] \left[ c_{1} \cdots c_{5} \right]} \varepsilon_{c_{1} \cdots c_{5} b_{1} \cdots b_{5} e} \bm{P}^{e} + \delta_{d b_{1} \cdots b_{5}}^{a_{1} \cdots a_{5} e} \bm{J}_{\phantom{d}e}^{d} + \nonumber \\
& - \frac{1}{3!3!5!} \varepsilon_{c_{1} \cdots c_{11}} \delta_{d_{1} d_{2} d_{3} b_{1} \cdots b_{5}}^{a_{1} \cdots a_{5} c_{4} c_{5} c_{6}} \eta^{\left[ c_{1} c_{2} c_{3} \right] \left[ d_{1} d_{2} d_{3} \right]} \bm{Z}^{c_{7} \cdots c_{11}}, \\
\left[ \bm{P}_{a}, \bm{Q} \right] = & - \frac{1}{2} \Gamma_{a} \bm{Q},\\
\left[ \bm{J}_{ab}, \bm{Q} \right] = & - \frac{1}{2} \Gamma_{ab} \bm{Q}, \\
\left[ \bm{Z}_{abcde}, \bm{Q} \right] = & - \frac{1}{2} \Gamma_{abcde} \bm{Q}, \\
\left\{ \bm{Q}, \bar{\bm{Q}} \right\} = & \frac{1}{8} \left( \Gamma^{a} \bm{P}_{a} - \frac{1}{2} \Gamma^{ab} \bm{J}_{ab} + \frac{1}{5!} \Gamma^{abcde} \bm{Z}_{abcde} \right).
\end{align}
\caption{\label{tab:osp321}Relaciones de (anti)conmutaci\'{o}n de la super\'{a}lgebra $\mathfrak{osp} \left( 32|1 \right)$. Aqu\'{\i} $\Gamma_{a}$ son matrices de Dirac en $d=11$ (ver Ap\'{e}ndice~\ref{Ap:Cliff}).}
\end{table}

Consideremos ahora el \'{a}lgebra $\mathfrak{osp} \left( 32|1 \right)$. Una base conveniente para esta \'{a}lgebra es dada por $\left\{ \bm{P}_{a}, \bm{J}_{ab}, \bm{Z}_{abcde}, \bm{Q} \right\}$, donde $\bm{P}_{a}$ son \textit{boosts} de AdS, $\bm{J}_{ab}$ son rotaciones de Lorentz, $\bm{Z}_{abcde}$ es un tensor antisim\'{e}trico de cinco \'{\i}ndices y $\bm{Q}$ es una carga spinorial de Majorana con 32 componentes. Las relaciones de (anti)conmutaci\'{o}n est\'{a}n consignadas en el cuadro~\ref{tab:osp321} (ver Ap\'{e}ndice~\ref{Ap:Delta} para notaci\'{o}n y convenciones).

Como toda super\'{a}lgebra, $\mathfrak{osp}\left(  32|1\right)  $ puede ser
separada en tres subespacios, $\mathfrak{osp}\left(  32|1\right)  =V_{0}\oplus
V_{1}\oplus V_{2}$, correspondiendo $V_{0}$ a la sub\'{a}lgebra de Lorentz
(generada por $\bm{J}_{ab}$), $V_{1}$ al subespacio fermi\'{o}nico
(generado por $\bm{Q}$) y $V_{2}$ al resto de los generadores
bos\'{o}nicos, $\bm{P}_{a}$ y $\bm{Z}_{abcde}$. Resulta directo
verificar que esta separaci\'{o}n en subespacios satisface~(\ref{SuperV0V0=V0})--(\ref{SuperV2V2=V0+V2}).

\begin{table}
\begin{center}
\begin{tabular}[c]{c r @{$\; = \;$} l}
\hline
\hline
Subespacios de $\mathfrak{G}_{\text{R}}$ & \multicolumn{2}{c}{Generadores} \\
\hline
\multirow{3}*{$S_{0} \otimes V_{0}$}
  & $\bm{J}_{ab}$
  & $\lambda_{0} \bm{J}_{ab}^{\left(  \mathfrak{osp} \right) }$ \\
  & $\bm{Z}_{ab}$
  & $\lambda_{2} \bm{J}_{ab}^{\left(  \mathfrak{osp} \right) }$ \\
  & $\bm{0}$
  & $\lambda_{3} \bm{J}_{ab}^{\left( \mathfrak{osp} \right) }$ \\
\hline
\multirow{2}*{$S_{1} \otimes V_{1}$}
  & $\bm{Q}$ & $\lambda_{1} \bm{Q}^{\left(  \mathfrak{osp} \right) }$ \\
  & $\bm{0}$ & $\lambda_{3} \bm{Q}^{\left( \mathfrak{osp} \right) }$ \\
\hline
\multirow{4}*{$S_{2} \otimes V_{2}$}
  & $\bm{P}_{a}$ & $\lambda_{2} \bm{P}_{a}^{\left(  \mathfrak{osp} \right) }$ \\
  & $\bm{Z}_{abcde}$ & $\lambda_{2} \bm{Z}_{abcde}^{\left(  \mathfrak{osp} \right) }$ \\
  & $\bm{0}$ & $\lambda_{3} \bm{P}_{a}^{\left( \mathfrak{osp} \right) }$ \\
  & $\bm{0}$ & $\lambda_{3} \bm{Z}_{abcde}^{\left( \mathfrak{osp} \right) }$ \\
\hline
\hline
\end{tabular}
\end{center}
\caption{\label{tab:mosp}El \'{A}lgebra~M puede ser considerada como una $S_{\mathrm{E}}^{(2)}$-Expansi\'{o}n de $\mathfrak{osp} \left( 32|1 \right)$. La tabla muestra la relaci\'{o}n entre los generadores de ambas \'{a}lgebras. Los tres niveles corresponden a las tres columnas en la fig.~\ref{fig:MAlg} o, alternativamente, a los tres subconjuntos en que $S_{\mathrm{E}}^{(2)}$ ha sido particionado.}
\end{table}

\begin{figure}
\begin{center}
\includegraphics[width=\textwidth]{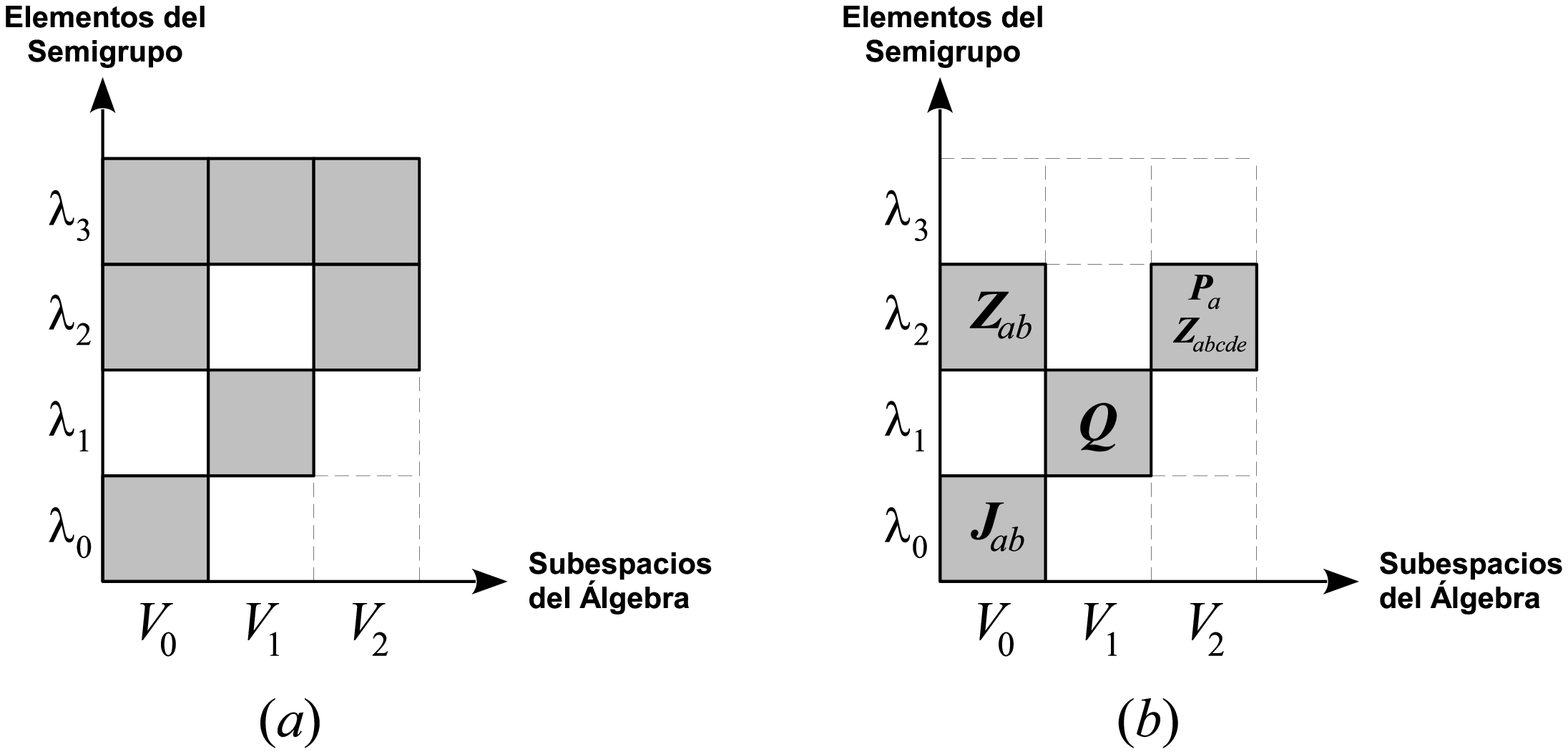}
\end{center}
\caption{\label{fig:MAlg}El \'{A}lgebra~M como una $S_{\mathrm{E}}^{\left(  2 \right)  }%
$-expansi\'{o}n de $\mathfrak{osp} \left(  32|1 \right)  $. (\textit{a})
Sub\'{a}lgebra resonante obtenida a partir del \'{a}lgebra $S_{\mathrm{E}%
}^{\left(  2 \right)  }$-expandida $\mathfrak{G} = S_{\mathrm{E}}^{\left(  2
\right)  } \otimes\mathfrak{osp} \left(  32|1 \right)  $. (\textit{b}) El
\'{a}lgebra~M propiamente tal es obtenida mediante el $0_{S}$-forzamiento de
la sub\'{a}lgebra resonante.}
\end{figure}

Para obtener el \'{a}lgebra~M uno debe proceder como en la
secci\'{o}n~\ref{sec:ExpSuper} con $S=S_{\mathrm{E}}^{\left(  2\right)  }$; es
decir, usar la partici\'{o}n resonante~(\ref{ParResSuper}) e imponer la
condici\'{o}n de $0_{S}$-forzamiento $\lambda_{3}\bm{T}_{A}=\bm{0}$. La relaci\'{o}n entre los generadores de $\mathfrak{osp} \left(
32|1\right)  $ y del \'{a}lgebra~M se muestra en el cuadro~\ref{tab:mosp}, en tanto que la fig.~\ref{fig:MAlg} muestra el esquema del \'{a}lgebra
$S$-expandida y su $0_{S}$-forzamiento.

\begin{table}
\begin{eqnarray}
\left[  \bm{J}^{ab},\bm{J}_{cd}\right] & = & \delta
_{ecd}^{abf}\bm{J}_{\phantom{e}f}^{e}, \\
\left[  \bm{J}^{ab},\bm{P}_{c}\right] & = & \delta_{ec}%
^{ab}\bm{P}^{e},\\
\left[  \bm{J}^{ab},\bm{Z}_{cd}\right] & = & \delta
_{ecd}^{abf}\bm{Z}_{\phantom{e}f}^{e},\\
\left[  \bm{J}^{ab},\bm{Z}_{c_{1}\cdots c_{5}}\right]
& = & \frac{1}{4!}\delta_{dc_{1}\cdots c_{5}}^{abe_{1}\cdots e_{4}}\bm{Z}%
_{\phantom{d}e_{1}\cdots e_{4}}^{d}, \\
\left[  \bm{J}_{ab},\bm{Q}\right] & = & -\frac{1}{2}%
\Gamma_{ab}\bm{Q}, \\
\left[  \bm{P}_{a},\bm{P}_{b}\right] & = & \bm{0},\\
\left[  \bm{P}_{a},\bm{Z}_{bc}\right] & = & \bm{0}, \\
\left[  \bm{P}_{a},\bm{Z}_{b_{1}\cdots b_{5}}\right]
& = & \bm{0},\\
\left[  \bm{Z}_{ab},\bm{Z}_{cd}\right] & = & \bm{0},\\
\left[  \bm{Z}_{ab},\bm{Z}_{c_{1}\cdots c_{5}}\right]
& = & \bm{0},\\
\left[  \bm{Z}_{a_{1}\cdots a_{5}},\bm{Z}_{b_{1}\cdots b_{5}%
}\right] & = & \bm{0}, \\
\left[  \bm{P}_{a},\bm{Q}\right] & = & \bm{0}, \label{PQM} \\
\left[  \bm{Z}_{ab},\bm{Q}\right] & = & \bm{0},\\
\left[  \bm{Z}_{abcde},\bm{Q}\right] & = & \bm{0}, \label{PZ5M} \\
\left\{  \bm{Q},\bar{\bm{Q}}\right\} & = & \frac{1}{8}\left(
\Gamma^{a}\bm{P}_{a}-\frac{1}{2}\Gamma^{ab}\bm{Z}_{ab}%
+\frac{1}{5!}\Gamma^{abcde}\bm{Z}_{abcde}\right)  .
\end{eqnarray}
\caption{\label{tab:MAlg}Las relaciones de (anti)conmutaci\'{o}n del \'{a}lgebra~M. Aqu\'{\i} $\Gamma_{a}$ son matrices de Dirac en $d=11$ (ver Ap\'{e}ndice~\ref{Ap:Cliff}).}
\end{table}

El \'{a}lgebra resultante se muestra en el cuadro~\ref{tab:MAlg}.

El \'{a}lgebra $S_{\mathrm{E}}^{\left(  2\right)  }$-expandida $\mathfrak{G}%
=S_{\mathrm{E}}^{\left(  2\right)  }\otimes\mathfrak{osp}\left(  32|1\right)
$ contiene una gran cantidad de generadores no incluidos en el \'{a}lgebra~M.
Los sectores indicados en la parte (\textit{a}) de la fig.~\ref{fig:MAlg} son eliminados al tomar la sub\'{a}lgebra resonante. Finalmente, el $\lambda_{3}$-forzamiento cumple un doble rol: elimina la sub\'{a}lgebra $\lambda_{3}\otimes\mathfrak{osp}\left(  32|1\right)  $ [que es isomorfa a
$\mathfrak{osp}\left(  32|1\right)  $] y abelianiza los conmutadores entre los generadores bos\'{o}nicos $\bm{P}_{a}$, $\bm{Z}_{ab}$ y
$\bm{Z}_{abcde}$, as\'{\i} como los de \'{e}stos con el generador fermi\'{o}nico $\bm{Q}$.

\subsection{\label{sec:DauFre}Superalgebra $\mathcal{N}=2$ similar a
D'Auria--Fr\'{e}}

En la secci\'{o}n anterior, el \'{a}lgebra~M fue obtenida como una $S$-expansi\'{o}n del \'{a}lgebra ortosimpl\'{e}ctica $\mathfrak{osp}\left( 32|1\right)  $ a trav\'{e}s del semigrupo $S=S_{\mathrm{E}}^{\left(  2\right)}$ (ver def.~\ref{def:SEN}). En esta secci\'{o}n exploramos los resultados de escoger $S=S_{\mathrm{E}}^{\left(  3\right)  }$, dejando todos los dem\'{a}s aspectos de la expansi\'{o}n (incluido el $0_{S}$-forzamiento) inalterados.

Al escoger la partici\'{o}n resonante~(\ref{ParResSuper}) y efectuar el $\lambda_{4}$-forzamiento de la sub\'{a}lgebra resonante resultante, obtenemos una super\'{a}lgebra muy similar al \'{a}lgebra~M, pero con un generador fermi\'{o}nico extra, $\bm{Q}^{\prime} = \lambda_{3} \bm{Q}^{\left( \mathfrak{osp} \right)}$. Este generador conmuta con todos los elementos de la super\'{a}lgebra (excepto con los generadores del \'{a}lgebra de Lorentz, lo cual es de esperar debido a su naturaleza spinorial), y anticonmuta adem\'{a}s con $\bm{Q}$ y consigo mismo. Otra diferencia con el \'{a}lgebra~M concierne los conmutadores del generador fermi\'{o}nico $\bm{Q}$ con los generadores bos\'{o}nicos $\bm{P}_{a}$, $\bm{Z}_{ab}$ y $\bm{Z}_{abcde}$, los cuales ya no son nulos [comparar con~(\ref{PQM})--(\ref{PZ5M})]:
\begin{eqnarray}
\left[  \bm{P}_{a},\bm{Q}\right]   &  = & -\frac{1}{2}\Gamma
_{a}\bm{Q}^{\prime},\\
\left[  \bm{Z}_{ab},\bm{Q}\right]   &  = & -\frac{1}{2}%
\Gamma_{ab}\bm{Q}^{\prime},\\
\left[  \bm{Z}_{abcde},\bm{Q}\right]   &  = & -\frac{1}{2}%
\Gamma_{abcde}\bm{Q}^{\prime}.
\end{eqnarray}

La relaci\'{o}n entre los generadores de la Super\'{a}lgebra de esta secci\'{o}n y los de $\mathfrak{osp} \left( 32|1 \right)$ se muestra en el cuadro~\ref{tab:dfosp}.

\begin{table}
\begin{center}
\begin{tabular}[c]{c r @{$\; = \;$} l}
\hline
\hline
Subespacios de $\mathfrak{G}_{\text{R}}$ & \multicolumn{2}{c}{Generadores} \\
\hline
\multirow{3}*{$S_{0} \otimes V_{0}$}
  & $\bm{J}_{ab}$
  & $\lambda_{0} \bm{J}_{ab}^{\left(  \mathfrak{osp} \right) }$ \\
  & $\bm{Z}_{ab}$
  & $\lambda_{2} \bm{J}_{ab}^{\left(  \mathfrak{osp} \right) }$ \\
  & $\bm{0}$
  & $\lambda_{4} \bm{J}_{ab}^{\left( \mathfrak{osp} \right) }$ \\
\hline
\multirow{3}*{$S_{1} \otimes V_{1}$}
  & $\bm{Q}$
  & $\lambda_{1} \bm{Q}^{\left(  \mathfrak{osp} \right) }$ \\
  & $\bm{Q}^{\prime}$
  & $\lambda_{3} \bm{Q}^{\left(  \mathfrak{osp} \right) }$ \\
  & $\bm{0}$
  & $\lambda_{4} \bm{Q}^{\left( \mathfrak{osp} \right) }$ \\
\hline
\multirow{4}*{$S_{2} \otimes V_{2}$}
  & $\bm{P}_{a}$ & $\lambda_{2} \bm{P}_{a}^{\left(  \mathfrak{osp} \right) }$ \\
  & $\bm{Z}_{abcde}$ & $\lambda_{2} \bm{Z}_{abcde}^{\left(  \mathfrak{osp} \right) }$ \\
  & $\bm{0}$ & $\lambda_{4} \bm{P}_{a}^{\left( \mathfrak{osp} \right) }$ \\
  & $\bm{0}$ & $\lambda_{4} \bm{Z}_{abcde}^{\left( \mathfrak{osp} \right) }$ \\
\hline
\hline
\end{tabular}
\end{center}
\caption{\label{tab:dfosp}Relaci\'{o}n entre los generadores de $\mathfrak{osp} \left( 32|1 \right)$ y los de la Super\'{a}lgebra extendida ($\mathcal{N} = 2$) similar a las de D'Auria y Fr\'{e}.}
\end{table}

\begin{figure}
\begin{center}
\includegraphics[width=\textwidth]{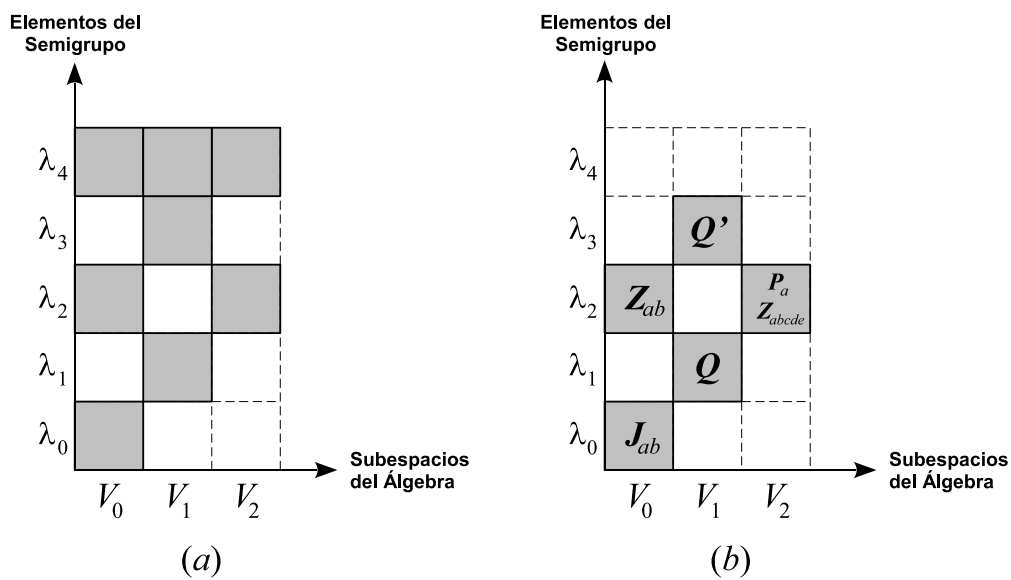}
\end{center}
\caption{\label{fig:DauFre}(\textit{a}) Una super\'{a}lgebra extendida ($\mathcal{N} = 2$) es obtenida como una sub\'{a}lgebra resonante del \'{a}lgebra $S_{\mathrm{E}}^{(3)}$-expandida $\mathfrak{G}=S_{\mathrm{E}}^{(3)} \otimes \mathfrak{osp} \left( 32|1 \right)$. (\textit{b}) El $0_{S}$-forzamiento de la Super\'{a}lgebra en (\textit{a}) elimina el subespacio $\lambda_{4} \otimes \mathfrak{osp} \left( 32|1 \right)$ y abelianiza algunos conmutadores. La Super\'{a}lgebra resultante es muy similar a las consideradas por D'Auria y Fr\'{e} en~\cite{DAu82}, como se explica en el texto.}
\end{figure}

La super\'{a}lgebra de esta secci\'{o}n es casi id\'{e}ntica a las introducidas por D'Auria y Fr\'{e} en~\cite{DAu82}; tiene el mismo n\'{u}mero y tipo de generadores que aquellas, y s\'{o}lo difiere por factores num\'{e}ricos en las relaciones de conmutaci\'{o}n. En la Ref.~\cite{Ban04b} se muestra c\'{o}mo las Super\'{a}lgebras de D'Auria--Fr\'{e} corresponden a los casos $s=3/2$ y $s=-1$ de una familia de Super\'{a}lgebras $\tilde{\mathfrak{E}} \left( s \right)$ parametrizadas por un n\'{u}mero real $s$. La super\'{a}lgebra presentada en esta secci\'{o}n corresponde al caso $s=0$.

El proceso de $S$-expansi\'{o}n llevado a cabo para obtener esta super\'{a}lgebra est\'{a} graficado en la fig.~\ref{fig:DauFre}, donde se muestran (\textit{a}) la sub\'{a}lgebra resonante obtenida directamente de $\mathfrak{osp}\left(  32|1\right)  $ y (\textit{b}) su $\lambda_{4}$-forzamiento.

\subsection{\label{sec:newsuper}Nueva Super\'{a}lgebra con $\mathcal{N}=2$}

Esta secci\'{o}n muestra un ejemplo de $S$-expansi\'{o}n con un semigrupo
distinto a aquel utilizado para reproducir los resultados de~\cite{deAz02}, vale decir, $S_{\mathrm{E}}^{\left(  N\right)  }$. Consideraremos una $S$-expansi\'{o}n del \'{a}lgebra ortosimpl\'{e}ctica $\mathfrak{osp}\left( 32|1\right)$ con $S=\mathbb{Z}_{4}$, el grupo c\'{\i}clico de cuatro
elementos. Esta elecci\'{o}n se hace pensando en la sencillez del ejemplo y en la posibilidad de encontrar una partici\'{o}n resonante de $S$. El caso
$S=\mathbb{Z}_{2}$ es trivial, en el sentido que la sub\'{a}lgebra resonante es isomorfa a $\mathfrak{osp}\left(  32|1\right)  $. El caso $S=\mathbb{Z}_{3}$ parece no poseer una partici\'{o}n resonante. Luego, $S=\mathbb{Z}_{4}$ corresponde al caso no trivial m\'{a}s simple.

\begin{table}
\begin{center}
\begin{tabular}{cccc}
\multicolumn{4}{c}{$S_{\mathrm{E}}^{\left( 2 \right)}$} \\
\hline
0 & 1 & 2 & 3 \\
1 & 2 & 3 & 3 \\
2 & 3 & 3 & 3 \\
3 & 3 & 3 & 3
\end{tabular}
\qquad
\begin{tabular}{cccc}
\multicolumn{4}{c}{$\phantom{S_{\mathrm{E}}^{\left( 2 \right)}} \mathbb{Z}_{4} \phantom{S_{\mathrm{E}}^{\left( 2 \right)}}$} \\
\hline
0 & 1 & 2 & 3 \\
1 & 2 & 3 & 0 \\
2 & 3 & 0 & 1 \\
3 & 0 & 1 & 2
\end{tabular}
\end{center}
\caption{\label{tab:Z4SE2}Tablas de multiplicar para los semigrupos abelianos $S_{\mathrm{E}}^{\left( 2 \right)}$ y $\mathbb{Z}_{4}$. La similitud entre ambos semigrupos es patente en estas tablas; los \'{u}nicos productos diferentes son aquellos que ocupan el tri\'{a}ngulo inferior derecho.}
\end{table}

La ley de multiplicaci\'{o}n para $\mathbb{Z}_{4}=\left\{ \lambda_{0}, \lambda_{1}, \lambda_{2}, \lambda_{3} \right\}$ es
\begin{equation}
\lambda_{\alpha}\lambda_{\beta}=\lambda_{\left(  \alpha+\beta\right)
\operatorname{mod}4}.
\end{equation}
Esta regla no es completamente distinta a la de $S_{\mathrm{E}}^{\left( 2 \right)  }$ (ver cuadro~\ref{tab:Z4SE2}), por lo que cabr\'{\i}a esperar resultados vagamente similares a
los de las secciones anteriores.

Una partici\'{o}n resonante de $\mathbb{Z}_{4}$ es dada por
\begin{align}
S_{0}  &  =\left\{  \lambda_{0},\lambda_{2}\right\}  ,\\
S_{1}  &  =\left\{  \lambda_{1},\lambda_{3}\right\}  ,\\
S_{2}  &  =\left\{  \lambda_{0},\lambda_{2}\right\}  .
\end{align}

Como $\mathbb{Z}_{4}$ no tiene un elemento cero, no es posible realizar un
$0_{S}$-forzamien\-to; la sub\'{a}lgebra resonante constituye por lo tanto
nuestro resultado final.

La nueva super\'{a}lgebra tiene dos generadores fermi\'{o}nicos
$\bm{Q}$ y $\bm{Q}^{\prime}$, al igual que la super\'{a}lgebra
de la secci\'{o}n~\ref{sec:DauFre}. A diferencia tanto de esta \'{u}ltima como
del \'{a}lgebra~M, contamos ahora con $11+\binom{11}{5}=473$ generadores
bos\'{o}nicos adicionales, $\bm{P}_{a}^{\prime}$ y $\bm{Z}%
_{abcde}^{\prime}$. Reflejando fielmente la estructura de $\mathbb{Z}_{4}$, no
hay ning\'{u}n generador abeliano en el \'{a}lgebra, ni siquiera las antiguas
cargas centrales tensoriales $\bm{Z}_{ab}$ y $\bm{Z}_{abcde}$.

\begin{table}
\begin{center}
\begin{tabular}[c]{c r @{$\; = \;$} l}
\hline
\hline
Subespacios de $\mathfrak{G}_{\text{R}}$ & \multicolumn{2}{c}{Generadores} \\
\hline
\multirow{2}*{$S_{0} \otimes V_{0}$}
  & $\bm{J}_{ab}$
  & $\lambda_{0} \bm{J}_{ab}^{\left(  \mathfrak{osp} \right) }$ \\
  & $\bm{Z}_{ab}$
  & $\lambda_{2} \bm{J}_{ab}^{\left(  \mathfrak{osp} \right) }$ \\
\hline
\multirow{2}*{$S_{1} \otimes V_{1}$}
  & $\bm{Q}$
  & $\lambda_{1} \bm{Q}^{\left(  \mathfrak{osp} \right) }$ \\
  & $\bm{Q}^{\prime}$
  & $\lambda_{3} \bm{Q}^{\left(  \mathfrak{osp} \right) }$ \\
\hline
\multirow{4}*{$S_{2} \otimes V_{2}$}
  & $\bm{P}_{a}^{\prime}$
  & $\lambda_{0} \bm{P}_{a}^{\left(  \mathfrak{osp} \right) }$ \\
  & $\bm{Z}_{abcde}^{\prime}$
  & $\lambda_{0} \bm{Z}_{abcde}^{\left(  \mathfrak{osp} \right) }$ \\
  & $\bm{P}_{a}$
  & $\lambda_{2} \bm{P}_{a}^{\left(  \mathfrak{osp} \right) }$ \\
  & $\bm{Z}_{abcde}$
  & $\lambda_{2} \bm{Z}_{abcde}^{\left(  \mathfrak{osp} \right) }$ \\
\hline
\hline
\end{tabular}
\end{center}
\caption{\label{tab:Z4}Relaci\'{o}n entre los generadores de $\mathfrak{osp} \left( 32|1 \right)$ y los del \'{a}lgebra obtenida como una sub\'{a}lgebra resonante de $\mathfrak{G} = \mathbb{Z}_{4} \otimes \mathfrak{osp} \left( 32|1 \right)$.}
\end{table}

\begin{figure}
\begin{center}
\includegraphics[width=.5\textwidth]{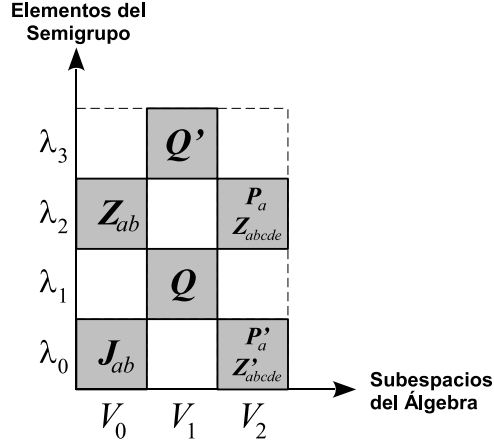}
\end{center}
\caption{\label{fig:Z4}Una nueva super\'{a}lgebra con $\mathcal{N}=2$ en $d=11$ es obtenida directamente como una sub\'{a}lgebra resonante del \'{a}lgebra $\mathbb{Z}_{4}$-expandida $\mathfrak{G} = \mathbb{Z}_{4} \otimes \mathfrak{osp} \left( 32|1 \right)$.}
\end{figure}

El cuadro~\ref{tab:Z4} muestra la relaci\'{o}n entre los generadores de esta
nueva super\'{a}lgebra y los de $\mathfrak{osp}\left(  32|1\right)  $. Las
relaciones de conmutaci\'{o}n, citadas m\'{a}s adelante, pueden leerse a
partir de la figura~\ref{fig:Z4}.

Deteng\'{a}monos un momento en las relaciones de (anti)conmutaci\'{o}n. Para
aligerar la notaci\'{o}n, \'{e}stas ser\'{a}n escritas s\'{o}lo en forma
esquem\'{a}tica. Su forma detallada puede recuperarse f\'{a}cilmente
comparando con $\mathfrak{osp}\left(  32|1\right)  $.

Los anticonmutadores entre los generadores fermi\'{o}nicos son
\begin{align}
\left\{  \bm{Q},\bm{Q}\right\}   &  \sim\bm{P}%
+\bm{Z}_{2}+\bm{Z}_{5},\\
\left\{  \bm{Q}^{\prime},\bm{Q}^{\prime}\right\}   &
\sim\bm{P}+\bm{Z}_{2}+\bm{Z}_{5},\\
\left\{  \bm{Q},\bm{Q}^{\prime}\right\}   &  \sim
\bm{P}^{\prime}+\bm{J}+\bm{Z}_{5}^{\prime}.
\end{align}

Todos los generadores son tensores o spinores de Lorentz, por lo cual sus
relaciones de conmutaci\'{o}n con $\bm{J}_{ab}$ son las usuales. Sin
embargo, hay \emph{dos} generadores que pueden interpretarse como \textit{boosts} de AdS, los cuales no conmutan
entre s\'{\i}:
\begin{align}
\left[  \bm{P},\bm{P}\right]   &  \sim\bm{J},\\
\left[  \bm{P}^{\prime},\bm{P}^{\prime}\right]   &
\sim\bm{J},\\
\left[  \bm{P},\bm{P}^{\prime}\right]   &  \sim\bm{Z}%
_{2}.
\end{align}

Las `cargas centrales' $\bm{Z}_{ab}$, $\bm{Z}_{abcde}$ y
$\bm{Z}_{abcde}^{\prime}$ ya no son abelianas:
\begin{align}
\left[  \bm{Z}_{2},\bm{Z}_{2}\right]   &  \sim\bm{J}%
,\\
\left[  \bm{Z}_{2},\bm{Z}_{5}\right]   &  \sim\bm{Z}%
_{5}^{\prime},\\
\left[  \bm{Z}_{2},\bm{Z}_{5}^{\prime}\right]   &
\sim\bm{Z}_{5},
\end{align}
\begin{align}
\left[  \bm{Z}_{5},\bm{Z}_{5}\right]   &  \sim\bm{P}%
^{\prime}+\bm{J}+\bm{Z}_{5}^{\prime},\\
\left[  \bm{Z}_{5},\bm{Z}_{5}^{\prime}\right]   &
\sim\bm{P}+\bm{Z}_{2}+\bm{Z}_{5},\\
\left[  \bm{Z}_{5}^{\prime},\bm{Z}_{5}^{\prime}\right]   &
\sim\bm{P}^{\prime}+\bm{J}+\bm{Z}_{5}^{\prime}.
\end{align}

Los conmutadores bos\'{o}nicos/fermi\'{o}nicos tienen la forma
\begin{align}
\left[  \bm{P},\bm{Q}\right]   &  \sim\bm{Q}^{\prime
},\\
\left[  \bm{Z}_{2},\bm{Q}\right]   &  \sim\bm{Q}%
^{\prime},\\
\left[  \bm{Z}_{5},\bm{Q}\right]   &  \sim\bm{Q}%
^{\prime},\\
\left[  \bm{P},\bm{Q}^{\prime}\right]   &  \sim\bm{Q}%
,\\
\left[  \bm{Z}_{2},\bm{Q}^{\prime}\right]   &  \sim
\bm{Q},\\
\left[  \bm{Z}_{5},\bm{Q}^{\prime}\right]   &  \sim
\bm{Q},
\end{align}%
\begin{align}
\left[  \bm{P}^{\prime},\bm{Q}\right]   &  \sim\bm{Q}%
,\\
\left[  \bm{J},\bm{Q}\right]   &  \sim\bm{Q},\\
\left[  \bm{Z}_{5}^{\prime},\bm{Q}\right]   &  \sim
\bm{Q},\\
\left[  \bm{P}^{\prime},\bm{Q}^{\prime}\right]   &
\sim\bm{Q}^{\prime},\\
\left[  \bm{J},\bm{Q}^{\prime}\right]   &  \sim\bm{Q}%
^{\prime},\\
\left[  \bm{Z}_{5}^{\prime},\bm{Q}^{\prime}\right]   &
\sim\bm{Q}^{\prime}.
\end{align}

La estructura c\'{\i}clica de $\mathbb{Z}_{4}$ es f\'{a}cilmente reconocible
en el patr\'{o}n de estas relaciones de conmutaci\'{o}n. No habiendo
generadores abelianos, el rol que cada uno de ellos cumple dentro del
\'{a}lgebra es m\'{a}s `equilibrado' que en el caso del \'{a}lgebra~M, donde,
por ejemplo, los generadores de Lorentz tienen claramente un papel especial.
Sin embargo, la simetr\'{\i}a no es total, pues sigue siendo posible
distinguir inequ\'{\i}vocamente a $\bm{J}_{ab}$ de $\bm{Z}%
_{ab}$. Por otro lado, los roles de $\bm{P}_{a}$ y $\bm{P}%
_{a}^{\prime}$, o de $\bm{Z}_{abcde}$ y $\bm{Z}_{abcde}%
^{\prime}$, son completamente sim\'{e}tricos.

Los tres \'{u}ltimos ejemplos (el \'{a}lgebra~M y las dos super\'{a}lgebras
extendidas) muestran como, a partir de una misma \'{a}lgebra $\mathfrak{g}$,
diferentes \'{a}lgebras $S$-expandidas son obtenidas para distintas elecciones
de semigrupo $S$ y distintas alternativas de particionamiento resonante.

\section[Tensores Invariantes para \'{a}lgebras $S$-ex\-pandidas]%
{\label{sec:TenInvSExp}Tensores Invariantes para \'{a}lgebras $S$-ex\-pandidas
\sectionmark{Tensores Invariantes para $S$-expansiones}}
\sectionmark{Tensores Invariantes para $S$-expansiones}

El problema de encontrar todos los tensores invariantes para un \'{a}lgebra de Lie\footnote{Para fijar ideas, en esta secci\'{o}n nos referimos expl\'{\i}citamente a s\'{u}peralgebras; sin embargo, todos los resultados siguen siendo v\'{a}lidos para el caso de \'{a}lgebras de Lie ordinarias.} $\mathfrak{g}$ arbitraria permanece abierto hasta hoy. \'{E}ste es no solo un importante problema matem\'{a}tico, sino tambi\'{e}n uno con relevancia f\'{\i}sica, porque un tensor invariante es un ingrediente clave en la construcci\'{o}n de lagrangeanos de CS o transgresiones (ver cap\'{\i}tulos~\ref{ch:CS} y~\ref{ch:trans}). Dada un \'{a}lgebra, la elecci\'{o}n de tensor invariante da forma a la teor\'{\i}a, fijando las clases de interacciones que ocurrir\'{a}n entre las distintas componentes de los campos independientes.

Un procedimiento est\'{a}ndar para obtener un tensor invariante de rango $r$ es usar la (s\'{u}per)traza del producto de $r$ generadores en alguna representaci\'{o}n matricial del \'{a}lgebra. Sin embargo, este procedimiento tiene una limitaci\'{o}n importante para el caso de \'{a}lgebras $0_{S}$-forzadas y, por este motivo, se vuelve importante considerar teoremas que provean de tensores invariantes para estas \'{a}lgebras.

\begin{theorem}
\label{th:TenInvSxg}Sea $S$ un semigrupo abeliano finito con $n$-selector
$K_{\alpha_{1}\cdots\alpha_{n}}^{\phantom{\alpha_{1} \cdots \alpha_{n}}\gamma
}$, $\mathfrak{g}$ una super\'{a}lgebra de Lie de base $\left\{
\bm{T}_{A}\right\}  $, y sea $\left\vert \bm{T}_{A_{1}}%
\cdots\bm{T}_{A_{n}}\right\vert $ un tensor invariante de rango $n$
para $\mathfrak{g}$. Denotando por $\left\{  \bm{T}_{\left(
A,\alpha\right)  }\right\}  $ a los generadores del \'{a}lgebra $S$-expandida
$\mathfrak{G}=S\otimes\mathfrak{g}$, se cumple que la expresi\'{o}n
\begin{equation}
\left\vert \bm{T}_{\left(  A_{1},\alpha_{1}\right)  }\cdots
\bm{T}_{\left(  A_{n},\alpha_{n}\right)  }\right\vert \equiv
\alpha_{\gamma}K_{\alpha_{1}\cdots\alpha_{n}}%
^{\phantom{\alpha_{1} \cdots \alpha_{n}}\gamma}\left\vert \bm{T}%
_{A_{1}}\cdots\bm{T}_{A_{n}}\right\vert , \label{TenInvSxg}%
\end{equation}
donde $\alpha_{\gamma}$ son constantes arbitrarias, corresponde a un tensor
invariante para $\mathfrak{G}$.
\end{theorem}

\begin{proof}
La condici\'{o}n de invariancia de $\left\vert \bm{T}_{A_{1}}%
\cdots\bm{T}_{A_{n}}\right\vert $ bajo $\mathfrak{g}$ puede escribirse
en la forma
\begin{equation}
\sum_{p=1}^{n}X_{A_{0}\cdots A_{n}}^{\left(  p\right)  }=0, \label{InvCondg}%
\end{equation}
con
\begin{align}
X_{A_{0}\cdots A_{n}}^{\left(  p\right)  } = & \left(  -1\right)
^{\mathfrak{q}\left(  A_{0}\right)  \left(  \mathfrak{q}\left(  A_{1}\right)
+\dotsb+\mathfrak{q}\left(  A_{p-1}\right)  \right)  }C_{A_{0}A_{p}%
}^{\phantom{A_{0} A_{p}}B}\times\nonumber\\
&  \times\left\vert \bm{T}_{A_{1}}\cdots\bm{T}_{A_{p-1}%
}\bm{T}_{B}\bm{T}_{A_{p+1}}\cdots\bm{T}_{A_{n}%
}\right\vert , \label{Xp-g}%
\end{align}
donde $\mathfrak{q}\left(  A\right)  $ denota el grado de $\bm{T}_{A}$
(1 para Fermi y 0 para Bose).

Definamos ahora el an\'{a}logo de~(\ref{Xp-g}) para el \'{a}lgebra
$S$-expandida $\mathfrak{G}$, es decir,
\begin{align}
X_{\left(  A_{0},\alpha_{0}\right)  \cdots\left(  A_{n},\alpha_{n}\right)
}^{\left(  p\right)  } = & \left(  -1\right)  ^{\mathfrak{q}\left(
A_{0},\alpha_{0}\right)  \left(  \mathfrak{q}\left(  A_{1},\alpha_{1}\right)
+\dotsb+\mathfrak{q}\left(  A_{p-1},\alpha_{p-1}\right)  \right)  }%
\times\nonumber\\
&  \times C_{\left(  A_{0},\alpha_{0}\right)  \left(  A_{p},\alpha_{p}\right)
}%
^{\phantom{\left( A_{0}, \alpha_{0} \right) \left( A_{p}, \alpha_{p} \right)}\left(
B,\beta\right)  }\left\vert \bm{T}_{\left(  A_{1},\alpha_{1}\right)
}\cdots\bm{T}_{\left(  A_{p-1},\alpha_{p-1}\right)  }\right.
\nonumber\\
&  \left.  \times\bm{T}_{\left(  B,\beta\right)  }\bm{T}%
_{\left(  A_{p+1},\alpha_{p+1}\right)  }\cdots\bm{T}_{\left(
A_{n},\alpha_{n}\right)  }\right\vert , \label{Xp-Sxg}%
\end{align}
donde $\left\vert \bm{T}_{\left(  A_{1},\alpha_{1}\right)  }%
\cdots\bm{T}_{\left(  A_{n},\alpha_{n}\right)  }\right\vert $ est\'{a}
dado por~(\ref{TenInvSxg}).

Usando el hecho que $\mathfrak{q}\left(  A,\alpha\right)  =\mathfrak{q}\left(
A\right)  $ y reemplazando tanto~(\ref{TenInvSxg}) como las constantes de
estructura~(\ref{C=KC}) para $\mathfrak{G}$ en~(\ref{Xp-Sxg}), encontramos
\begin{equation}
X_{\left(  A_{0},\alpha_{0}\right)  \cdots\left(  A_{n},\alpha_{n}\right)
}^{\left(  p\right)  }=\alpha_{\gamma}K_{\alpha_{0}\cdots\alpha_{n}%
}^{\phantom{\alpha_{0} \cdots \alpha_{n}}\gamma}X_{A_{0}\cdots A_{n}}^{\left(
p\right)  }. \label{Xp-Sxg2}%
\end{equation}

De~(\ref{InvCondg}) concluimos inmediatamente que
\begin{equation}
\sum_{p=1}^{n}X_{\left(  A_{0},\alpha_{0}\right)  \cdots\left(  A_{n}%
,\alpha_{n}\right)  }^{\left(  p\right)  }=0. \label{Xp-Sxg3}%
\end{equation}
Por lo tanto, $\left\vert \bm{T}_{\left(  A_{1},\alpha_{1}\right)
}\cdots\bm{T}_{\left(  A_{n},\alpha_{n}\right)  }\right\vert
=\alpha_{\gamma}K_{\alpha_{1}\cdots\alpha_{n}}%
^{\phantom{\alpha_{1} \cdots \alpha_{n}}\gamma}\left\vert \bm{T}%
_{A_{1}}\cdots\bm{T}_{A_{n}}\right\vert $ es un tensor invariante para
$\mathfrak{G}$.
\end{proof}

Vale la pena notar que, en general, la expresi\'{o}n
\begin{equation}
\left\vert \bm{T}_{\left(  A_{1},\alpha_{1}\right)  }\cdots
\bm{T}_{\left(  A_{n},\alpha_{n}\right)  }\right\vert =\sum_{m=0}%
^{M}\alpha_{\gamma}^{\beta_{1}\cdots\beta_{m}}K_{\beta_{1}\cdots\beta
_{m}\alpha_{1}\cdots\alpha_{n}}%
^{\phantom{\beta_{1} \cdots \beta_{m} \alpha_{1} \cdots \alpha_{n}}\gamma
}\left\vert \bm{T}_{A_{1}}\cdots\bm{T}_{A_{n}}\right\vert ,
\label{UltraTenInvSxg}%
\end{equation}
donde $M$ es el n\'{u}mero de elementos de $S$ y $\alpha_{\gamma}^{\beta
_{1}\cdots\beta_{m}}$ son constantes arbitrarias, es tambi\'{e}n un tensor
invariante para $\mathfrak{G}$. Un ejemplo de~(\ref{UltraTenInvSxg}) es
provisto por la supertraza. Una representaci\'{o}n matricial para
$\bm{T}_{\left(  A,\alpha\right)  }$ es inducida por una
representaci\'{o}n para $\bm{T}_{A}$ cuando usamos la
expresi\'{o}n~(\ref{lambda=K}) para los elementos de $S$, $\bm{T}%
_{\left(  A,\alpha\right)  }=\left(  \lambda_{\alpha}\right)  _{\mu
}^{\phantom{\mu}\nu}\bm{T}_{A}$. En esta representaci\'{o}n, tenemos que
\begin{equation}
\operatorname*{STr}\left(  \bm{T}_{\left(  A_{1},\alpha_{1}\right)
}\cdots\bm{T}_{\left(  A_{n},\alpha_{n}\right)  }\right)
=K_{\gamma\alpha_{1}\cdots\alpha_{n}}%
^{\phantom{\gamma \alpha_{1} \cdots \alpha_{n}}\gamma}\operatorname{Str}%
\left(  \bm{T}_{A_{1}}\cdots\bm{T}_{A_{n}}\right)  ,
\label{STr-Str}%
\end{equation}
donde $\operatorname*{STr}$ es la supertraza para $\bm{T}_{\left(  A,\alpha\right)  }$ y $\operatorname{Str}$ la supertraza para
$\bm{T}_{A}$. La ec.~(\ref{STr-Str}) corresponde al caso
$\alpha_{\gamma}^{\beta}=\delta_{\gamma}^{\beta}$ de~(\ref{UltraTenInvSxg}),
con todos los dem\'{a}s $\alpha$'s iguales a cero.

Aunque la expresi\'{o}n~(\ref{UltraTenInvSxg}) puede parecer m\'{a}s general
que~(\ref{TenInvSxg}), \'{e}ste no es el caso. Recurriendo a la asociatividad
y la clausura de la multiplicaci\'{o}n del semigrupo es siempre posible
reducir~(\ref{UltraTenInvSxg}) a~(\ref{TenInvSxg}), la cual resulta entonces
ser m\'{a}s `fundamental'.

Dado un tensor invariante para un \'{a}lgebra, sus componentes valuadas sobre
una sub\'{a}lgebra proporcionan un tensor invariante para la sub\'{a}lgebra
(siempre que no se anulen id\'{e}nticamente). Esta simple observaci\'{o}n nos
provee de un tensor invariante para cada sub\'{a}lgebra resonante
$\mathfrak{G}_{\mathrm{R}}$. Dada una partici\'{o}n resonante de $S$,
$S=\bigcup_{p\in I}S_{p}$, y denotando la base del subespacio $V_{p}$ por
$\left\{  \bm{T}_{a_{p}}\right\}  $, las componentes
de~(\ref{TenInvSxg}) valuadas en $\mathfrak{G}_{\mathrm{R}}$ est\'{a}n dadas
por
\begin{equation}
\left\vert \bm{T}_{\left(  a_{p_{1}},\alpha_{p_{1}}\right)  }%
\cdots\bm{T}_{\left(  a_{p_{n}},\alpha_{p_{n}}\right)  }\right\vert
=\alpha_{\gamma}K_{\alpha_{p_{1}}\cdots\alpha_{p_{n}}}%
^{\phantom{\alpha_{p_{1}} \cdots \alpha_{p_{n}}}\gamma}\left\vert
\bm{T}_{a_{p_{1}}}\cdots\bm{T}_{a_{p_{n}}}\right\vert ,
\label{TenInvResSubAlg}%
\end{equation}
con $\lambda_{\alpha_{p}}\in S_{p}$. Estas componentes forman un tensor
invariante para la sub\'{a}lgebra resonante $\mathfrak{G}_{\mathrm{R}}%
=\sum_{p\in I}S_{p}\otimes V_{p}$. Dado que $S$ es cerrado bajo el
producto~(\ref{tms}), para cada elecci\'{o}n de $\alpha_{p_{1}},\dotsc
,\alpha_{p_{n}}$ siempre existe un valor de $\gamma$ tal que $K_{\alpha
_{p_{1}}\cdots\alpha_{p_{n}}}%
^{\phantom{\alpha_{p_{1}} \cdots \alpha_{p_{n}}}\gamma}=1$ y, por lo tanto, el
tensor invariante~(\ref{TenInvResSubAlg}) es gen\'{e}ricamente distinto de
cero (asumiendo $\alpha_{\gamma}\neq0$).

Habiendo obtenido tensores invariantes para las \'{a}lgebras $S$-expandidas y
para sus sub\'{a}lgebras resonantes, pasamos a considerar el problema de
encontrar un tensor invariante para \'{a}lgebras obtenidas como el $0_{S}%
$-forzamiento de un \'{a}lgebra $S$-expandida (o de su sub\'{a}lgebra
resonante). Por supuesto, el problema radica en que un \'{a}lgebra $0_{S}%
$-forzada no corresponde a una sub\'{a}lgebra, de modo que las componentes
correspondientes de~(\ref{TenInvSxg}) o~(\ref{TenInvResSubAlg}) no
proporcionan un tensor invariante v\'{a}lido. Una soluci\'{o}n se ofrece por
medio del siguiente teorema, el cual entrega una expresi\'{o}n general para un
tensor invariante de un \'{a}lgebra $0_{S}$-forzada.

\begin{theorem}
\label{th:TenInv0F}Sea $S$ un semigrupo abeliano finito con elementos no nulos $\lambda_{i},i=0,\dotsc,N$, y un elemento cero $\lambda_{N+1}=0_{S}$. Sea $\mathfrak{g}$ una super\'{a}lgebra de Lie con base $\left\{ \bm{T}_{A} \right\}$ y sea $\left\vert \bm{T}_{A_{1}} \cdots \bm{T}_{A_{n}} \right\vert$ un tensor invariante para $\mathfrak{g}$. La expresi\'{o}n
\begin{equation}
\left\vert \bm{T}_{\left(  A_{1},i_{1}\right)  }\cdots\bm{T}%
_{\left(  A_{n},i_{n}\right)  }\right\vert \equiv\alpha_{j}K_{i_{1}\cdots
i_{n}}^{\phantom{i_{1} \cdots i_{n}}j}\left\vert \bm{T}_{A_{1}}%
\cdots\bm{T}_{A_{n}}\right\vert , \label{TenInv0F}%
\end{equation}
donde $\alpha_{j}$ son constantes arbitrarias, corresponde a un tensor
invariante para el \'{a}lgebra $0_{S}$-forzada obtenida a partir del
\'{a}lgebra $S$-expandida $\mathfrak{G}=S\otimes\mathfrak{g}$.
\end{theorem}

\begin{proof}
La demostraci\'{o}n de este teorema sigue las mismas l\'{\i}neas que la del
teorema~\ref{th:TenInvSxg}.

Definamos el an\'{a}logo de~(\ref{Xp-g}) para el $0_{S}$-forzamiento del
\'{a}lgebra $S$-expandida $\mathfrak{G}$, es decir,
\begin{align}
Y_{\left(  A_{0},i_{0}\right)  \cdots\left(  A_{n},i_{n}\right)  }^{\left(
p\right)  } = & \left(  -1\right)  ^{\mathfrak{q}\left(  A_{0},i_{0}\right)
\left(  \mathfrak{q}\left(  A_{1},i_{1}\right)  +\dotsb+\mathfrak{q}\left(
A_{p-1},i_{p-1}\right)  \right)  }\times\nonumber\\
&  \times C_{\left( A_{0}, i_{0} \right) \left( A_{p}, i_{p} \right)}^{\phantom{\left( A_{0}, i_{0} \right) \left( A_{p}, i_{p} \right)} \left(B,j\right)  }\left\vert \bm{T}_{\left(  A_{1},i_{1}\right)  }%
\cdots\bm{T}_{\left(  A_{p-1},i_{p-1}\right)  }\right. \nonumber\\
&  \left.  \times\bm{T}_{\left(  B,j\right)  }\bm{T}_{\left(
A_{p+1},i_{p+1}\right)  }\cdots\bm{T}_{\left(  A_{n},i_{n}\right)
}\right\vert . \label{Yp}%
\end{align}
donde $\left\vert \bm{T}_{\left(  A_{1},i_{1}\right)  }\cdots
\bm{T}_{\left(  A_{n},i_{n}\right)  }\right\vert $ est\'{a} dado
por~(\ref{TenInv0F}).

Usando el hecho que $\mathfrak{q}\left(  A,\alpha\right)  =\mathfrak{q}\left(
A\right)  $ y reemplazando tanto~(\ref{TenInv0F}) como las constantes de
estructura~(\ref{C=KC0F}) para el $0_{S}$.forzamiento de $\mathfrak{G}$
en~(\ref{Yp}), encontramos
\begin{equation}
Y_{\left(  A_{0},i_{0}\right)  \cdots\left(  A_{n},i_{n}\right)  }^{\left(
p\right)  }=\alpha_{k}K_{i_{0}i_{p}}^{\phantom{i_{0} i_{p}}j}K_{i_{1}\cdots
i_{p-1}ji_{p+1}\cdots i_{n}}%
^{\phantom{i_{1} \cdots i_{p-1} j i_{p+1} \cdots i_{n} }k}X_{A_{0}\cdots
A_{n}}^{\left(  p\right)  }. \label{Yp1}%
\end{equation}
El producto $K_{i_{0} i_{p}}^{\phantom{i_{0} i_{p}}j} K_{i_{1} \cdots i_{p-1} j i_{p+1} \cdots i_{n}}^{\phantom{i_{1} \cdots i_{p-1} j i_{p+1} \cdots i_{n}}k}$ puede ser reescrito en la forma
\begin{align*}
K_{i_{0}i_{p}}^{\phantom{i_{0} i_{p}}j}K_{i_{1}\cdots i_{p-1}ji_{p+1}\cdots
i_{n}}^{\phantom{i_{1} \cdots i_{p-1} j i_{p+1} \cdots i_{n} }k}
= & K_{i_{0}i_{p}}^{\phantom{i_{0} i_{p}}\gamma}K_{i_{1}\cdots i_{p-1}\gamma
i_{p+1}\cdots i_{n}}%
^{\phantom{i_{1} \cdots i_{p-1} \gamma i_{p+1} \cdots i_{n} }k}+\\
&  -K_{i_{0}i_{p}}^{\phantom{i_{0} i_{p}}N+1}K_{i_{1}\cdots i_{p-1}\left(
N+1\right)  i_{p+1}\cdots i_{n}}%
^{\phantom{i_{1} \cdots i_{p-1} \left( N+1 \right) i_{p+1} \cdots i_{n} }k}\\
&  =K_{i_{0} \cdots i_{n}}^{\phantom{i_{0} \cdots i_{n}}k}-K_{i_{0}i_{p}%
}^{\phantom{i_{0} i_{p}}N+1}K_{i_{1}\cdots i_{p-1}\left(  N+1\right)
i_{p+1}\cdots i_{n}}%
^{\phantom{i_{1} \cdots i_{p-1} \left( N+1 \right) i_{p+1} \cdots i_{n} }k}.
\end{align*}
Notando ahora que $K_{i_{1}\cdots i_{p-1}\left(  N+1\right)  i_{p+1}\cdots
i_{n}}^{\phantom{i_{1} \cdots i_{p-1} \left( N+1 \right) i_{p+1} \cdots i_{n} }k}=0$, encontramos
\begin{equation}
K_{i_{0}i_{p}}^{\phantom{i_{0} i_{p}}j}K_{i_{1}\cdots i_{p-1}ji_{p+1}\cdots
i_{n}}^{\phantom{i_{1} \cdots i_{p-1} j i_{p+1} \cdots i_{n} }k}%
=K_{i_{0} \cdots i_{n}}^{\phantom{i_{0} \cdots i_{n}}k}. \label{Yp2}%
\end{equation}
Reemplazando~(\ref{Yp2}) en~(\ref{Yp1}) hallamos
\begin{equation}
Y_{\left(  A_{0},i_{0}\right)  \cdots\left(  A_{n},i_{n}\right)  }^{\left(
p\right)  }=\alpha_{k}K_{i_{0}\cdots i_{n}}^{\phantom{i_{0} \cdots i_{n} }k} X_{A_{0}\cdots A_{n}}^{\left(  p\right)  },
\end{equation}
de donde concluimos inmediatamente que
\begin{equation}
\sum_{p=1}^{n}Y_{\left(  A_{0},i_{0}\right)  \cdots\left(  A_{n},i_{n}\right)
}^{\left(  p\right)  }=0.
\end{equation}
Por lo tanto, $\left\vert \bm{T}_{\left(  A_{1},i_{1}\right)  }%
\cdots\bm{T}_{\left(  A_{n},i_{n}\right)  }\right\vert =\alpha
_{j}K_{i_{1}\cdots i_{n}}^{\phantom{i_{1} \cdots i_{n}}j}\left\vert
\bm{T}_{A_{1}}\cdots\bm{T}_{A_{n}}\right\vert $ es un tensor
invariante para el \'{a}lgebra $0_{S}$-forzada obtenida de $\mathfrak{G}$.
\end{proof}

El teorema~\ref{th:TenInv0F} tambi\'{e}n es v\'{a}lido para el \'{a}lgebra
$0_{S}$-forzada de una sub\'{a}lgebra resonante $\mathfrak{G}_{\mathrm{R}}$.
La demostraci\'{o}n es an\'{a}loga a la que hemos presentado. En
consecuencia,
\begin{equation}
\left\vert \bm{T}_{\left(  a_{p_{1}},i_{p_{1}}\right)  }%
\cdots\bm{T}_{\left(  a_{p_{n}},i_{p_{n}}\right)  }\right\vert
=\alpha_{j}K_{i_{p_{1}}\cdots i_{p_{n}}}%
^{\phantom{i_{p_{1}} \cdots i_{p_{n}}}j}\left\vert \bm{T}_{a_{p_{1}}%
}\cdots\bm{T}_{a_{p_{n}}}\right\vert ,
\end{equation}
con $\lambda_{i_{p}}\in S_{p}$, es un tensor invariante para el \'{a}lgebra
$0_{S}$-forzada de una sub\'{a}lgebra resonante $\mathfrak{G}_{\mathrm{R}%
}=\sum_{p\in I}S_{p}\otimes V_{p}$.

La utilidad del teorema~\ref{th:TenInv0F} proviene del hecho que, en general,
la (s\'{u}per)\-traza en la representaci\'{o}n adjunta de un \'{a}lgebra
$0_{S}$-forzada tiene un n\'{u}mero menor de componentes no nulas
que~(\ref{TenInv0F}).

En efecto, usando la representaci\'{o}n adjunta dada por las constantes de
estructura~(\ref{C=KC0F}), encontramos
\begin{align}
\operatorname*{STr}\left(  \bm{T}_{\left(  A_{1},i_{1}\right)  }%
\cdots\bm{T}_{\left(  A_{n},i_{n}\right)  }\right) = &
K_{j_{1}i_{1}}^{\phantom{j_{1}i_{1}}j_{2}} K_{j_{2}i_{2}}%
^{\phantom{j_{2}i_{2}}j_{3}} \cdots K_{j_{n-1}i_{n-1}}%
^{\phantom{j_{n-1}i_{n-1}}j_{n}} \times\nonumber\\
&  \times K_{j_{n}i_{n}}^{\phantom{j_{n}i_{n}}j_{1}}\operatorname*{Str}\left(
\bm{T}_{A_{1}}\cdots\bm{T}_{A_{n}}\right)  . \label{noname}
\end{align}
Como
\begin{align}
K_{j_{1}i_{1}\cdots i_{n}}^{\phantom{j_{1} i_{1} \cdots i_{n}}j_{1}}  &
=K_{j_{1}i_{1}}^{\phantom{j_{1}i_{1}}\gamma_{2}}K_{\gamma_{2}i_{2}%
}^{\phantom{\gamma_{2}i_{2}}\gamma_{3}}\cdots K_{\gamma_{n-1}i_{n-1}%
}^{\phantom{\gamma_{n-1}i_{n-1}}\gamma_{n}}K_{\gamma_{n}i_{n}}%
^{\phantom{\gamma_{n}i_{n}}j_{1}}\nonumber\\
&  =K_{j_{1}i_{1}}^{\phantom{j_{1}i_{1}}j_{2}}K_{j_{2}i_{2}}%
^{\phantom{j_{2}i_{2}}j_{3}}\cdots K_{j_{n-1}i_{n-1}}%
^{\phantom{j_{n-1}i_{n-1}}j_{n}}K_{j_{n}i_{n}}^{\phantom{j_{n}i_{n}}j_{1}},
\end{align}
la ec.~(\ref{noname}) resulta ser equivalente a
\begin{equation}
\operatorname*{STr}\left(  \bm{T}_{\left(  A_{1},i_{1}\right)  }%
\cdots\bm{T}_{\left(  A_{n},i_{n}\right)  }\right)  =K_{j_{1}%
i_{1}\cdots i_{n}}^{\phantom{j_{1} i_{1} \cdots i_{n}}j_{1}}%
\operatorname*{Str}\left(  \bm{T}_{A_{1}}\cdots\bm{T}_{A_{n}%
}\right)  .
\end{equation}

En general, esta expresi\'{o}n tiene menos componentes que~(\ref{TenInv0F}).
Para visualizar esto es conveniente considerar el caso cuando el
semigrupo $S$ contiene un elemento identidad $\lambda_{0}=e$, y cada
$\lambda_{i}$ aparece s\'{o}lo una vez en cada fila y en cada columna de la
tabla de multiplicaci\'{o}n del grupo (i.e. para todo $\lambda_{i},\lambda
_{j}\neq e$, tenemos $\lambda_{i}\lambda_{j}\neq\lambda_{i}$ y $\lambda
_{i}\lambda_{j}\neq\lambda_{j}$). En este caso, $K_{j_{1}i_{1}\cdots i_{n}%
}^{\phantom{j_{1} i_{1} \cdots i_{n}}j_{1}}=K_{i_{1}\cdots i_{n}%
}^{\phantom{i_{1} \cdots i_{n}}0}$, y las \'{u}nicas componentes no nulas de
la supertraza son
\begin{equation}
\operatorname*{STr}\left(  \bm{T}_{\left(  A_{1},i_{1}\right)  }%
\cdots\bm{T}_{\left(  A_{n},i_{n}\right)  }\right)  =K_{i_{1}\cdots
i_{n}}^{\phantom{i_{1} \cdots i_{n}}0}\operatorname*{Str}\left(
\bm{T}_{A_{1}}\cdots\bm{T}_{A_{n}}\right)  ,
\end{equation}
lo cual corresponde a la elecci\'{o}n $\alpha_{j}=\delta_{j}^{0}$
en~(\ref{TenInv0F}). Todos los otros $\alpha_{j}$ han sido fijados iguales a cero.

El semigrupo $S=S_{\mathrm{E}}^{\left(  N\right)  }$ usado para reproducir las
expansiones de MC de la Ref.~\cite{deAz02} cumple con las condiciones
se\~{n}aladas en el p\'{a}rrafo anterior. En este caso, $K_{i_{1}\cdots i_{n}}^{\phantom{i_{1} \cdots i_{n}}0}=\delta_{i_{1}+\dotsb+i_{n}}^{0}$ y, por lo
tanto, la \'{u}nica componente no nula de la supertraza es
\begin{equation}
\operatorname*{STr}\left(  \bm{T}_{\left(  A_{1},0\right)  }%
\cdots\bm{T}_{\left(  A_{n},0\right)  }\right)  =\operatorname*{Str}%
\left(  \bm{T}_{A_{1}}\cdots\bm{T}_{A_{n}}\right)  .
\end{equation}

La superioridad del tensor invariante~(\ref{TenInv0F}) con respecto a la
(s\'{u}per)tra\-za es ahora evidente; en el caso $\mathfrak{G}=S_{\mathrm{E}%
}^{\left(  N\right)  }\otimes\mathfrak{g}$, la (s\'{u}per)traza proporciona
s\'{o}lo una repetici\'{o}n trivial del tensor invariante para $\mathfrak{g}$,
y s\'{o}lo una parte de ella en el caso de una sub\'{a}lgebra resonante.

En el cap\'{\i}tulo~\ref{ch:TGFTMAlg} usaremos los teoremas de esta secci\'{o}n para escribir una acci\'{o}n transgresora en $d=11$ para el \'{a}lgebra~M, obtenida como el $0_{S}$-forzamiento del \'{a}lgebra $S_{\mathrm{E}}^{\left( 2 \right)}$-expandida $\mathfrak{G} = S_{\mathrm{E}}^{\left( 2 \right)} \otimes \mathfrak{osp} \left( 32|1 \right)$.

\chapter{\label{ch:TGFTMAlg}Teor\'{\i}a de Gauge para el \'{a}lgebra~M en $\lowercase{d}=11$}
\chaptermark{Teor\'{\i}a de Gauge para el \'{a}lgebra~M}

En este cap\'{\i}tulo convergen los dos cauces principales descritos en el transcurso de la Tesis; las formas de transgresi\'{o}n y la $S$-expansi\'{o}n de \'{a}lgebras de Lie.

Las formas de transgresi\'{o}n proveen de un marco general para una
teor\'{\i}a de gauge con un lagrangeano completamente invariante. Conocemos
las ecuaciones de movimiento, las condiciones de borde, las cargas conservadas
y hemos introducido adem\'{a}s un m\'{e}todo de separaci\'{o}n en subespacios
que ayuda a interpretar f\'{\i}sicamente el lagrangeano.

El m\'{e}todo de $S$-expansi\'{o}n de \'{a}lgebras de Lie, y en particular los
teoremas demostrados en la secci\'{o}n~\ref{sec:TenInvSExp}, provee de un
tensor invariante para un \'{a}lgebra $S$-expandida, completando as\'{\i} la
corta lista de elementos que definen una teor\'{\i}a de gauge transgresora.

En las secciones siguientes presentamos una teor\'{\i}a de gauge para el
\'{a}lgebra~M (a menudo denotada sencillamente por $\mathfrak{M}$) en $d=11$
cuyo lagrangeano es una forma de transgresi\'{o}n. El tensor invariante
utilizado proviene de considerar $\mathfrak{M}$ como una $S$-expansi\'{o}n del
\'{a}lgebra ortosimpl\'{e}ctica $\mathfrak{osp}\left(  32|1\right)  $ (ver secci\'{o}n~\ref{sec:MAlgSExp}).

Este cap\'{\i}tulo est\'{a} basado en la Ref.~\cite{Iza06c}.

\section{\label{sec:MAlg}El \'{A}lgebra~M}

El \'{a}lgebra~M~\cite{Tow95} es la extensi\'{o}n maximal en $d=11$ del
\'{a}lgebra de s\'{u}per Poincar\'{e} en $d=4$,
\begin{equation}
\left\{  \bm{Q},\bar{\bm{Q}}\right\}  =2\gamma^{a}%
\bm{P}_{a}.
\end{equation}
El anticonmutador entre generadores fermi\'{o}nicos en $\mathfrak{M}$ tiene la
forma
\begin{equation}
\left\{  \bm{Q},\bar{\bm{Q}}\right\}  =\frac{1}{8}\left(
\Gamma^{a}\bm{P}_{a}-\frac{1}{2}\Gamma^{ab}\bm{Z}_{ab}%
+\frac{1}{5!}\Gamma^{abcde}\bm{Z}_{abcde}\right)  , \label{qqm}%
\end{equation}
donde $\bm{P}_{a}$ son traslaciones y $\bm{Z}_{ab}$ y
$\bm{Z}_{abcde}$ son cargas centrales. Estas `cargas centrales' son en
realidad tensores de Lorentz, como muestran las relaciones de conmutaci\'{o}n
\begin{align}
\left[  \bm{J}^{ab},\bm{Z}_{cd}\right]   &  =\delta
_{ecd}^{abf}\bm{Z}_{\phantom{e}f}^{e},\\
\left[  \bm{J}^{ab},\bm{Z}_{c_{1}\cdots c_{5}}\right]   &
=\frac{1}{4!}\delta_{dc_{1}\cdots c_{5}}^{abe_{1}\cdots e_{4}}\bm{Z}%
_{\phantom{d}e_{1}\cdots e_{4}}^{d}.
\end{align}

El calificativo de `maximal' dado a $\mathfrak{M}$ se justifica mediante el
siguiente argumento. Un spinor de Dirac en $d=11$ tiene 32 componentes. El
anticonmutador entre dos generadores fermi\'{o}nicos tiene la
forma\footnote{El factor de 2 es puramente convencional.} $\left\{
\bar{\bm{Q}}_{\alpha},\bar{\bm{Q}}_{\beta}\right\}
=2\bm{P}_{\alpha\beta}$, donde $\bm{P}_{\alpha\beta}$ es una
matriz sim\'{e}trica de $32\times32$. Esto significa que puede ser expandida
en la base de las matrices de Dirac (ver Ap\'{e}ndice~\ref{Ap:DiracDirac}), usando s\'{o}lo aquellas que son sim\'{e}tricas: $\Gamma_{a}$, $\Gamma_{ab}$ y $\Gamma_{abcde}$. La
expansi\'{o}n de $\bm{P}_{\alpha\beta}$ es precisamente lo que se
muestra en el lado derecho de la ec.~(\ref{qqm}). Las componentes de $\bm{P}_{\alpha
\beta}$ deben ser tensores de Lorentz por consistencia, pero aparte de esto
deben ser abelianos para reproducir el comportamiento de $\bm{P}_{a}$
en $d=4$.

La lista completa con todas las relaciones de (anti)conmutaci\'{o}n del \'{a}lgebra~M aparece en el cuadro~\ref{tab:MAlg} (p\'{a}g.~\pageref{tab:MAlg}).

Con 583 generadores bos\'{o}nicos (11 traslaciones $\bm{P}_{a}$, 55
rotaciones de Lorentz $\bm{J}_{ab}$, 55 componentes de la carga
central $\bm{Z}_{ab}$ y 462 de la carga central $\bm{Z}%
_{abcde}$), el \'{a}lgebra~M puede ser considerada como una expansi\'{o}n del
\'{a}lgebra ortosimpl\'{e}ctica $\mathfrak{osp}\left(  32|1\right)  $, que
posee s\'{o}lo 528 (ver cap\'{\i}tulo~\ref{ch:expansion} y las Refs.~\cite{deAz02,Iza06b}).

Esta relaci\'{o}n entre \'{a}lgebras es esencial para escribir un lagrangeano,
pues nos permite extraer un tensor invariante para $\mathfrak{M}$ a partir de
un tensor invariante para $\mathfrak{osp}\left(  32|1\right)  $.

\section{\label{sec:MAlgTenInv}El Tensor Invariante}

\subsection{\label{sec:TIrel}Relaci\'{o}n entre tensores invariantes para dos \'{a}l\-gebras}

Como se explica en detalle en la secci\'{o}n~\ref{sec:MAlgSExp}, el
\'{a}lgebra~M corresponde a una $S_{\mathrm{E}}^{\left(  2\right)  }%
$-expansi\'{o}n de $\mathfrak{osp}\left(  32|1\right)  $, donde $S_{\mathrm{E}%
}^{\left(  2\right)  }$ es el semigrupo de elementos $S_{\mathrm{E}}^{\left(
2\right)  }=\left\{  \lambda_{0},\lambda_{1},\lambda_{2},\lambda_{3}\right\}
$, con el producto (ver def.~\ref{def:SEN})
\begin{equation}
\lambda_{\alpha} \lambda_{\beta} =
\left\{
\begin{array}[c]{cl}
\lambda_{\alpha+\beta}, & \text{cuando } \alpha+\beta \leq 2 \\
\lambda_{3}, & \text{cuando } \alpha+\beta \geq 3
\end{array}
\right. .
\end{equation}

La relaci\'{o}n entre los generadores de ambas \'{a}lgebras est\'{a} dada en
el cuadro~\ref{tab:mosp} (p\'{a}g.~\pageref{tab:mosp}).

De acuerdo al Teorema~\ref{th:TenInv0F} (ver secci\'{o}n~\ref{sec:TenInvSExp},
p\'{a}g.~\pageref{th:TenInv0F}), un tensor sim\'{e}trico\footnote{El
Teorema~\ref{th:TenInv0F} proporciona un tensor invariante general, no
necesariamente sim\'{e}trico. Si el tensor invariante del \'{a}lgebra original
es adem\'{a}s sim\'{e}trico, entonces tambi\'{e}n lo ser\'{a} el del
\'{a}lgebra $S$-expandida (o el de su $0_{S}$-forzamiento). Esto es una
consecuencia directa de la abelianidad del semigrupo $S$.} invariante para
$\mathfrak{M}$ es encontrado en la expresi\'{o}n
\begin{equation}
\left\langle \bm{T}_{\left(  A_{1},i_{1}\right)  }\cdots
\bm{T}_{\left(  A_{n},i_{n}\right)  }\right\rangle =\alpha_{j}%
K_{i_{1}\cdots i_{n}}^{\phantom{i_{1} \cdots i_{n}}j}\left\langle
\bm{T}_{A_{1}}\cdots\bm{T}_{A_{n}}\right\rangle ,
\label{TenInv0F2}%
\end{equation}
donde $\alpha_{j}$, $j=0,1,2$, son constantes arbitrarias y $K_{i_{1}\cdots
i_{n}}^{\phantom{i_{1} \cdots i_{n}}j}$ es el $n$-selector para el semigrupo
$S=S_{\mathrm{E}}^{\left(  2\right)  }$. Este $n$-selector tiene la forma
\begin{equation}
K_{i_{1}\cdots i_{n}}^{\phantom{i_{1} \cdots i_{n}}j}=\delta_{i_{1}%
+\dotsb+i_{n}}^{j}, \label{KSEN}%
\end{equation}
donde $\delta_{\sigma}^{\rho}$ es la delta de Kronecker. Reemplazando
el $n$-selector~(\ref{KSEN}) en~(\ref{TenInv0F2}), encontramos que las
\'{u}nicas componentes no nulas del tensor invariante para el \'{a}lgebra~M
son
\begin{align}
\left\langle \bm{J}_{a_{1}b_{1}}\cdots\bm{J}_{a_{6}b_{6}%
}\right\rangle _{\mathrm{M}}  &  =\alpha_{0}\left\langle \bm{J}%
_{a_{1}b_{1}}\cdots\bm{J}_{a_{6}b_{6}}\right\rangle _{\mathfrak{osp}%
},\label{itmalg1}\\
\left\langle \bm{J}_{a_{1}b_{1}}\cdots\bm{J}_{a_{5}b_{5}%
}\bm{P}_{c}\right\rangle _{\mathrm{M}}  &  =\alpha_{2}\left\langle
\bm{J}_{a_{1}b_{1}}\cdots\bm{J}_{a_{5}b_{5}}\bm{P}%
_{c}\right\rangle _{\mathfrak{osp}},\\
\left\langle \bm{J}_{a_{1}b_{1}}\cdots\bm{J}_{a_{5}b_{5}%
}\bm{Z}_{cd}\right\rangle _{\mathrm{M}}  &  =\alpha_{2}\left\langle
\bm{J}_{a_{1}b_{1}}\cdots\bm{J}_{a_{5}b_{5}}\bm{J}%
_{cd}\right\rangle _{\mathfrak{osp}},\\
\left\langle \bm{J}_{a_{1}b_{1}}\cdots\bm{J}_{a_{5}b_{5}%
}\bm{Z}_{c_{1}\cdots c_{5}}\right\rangle _{\mathrm{M}}  &  =\alpha
_{2}\left\langle \bm{J}_{a_{1}b_{1}}\cdots\bm{J}_{a_{5}b_{5}%
}\bm{Z}_{c_{1}\cdots c_{5}}\right\rangle _{\mathfrak{osp}},\\
\left\langle \bm{QJ}_{a_{1}b_{1}}\cdots\bm{J}_{a_{4}b_{4}}%
\bar{\bm{Q}}\right\rangle _{\mathrm{M}}  &  =\alpha_{2}\left\langle
\bm{QJ}_{a_{1}b_{1}}\cdots\bm{J}_{a_{4}b_{4}}\bar
{\bm{Q}}\right\rangle _{\mathfrak{osp}}, \label{itmalg5}%
\end{align}
donde $\alpha_{0}$ y $\alpha_{2}$ son constantes arbitrarias.

Es interesante notar que este tensor invariante, a\'{u}n teniendo m\'{a}s
componentes no nulas que la supertraza [la cu\'{a}l consistir\'{\i}a s\'{o}lo
de~(\ref{itmalg1}), ver secci\'{o}n~\ref{sec:TenInvSExp}], es sin embargo mucho m\'{a}s peque\~{n}o que aquel para $\mathfrak{osp}\left(  32|1\right)  $. Estas es una caracter\'{\i}stica
com\'{u}n de las \'{a}lgebras $0_{S}$-forzadas, y ejerce una profunda
influencia en la din\'{a}mica producida por la teor\'{\i}a, como veremos en la
secci\'{o}n~\ref{sec:MAlgDyn}.

\subsection{\label{sec:TIosp}Tensor Invariante para $\mathfrak{osp}\left(
32|1\right)  $}

Para obtener un tensor invariante para el \'{a}lgebra~M es necesario partir
reconsiderando $\mathfrak{osp}\left(  32|1\right)  $.

El supergrupo $\mathrm{OSp}\left(  32|1\right)  $ es definido como aquel que
deja invariante una m\'{e}trica de la forma
\begin{equation}
G =
\left[
\begin{array}[c]{cc}
C_{\alpha\beta} & 0 \\
0 & 1
\end{array}
\right] ,
\end{equation}
con $\alpha,\beta=1,\dotsc,32$ y $C_{\beta\alpha}=-C_{\alpha\beta}$. El
subgrupo\ que deja invariante la m\'{e}trica antisim\'{e}trica $C_{\alpha
\beta}$ es $\mathrm{Sp}\left(  32\right)  $. El \'{a}lgebra $\mathfrak{sp}%
\left(  32\right)  $ consiste de todas las matrices \emph{sim\'{e}tricas} de
$32\times32$, y su rango es $\frac{1}{2} \times 32 \times 33 = 528$. Una base para
$\mathfrak{sp}\left(  32\right)  $ es provista por las matrices de Dirac
$\Gamma_{a}$, $\Gamma_{ab}$ y $\Gamma_{abcde}$ en $d=11$; todas ellas son
sim\'{e}tricas (cuando se les baja un \'{\i}ndice con $C_{\alpha\beta}$; ver
cuadro~\ref{tab:Dirac} en el Ap\'{e}ndice~\ref{Ap:DiracSym}), linealmente
independientes, y su n\'{u}mero total es el correcto:
\begin{equation}
11+\binom{11}{2}+\binom{11}{5}=528.
\end{equation}

Una representaci\'{o}n en t\'{e}rminos de supermatrices para la
super\'{a}lgebra $\mathfrak{osp}\left(  32|1\right)  $ es
\begin{align}
\bm{P}_{a}  &  =\left[
\begin{array}
[c]{cc}%
\frac{1}{2}\left(  \Gamma_{a}\right)  _{\phantom{\alpha}\beta}^{\alpha} & 0\\
0 & 0
\end{array}
\right]  ,\\
\bm{J}_{ab}  &  =\left[
\begin{array}
[c]{cc}%
\frac{1}{2}\left(  \Gamma_{ab}\right)  _{\phantom{\alpha}\beta}^{\alpha} & 0\\
0 & 0
\end{array}
\right]  ,\\
\bm{Z}_{abcde}  &  =\left[
\begin{array}
[c]{cc}%
\frac{1}{2}\left(  \Gamma_{abcde}\right)  _{\phantom{\alpha}\beta}^{\alpha} &
0\\
0 & 0
\end{array}
\right]  ,\\
\bm{Q}^{\gamma}  &  =\left[
\begin{array}
[c]{cc}%
0 & C^{\gamma\alpha}\\
\delta_{\beta}^{\gamma} & 0
\end{array}
\right]  .
\end{align}

La supertraza supersimetrizada del producto de seis de estas supermatrices
proporciona un tensor sim\'{e}trico invariante de rango apropiado para la
construcci\'{o}n de una acci\'{o}n transgresora en $d=11$.

Para facilitar la supersimetrizaci\'{o}n es conveniente `amarrar' los
generadores del \'{a}lgebra a par\'{a}metros apropiados, i.e. n\'{u}meros
reales (conmutantes) en el caso de los generadores bos\'{o}nicos y n\'{u}meros
de Grassmann (anticonmutantes) en el caso de los generadores fermi\'{o}nicos.
Este sencillo truco reduce el problema de la supersimetrizaci\'{o}n al de una
simetrizaci\'{o}n ordinaria.

Sean $A_{1},\dotsc,A_{6}$ seis supermatrices de la forma
\begin{equation}
A=\left[
\begin{array}
[c]{cc}%
A_{\phantom{\alpha}\beta}^{\alpha} & A^{\alpha}\\
A_{\beta} & 0
\end{array}
\right]  .
\end{equation}
Denotaremos por $\left\{  A_{1}\cdots A_{6}\right\}  $ el producto
completamente simetrizado de estas seis supermatrices. Este producto puede
calcularse usando la f\'{o}rmula iterativa
\begin{align}
\left\{  A_{1}\right\}   &  =A_{1},\\
\left\{  A_{1}\cdots A_{n}\right\}   &  =\frac{1}{n}\sum_{p=1}^{n}%
A_{p}\left\{  A_{1}\cdots\hat{A}_{p}\cdots A_{n}\right\}  ,
\label{simetrizame}%
\end{align}
donde la notaci\'{o}n $\hat{A}_{p}$ indica que la matriz $A_{p}$ debe ser omitida.

La supertraza supersimetrizada $\left\langle A_{1}\cdots A_{6}\right\rangle $
corresponde entonces a la supertraza del producto completamente simetrizado
$\left\{  A_{1}\cdots A_{6}\right\}  $:
\begin{equation}
\left\langle A_{1}\cdots A_{6}\right\rangle =\operatorname{STr}\left\{
A_{1}\cdots A_{6}\right\}  .
\end{equation}

Cuando las seis matrices $A_{1},\dotsc,A_{6}$ son bos\'{o}nicas, la supertraza
se reduce trivialmente a la traza\footnote{Nos referimos aqu\'{\i}, por
supuesto, a la traza del bloque superior izquierdo de la supermatriz. La
notaci\'{o}n es un tanto inexacta, pero confiamos en que no produce
confusi\'{o}n.}
\begin{equation}
\left\langle A_{1}\cdots A_{6}\right\rangle =\operatorname{Tr}\left\{
A_{1}\cdots A_{6}\right\}  .
\end{equation}
En este caso, la propiedad c\'{\i}clica de la traza puede ser usada para
simplificar ligeramente esta expresi\'{o}n:
\begin{equation}
\left\langle A_{1}\cdots A_{6}\right\rangle =\operatorname{Tr}\left(  \left\{
A_{1}\cdots A_{5}\right\}  A_{6}\right)  .
\end{equation}

Cuando dos de las matrices son fermi\'{o}nicas, $A_{5}=\bm{\chi}%
=\bar{\chi}_{\alpha}\bm{Q}^{\alpha}$, $A_{6}=\bm{\zeta}%
=\bar{\zeta}_{\alpha}\bm{Q}^{\alpha}$, entonces es posible demostrar
[usando~(\ref{simetrizame})] que la supertraza supersimetrizada puede
calcularse a partir de la expresi\'{o}n
\begin{equation}
\left\langle \bm{\chi\zeta}A_{1}\cdots A_{4}\right\rangle =-\frac
{2}{5}\bar{\chi}\left\{  A_{1}\cdots A_{4}\right\}  \zeta.
\end{equation}

Los dos casos considerados cubren todas las posibilidades que necesitamos de
acuerdo a las ecs.~(\ref{itmalg1})--(\ref{itmalg5}). De estas ecuaciones
resulta tambi\'{e}n claro que es necesario calcular el producto simetrizado de
hasta cinco matrices bos\'{o}nicas de la forma $A=A^{ab}\Gamma_{ab}$.

Ocupando la ec.~(\ref{simetrizame}) y la f\'{o}rmula~(\ref{ABD}) para el
producto de matrices de Dirac contra\'{\i}das con tensores arbitrarios es
posible calcular los siguientes productos simetrizados (con $A_{k}=A_{k}%
^{ab}\Gamma_{ab}$):
\begin{equation}
\left\{  A_{1}A_{2}\right\}  =2\operatorname{Tr}\left(  A_{1}A_{2}\right)
\mathbbm{1}+A_{1}A_{2}\Gamma_{\left[  4\right]  }, \label{s2}%
\end{equation}%
\begin{equation}
\left\{  A_{1}A_{2}A_{3}\right\}  =\sum_{\left\langle ijk\right\rangle
}\left(  \operatorname{Tr}\left(  A_{i}A_{j}\right)  A_{k}-\frac{4}{3}\left[
A_{i}A_{j}A_{k}\right]  \right)  \Gamma_{\left[  2\right]  }+A_{1}A_{2}%
A_{3}\Gamma_{\left[  6\right]  }, \label{s3}%
\end{equation}%
\begin{align}
\left\{  A_{1}\cdots A_{4}\right\} = & \sum_{\left\langle ijkl\right\rangle
}\left(  \frac{1}{2}\operatorname{Tr}\left(  A_{i}A_{j}\right)
\operatorname{Tr}\left(  A_{k}A_{l}\right)  -\frac{2}{3}\operatorname{Tr}%
\left(  A_{i}A_{j}A_{k}A_{l}\right)  \right)  \mathbbm{1}+\nonumber\\
&  +\sum_{\left\langle ijkl\right\rangle }\left(  \frac{1}{2}\operatorname{Tr}%
\left(  A_{i}A_{j}\right)  A_{k}A_{l}-\frac{4}{3}A_{i}\left[  A_{j}A_{k}%
A_{l}\right]  \right)  \Gamma_{\left[  4\right]  }+ \nonumber \\
&  +\left(  A_{1}\cdots A_{4}\right)  \Gamma_{\left[  8\right]  }, \label{s4}%
\end{align}%
\begin{align}
\left\{  A_{1}\cdots A_{5}\right\} = & \sum_{\left\langle ijklm\right\rangle
}\left(  \frac{1}{2}\operatorname{Tr}\left(  A_{i}A_{j}\right)
\operatorname{Tr}\left(  A_{k}A_{l}\right)  A_{m}-\frac{2}{3}\operatorname{Tr}%
\left(  A_{i}A_{j}A_{k}A_{l}\right)  A_{m}+\right. \nonumber\\
&  \left.  -\frac{4}{3}\operatorname{Tr}\left(  A_{i}A_{j}\right)  \left[
A_{k}A_{l}A_{m}\right]  +\frac{32}{15}\left[  A_{i}A_{j}A_{k}A_{l}%
A_{m}\right]  \right)  \Gamma_{\left[  2\right]  }+\nonumber\\
&  +\sum_{\left\langle ijklm\right\rangle }\left(  \frac{1}{6}%
\operatorname{Tr}\left(  A_{i}A_{j}\right)  A_{k}A_{l}A_{m}-\frac{2}{3}%
A_{i}A_{j}\left[  A_{k}A_{l}A_{m}\right]  \right)  \Gamma_{\left[  6\right]
}+\nonumber\\
&  +\left(  A_{1}\cdots A_{5}\right)  \Gamma_{\left[  10\right]  }. \label{s5}%
\end{align}

Para simplificar la escritura de las ecs.~(\ref{s2})--(\ref{s5}) hemos omitido
todos los \'{\i}ndices de Lorentz, los cuales est\'{a}n contra\'{\i}dos en
orden can\'{o}nico (i.e. $A_{i}A_{j}\Gamma_{\left[  4\right]  }=A_{i}%
^{ab}A_{j}^{cd}\Gamma_{abcd}$), y hemos introducido las abreviaciones
\begin{align}
\operatorname{Tr}\left(  A_{i_{1}}\cdots A_{i_{n}}\right)   &  =\left(
A_{i_{1}}\right)  _{\phantom{c_{1}}c_{2}}^{c_{1}}\left(  A_{i_{1}}\right)
_{\phantom{c_{2}}c_{3}}^{c_{2}}\cdots\left(  A_{i_{n}}\right)
_{\phantom{c_{n}}c_{1}}^{c_{n}},\\
\left[  A_{i_{1}}\cdots A_{i_{n}}\right]  ^{ab}  &  =\left(  A_{i_{1}}\right)
_{\phantom{a}c_{1}}^{a}\left(  A_{i_{2}}\right)  _{\phantom{c_{1}}c_{2}%
}^{c_{1}}\cdots\left(  A_{i_{n}}\right)  ^{c_{n-1}b}.
\end{align}
El s\'{\i}mbolo $\left\langle i_{1}\cdots i_{n}\right\rangle $ usado en las
sumatorias indica que debe sumarse sobre $i_{1},\dotsc,i_{n}=1,\dotsc,n$, con
la restricci\'{o}n de que todos los $i_{k}$ sean \emph{distintos}. Esto
representa una manera sencilla de implementar la simetrizaci\'{o}n en
$i_{1}\cdots i_{n}$.

Usando los productos simetrizados~(\ref{s2})--(\ref{s5}) y las propiedades de
las matrices de Dirac en $d=11$ (ver Ap\'{e}ndice~\ref{Ap:Cliff}) uno obtiene
las siguientes componentes del tensor invariante para la super\'{a}lgebra
$\mathfrak{osp}\left(  32|1\right)  $:
\begin{equation}
\left\langle \bm{J}^{5}\bm{P}\right\rangle _{\mathfrak{osp}%
}=\frac{1}{2}\varepsilon_{a_{1}\cdots a_{11}}L_{1}^{a_{1}a_{2}}\cdots
L_{5}^{a_{9}a_{10}}B_{1}^{a_{11}}, \label{L5B1}%
\end{equation}
\begin{align}
\left\langle \bm{J}^{6}\right\rangle _{\mathfrak{osp}} = & \frac{1}{3} \sum_{\left\langle i_{1}\cdots i_{6}\right\rangle }\left[  \frac{1}{4}\operatorname*{Tr}\left(  L_{i_{1}}L_{i_{2}}\right)  \operatorname*{Tr} \left(  L_{i_{3}}L_{i_{4}}\right)  \operatorname*{Tr}\left(  L_{i_{5}} L_{i_{6}} \right)  +\right. \nonumber\\
&  \left.  -\operatorname*{Tr}\left(  L_{i_{1}}L_{i_{2}}L_{i_{3}}L_{i_{4}%
}\right)  \operatorname*{Tr}\left(  L_{i_{5}}L_{i_{6}}\right)  +\frac{16}%
{15}\operatorname*{Tr}\left(  L_{i_{1}}L_{i_{2}}L_{i_{3}}L_{i_{4}}L_{i_{5}%
}L_{i_{6}}\right)  \right]  , \label{L6}%
\end{align}
\begin{align}
\left\langle \bm{J}^{5}\bm{Z}\right\rangle _{\mathfrak{osp}}
= & \frac{1}{3}\varepsilon_{a_{1}\cdots a_{11}}\sum_{\left\langle i_{1}\cdots
i_{5}\right\rangle }\left[  -\frac{5}{4}L_{i_{1}}^{a_{1}a_{2}}\cdots L_{i_{4}%
}^{a_{7}a_{8}}\left(  L_{i_{5}}\right)  _{bc}B_{5}^{bca_{9}a_{10}a_{11}%
}+\right. \nonumber\\
&  +10L_{i_{1}}^{a_{1}a_{2}}L_{i_{2}}^{a_{3}a_{4}}L_{i_{3}}^{a_{5}a_{6}%
}\left(  L_{i_{4}}\right)  _{\phantom{a_{7}}b}^{a_{7}}\left(  L_{i_{5}%
}\right)  _{\phantom{a_{8}}c}^{a_{8}}B_{5}^{bca_{9}a_{10}a_{11}}+\nonumber\\
&  +\frac{1}{4}L_{i_{1}}^{a_{1}a_{2}}L_{i_{2}}^{a_{3}a_{4}}L_{i_{3}}%
^{a_{5}a_{6}}B_{5}^{a_{7}\cdots a_{11}}\operatorname*{Tr}\left(  L_{i_{4}%
}L_{i_{5}}\right)  +\nonumber\\
&  \left.  -L_{i_{1}}^{a_{1}a_{2}}L_{i_{2}}^{a_{3}a_{4}}\left[  L_{i_{3}%
}L_{i_{4}}L_{i_{5}}\right]  ^{a_{5}a_{6}}B_{5}^{a_{7}\cdots a_{11}}\right]  ,
\label{L5B5}
\end{align}
\begin{align}
\left\langle \bm{QJ}^{4}\bar{\bm{Q}}\right\rangle
_{\mathfrak{osp}} = & -\frac{1}{240}\varepsilon_{a_{1}\cdots a_{8}abc}%
L_{1}^{a_{1}a_{2}}\cdots L_{4}^{a_{7}a_{8}}\left(  \bar{\chi}\Gamma^{abc}%
\zeta\right)  +\nonumber\\
&  +\frac{1}{60}\sum_{\left\langle i_{1}\cdots i_{4}\right\rangle }\left[
\frac{3}{4}\operatorname*{Tr}\left(  L_{i_{1}}L_{i_{2}}\right)  L_{i_{3}%
}^{a_{1}a_{2}}L_{i_{4}}^{a_{3}a_{4}}\left(  \bar{\chi}\Gamma_{a_{1}\cdots
a_{4}}\zeta\right)  +\right. \nonumber\\
&  -2L_{i_{1}}^{a_{1}a_{2}}\left[  L_{i_{2}}L_{i_{3}}L_{i_{4}}\right]
^{a_{3}a_{4}}\left(  \bar{\chi}\Gamma_{a_{1}\cdots a_{4}}\zeta\right)
+\nonumber\\
&  \left.  +\frac{3}{4}\operatorname*{Tr}\left(  L_{i_{1}}L_{i_{2}}\right)
\operatorname*{Tr}\left(  L_{i_{3}}L_{i_{4}}\right)  \bar{\chi}\zeta
-\operatorname*{Tr}\left(  L_{i_{1}}L_{i_{2}}L_{i_{3}}L_{i_{4}}\right)
\bar{\chi}\zeta\right]  , \label{L4FF}%
\end{align}
donde, para hacer m\'{a}s llevadera la vida, hemos introducido las
abreviaciones
\begin{align}
\left\langle \bm{J}^{5}\bm{P}\right\rangle _{\mathfrak{osp}}
&  =L_{1}^{a_{1}b_{1}}\cdots L_{5}^{a_{5}b_{5}}B_{1}^{c}\left\langle
\bm{J}_{a_{1}b_{1}}\cdots\bm{J}_{a_{5}b_{5}}\bm{P}%
_{c}\right\rangle _{\mathfrak{osp}},\label{J5Posp}\\
\left\langle \bm{J}^{6}\right\rangle _{\mathfrak{osp}}  &
=L_{1}^{a_{1}b_{1}}\cdots L_{6}^{a_{6}b_{6}}\left\langle \bm{J}%
_{a_{1}b_{1}}\cdots\bm{J}_{a_{6}b_{6}}\right\rangle _{\mathfrak{osp}%
},\\
\left\langle \bm{J}^{5}\bm{Z}\right\rangle _{\mathfrak{osp}}
&  =L_{1}^{a_{1}b_{1}}\cdots L_{5}^{a_{5}b_{5}}B_{5}^{c_{1}\cdots c_{5}%
}\left\langle \bm{J}_{a_{1}b_{1}}\cdots\bm{J}_{a_{5}b_{5}%
}\bm{Z}_{c_{1}\cdots c_{5}}\right\rangle _{\mathfrak{osp}},\\
\left\langle \bm{QJ}^{4}\bar{\bm{Q}}\right\rangle
_{\mathfrak{osp}}  &  =L_{1}^{a_{1}b_{1}}\cdots L_{4}^{a_{4}b_{4}}\bar{\chi
}_{\alpha}\zeta^{\beta}\left\langle \bm{Q}^{\alpha}\bm{J}%
_{a_{1}b_{1}}\cdots\bm{J}_{a_{4}b_{4}}\bar{\bm{Q}}_{\beta
}\right\rangle _{\mathfrak{osp}}. \label{QJ4Qosp}%
\end{align}
Aqu\'{\i} $L_{i}^{ab}$ son tensores de Lorentz antisim\'{e}tricos en $ab$, en
tanto que $\chi$ y $\zeta$ son spinores de Majorana.

\subsection{\label{sec:TIM}Tensor Invariante para el \'{a}lgebra~M}

A partir de las componentes~(\ref{J5Posp})--(\ref{QJ4Qosp}) del tensor
invariante para $\mathfrak{osp}\left(  32|1\right)  $ resulta directo escribir
las componentes no nulas del tensor invariante inducido para el \'{a}lgebra~M.
Recurriendo a~(\ref{itmalg1})--(\ref{itmalg5}), encontramos
\begin{align}
\left\langle \bm{J}^{6}\right\rangle _{\mathrm{M}} = & \frac{1}%
{3}\alpha_{0}\sum_{\left\langle i_{1}\cdots i_{6}\right\rangle }\left[
\frac{1}{4}\operatorname*{Tr}\left(  L_{i_{1}}L_{i_{2}}\right)
\operatorname*{Tr}\left(  L_{i_{3}}L_{i_{4}}\right)  \operatorname*{Tr}\left(
L_{i_{5}}L_{i_{6}}\right)  +\right. \nonumber\\
&  \left.  -\operatorname*{Tr}\left(  L_{i_{1}}L_{i_{2}}L_{i_{3}}L_{i_{4}%
}\right)  \operatorname*{Tr}\left(  L_{i_{5}}L_{i_{6}}\right)  +\frac{16}%
{15}\operatorname*{Tr}\left(  L_{i_{1}}L_{i_{2}}L_{i_{3}}L_{i_{4}}L_{i_{5}%
}L_{i_{6}}\right)  \right]  , \label{J6M}%
\end{align}
\begin{equation}
\left\langle \bm{J}^{5}\bm{P}\right\rangle _{\mathrm{M}}%
=\frac{1}{2}\alpha_{2}\varepsilon_{a_{1}\cdots a_{11}}L_{1}^{a_{1}a_{2}}\cdots
L_{5}^{a_{9}a_{10}}B_{1}^{a_{11}}, \label{J5PM}%
\end{equation}
\begin{align}
\left\langle \bm{J}^{5}\bm{Z}_{2}\right\rangle _{\mathrm{M}}
= & \alpha_{2}\sum_{\left\langle i_{1}\cdots i_{5}\right\rangle }\left[
\frac{1}{2}\operatorname*{Tr}\left(  L_{i_{1}}L_{i_{2}}\right)
\operatorname*{Tr}\left(  L_{i_{3}}L_{i_{4}}\right)  \operatorname*{Tr}\left(
L_{i_{5}}B_{2}\right)  +\right. \nonumber\\
&  -\frac{4}{3}\operatorname*{Tr}\left(  L_{i_{1}}L_{i_{2}}\right)
\operatorname*{Tr}\left(  L_{i_{3}}L_{i_{4}}L_{i_{5}}B_{2}\right)
+\nonumber\\
&  -\frac{2}{3}\operatorname*{Tr}\left(  L_{i_{1}}L_{i_{2}}L_{i_{3}}L_{i_{4}%
}\right)  \operatorname*{Tr}\left(  L_{i_{5}}B_{2}\right)  +\nonumber\\
&  \left.  +\frac{32}{15}\operatorname*{Tr}\left(  L_{i_{1}}L_{i_{2}}L_{i_{3}%
}L_{i_{4}}L_{i_{5}}B_{2}\right)  \right]  , \label{J5Z2M}
\end{align}
\begin{align}
\left\langle \bm{J}^{5}\bm{Z}_{5}\right\rangle _{\mathrm{M}}
= & \frac{1}{3}\alpha_{2}\varepsilon_{a_{1}\cdots a_{11}}\sum_{\left\langle
i_{1}\cdots i_{5}\right\rangle }\left[  -\frac{5}{4}L_{i_{1}}^{a_{1}a_{2}%
}\cdots L_{i_{4}}^{a_{7}a_{8}}\left(  L_{i_{5}}\right)  _{bc}B_{5}%
^{bca_{9}a_{10}a_{11}}+\right. \nonumber\\
&  +10L_{i_{1}}^{a_{1}a_{2}}L_{i_{2}}^{a_{3}a_{4}}L_{i_{3}}^{a_{5}a_{6}%
}\left(  L_{i_{4}}\right)  _{\phantom{a_{7}}b}^{a_{7}}\left(  L_{i_{5}%
}\right)  _{\phantom{a_{8}}c}^{a_{8}}B_{5}^{bca_{9}a_{10}a_{11}}+\nonumber\\
&  +\frac{1}{4}\operatorname*{Tr}\left(  L_{i_{1}}L_{i_{2}}\right)  L_{i_{3}%
}^{a_{1}a_{2}}L_{i_{4}}^{a_{3}a_{4}}L_{i_{5}}^{a_{5}a_{6}}B_{5}^{a_{7}\cdots
a_{11}}+\nonumber\\
&  \left.  -L_{i_{1}}^{a_{1}a_{2}}L_{i_{2}}^{a_{3}a_{4}}\left[  L_{i_{3}%
}L_{i_{4}}L_{i_{5}}\right]  ^{a_{5}a_{6}}B_{5}^{a_{7}\cdots a_{11}}\right]  ,
\label{J5Z5M}
\end{align}
\begin{align}
\left\langle \bm{QJ}^{4}\bar{\bm{Q}}\right\rangle
_{\mathrm{M}} = & -\frac{\alpha_{2}}{240}\varepsilon_{a_{1}\cdots a_{8}%
abc}L_{1}^{a_{1}a_{2}}\cdots L_{4}^{a_{7}a_{8}}\left(  \bar{\chi}\Gamma
^{abc}\zeta\right)  +\nonumber\\
&  +\frac{\alpha_{2}}{60}\sum_{\left\langle i_{1}\cdots i_{4}\right\rangle
}\left[  \frac{3}{4}\operatorname*{Tr}\left(  L_{i_{1}}L_{i_{2}}\right)
L_{i_{3}}^{a_{1}a_{2}}L_{i_{4}}^{a_{3}a_{4}}\left(  \bar{\chi}\Gamma
_{a_{1}\cdots a_{4}}\zeta\right)  +\right. \nonumber\\
&  -2L_{i_{1}}^{a_{1}a_{2}}\left[  L_{i_{2}}L_{i_{3}}L_{i_{4}}\right]
^{a_{3}a_{4}}\left(  \bar{\chi}\Gamma_{a_{1}\cdots a_{4}}\zeta\right)
+\nonumber\\
&  \left.  +\frac{3}{4}\operatorname*{Tr}\left(  L_{i_{1}}L_{i_{2}}\right)
\operatorname*{Tr}\left(  L_{i_{3}}L_{i_{4}}\right)  \bar{\chi}\zeta
-\operatorname*{Tr}\left(  L_{i_{1}}L_{i_{2}}L_{i_{3}}L_{i_{4}}\right)
\bar{\chi}\zeta\right]  , \label{QJ4QM}
\end{align}
donde las abreviaciones correspondientes son ahora
\begin{align}
\left\langle \bm{J}^{6}\right\rangle _{\mathrm{M}}  &  =L_{1}%
^{a_{1}b_{1}}\cdots L_{6}^{a_{6}b_{6}}\left\langle \bm{J}_{a_{1}b_{1}%
}\cdots\bm{J}_{a_{6}b_{6}}\right\rangle _{\mathrm{M}},\label{ab1}\\
\left\langle \bm{J}^{5}\bm{P}\right\rangle _{\mathrm{M}}  &
=L_{1}^{a_{1}b_{1}}\cdots L_{5}^{a_{5}b_{5}}B_{1}^{c}\left\langle
\bm{J}_{a_{1}b_{1}}\cdots\bm{J}_{a_{5}b_{5}}\bm{P}%
_{c}\right\rangle _{\mathrm{M}},\\
\left\langle \bm{J}^{5}\bm{Z}_{2}\right\rangle _{\mathrm{M}}
&  =L_{1}^{a_{1}b_{1}}\cdots L_{5}^{a_{5}b_{5}}B_{2}^{cd}\left\langle
\bm{J}_{a_{1}b_{1}}\cdots\bm{J}_{a_{65}b_{5}}\bm{Z}%
_{cd}\right\rangle _{\mathrm{M}},\\
\left\langle \bm{J}^{5}\bm{Z}_{5}\right\rangle _{\mathrm{M}}
&  =L_{1}^{a_{1}b_{1}}\cdots L_{5}^{a_{5}b_{5}}B_{5}^{c_{1}\cdots c_{5}%
}\left\langle \bm{J}_{a_{1}b_{1}}\cdots\bm{J}_{a_{5}b_{5}%
}\bm{Z}_{c_{1}\cdots c_{5}}\right\rangle _{\mathrm{M}},\\
\left\langle \bm{QJ}^{4}\bar{\bm{Q}}\right\rangle
_{\mathrm{M}}  &  =L_{1}^{a_{1}b_{1}}\cdots L_{4}^{a_{4}b_{4}}\bar{\chi
}_{\alpha}\zeta^{\beta}\left\langle \bm{Q}^{\alpha}\bm{J}%
_{a_{1}b_{1}}\cdots\bm{J}_{a_{4}b_{4}}\bar{\bm{Q}}_{\beta
}\right\rangle _{\mathrm{M}}. \label{ab5}%
\end{align}

En la secci\'{o}n~\ref{sec:MAlgLag} construimos una acci\'{o}n transgresora
para el \'{a}lgebra~M usando el tensor invariante~(\ref{J6M})--(\ref{QJ4QM}).

\subsection{\label{sec:TIrelax}Relajando Constantes de Acoplamiento}

El tensor invariante para el \'{a}lgebra~M expuesto en las ecs.~(\ref{J6M}%
)--(\ref{QJ4QM}) fue calculado a partir de la supertraza supersimetrizada del
producto de seis generadores de $\mathfrak{osp}\left(  32|1\right)  $. En
particular, hemos utilizado matrices de Dirac en $d=11$ para representar el
sector bos\'{o}nico, de manera que las componentes puramente bos\'{o}nicas del
tensor invariante corresponden a su traza simetrizada.

Es posible obtener nuevos tensores invariantes considerando productos
simetrizados de trazas, como en la expresi\'{o}n $\left\langle \bm{F}%
^{p}\right\rangle \left\langle \bm{F}^{n-p}\right\rangle $. Para
barrer todas las posibilidades uno debe considerar las particiones de seis
(que es el orden deseado para el tensor invariante). Un an\'{a}lisis sencillo
muestra que, aparte de la partici\'{o}n $6=6$ ya considerada, los \'{u}nicos
otros casos que hacen una contribuci\'{o}n no nula son $6=4+2$ y $6=2+2+2$. De
este modo, nos vemos conminados a considerar la combinaci\'{o}n lineal
\begin{equation}
\left\langle \cdots\right\rangle _{\mathrm{M}}^{\prime}=\left\langle
\cdots\right\rangle _{6=6}+\beta_{4+2}\left\langle \cdots\right\rangle
_{6=4+2}+\beta_{2+2+2}\left\langle \cdots\right\rangle _{6=2+2+2},
\end{equation}
donde $\beta_{4+2}$ y $\beta_{2+2+2}$ son nuevas constantes arbitrarias (el
coeficiente en frente de $\left\langle \cdots\right\rangle _{6=6}$ puede ser
normalizado a 1 sin p\'{e}rdida de generalidad).

El sorprendente resultado de realizar este ejercicio es que \emph{ning\'{u}n
t\'{e}rmino nuevo aparece en el tensor invariante}; en vez de eso, la
estructura r\'{\i}gida encontrada en~(\ref{J6M})--(\ref{QJ4QM}) se ve relajada
a otra que toma en cuenta las nuevas constantes de acoplamiento $\beta_{4+2}$
y $\beta_{2+2+2}$. Manipulando estas constantes uno encuentra que hay
distintos sectores que son por s\'{\i} mismos invariantes bajo $\mathfrak{M}$, de
modo que es perfectamente razonable asociarlos con acoplamientos diferentes.

El nuevo tensor invariante tiene la forma
\begin{align}
\left\langle \bm{J}^{6}\right\rangle _{\mathrm{M}}^{\prime}
= & \frac{1}{3}\alpha_{0}\sum_{\left\langle i_{1}\cdots i_{6}\right\rangle
}\left[  \frac{1}{4}\gamma_{5}\operatorname*{Tr}\left(  L_{i_{1}}L_{i_{2}%
}\right)  \operatorname*{Tr}\left(  L_{i_{3}}L_{i_{4}}\right)
\operatorname*{Tr}\left(  L_{i_{5}}L_{i_{6}}\right)  +\right. \nonumber\\
&  \left.  -\kappa_{15}\operatorname*{Tr}\left(  L_{i_{1}}L_{i_{2}}L_{i_{3}%
}L_{i_{4}}\right)  \operatorname*{Tr}\left(  L_{i_{5}}L_{i_{6}}\right)
+\frac{16}{15}\operatorname*{Tr}\left(  L_{i_{1}}L_{i_{2}}L_{i_{3}}L_{i_{4}%
}L_{i_{5}}L_{i_{6}}\right)  \right]  , \label{J6M2}
\end{align}
\begin{equation}
\left\langle \bm{J}^{5}\bm{P}\right\rangle _{\mathrm{M}%
}^{\prime}=\frac{1}{2}\alpha_{2}\varepsilon_{a_{1}\cdots a_{11}} L_{1}^{a_{1}a_{2}} \cdots L_{5}^{a_{9}a_{10}}B_{1}^{a_{11}}, \label{J5PM2}
\end{equation}
\begin{align}
\left\langle \bm{J}^{5}\bm{Z}_{2}\right\rangle _{\mathrm{M}%
}^{\prime} = & \alpha_{2}\sum_{\left\langle i_{1}\cdots i_{5}\right\rangle
}\left[  \frac{1}{2}\gamma_{5}\operatorname*{Tr}\left(  L_{i_{1}}L_{i_{2}%
}\right)  \operatorname*{Tr}\left(  L_{i_{3}}L_{i_{4}}\right)
\operatorname*{Tr}\left(  L_{i_{5}}B_{2}\right)  +\right. \nonumber\\
&  -\frac{4}{3}\kappa_{15}\operatorname*{Tr}\left(  L_{i_{1}}L_{i_{2}}\right)
\operatorname*{Tr}\left(  L_{i_{3}}L_{i_{4}}L_{i_{5}}B_{2}\right)
+\nonumber\\
&  -\frac{2}{3}\kappa_{15}\operatorname*{Tr}\left(  L_{i_{1}}L_{i_{2}}%
L_{i_{3}}L_{i_{4}}\right)  \operatorname*{Tr}\left(  L_{i_{5}}B_{2}\right)
+\nonumber\\
&  \left.  +\frac{32}{15}\operatorname*{Tr}\left(  L_{i_{1}}L_{i_{2}}L_{i_{3}%
}L_{i_{4}}L_{i_{5}}B_{2}\right)  \right]  , \label{J5Z2M2}
\end{align}
\begin{align}
\left\langle \bm{J}^{5}\bm{Z}_{5}\right\rangle _{\mathrm{M}%
}^{\prime} = & \frac{1}{3}\alpha_{2}\varepsilon_{a_{1}\cdots a_{11}}%
\sum_{\left\langle i_{1}\cdots i_{5}\right\rangle }\left[  -\frac{5}%
{4}L_{i_{1}}^{a_{1}a_{2}}\cdots L_{i_{4}}^{a_{7}a_{8}}\left(  L_{i_{5}%
}\right)  _{bc}B_{5}^{bca_{9}a_{10}a_{11}}+\right. \nonumber\\
&  +10L_{i_{1}}^{a_{1}a_{2}}L_{i_{2}}^{a_{3}a_{4}}L_{i_{3}}^{a_{5}a_{6}%
}\left(  L_{i_{4}}\right)  _{\phantom{a_{7}}b}^{a_{7}}\left(  L_{i_{5}%
}\right)  _{\phantom{a_{8}}c}^{a_{8}}B_{5}^{bca_{9}a_{10}a_{11}}+\nonumber\\
&  +\frac{1}{4}\kappa_{15}\operatorname*{Tr}\left(  L_{i_{1}}L_{i_{2}}\right)
L_{i_{3}}^{a_{1}a_{2}}L_{i_{4}}^{a_{3}a_{4}}L_{i_{5}}^{a_{5}a_{6}}B_{5}%
^{a_{7}\cdots a_{11}}+\nonumber\\
&  \left.  -L_{i_{1}}^{a_{1}a_{2}}L_{i_{2}}^{a_{3}a_{4}}\left[  L_{i_{3}%
}L_{i_{4}}L_{i_{5}}\right]  ^{a_{5}a_{6}}B_{5}^{a_{7}\cdots a_{11}}\right]  ,
\label{J5Z5M2}
\end{align}
\begin{align}
\left\langle \bm{QJ}^{4}\bar{\bm{Q}}\right\rangle
_{\mathrm{M}}^{\prime} = & -\frac{\alpha_{2}}{240}\varepsilon_{a_{1}\cdots
a_{8}abc}L_{1}^{a_{1}a_{2}}\cdots L_{4}^{a_{7}a_{8}}\left(  \bar{\chi}%
\Gamma^{abc}\zeta\right)  +\nonumber\\
&  +\frac{\alpha_{2}}{60}\sum_{\left\langle i_{1}\cdots i_{4}\right\rangle} \left\{  \frac{3}{4}\kappa_{9}\operatorname*{Tr}\left(  L_{i_{1}}L_{i_{2}%
}\right)  L_{i_{3}}^{a_{1}a_{2}}L_{i_{4}}^{a_{3}a_{4}}\left(  \bar{\chi}%
\Gamma_{a_{1}\cdots a_{4}}\zeta\right)  +\right. \nonumber\\
&  -2L_{i_{1}}^{a_{1}a_{2}}\left[  L_{i_{2}}L_{i_{3}}L_{i_{4}}\right]
^{a_{3}a_{4}}\left(  \bar{\chi}\Gamma_{a_{1}\cdots a_{4}}\zeta\right)
+\nonumber\\
&   +\frac{3}{4}\left(  5\gamma_{9}-4\right)  \operatorname*{Tr}\left(
L_{i_{1}}L_{i_{2}}\right)  \operatorname*{Tr}\left(  L_{i_{3}}L_{i_{4}%
}\right)  \bar{\chi}\zeta + \nonumber \\
& \left. -\kappa_{3}\operatorname*{Tr}\left(  L_{i_{1}%
}L_{i_{2}}L_{i_{3}}L_{i_{4}}\right)  \bar{\chi}\zeta \right\}  , \label{QJ4QM2}%
\end{align}
donde hemos utilizado las mismas abreviaciones que en la
secci\'{o}n~\ref{sec:TIM} [cf.~ecs.~(\ref{ab1})--(\ref{ab5})].

Las constantes $\kappa_{n}$ y $\gamma_{n}$ que aparecen en~(\ref{J6M2}%
)--(\ref{QJ4QM2}) no constituyen, como pudiera parecer a simple vista, una
secuencia infinita de constantes de acoplamiento arbitrarias, sino que pueden ser
expresadas en t\'{e}rminos de $\beta_{4+2}$ y $\beta_{2+2+2}$ (que s\'{\i} son
constantes de acoplamiento arbitrarias) en la forma\footnote{Aqu\'{\i}
$\mathbbm{1}$ denota la matriz identidad de $32\times32$, de donde
$\operatorname*{Tr}\left(  \mathbbm{1}\right)  =32$.}
\begin{align}
\kappa_{n}  &  =1+\frac{1}{n}\beta_{4+2}\operatorname*{Tr}\left(
\mathbbm{1}\right)  ,\\
\gamma_{n}  &  =\kappa_{n}+\frac{1}{15}\beta_{2+2+2}\left[  \operatorname*{Tr}%
\left(  \mathbbm{1}\right)  \right]  ^{2}.
\end{align}
Las relaciones inversas son
\begin{align}
\beta_{4+2}  &  =\frac{1}{\operatorname*{Tr}\left(  \mathbbm{1}\right)
}n\left(  \kappa_{n}-1\right)  ,\\
\beta_{2+2+2}  &  =\frac{15}{\left[  \operatorname*{Tr}\left(  \mathbbm{1}%
\right)  \right]  ^{2}}\left(  \gamma_{n}-\kappa_{n}\right)  .
\end{align}

Una manera equivalente de plantear el problema es olvidar las constantes
de acoplamiento originales $\beta_{4+2}$ y $\beta_{2+2+2}$ y exigir que las
nuevas constantes $\kappa_{n}$ y $\gamma_{n}$ satisfagan las relaciones
\begin{align}
\kappa_{m}  &  =1+\frac{n}{m}\left(  \kappa_{n}-1\right)  ,\label{km}\\
\gamma_{m}  &  =\gamma_{n}+\left(  \frac{n}{m}-1\right)  \left(  \kappa
_{n}-1\right)  . \label{gm}%
\end{align}
En este enfoque, uno puede fijar el valor de una constante de la familia
$\kappa$ y una constante de la familia $\gamma$; el valor de todas las
dem\'{a}s queda un\'{\i}vocamente determinado a partir de~(\ref{km}%
)--(\ref{gm}).

Vale la pena notar tambi\'{e}n que
\begin{align}
\beta_{4+2}  &  =0\qquad\Leftrightarrow\qquad\kappa_{n}=1,\\
\beta_{2+2+2}  &  =0\qquad\Leftrightarrow\qquad\gamma_{n}=\kappa_{n}.
\end{align}

El efecto de este nuevo tensor invariante en la teor\'{\i}a de gauge para
$\mathfrak{M}$ que describimos en este cap\'{\i}tulo concierne la forma
espec\'{\i}fica de ciertos tensores usados para escribir el lagrangeano y las
ecuaciones de movimiento [cf.~ecs.~(\ref{H1})--(\ref{Rcal}) y~(\ref{L2})--(\ref{H7})]. La estructura general de
la teor\'{\i}a no es modificada.

\section{\label{sec:MAlgLag}La Acci\'{o}n}

El lagrangeano invariante de gauge que utilizamos corresponde a una forma de
transgresi\'{o}n para las conexiones (valuadas en $\mathfrak{M}$)
$\bm{A}$ y $\bar{\bm{A}}$,
\begin{align}
\bm{A}  &  =\bm{\omega}+\bm{e}+\bm{b}%
_{2}+\bm{b}_{5}+\bar{\bm{\psi}},\label{AM}\\
\bar{\bm{A}}  &  =\bar{\bm{\omega}}+\bar{\bm{e}}%
+\bar{\bm{b}}_{2}+\bar{\bm{b}}_{5}+\bar{\bm{\chi}}.
\label{AbM}%
\end{align}
Cada t\'{e}rmino en los que han sido descompuestas $\bm{A}$ y
$\bar{\bm{A}}$ hace referencia a un subespacio distinto de
$\mathfrak{M}$. En concreto,
\begin{align}
\bm{\omega}  &  =\frac{1}{2}\omega^{ab}\bm{J}_{ab},\\
\bm{e}  &  =\frac{1}{\ell}e^{a}\bm{P}_{a},\\
\bm{b}_{2}  &  =\frac{1}{2}b_{2}^{ab}\bm{Z}_{ab},\\
\bm{b}_{5}  &  =\frac{1}{5!}b_{5}^{abcde}\bm{Z}_{abcde},\\
\bar{\bm{\psi}}  &  =\bar{\psi}_{\alpha}\bm{Q}^{\alpha},
\end{align}
y de modo similar para $\bar{\bm{A}}$. Como es usual, identificamos
$e^{a}$ y $\omega^{ab}$ con el vielbein y la conexi\'{o}n de spin en una
formulaci\'{o}n de primer orden de gravedad. Los campos bos\'{o}nicos
$b_{2}^{ab}$ y $b_{5}^{abcde}$ son requeridos por el \'{a}lgebra, y no tienen
an\'{a}logo en $d=4$. Por \'{u}ltimo, $\psi$ es un spinor de Majorana de 32
componentes. Es importante recalcar que todos estos campos son componentes de
una conexi\'{o}n de gauge valuada en $\mathfrak{M}$. Las transformaciones de
gauge para estos campos son deducidas de la ley general para una conexi\'{o}n,
ec.~(\ref{deltaAinf}).

La curvatura $\bm{F}=\mathrm{d}\bm{A}+\bm{A}^{2}$
asociada a $\bm{A}$ tiene la forma
\begin{equation}
\bm{F}=\bm{R}+\bm{F}_{P}+\bm{F}_{2}%
+\bm{F}_{5}+\mathrm{D}_{\omega}\bar{\bm{\psi}}, \label{FM}%
\end{equation}
donde, al igual que con $\bm{A}$, cada t\'{e}rmino hace referencia a
un subespacio distinto:
\begin{align}
\bm{R}  &  =\frac{1}{2}R^{ab}\bm{J}_{ab},\\
\bm{F}_{P}  &  =\left(  \frac{1}{\ell}T^{a}+\frac{1}{16}\bar{\psi
}\Gamma^{a}\psi\right)  \bm{P}_{a},\\
\bm{F}_{2}  &  =\frac{1}{2}\left(  \mathrm{D}_{\omega}b^{ab}-\frac
{1}{16}\bar{\psi}\Gamma^{ab}\psi\right)  \bm{Z}_{ab},\\
\bm{F}_{5}  &  =\frac{1}{5!}\left(  \mathrm{D}_{\omega}b^{a_{1}\cdots
a_{5}}+\frac{1}{16}\bar{\psi}\Gamma^{a_{1}\cdots a_{5}}\psi\right)
\bm{Z}_{a_{1}\cdots a_{5}},\\
\mathrm{D}_{\omega}\bar{\bm{\psi}}  &  =\mathrm{D}_{\omega}\bar{\psi
}\bm{Q}.
\end{align}
La derivada covariante de Lorentz de un spinor tiene la forma usual,
\begin{align}
\mathrm{D}_{\omega}\psi &  =\mathrm{d}\psi+\frac{1}{4}\omega^{ab}\Gamma
_{ab}\psi,\\
\mathrm{D}_{\omega}\bar{\psi}  &  =\mathrm{d}\bar{\psi}-\frac{1}{4}\omega
^{ab}\bar{\psi}\Gamma_{ab}.
\end{align}

\subsection{Simetr\'{\i}as}

La acci\'{o}n transgresora es invariante bajo las transformaciones de gauge
infinitesimales [cf.~ecs.~(\ref{dgA})--(\ref{dgAb})]%
\begin{align}
\delta_{\mathrm{gauge}}\bm{A}  &  =-\mathrm{D}\bm{\lambda},\\
\delta_{\mathrm{gauge}}\bar{\bm{A}}  &  =-\mathrm{\bar{D}%
}\bm{\lambda},
\end{align}
donde
\begin{equation}
\bm{\lambda} = \frac{1}{2} \lambda^{ab} \bm{J}_{ab} + \frac{1}{\ell} \kappa^{a} \bm{P}_{a} + \frac{1}{2} \kappa^{ab} \bm{Z}_{ab} + \frac{1}{5!} \kappa^{abcde} \bm{Z}_{abcde} + \bar{\varepsilon} \bm{Q}
\end{equation}
es una 0-forma valuada en $\mathfrak{M}$.

En esta secci\'{o}n escribimos expl\'{\i}citamente la variaci\'{o}n de cada una de las componentes de la conexi\'{o}n~(\ref{AM}) bajo transformaciones de gauge generadas por los distintos elementos del \'{a}lgebra~M.

\begin{itemize}
\item Traslaciones: $\bm{\lambda}=\left(  1/\ell\right)  \kappa^{a}\bm{P}_{a}$,
\begin{align}
\delta e^{a}  &  =-\mathrm{D}_{\omega}\kappa^{a},\\
\delta b_{2}^{ab}  &  =0,\\
\delta b_{5}^{abcde}  &  =0,\\
\delta\omega^{ab}  &  =0,\\
\delta\psi &  =0.
\end{align}

\item $\bm{Z}_{2}$: $\bm{\lambda}=\left(  1/2\right)
\kappa^{ab}\bm{Z}_{ab}$,
\begin{align}
\delta e^{a}  &  =0,\\
\delta b_{2}^{ab}  &  =-\mathrm{D}_{\omega}\kappa^{ab},\\
\delta b_{5}^{abcde}  &  =0,\\
\delta\omega^{ab}  &  =0,\\
\delta\psi &  =0.
\end{align}

\item $\bm{Z}_{5}$: $\bm{\lambda}=\left(  1/5!\right)
\kappa^{abcde}\bm{Z}_{abcde}$,
\begin{align}
\delta e^{a}  &  =0,\\
\delta b_{2}^{ab}  &  =0,\\
\delta b_{5}^{abcde}  &  =-\mathrm{D}_{\omega}\kappa^{abcde},\\
\delta\omega^{ab}  &  =0,\\
\delta\psi &  =0.
\end{align}

\item Rotaciones de Lorentz\footnote{La ley de transformaci\'{o}n general para un tensor completamente antisim\'{e}trico de $p$ \'{\i}ndices es $\delta b_{p}^{a_{1} \cdots a_{p}} = \left( -1 \right)^{p+1} p \lambda_{\phantom{[a_{1}}c}^{[a_{1}} b_{p}^{a_{2} \cdots a_{p}]c}$.}: $\bm{\lambda}=\left(  1/2\right)
\lambda^{ab}\bm{J}_{ab}$,
\begin{align}
\delta e^{a}  &  =\lambda_{\phantom{a}b}^{a}e^{b},\\
\delta b_{2}^{ab}  &  =-2\lambda_{\phantom{[a}c}^{[a}b_{2}^{b]c},\\
\delta b_{5}^{abcde} & = 5 \lambda_{\phantom{[a}f}^{[a} b_{5}^{bcde]f} , \\
\delta\omega^{ab}  &  =-\mathrm{D}_{\omega}\lambda^{ab},\\
\delta\psi &  =\frac{1}{4}\lambda^{ab}\Gamma_{ab}\psi.
\end{align}

\item Supersimetr\'{\i}a: $\bm{\lambda}=\bar{\varepsilon}\bm{Q}$,
\begin{align}
\delta e^{a}  &  =\frac{\ell}{8}\bar{\varepsilon}\Gamma^{a}\psi,\\
\delta b_{2}^{ab}  &  =-\frac{1}{8}\bar{\varepsilon}\Gamma^{ab}\psi,\\
\delta b_{5}^{abcde} & = \frac{1}{8} \bar{\varepsilon} \Gamma^{abcde} \psi, \\
\delta\omega^{ab}  &  =0,\\
\delta\psi &  =-\mathrm{D}_{\omega}\varepsilon.
\end{align}

\end{itemize}

En su simplicidad, estas leyes de transformaci\'{o}n muestran de manera
transparente el rol de las distintas componentes de $\bm{A}$. Todos
los campos son campos de gauge y tensores (o spinores) de Lorentz. Las
transformaciones de supersimetr\'{\i}a act\'{u}an del modo usual sobre $e^{a}%
$, $b_{2}^{ab}$ y $b_{5}^{abcde}$, en tanto que dejan invariante la
conexi\'{o}n de spin $\omega^{ab}$.

Debe destacarse que la acci\'{o}n es completamente invariante bajo estas
transformaciones, sin que sea necesario recurrir a campos auxiliares. El
\'{a}lgebra satisfecha por estas transformaciones es, por construcci\'{o}n, el
\'{a}lgebra~M, la cual se cierra sin necesidad de usar las ecuaciones de
movimiento. Estas caracter\'{\i}sticas son consecuencia directa de la
elecci\'{o}n del lagrangeano como una forma de transgresi\'{o}n.

\subsection{El Lagrangeano}

El lagrangeano invariante para $\mathfrak{M}$ puede escribirse
como\footnote{La constante adimensional $k$ usualmente escrita frente a una
forma de transgresi\'{o}n cuando ella es utilizada como lagrangeano [ver,
e.g., ec.~(\ref{AcTra})] ha sido absorbida aqu\'{\i} en las constantes
$\alpha_{0}$ y $\alpha_{2}$ del tensor invariante~(\ref{J6M})--(\ref{QJ4QM}%
).}
\begin{equation}
L_{\mathrm{M}}^{\left(  11\right)  }\left(  \bm{A},\bar{\bm{A}%
}\right)  =\mathcal{Q}_{\bm{A}\leftarrow\bar{\bm{A}}}^{\left(
11\right)  }, \label{LM11}%
\end{equation}
donde $\mathcal{Q}_{\bm{A}\leftarrow\bar{\bm{A}}}^{\left(
11\right)  }$ es una forma de transgresi\'{o}n. Para obtener una versi\'{o}n
m\'{a}s expl\'{\i}cita de este lagrangeano usamos el m\'{e}todo de
separaci\'{o}n en subespacios desarrollado en el
cap\'{\i}tulo~\ref{ch:metsepsub}. Como primer paso, introducimos la
conexi\'{o}n intermedia $\tilde{\bm{A}}=\bar{\bm{\omega}}$.
Esto nos permite separar el lagrangeano~(\ref{LM11}) en la forma
\begin{equation}
L_{\mathrm{M}}^{\left(  11\right)  }\left(  \bm{A},\bar{\bm{A}%
}\right)  =\mathcal{Q}_{\bm{A}\leftarrow\bar{\bm{\omega}}%
}^{\left(  11\right)  }+\mathcal{Q}_{\bar{\bm{\omega}}\leftarrow
\bar{\bm{A}}}^{\left(  11\right)  }+\mathrm{d}\mathcal{Q}%
_{\bm{A}\leftarrow\bar{\bm{\omega}}\leftarrow\bar
{\bm{A}}}^{\left(  10\right)  }.
\end{equation}
La transgresi\'{o}n $\mathcal{Q}_{\bm{A}\leftarrow\bar
{\bm{\omega}}}^{\left(  11\right)  }$ puede ser separada a su vez a
trav\'{e}s de la conexi\'{o}n intermedia $\tilde{\bm{A}}%
=\bm{\omega}$:
\begin{equation}
\mathcal{Q}_{\bm{A}\leftarrow\bar{\bm{\omega}}}^{\left(
11\right)  }=\mathcal{Q}_{\bm{A}\leftarrow\bm{\omega}%
}^{\left(  11\right)  }+\mathcal{Q}_{\bm{\omega}\leftarrow
\bar{\bm{\omega}}}^{\left(  11\right)  }+\mathrm{d}\mathcal{Q}%
_{\bm{A}\leftarrow\bm{\omega}\leftarrow\bar{\bm{\omega
}}}^{\left(  10\right)  }.
\end{equation}
Estas dos iteraciones del m\'{e}todo nos conducen a la siguiente expresi\'{o}n
para el lagrangeano:
\begin{equation}
L_{\mathrm{M}}^{\left(  11\right)  }\left(  \bm{A},\bar{\bm{A}%
}\right)  =\mathcal{Q}_{\bm{A}\leftarrow\bm{\omega}}^{\left(
11\right)  }-\mathcal{Q}_{\bar{\bm{A}}\leftarrow\bar
{\bm{\omega}}}^{\left(  11\right)  }+\mathcal{Q}_{\bm{\omega
}\leftarrow\bar{\bm{\omega}}}^{\left(  11\right)  }+\mathrm{d}%
\mathcal{B}_{\mathrm{M}}^{\left(  10\right)  }, \label{LM11b}%
\end{equation}
con
\begin{equation}
\mathcal{B}_{\mathrm{M}}^{\left(  10\right)  }=\mathcal{Q}_{\bm{A}%
\leftarrow\bm{\omega}\leftarrow\bar{\bm{\omega}}}^{\left(
10\right)  }+\mathcal{Q}_{\bm{A}\leftarrow\bar{\bm{\omega}%
}\leftarrow\bar{\bm{A}}}^{\left(  10\right)  }. \label{BM10}%
\end{equation}

Los dos primeros t\'{e}rminos en~(\ref{LM11b}) tienen exactamente la misma
forma (con los reemplazos evidentes), de modo que basta con concentrarse en
analizar uno de ellos. M\'{a}s adelante mostraremos que el tercer t\'{e}rmino
carece de relaci\'{o}n con los dos primeros; en particular, es posible
anularlo sin afectar al resto. El t\'{e}rmino de borde~(\ref{BM10}) puede ser
escrito de manera m\'{a}s expl\'{\i}cita volviendo a la ec.~(\ref{borde})\ y
sustituyendo las conexiones y las curvaturas correspondientes. El resultado,
sin embargo, no es particularmente inspirador y, dado que su forma
expl\'{\i}cita no es necesaria para escribir las condiciones de borde, no nos
ocuparemos m\'{a}s de \'{e}l.

Examinemos ahora la transgresi\'{o}n $\mathcal{Q}_{\bm{A}%
\leftarrow\bm{\omega}}^{\left(  11\right)  }$. El m\'{e}todo de
separaci\'{o}n en subespacios introducido en el
cap\'{\i}tulo~\ref{ch:metsepsub} puede volver a ser usado para obtener una
expresi\'{o}n cerrada para ella. Con este objetivo introducimos el siguiente
conjunto de conexiones intermedias:
\begin{align}
\bm{A}_{0} & = \bm{\omega}, \\
\bm{A}_{1} & = \bm{\omega} + \bm{e}, \\
\bm{A}_{2} & = \bm{\omega} + \bm{e} + \bm{b}_{2}, \\
\bm{A}_{3} & = \bm{\omega} + \bm{e} + \bm{b}_{2} + \bm{b}_{5}, \\
\bm{A}_{4} & = \bm{\omega} + \bm{e} + \bm{b}_{2} + \bm{b}_{5} + \bar{\bm{\psi}}.
\end{align}
La identidad triangular~(\ref{treqdef}) nos permite separar la transgresi\'{o}n
$\mathcal{Q}_{\bm{A}_{4}\leftarrow\bm{A}_{0}}^{\left(
11\right)  }$ de acuerdo al patr\'{o}n
\begin{align}
\mathcal{Q}_{\bm{A}_{4}\leftarrow\bm{A}_{0}}^{\left(
11\right)  }  &  =\mathcal{Q}_{\bm{A}_{4}\leftarrow\bm{A}_{3}%
}^{\left(  11\right)  }+\mathcal{Q}_{\bm{A}_{3}\leftarrow
\bm{A}_{0}}^{\left(  11\right)  }+\mathrm{d}\mathcal{Q}%
_{\bm{A}_{4}\leftarrow\bm{A}_{3}\leftarrow\bm{A}_{0}%
}^{\left(  10\right)  },\label{part1}\\
\mathcal{Q}_{\bm{A}_{3}\leftarrow\bm{A}_{0}}^{\left(
11\right)  }  &  =\mathcal{Q}_{\bm{A}_{3}\leftarrow\bm{A}_{2}%
}^{\left(  11\right)  }+\mathcal{Q}_{\bm{A}_{2}\leftarrow
\bm{A}_{0}}^{\left(  11\right)  }+\mathrm{d}\mathcal{Q}%
_{\bm{A}_{3}\leftarrow\bm{A}_{2}\leftarrow\bm{A}_{0}%
}^{\left(  10\right)  },\label{part2}\\
\mathcal{Q}_{\bm{A}_{2}\leftarrow\bm{A}_{0}}^{\left(
11\right)  }  &  =\mathcal{Q}_{\bm{A}_{2}\leftarrow\bm{A}_{1}%
}^{\left(  11\right)  }+\mathcal{Q}_{\bm{A}_{1}\leftarrow
\bm{A}_{0}}^{\left(  11\right)  }+\mathrm{d}\mathcal{Q}%
_{\bm{A}_{2}\leftarrow\bm{A}_{1}\leftarrow\bm{A}_{0}%
}^{\left(  10\right)  }. \label{part3}%
\end{align}

Evaluando expl\'{\i}citamente cada uno de los t\'{e}rminos en~(\ref{part1})--(\ref{part3}) uno encuentra el lagrangeano
\begin{equation}
\mathcal{Q}_{\bm{A}_{4}\leftarrow\bm{A}_{0}}^{\left(
11\right)  }=6\left[  H_{a}e^{a}+\frac{1}{2}H_{ab}b_{2}^{ab}+\frac{1}%
{5!}H_{abcde}b_{5}^{abcde}-\frac{5}{2}\bar{\psi}\mathcal{R}\mathrm{D}_{\omega
}\psi\right]  . \label{q40}%
\end{equation}

Los tres t\'{e}rminos de borde presentes en~(\ref{part1})--(\ref{part3}) se
anulan id\'{e}nticamente; esto es una consecuencia de la forma particular del
tensor invariante escogido [cf.~ecs.~(\ref{itmalg1})--(\ref{itmalg5})].

Los tensores $H_{a}$, $H_{ab}$, $H_{abcde}$ y $\mathcal{R}$ que aparecen
en~(\ref{part1})--(\ref{part3}) est\'{a}n definidos mediante las expresiones
\begin{align}
H_{a} & \equiv \left\langle \bm{R}^{5} \bm{P}_{a} \right\rangle_{\mathrm{M}}, \label{H1} \\
H_{ab} & \equiv \left\langle \bm{R}^{5} \bm{Z}_{ab} \right\rangle_{\mathrm{M}}, \label{H2} \\
H_{abcde} & \equiv \left\langle \bm{R}^{5} \bm{Z}_{abcde} \right\rangle_{\mathrm{M}}, \label{H5} \\
\mathcal{R}_{\phantom{\alpha}\beta}^{\alpha} & \equiv \left\langle \bm{Q}^{\alpha} \bm{R}^{4} \bar{\bm{Q}}_{\beta} \right\rangle_{\mathrm{M}}, \label{Rcal}
\end{align}
donde $\bm{R}=\left(  1/2\right)  R^{ab}\bm{J}_{ab}$ es la
curvatura de Lorentz. Versiones expl\'{\i}citas de~(\ref{H1})--(\ref{Rcal})
pueden ser obtenidas sin demasiada dificultad recurriendo al tensor
invariante~(\ref{itmalg1})--(\ref{itmalg5}). Efectuando los reemplazos
correspondientes, encontramos
\begin{align}
H_{a} = & \frac{\alpha_{2}}{64}R_{a}^{\left(  5\right)  },\label{H1ex}\\
H_{ab} = & \alpha_{2}\left[  \frac{5}{2}\left(  R^{4}-\frac{3}{4}R^{2}%
R^{2}\right)  R_{ab}+5R^{2}R_{ab}^{3}-8R_{ab}^{5}\right]  ,\label{H2ex}\\
H_{abcde} = & -\frac{5}{16}\alpha_{2}\left[  5R_{[ab}R_{cde]}^{\left(
4\right)  }-40R_{\phantom{f}[a}^{f}R_{\phantom{g}b}^{g}R_{cde]fg}^{\left(
3\right)  }+\right. \nonumber\\
&  \left.  -R^{2}R_{abcde}^{\left(  3\right)  }+4R_{abcdefg}^{\left(
2\right)  }\left(  R^{3}\right)  ^{fg}\right]  ,\label{H5ex}\\
\mathcal{R} = & -\frac{\alpha_{2}}{40}\left\{  \left(  R^{4}-\frac{3}{4}%
R^{2}R^{2}\right)  \mathbbm{1}+\frac{1}{96}R_{abc}^{\left(  4\right)
}\Gamma^{abc}+\right. \nonumber\\
&  \left.  -\frac{3}{4}\left[  R^{2}R^{ab}-\frac{8}{3}\left(  R^{3}\right)
^{ab}\right]  R^{cd}\Gamma_{abcd}\right\}  . \label{Rcalex}%
\end{align}

Aqu\'{\i} hemos usado las abreviaciones\footnote{La antisimetr\'{\i}a de $R^{ab}$ implica que la traza del producto de un n\'{u}mero impar de curvaturas de Lorentz se anula: $R^{2n+1}=0$.}
\begin{align}
R^{n}  &  =R_{\phantom{a_{1}}a_{2}}^{a_{1}}\cdots R_{\phantom{a_{n}}a_{1}%
}^{a_{n}},\label{Rabv1}\\
R_{ab}^{n}  &  =R_{ac_{1}}R_{\phantom{c_{1}}c_{2}}^{c_{1}}\cdots
R_{\phantom{c_{n-1}}b}^{c_{n-1}},\label{Rabv2}\\
R_{a_{1}\cdots a_{d-2n}}^{\left(  n\right)  }  &  =\varepsilon_{a_{1}\cdots
a_{d-2n}b_{1}\cdots b_{2n}}R^{b_{1}b_{2}}\cdots R^{b_{2n-1}b_{2n}}.
\label{Rabv3}%
\end{align}

En la secci\'{o}n~\ref{sec:MAlgDyn} comentamos acerca de la din\'{a}mica
producida por este lagrangeano; por ahora, notamos que en \'{e}l no aparece
ninguna derivada de $e^{a}$, $b_{2}^{ab}$ \'{o} $b_{5}^{abcde}$. Esto es una
consecuencia del tensor invariante utilizado (y, a la vez, del $0_{S}%
$-forzamiento), el cual no contiene componentes no nulas de la forma
$\left\langle \bm{J}^{3}\bm{PZ}_{2}\right\rangle $, etc.

La \'{u}ltima contribuci\'{o}n al lagrangeano~(\ref{LM11b}) proviene del
t\'{e}rmino `especular' $\mathcal{Q}_{\bm{\omega}\leftarrow
\bar{\bm{\omega}}}^{\left(  11\right)  }$. Recurriendo a la
definici\'{o}n de forma de transgresi\'{o}n y a la forma del tensor invariante
para $\mathfrak{M}$, resulta directo escribir la expresi\'{o}n
\begin{equation}
\mathcal{Q}_{\bm{\omega}\leftarrow\bar{\bm{\omega}}}^{\left(
11\right)  }=3\int_{0}^{1} dt \theta^{ab}L_{ab}\left(  t\right)
,\label{qww1}%
\end{equation}
donde
\begin{align}
L_{ab}\left(  t\right)   &  =\left\langle \bm{R}_{t}^{5}%
\bm{J}_{ab}\right\rangle _{\mathrm{M}},\\
\bm{R}_{t} &  =\frac{1}{2}R_{t}^{ab}\bm{J}_{ab},\\
R_{t}^{ab} &  =\bar{R}^{ab}+t\mathrm{D}_{\bar{\omega}}\theta
^{ab}+t^{2}\theta_{\phantom{a}c}^{a}\theta^{cb}.
\end{align}
Una versi\'{o}n expl\'{\i}cita de $L_{ab}\left(  t\right)  $, obtenida
sustituyendo las componentes relevantes del tensor invariante en~(\ref{qww1}),
es
\begin{equation}
L_{ab}\left(  t\right)  =\alpha_{0}\left[  \frac{5}{2}\left(  R_{t}^{4}%
-\frac{3}{4}R_{t}^{2}R_{t}^{2}\right)  \left(  R_{t}\right)  _{ab}+5R_{t}%
^{2}\left(  R_{t}^{3}\right)  _{ab}-8\left(  R_{t}^{5}\right)  _{ab}\right]
.\label{Lab(t)}%
\end{equation}

Como se observa en~(\ref{Lab(t)}), $\mathcal{Q}_{\bm{\omega} \leftarrow \bar{\bm{\omega}}}^{\left( 11 \right)}$ es proporcional a $\alpha_{0}$, a diferencia de todos los dem\'{a}s t\'{e}rminos, que son proporcionales a $\alpha_{2}$. Esto es una consecuencia directa de la elecci\'{o}n de tensor invariante. Siendo el \'{u}nico sector del lagrangeano no relacionado con $\alpha_{2}$, puede ser omitido simplemente escogiendo $\alpha_{0}=0$. Esta independencia tambi\'{e}n significa que $\mathcal{Q}_{\bm{\omega} \leftarrow \bar{\bm{\omega}}}^{\left( 11 \right)}$ es por s\'{\i} solo invariante bajo el \'{a}lgebra~M. Esto est\'{a} relacionado con el hecho que este t\'{e}rmino corresponde a la \'{u}nica componente del lagrangeano que sobrevive cuando se ocupa directamente la supertraza del \'{a}lgebra~M para construir el tensor invariante.

A causa de su forma, $\mathcal{Q}_{\bm{\omega}\leftarrow
\bar{\bm{\omega}}}^{\left(  11\right)  }$ aparentemente contiene un
t\'{e}rmino de interacci\'{o}n entre $\bm{\omega}$ y $\bar
{\bm{\omega}}$ no confinado al borde del espacio-tiempo. Esto no es
m\'{a}s que una ilusi\'{o}n; para visualizarlo, basta con usar la identidad
triangular con $\tilde{\bm{A}}=0$,
\begin{equation}
\mathcal{Q}_{\bm{\omega}\leftarrow\bar{\bm{\omega}}}^{\left(
11\right)  }=\mathcal{Q}_{\bm{\omega}\leftarrow0}^{\left(  11\right)
}-\mathcal{Q}_{\bar{\bm{\omega}}\leftarrow0}^{\left(  11\right)
}+\mathrm{d}\mathcal{Q}_{\bm{\omega}\leftarrow\bm{A}%
_{3}\leftarrow\bar{\bm{\omega}}}^{\left(  10\right)  }.
\end{equation}
Aqu\'{\i} $\mathcal{Q}_{\bm{\omega}\leftarrow0}^{\left(  11\right)  }$
y $\mathcal{Q}_{\bar{\bm{\omega}}\leftarrow0}^{\left(  11\right)  }$
corresponden a dos lagrangeanos de CS independientes de `gravedad
ex\'{o}tica', mientras que $\mathcal{Q}_{\bm{\omega}\leftarrow
\bm{A}_{3}\leftarrow\bar{\bm{\omega}}}^{\left(  10\right)  }$
es el t\'{e}rmino de borde que contiene toda la interacci\'{o}n entre
$\bm{\omega}$ y $\bar{\bm{\omega}}$.

\section{\label{sec:MAlgDyn}La Din\'{a}mica}

En esta secci\'{o}n estudiamos las ecuaciones de movimiento para el lagrangeano~(\ref{LM11}), comentando brevemente acerca del problema del vac\'{\i}o.

\subsection{Ecuaciones de movimiento}

Las ecuaciones de movimiento para $\bm{A}$ y $\bar{\bm{A}}$
son completamente an\'{a}logas [cf.~ecs.~(\ref{FnGa1})--(\ref{FnGa2})], por lo
cual en esta secci\'{o}n ser\'{a}n presentadas s\'{o}lo para $\bm{A}$.
La forma general de las ecuaciones de movimiento para $\bm{A}$ es
\begin{equation}
\left\langle \bm{F}^{5}\bm{G}_{A}\right\rangle _{\mathrm{M}%
}=0,
\end{equation}
donde $\left\{ \bm{G}_{A} \right\} = \left\{ \bm{P}_{a}, \bm{Z}_{ab}, \bm{Z}_{abcde}, \bm{J}_{ab}, \bm{Q} \right\}$ es una base para $\mathfrak{M}$, $\left\langle \cdots \right\rangle_{\mathrm{M}}$ es el tensor sim\'{e}trico invariante~(\ref{J5PM})--(\ref{QJ4QM}) y $\bm{F}$ es la curvatura~(\ref{FM}).

Las ecuaciones de movimiento asociadas a las variaciones de $e^{a}$,
$b_{2}^{ab}$, $b_{5}^{abcde}$ y $\psi$ est\'{a}n dadas por
\begin{align}
H_{a}  &  =0,\label{H1=0}\\
H_{ab}  &  =0,\label{H2=0}\\
H_{abcde}  &  =0,\label{H5=0}\\
\mathcal{R}\mathrm{D}_{\omega}\psi &  =0. \label{RcalDpsi=0}%
\end{align}
Las expresiones expl\'{\i}citas para $H_{a}$, $H_{ab}$, $H_{abcde}$ y
$\mathcal{R}$ se encuentran en las ecs.~(\ref{H1ex})--(\ref{Rcalex}). Estas
ecuaciones de movimiento pueden deducirse f\'{a}cilmente de la forma del trozo
correspondiente del lagrangeano, ec.~(\ref{q40}).

La ecuaci\'{o}n de movimiento asociada a la variaci\'{o}n de $\omega^{ab}$ es
m\'{a}s complicada, y puede ser escrita en la forma
\begin{align}
L_{ab}-10\left(  \mathrm{D}_{\omega}\bar{\psi}\right)  \mathcal{Z}_{ab}\left(
\mathrm{D}_{\omega}\psi\right)  +5H_{abc}\left(  T^{c}+\frac{1}{16}\bar{\psi
}\Gamma^{c}\psi\right)  +  & \nonumber\\
+\frac{5}{2}H_{abcd}\left(  \mathrm{D}_{\omega}b^{cd}-\frac{1}{16}\bar{\psi
}\Gamma^{cd}\psi\right)  +\frac{1}{24}H_{abc_{1}\cdots c_{5}}\left(
\mathrm{D}_{\omega}b^{c_{1}\cdots c_{5}}+\frac{1}{16}\bar{\psi}\Gamma
^{c_{1}\cdots c_{5}}\psi\right)   &  =0, \label{DeltaOmega}%
\end{align}
donde hemos definido
\begin{align}
L_{ab}  &  \equiv\left\langle \bm{R}^{5}\bm{J}_{ab}%
\right\rangle _{\mathrm{M}},\label{L2}\\
\left(  \mathcal{Z}_{ab}\right)  _{\phantom{\alpha}\beta}^{\alpha}  &
\equiv\left\langle \bm{Q}^{\alpha}\bm{R}^{3}\bm{J}%
_{ab}\bar{\bm{Q}}_{\beta}\right\rangle _{\mathrm{M}},\label{Zcal}\\
H_{abc}  &  \equiv\left\langle \bm{R}^{4}\bm{J}_{ab}%
\bm{P}_{c}\right\rangle _{\mathrm{M}},\label{H3}\\
H_{abcd}  &  \equiv\left\langle \bm{R}^{4}\bm{J}%
_{ab}\bm{Z}_{cd}\right\rangle _{\mathrm{M}},\label{H4}\\
H_{abcdefg}  &  \equiv\left\langle \bm{R}^{4}\bm{J}%
_{ab}\bm{Z}_{cdefg}\right\rangle _{\mathrm{M}}. \label{H7}%
\end{align}
Estos tensores satisfacen las identidades
\begin{align}
H_{c}  &  =\frac{1}{2}R^{ab}H_{abc},\\
H_{cd}  &  =\frac{1}{2}R^{ab}H_{abcd},\\
H_{cdefg}  &  =\frac{1}{2}R^{ab}H_{abcdefg},\\
\mathcal{R}_{\phantom{\alpha}\beta}^{\alpha}  &  =\frac{1}{2}R^{ab}\left(
\mathcal{Z}_{ab}\right)  _{\phantom{\alpha}\beta}^{\alpha}.
\end{align}

Versiones expl\'{\i}citas para~(\ref{L2})--(\ref{H7}) pueden ser obtenidas recurriendo a las componentes relevantes del tensor invariante~(\ref{J5PM})--(\ref{QJ4QM}). El resultado es
\begin{equation}
L_{ab}=\alpha_{0}\left[  \frac{5}{2}\left(  R^{4}-\frac{3}{4}R^{2}%
R^{2}\right)  R_{ab}+5R^{2}R_{ab}^{3}-8R_{ab}^{5}\right]  ,
\end{equation}
\begin{align}
\mathcal{Z}_{ab} = & \frac{\alpha_{2}}{40}\left\{  2\left(  R_{ab}^{3}%
-\frac{3}{4}R^{2}R_{ab}\right)  \mathbbm{1}-\frac{1}{48}R_{abcde}^{\left(
3\right)  }\Gamma^{cde}+\right. \nonumber\\
&  -\frac{3}{4}\left(  R_{ab}R^{cd}-\frac{1}{2}R^{2}\delta_{ab}^{cd}\right)
R^{ef}\Gamma_{cdef}+\nonumber\\
&  \left.  -\left[  \delta_{ab}^{cg}R_{gh}R^{hd}R^{ef}-R_{\phantom{c}a}%
^{c}R_{\phantom{d}b}^{d}R^{ef}+\frac{1}{2}\delta_{ab}^{ef}\left(
R^{3}\right)  ^{cd}\right]  \Gamma_{cdef}\right)  ,
\end{align}
\begin{equation}
H_{abc} = \frac{\alpha_{2}}{32} R_{abc}^{\left( 4 \right)},
\end{equation}
\begin{align}
H_{abcd} = & \alpha_{2}\delta_{ab}^{ef}\delta_{cd}^{gh}\left[  \frac{3}%
{4}R^{2}R_{ef}R_{gh}-R_{ef}^{3}R_{gh}-R_{ef}R_{gh}^{3}+\right. \nonumber\\
&  -\frac{4}{5}\left(  R_{eh}R_{fg}^{3}+R_{eh}^{3}R_{fg}-R_{eh}^{2}R_{fg}%
^{2}\right)  +\frac{1}{2}R^{2}R_{eh}R_{fg}+\nonumber\\
&  \left.  +\frac{1}{8}\eta_{\left[  ef\right]  \left[  gh\right]  }\left(
R^{4}-\frac{3}{4} R^{2}R^{2}\right)  -\eta_{fg}\left(  R^{2}%
R_{eh}^{2}-\frac{8}{5}R_{eh}^{4}\right)  \right]  ,
\end{align}
\begin{align}
H_{abc_{1}\cdots c_{5}} = & \frac{\alpha_{2}}{80}\delta_{c_{1}\cdots c_{5}%
}^{cdefg}\left(  -\frac{5}{3}R_{abcde}^{\left(  3\right)  }R_{fg}%
+10R_{abcdepq}^{\left(  2\right)  }R_{\phantom{p}f}^{p}R_{\phantom{q}g}%
^{q}+\right. \nonumber\\
&  -\frac{1}{6}R_{ab}R_{cdefg}^{\left(  3\right)  }+\frac{1}{4}R^{2}%
R_{abcdefg}^{\left(  2\right)  }-\frac{2}{3}R_{abcdefgpq}^{\left(  1\right)
}\left(  R^{3}\right)  ^{pq}+\nonumber\\
&  +\frac{1}{3}R_{\phantom{p}a}^{p}R_{\phantom{q}b}^{q}R_{cdefgpq}^{\left(
2\right)  }-\frac{1}{3}R_{\phantom{q}a}^{q}R_{bcdefgp}^{\left(  2\right)
}R_{\phantom{p}q}^{p}+\frac{1}{3}R_{\phantom{q}b}^{q}R_{acdefgp}^{\left(
2\right)  }R_{\phantom{p}q}^{p}+\nonumber\\
&  \left.  -\frac{10}{3}\eta_{ga}R_{bcdep}^{\left(  3\right)  }%
R_{\phantom{p}f}^{p}+\frac{10}{3}\eta_{gb}R_{acdep}^{\left(  3\right)
}R_{\phantom{p}f}^{p}-\frac{5}{24}\eta_{\left[  ab\right]  \left[  cd\right]
}R_{efg}^{\left(  4\right)  }\right)  .
\end{align}
Al escribir estos tensores hemos utilizado las abreviaciones~(\ref{Rabv1})--(\ref{Rabv3}).

\subsection{El Vac\'{\i}o y $d=4$}

Como primera aplicaci\'{o}n de las ecuaciones de movimiento~(\ref{H1=0}%
)--(\ref{DeltaOmega}) discutimos el problema del vac\'{\i}o. El
\textit{ansatz} que ocupamos en esta secci\'{o}n fue propuesto por primera vez
en las Refs.~\cite{Has03,Has05}.

Encontrar el vac\'{\i}o `verdadero' para una teor\'{\i}a de campos
transgresora es un problema no trivial muy interesante. El candidato natural
es $\bm{F}=0$; esta configuraci\'{o}n satisface las ecuaciones de
campo, es estable, sin carga y completamente invariante de gauge. Sin embargo,
cuando la dimensi\'{o}n del espacio-tiempo es $d=2n+1\geq5$, existe un
problema: las perturbaciones de primer orden sobre esta soluci\'{o}n son modos
cero\footnote{Este problema existe tambi\'{e}n para las teor\'{\i}as de CS
(recordemos que las ecuaciones de movimiento son id\'{e}nticas).}, como puede apreciarse
directamente de las ecuaciones de movimiento,
\begin{equation}
\left\langle \bm{F}^{n}\bm{G}_{A}\right\rangle =0.
\label{FnGA=0}%
\end{equation}
Esto significa que no hay grados de libertad propagantes locales.

Varias soluciones han sido ofrecidas para este problema. En la
Ref.~\cite{Hor97}, por ejemplo, se propone acoplar materia a la acci\'{o}n de
CS por medio de l\'{\i}neas de Wilson; esto resulta en perturbaciones no nulas
alrededor de la soluci\'{o}n $\bm{F}=0$. Una soluci\'{o}n alternativa
fue sugerida en las Refs.~\cite{Has03,Has05}, donde, sin agregar materia ni
modificar la acci\'{o}n en modo alguno, se abandona la soluci\'{o}n
$\bm{F}=0$ en favor de otra que satisfaga~(\ref{FnGA=0}) como un cero
simple, de manera de permitir que las perturbaciones de propaguen. Una
consecuencia impresionante de este \'{u}ltimo enfoque es que, escogiendo un
\textit{ansatz} apropiado, uno encuentra una configuraci\'{o}n de universo
tetradimensional como soluci\'{o}n.

En esta secci\'{o}n analizamos como el modelo propuesto en~\cite{Has03,Has05}
encaja dentro de esta teor\'{\i}a, que posee un lagrangeano distinto.
Tambi\'{e}n exploramos las consecuencias de permitir que la torsi\'{o}n
11-dimensional sea distinta de cero.

Consideremos el \textit{ansatz} geom\'{e}trico $M=X_{d+1}\times S^{10-d}$, donde
$X_{d+1}$ es el `producto combado' (`\textit{warped product}') de una variedad
$d$-dimensional $M_{d}$ y $\mathbb{R}$, y $S^{10-d}$ corresponde a una
variedad $\left(  10-d\right)  $-dimensional con curvatura constante y sin
torsi\'{o}n. La m\'{e}trica para este caso corresponde a la ec.~(4.3) de la
Ref.~\cite{Has05},
\begin{equation}
\mathrm{d}s^{2}=e^{-2\xi\left\vert z\right\vert }\left(  \mathrm{d}%
z^{2}+\tilde{g}_{\alpha\beta}^{\left(  d\right)  }\mathrm{d}x^{\alpha
}\mathrm{d}x^{\beta}\right)  +\gamma_{\kappa\lambda}^{\left(  10-d\right)
}\mathrm{d}y^{\kappa}\mathrm{d}y^{\lambda},
\end{equation}
donde $x^{\alpha}$ son las coordenadas locales en $M_{d}$, $z$ es una
coordenada a lo largo de $\mathbb{R}$, y $y^{\kappa}$ son las coordenadas
locales en $S^{10-d}$. S\'{o}lo por esta secci\'{o}n usamos el \'{\i}ndice $Z$
para el espacio tangente de $\mathbb{R}$; $a,b,c,\dotsc$ para el espacio
tangente de $M_{d}$ e $i,j,k,\dotsc$ para el espacio tangente de $S^{10-d}$.

Las componentes de la curvatura y la torsi\'{o}n para este \textit{ansatz} son
\begin{align}
R^{ab}  &  =\tilde{R}^{ab}-\xi^{2}\tilde{e}^{a}\tilde{e}^{b}+2\xi\theta\left(
z\right)  \left(  \tilde{e}^{a}\kappa^{b}-\tilde{e}^{b}\kappa^{a}\right)
-\kappa^{a}\kappa^{b},\\
R^{aZ}  &  =-2e^{\xi\left\vert z\right\vert }\xi\delta\left(  z\right)
E^{Z}\tilde{e}^{a}-2\xi\theta\left(  z\right)  \tilde{T}^{a}+\mathrm{D}%
_{\tilde{\omega}}\kappa^{a},\\
T^{a}  &  =\kappa^{a}E^{Z}+e^{-\xi\left\vert z\right\vert }\tilde{T}%
^{a},\label{T11}\\
T^{Z}  &  =-e^{-\xi\left\vert z\right\vert }\kappa^{a}\tilde{e}_{a}.
\end{align}
Aqu\'{\i} las cantidades del lado izquierdo son las componentes sobre
$M_{d} {\times_{\xi}} \mathbb{R}$ de la torsi\'{o}n y la curvatura de Lorentz. Las
cantidades con tilde del lado derecho, como $\tilde{R}^{ab}$, $\tilde{T}^{a}$
y $\tilde{e}^{a}$, denotan objetos pertenecientes a la variedad $M_{d}$.
$\kappa^{a}$ corresponde a la componente $k^{aZ}$ de la contorsi\'{o}n
11-dimensional. La funci\'{o}n escal\'{o}n de Heaviside y la delta de Dirac
son denotadas de manera usual como $\theta\left(  z\right)  $ y $\delta\left(
z\right)  $.

A partir de la ecuaci\'{o}n de movimiento $H_{a}=0$ [cf.~ec.~(\ref{H1=0})] es
posible demostrar, siguiendo los mismos argumentos que en~\cite{Has03,Has05},
que la \'{u}nica manera de tener grados de libertad propagantes es imponer
$d=4$.

En este caso ($d=4$), la ec. $H_{a}=0$ se desdobla en
\begin{align}
\xi e^{\xi\left\vert z\right\vert }\delta\left(  z\right)  E^{Z}%
\varepsilon_{abcd}\left(  \tilde{R}^{ab}-\xi^{2}\tilde{e}^{a}\tilde{e}%
^{b}\right)  \tilde{e}^{c}  &  =\mathcal{T}_{d},\\
\varepsilon_{abcd}\left(  \tilde{R}^{ab}-\xi^{2}\tilde{e}^{a}\tilde{e}%
^{b}\right)  \left(  \tilde{R}^{cd}-\xi^{2}\tilde{e}^{c}\tilde{e}^{d}\right)
&  =\mathcal{T},
\end{align}
con
\begin{align}
\mathcal{T}_{d} = & 2E^{Z}e^{\xi\left\vert z\right\vert }\xi\delta\left(
z\right)  \varepsilon_{abcd}\left(  \frac{1}{2}\kappa^{a}\kappa^{b}-\xi
\theta\left(  z\right)  \left(  \tilde{e}^{a}\kappa^{b}-\tilde{e}^{b}%
\kappa^{a}\right)  \right)  \tilde{e}^{c}+\nonumber\\
&  +\varepsilon_{abcd}\left[  \tilde{R}^{ab}-\xi^{2}\tilde{e}^{a}\tilde{e}%
^{b}+2\xi\theta\left(  z\right)  \left(  \tilde{e}^{a}\kappa^{b}-\tilde{e}%
^{b}\kappa^{a}\right)  -\kappa^{a}\kappa^{b}\right] \times \nonumber \\
& \times \left(  \frac{1}%
{2}\mathrm{D}_{\tilde{\omega}}\kappa^{c}-\xi\theta\left(  z\right)  \tilde
{T}^{c}\right)  ,
\end{align}%
\begin{align}
\mathcal{T} = & -4\varepsilon_{abcd}\left(  \tilde{R}^{ab}-\xi^{2}\tilde
{e}^{a}\tilde{e}^{b}+\xi\theta\left(  z\right)  \left(  \tilde{e}^{a}%
\kappa^{b}-\tilde{e}^{b}\kappa^{a}\right)  -\frac{1}{2}\kappa^{a}\kappa
^{b}\right)  \times\nonumber\\
&  \times\left(  \xi\theta\left(  z\right)  \left(  \tilde{e}^{c}\kappa
^{d}-\tilde{e}^{d}\kappa^{c}\right)  -\frac{1}{2}\kappa^{c}\kappa^{d}\right)
.
\end{align}

La primera de estas ecuaciones corresponde a las ecuaciones de Einstein, con
soporte limitado a $M_{4}$. El lado derecho de esta ecuaci\'{o}n contiene
acoplamientos entre gravedad y torsi\'{o}n. A\'{u}n imponiendo la torsi\'{o}n
4-dimensional $\tilde{T}^{a}$ igual a cero, las componentes restantes
$\kappa^{a}$ se camuflan como una especie de materia seg\'{u}n el punto de
vista de un observador 4-dimensional. La segunda ecuaci\'{o}n impone
relaciones extra entre la geometr\'{\i}a 4-dimensional y $\kappa^{a}$.

Las ecuaciones de movimiento~(\ref{H2=0})--(\ref{H5=0}) imponen a\'{u}n
m\'{a}s restricciones sobre la geometr\'{\i}a.

En contraste, las ecuaciones de movimiento~(\ref{RcalDpsi=0}%
)--(\ref{DeltaOmega}) relacionan la geometr\'{\i}a 4-dimensional con
$\kappa^{a}$, los fermiones y las cargas centrales.

De este modo se encuentra que existen demasiadas restricciones sobre la
geometr\'{\i}a como para reproducir Relatividad General en $d=4$ (para un
an\'{a}lisis de una situaci\'{o}n similar que ocurre en cinco dimensiones, ver
la Ref.~\cite{Ede06a}).

Existen varias maneras de lidiar con este problema. El exceso de restricciones
est\'{a} relacionado con la elecci\'{o}n de semigrupo hecha al construir el
\'{a}lgebra~M y tambi\'{e}n al procedimiento de $0_{S}$-forzamiento.
Escogiendo distintos semigrupos uno podr\'{\i}a estudiar distintas
\'{a}lgebras relacionadas con $\mathfrak{M}$ pero que careciesen de su rigidez din\'{a}mica.

\chapter{\label{ch:final}Conclusiones y Resultados Principales}

En este cap\'{\i}tulo final resumimos los resultados principales obtenidos en la Tesis.

Las formas de transgresi\'{o}n permiten definir teor\'{\i}as de gauge en dimensiones impares con un funcional de acci\'{o}n completamente invariante bajo transformaciones de gauge. La conexi\'{o}n con los lagrangeanos de CS es analizada en detalle, poniendo \'{e}nfasis en los aspectos relacionados con la p\'{e}rdida de la invariancia de gauge.

La f\'{o}rmula extendida de la homotop\'{\i}a de Cartan es utilizada como base para un m\'{e}todo de separaci\'{o}n en subespacios para lagrangeanos transgresores. El m\'{e}todo permite separar un lagrangeano en lagrangeanos parciales para los distintos subespacios del \'{a}lgebra de gauge. Las identidades involucradas, derivadas de la FEHC, son tambi\'{e}n de ayuda para clarificar la conexi\'{o}n entre transgresi\'{o}n y CS.

Como ejemplo de acci\'{o}n transgresora y de aplicaci\'{o}n del m\'{e}todo de separaci\'{o}n en subespacios rederivamos la acci\'{o}n finita para gravitaci\'{o}n en dimensiones impares introducida en~\cite{Mor04a}.

Introducimos un m\'{e}todo de expansi\'{o}n de \'{a}lgebras de Lie complementario al presentado por de~Azc\'{a}rraga, Izquierdo, Pic\'{o}n y Varela en~\cite{deAz02}. El m\'{e}todo hace uso extensivo de semigrupos abelianos discretos, generalizando de este modo el proceso de expansi\'{o}n en serie ocupado en~\cite{deAz02}. Explicamos c\'{o}mo obtener sub\'{a}lgebras y lo que hemos llamado `\'{a}lgebras forzadas' a partir de una \'{a}lgebra $S$-expandida. Damos teoremas generales que permiten extraer un tensor invariante para \'{a}lgebras $S$-expandidas y algunos de sus derivados en t\'{e}rminos de un tensor invariante para el \'{a}lgebra original.

En el cap\'{\i}tulo~\ref{ch:TGFTMAlg} ocupamos el m\'{e}todo de $S$-expansi\'{o}n para escribir una acci\'{o}n transgresora en $d=11$ para el \'{a}lgebra~M. Tambi\'{e}n hacemos uso del m\'{e}todo de separaci\'{o}n en subespacios para encontrar una forma expl\'{\i}cita para el lagrangeano. Estudiamos brevemente la din\'{a}mica producida por la teor\'{\i}a, encontrando que la forma del \'{a}lgebra y del tensor invariante conducen a una geometr\'{\i}a fuertemente constre\~{n}ida. Proponemos algunas alternativas para mejorar el comportamiento din\'{a}mico basadas en lo aprendido a trav\'{e}s del procedimiento de $S$-Expansi\'{o}n.

El trabajo desarrollado en esta Tesis puede ser ampliado en varias direcciones.

En el marco de las teor\'{\i}as de gauge transgresoras, existen muchos ejemplos interesantes, correspondientes a diferentes \'{a}lgebras y tensores invariantes, que pueden ser explorados y utilizados como base para modelos fenomenol\'{o}gicos, ya sea de objetos locales como agujeros negros y de gusano, como de cosmolog\'{\i}a.

La $S$-Expansi\'{o}n probablemente admita una generalizaci\'{o}n al caso de un semigrupo infinito con un n\'{u}mero contable de elementos. Esto permitir\'{\i}a hacer contacto con las \'{a}lgebras de Kac--Moody, de inter\'{e}s tanto para matem\'{a}ticos como para f\'{\i}sicos. Otra generalizaci\'{o}n imaginable consiste en admitir la posibilidad de un conjunto con elementos conmutantes y anticonmutantes que tome el lugar del semigrupo. Si esta idea resulta factible, proveer\'{\i}a de un medio de derivar super\'{a}lgebras a partir de \'{a}lgebras de Lie ordinarias.

\appendix

\chapter{\label{Ap:NC}Notaci\'{o}n y Convenciones}

La dimensi\'{o}n del espacio-tiempo es denotada por $d$. \'{E}sta es casi
siempre impar, $d=2n+1$.

\'{I}ndices latinos en may\'{u}sculas ($A,B,C,\dotsc$) se usan para denotar
conjuntos de \'{\i}ndices o \'{\i}ndices de conjuntos no especificados;
t\'{\i}picamente, generadores de un \'{a}lgebra de Lie arbitraria.

\'{I}ndices latinos en min\'{u}sculas a partir del comienzo del alfabeto
($a,b,c,\dotsc$) denotan \'{\i}ndices de Lorentz, i.e. \'{\i}ndices del
espacio tangente.

\'{I}ndices latinos en min\'{u}sculas a partir de la mitad del alfabeto
($i,j,k,\dotsc$) son usados en multitud de contextos distintos; para las
componentes no nulas de un semigrupo, para contar tensores de Lorentz, etc.

\'{I}ndices griegos a partir del comienzo del alfabeto ($\alpha,\beta
,\gamma,\dotsc$) son usados en dos contextos distintos: para las componentes
de un semigrupo gen\'{e}rico $S$ y para las componentes de spinores y matrices
de Dirac.

\'{I}ndices griegos a partir de la mitad del alfabeto ($\mu,\nu,\dotsc$) son
usados (escasamente) para componentes de vectores o tensores en la base coordenada.

Las \'{a}lgebras de Lie tienen nombres en fuente $\mathfrak{Fraktur}$:
$\mathfrak{osp}\left(  32|1\right)  $, $\mathfrak{M}$, $\mathfrak{so}\left(
n\right)  $.

Los objetos valuados en un \'{a}lgebra de Lie son escritos en negrita:
$\bm{A}$, $\bm{T}_{A}$.

El s\'{\i}mbolo $\wedge$ utilizado a menudo para denotar el producto entre formas diferenciales es frecuentemente omitido a lo largo de la Tesis.

La simetrizaci\'{o}n y antisimetrizaci\'{o}n son realizadas con peso 1, i.e. $Z_{\left[ ab \right]} = \frac{1}{2} \left( Z_{ab} - Z_{ba} \right) $.

\chapter{\label{Ap:CargasBig}Cargas Conservadas para la Acci\'{o}n Transgresora}
\chaptermark{Cargas Transgresoras}

\section{\label{Ap:Noether}El Teorema de Noether}

En este Ap\'{e}ndice damos una demostraci\'{o}n del Teorema de Noether en el
lenguaje de las formas diferenciales. El objetivo de este Ap\'{e}ndice es
proveer de un fundamento coherente para las f\'{o}rmulas usadas para calcular
corrientes conservadas para la acci\'{o}n transgresora (ver
Ap\'{e}ndice~\ref{Ap:Cargas}), as\'{\i} como clarificar la notaci\'{o}n y las
convenciones usadas.

Sea $L=L\left(  \phi\right)  $ una $d$-forma lagrangeana para un campo $\phi$
arbitrario (por supuesto, $\phi$ representa en general un conjunto de campos).
Asumiremos que este lagrangeano es invariante bajo difeomorfismos y bajo
transformaciones de gauge asociadas a una cierta \'{a}lgebra de Lie.

Bajo una variaci\'{o}n infinitesimal arbitraria $\phi\rightarrow\phi
+\delta\phi$, el lagrangeano cambia en
\begin{equation}
\delta L=E\left(  \phi\right)  \delta\phi+\mathrm{d}\Xi\left(  \phi,\delta
\phi\right)  , \label{deltaLgral}%
\end{equation}
donde $E\left(  \phi\right)  =0$ corresponden a las ecuaciones de movimiento y
$\Xi$ es una $\left(  d-1\right)  $-forma que depende de $\phi$ y de su
variaci\'{o}n $\delta\phi$.

Cuando la variaci\'{o}n $\delta\phi$ corresponde a una transformaci\'{o}n de
gauge, el lagrangeano variar\'{a} a lo m\'{a}s en una forma exacta,
\begin{equation}
\delta_{\mathrm{gauge}}L=\mathrm{d}\Omega.
\end{equation}
Esto significa que, cuando las ecuaciones de movimiento $E\left(  \phi\right)
=0$ son satisfechas, la corriente
\begin{equation}
\left.  \star J_{\mathrm{gauge}}\right.  =\Omega-\Xi_{\mathrm{gauge}}
\label{JgaugeDef}%
\end{equation}
es conservada \textit{on-shell}, i.e. $\mathrm{d}\left.  \star
J_{\mathrm{gauge}}\right.  =0$. La notaci\'{o}n $\Xi_{\mathrm{gauge}}$ usada
en~(\ref{JgaugeDef}) indica que debemos reemplazar en $\Xi\left(  \phi
,\delta\phi\right)  $ la transformaci\'{o}n de gauge co\-rrespondiente a $\phi$,
es decir, $\Xi_{\mathrm{gauge}}\equiv\Xi\left(  \phi,\delta_{\mathrm{gauge}%
}\phi\right)  $. La ec.~(\ref{JgaugeDef}) es la base para el c\'{a}lculo de
la corriente de gauge para la acci\'{o}n transgresora presentado en el
Ap\'{e}ndice~\ref{Ap:cgauge}.

Bajo un difeomorfismo infinitesimal $\delta x^{\mu}=\xi^{\mu}\left(  x\right)
$, la variaci\'{o}n funcional del lagrangeano puede escribirse en la forma
\begin{align}
\delta_{\mathrm{dif}}L  &  =-\pounds _{\xi}L\nonumber\\
&  =-\left(  \mathrm{dI}_{\xi}+\mathrm{I}_{\xi}\mathrm{d}\right)  L\nonumber\\
&  =-\mathrm{dI}_{\xi}L, \label{deltadiffL}%
\end{align}
donde hemos usado la identidad $\mathrm{d}L=0$. Reemplazando~(\ref{deltadiffL}%
) en~(\ref{deltaLgral}) encontramos que, cuando las ecuaciones de movimiento
$E\left(  \phi\right)  =0$ son satisfechas, la corriente
\begin{equation}
\left.  \star J_{\mathrm{dif}}\right.  =-\Xi_{\mathrm{dif}}-\mathrm{I}_{\xi}L
\label{JdifDef}%
\end{equation}
es conservada \textit{on-shell}, i.e. $\mathrm{d}\left.  \star J_{\mathrm{dif}%
}\right.  =0$. La notaci\'{o}n $\Xi_{\mathrm{dif}}$ usada en~(\ref{JdifDef})
indica que debemos reemplazar en $\Xi\left(  \phi,\delta\phi\right)  $ la
acci\'{o}n del difeomorfismo sobre $\phi$, es decir, $\Xi_{\mathrm{dif}}%
\equiv\Xi\left(  \phi,\delta_{\mathrm{dif}}\phi\right)  $. La
ec.~(\ref{JdifDef}) es la base para el c\'{a}lculo de la corriente de
difeomorfismos para la acci\'{o}n transgresora presentado en el
Ap\'{e}ndice~\ref{Ap:cdiff}.

\section[Corrientes conservadas para la acci\'{o}n transgresora]%
{\label{Ap:Cargas}Corrientes Conservadas para la Acci\'{o}n Transgresora
\sectionmark{Corrientes Transgresoras}}
\sectionmark{Corrientes Transgresoras}

En este Ap\'{e}ndice presentamos una deducci\'{o}n de las corrientes de
Noether asociadas a transformaciones de gauge y difeomorfismos para la
acci\'{o}n transgresora.

Con la notaci\'{o}n del Ap\'{e}ndice~\ref{Ap:Noether} y la
secci\'{o}n~\ref{sec:eomTr}, tenemos
\begin{equation}
\Xi=n\left(  n+1\right)  k\int_{0}^{1}dt\left\langle \delta\bm{A}%
_{t}\bm{\Theta F}_{t}^{n-1}\right\rangle . \label{Xi0}%
\end{equation}
Parte importante del c\'{a}lculo desarrollado en las secciones~\ref{Ap:cgauge}
y~\ref{Ap:cdiff} consiste en particularizar $\Xi$ para los casos de
transformaciones de gauge y difeomorfismos.

\subsection{\label{Ap:cgauge}Corriente de gauge}

Como se mostr\'{o} en el Ap\'{e}ndice~\ref{Ap:Noether}, la corriente de
Noether asociada a transformaciones de gauge tiene la forma
general\ [cf.~ec.~(\ref{JgaugeDef})]%
\begin{equation}
\left.  \star J_{\mathrm{gauge}}\right.  =\Omega-\Xi_{\mathrm{gauge}}.
\label{Jgaugecopy}%
\end{equation}

En el caso particular de la acci\'{o}n transgresora~(\ref{AcTra}), el primer
t\'{e}rmino en~(\ref{Jgaugecopy}) se anula id\'{e}nticamente (ver
secci\'{o}n~\ref{sec:symtra}):
\begin{equation}
\Omega=0. \label{W=0}%
\end{equation}

Para evaluar el segundo t\'{e}rmino partimos de la variaci\'{o}n de
$\bm{A}$ y $\bar{\bm{A}}$ bajo transformaciones de gauge
[cf.~ecs.~(\ref{dgA})--(\ref{dgAb})],
\begin{align}
\delta_{\mathrm{gauge}}\bm{A}  &  =-\mathrm{D}\bm{\lambda
},\label{dgA3}\\
\delta_{\mathrm{gauge}}\bar{\bm{A}}  &  =-\mathrm{\bar{D}%
}\bm{\lambda}. \label{dgAb3}%
\end{align}
Reemplazando~(\ref{dgA3}) y~(\ref{dgAb3}) en la definici\'{o}n de
$\bm{A}_{t}$ encontramos
\begin{equation}
\delta_{\mathrm{gauge}}\bm{A}_{t}=-\mathrm{D}_{t}\bm{\lambda},
\label{deltagaugeAt}%
\end{equation}
donde $\mathrm{D}_{t}$ denota la derivada covariante en la conexi\'{o}n
$\bm{A}_{t}$. Sustituyendo ahora~(\ref{dgAb3}) en~(\ref{Xi0})
hallamos
\begin{equation}
\Xi_{\mathrm{gauge}}=-n\left(  n+1\right)  k\int_{0}^{1}dt\left\langle
\mathrm{D}_{t}\bm{\lambda\Theta F}_{t}^{n-1}\right\rangle .
\end{equation}

Usando la identidad de Bianchi $\mathrm{D}_{t}\bm{F}_{t}=0$, la regla
de Leibniz para $\mathrm{D}_{t}$ y la propiedad de invariancia del tensor
sim\'{e}trico $\left\langle \cdots\right\rangle $, obtenemos
\begin{equation}
\Xi_{\mathrm{gauge}}=-n\left(  n+1\right)  k\mathrm{d}\int_{0}^{1}%
dt\left\langle \bm{\lambda\Theta F}_{t}^{n-1}\right\rangle +n\left(
n+1\right)  k\int_{0}^{1}dt\left\langle \bm{\lambda}\mathrm{D}%
_{t}\bm{\Theta F}_{t}^{n-1}\right\rangle .
\end{equation}

Ahora reemplazamos la identidad $\left(  d/dt\right)  \bm{F}%
_{t}=\mathrm{D}_{t}\bm{\Theta}$ e integramos por partes en $t$ en el
segundo t\'{e}rmino para obtener
\begin{equation}
\Xi_{\mathrm{gauge}}=-n\left(  n+1\right)  k\mathrm{d}\int_{0}^{1}%
dt\left\langle \bm{\lambda\Theta F}_{t}^{n-1}\right\rangle +\left(
n+1\right)  k\int_{0}^{1}dt\frac{d}{dt}\left\langle \bm{\lambda F}%
_{t}^{n}\right\rangle ,
\end{equation}
de donde nos vemos conducidos directamente a
\begin{equation}
\Xi_{\mathrm{gauge}}=-n\left(  n+1\right)  k\mathrm{d}\int_{0}^{1}%
dt\left\langle \bm{\lambda\Theta F}_{t}^{n-1}\right\rangle +\left(
n+1\right)  k\left(  \left\langle \bm{\lambda F}^{n}\right\rangle
-\left\langle \bm{\lambda}\bar{\bm{F}}^{n}\right\rangle
\right)  . \label{Xi-gauge}%
\end{equation}

Sustituyendo~(\ref{W=0}) y~(\ref{Xi-gauge}) en~(\ref{Jgaugecopy}) hallamos
finalmente la corriente de gauge
\begin{equation}
\left.  \star J_{\mathrm{gauge}}\right.  =n\left(  n+1\right)  k\mathrm{d}%
\int_{0}^{1}dt\left\langle \bm{\lambda\Theta F}_{t}^{n-1}\right\rangle
.
\end{equation}
Al escribir esta corriente hemos omitido t\'{e}rminos proporcionales a las
ecuaciones de movimiento, de modo que ella s\'{o}lo est\'{a} definida
\textit{on-shell}.

\subsection{\label{Ap:cdiff}Corriente de difeomorfismos}

Como se mostr\'{o} en el Ap\'{e}ndice~\ref{Ap:Noether}, la corriente de
Noether asociada a la invariancia bajo difeomorfismos de la acci\'{o}n tiene
la forma general\ [cf.~ec.~(\ref{JdifDef})]%
\begin{equation}
\left.  \star J_{\mathrm{dif}}\right.  =-\Xi_{\mathrm{dif}}-\mathrm{I}_{\xi
}L_{\mathrm{T}}^{\left(  2n+1\right)  }. \label{JdifDefcopy}%
\end{equation}

Para evaluar el primer t\'{e}rmino en~(\ref{JdifDefcopy}) partimos de la
variaci\'{o}n de $\bm{A}$ y $\bar{\bm{A}}$ bajo difeomorfismos
[cf.~ecs.~(\ref{ddA})--(\ref{ddAb})],
\begin{align}
\delta_{\mathrm{dif}}\bm{A}  &  =-\pounds _{\xi}\bm{A}%
,\label{ddA3}\\
\delta_{\mathrm{dif}}\bar{\bm{A}}  &  =-\pounds _{\xi}\bar
{\bm{A}}. \label{ddAb3}%
\end{align}
Reemplazando~(\ref{ddA3}) y~(\ref{ddAb3}) en la definici\'{o}n de
$\bm{A}_{t}$ encontramos
\begin{equation}
\delta_{\mathrm{dif}}\bm{A}_{t}=-\pounds _{\xi}\bm{A}_{t}.
\label{deltadifAt}%
\end{equation}
Sustituyendo~(\ref{deltadifAt}) en~(\ref{Xi0}) obtenemos%

\begin{equation}
\Xi_{\mathrm{dif}}=-n\left(  n+1\right)  k\int_{0}^{1}dt\left\langle
\pounds _{\xi}\bm{A}_{t}\bm{\Theta F}_{t}^{n-1}\right\rangle .
\end{equation}
Usando ahora la identidad
\begin{equation}
\pounds _{\xi}\bm{A}_{t}=\mathrm{I}_{\xi}\bm{F}_{t}%
+\mathrm{D}_{t}\mathrm{I}_{\xi}\bm{A}_{t}%
\end{equation}
podemos escribir
\begin{equation}
\Xi_{\mathrm{dif}}=-n\left(  n+1\right)  k\int_{0}^{1}dt\left\langle
\mathrm{I}_{\xi}\bm{F}_{t}\bm{\Theta F}_{t}^{n-1}\right\rangle
-n\left(  n+1\right)  k\int_{0}^{1}dt\left\langle \mathrm{D}_{t}%
\mathrm{I}_{\xi}\bm{A}_{t}\bm{\Theta F}_{t}^{n-1}\right\rangle
.
\end{equation}
Ocupando la regla de Leibniz para el operador de contracci\'{o}n
$\mathrm{I}_{\xi}$ y para la derivada covariante $\mathrm{D}_{t}$, m\'{a}s la
identidad de Bianchi $\mathrm{D}_{t}\bm{F}_{t}=0$ y las propiedades
del tensor sim\'{e}trico invariante $\left\langle \cdots\right\rangle $
hallamos
\begin{align*}
\Xi_{\mathrm{dif}} = & -\mathrm{I}_{\xi}L_{\mathrm{T}}^{\left(  2n+1\right)
}+\left(  n+1\right)  k\int_{0}^{1}dt\left\langle \mathrm{I}_{\xi
}\bm{\Theta F}_{t}^{n}\right\rangle +\\
&  -n\left(  n+1\right)  k\mathrm{d}\int_{0}^{1}dt\left\langle \mathrm{I}%
_{\xi}\bm{A}_{t}\bm{\Theta F}_{t}^{n-1}\right\rangle +\\
&  +n\left(  n+1\right)  k\int_{0}^{1}dt\left\langle \mathrm{I}_{\xi
}\bm{A}_{t}\mathrm{D}_{t}\bm{\Theta F}_{t}^{n-1}\right\rangle
.
\end{align*}
A continuaci\'{o}n hacemos uso de las identidades
\begin{align}
\frac{d}{dt}\bm{F}_{t}  &  =\mathrm{D}_{t}\bm{\Theta},\\
\frac{d}{dt}\mathrm{I}_{\xi}\bm{A}_{t}  &  =\mathrm{I}_{\xi
}\bm{\Theta},
\end{align}
en el segundo y cuarto t\'{e}rminos e integramos por partes en $t$ para
obtener
\begin{align}
\Xi_{\mathrm{dif}} = & -\mathrm{I}_{\xi}L_{\mathrm{T}}^{\left(  2n+1\right)
}+\left(  n+1\right)  k\int_{0}^{1}dt\frac{d}{dt}\left\langle \mathrm{I}_{\xi
}\bm{A}_{t}\bm{F}_{t}^{n}\right\rangle +\nonumber\\
&  -n\left(  n+1\right)  k\mathrm{d}\int_{0}^{1}dt\left\langle \mathrm{I}%
_{\xi}\bm{A}_{t}\bm{\Theta F}_{t}^{n-1}\right\rangle ,
\end{align}
de donde nos vemos conducidos directamente a
\begin{align}
\Xi_{\mathrm{dif}}+\mathrm{I}_{\xi}L_{\mathrm{T}}^{\left(  2n+1\right)  }
= & -n\left(  n+1\right)  k\mathrm{d}\int_{0}^{1}dt\left\langle \mathrm{I}_{\xi
}\bm{A}_{t}\bm{\Theta F}_{t}^{n-1}\right\rangle +\nonumber\\
&  +\left(  n+1\right)  k\left(  \left\langle \mathrm{I}_{\xi}\bm{AF}%
^{n}\right\rangle -\left\langle \mathrm{I}_{\xi}\bar{\bm{A}}%
\bar{\bm{F}}^{n}\right\rangle \right)  . \label{Xi-dif}%
\end{align}

Sustituyendo~(\ref{Xi-dif}) en~(\ref{JdifDefcopy}) hallamos la corriente de
difeomorfismos
\begin{equation}
\left.  \star J_{\mathrm{dif}}\right.  =n\left(  n+1\right)  k\mathrm{d}%
\int_{0}^{1}dt\left\langle \mathrm{I}_{\xi}\bm{A}_{t}%
\bm{\Theta F}_{t}^{n-1}\right\rangle .
\end{equation}
Al escribir esta corriente hemos omitido t\'{e}rminos proporcionales a las
ecuaciones de movimiento, de modo que ella s\'{o}lo est\'{a} definida
\textit{on-shell}.

\chapter{\label{Ap:Delta}La Delta de Kronecker generalizada}

La delta de Kronecker generalizada es definida mediante el determinante
\begin{equation}
\delta_{b_{1}\cdots b_{n}}^{a_{1}\cdots a_{n}}\equiv\det\left[
\begin{array}
[c]{cccc}%
\delta_{b_{1}}^{a_{1}} & \delta_{b_{2}}^{a_{1}} & \cdots & \delta_{b_{n}%
}^{a_{1}}\\
\delta_{b_{1}}^{a_{2}} & \delta_{b_{2}}^{a_{2}} & \cdots & \delta_{b_{n}%
}^{a_{2}}\\
\vdots & \vdots & \ddots & \vdots\\
\delta_{b_{1}}^{a_{n}} & \delta_{b_{2}}^{a_{n}} & \cdots & \delta_{b_{n}%
}^{a_{n}}%
\end{array}
\right]  , \label{deltadef}%
\end{equation}
con $1\leq n\leq d$, siendo $d$ el rango de los \'{\i}ndices (usualmente
identificado con la dimensionalidad del espacio-tiempo).

De su definici\'{o}n, la delta $\delta_{b_{1}\cdots b_{n}}^{a_{1}\cdots a_{n}%
}$ es un tensor de tipo $\left(  n,n\right)  $ totalmente antisim\'{e}trico,
i.e. $\delta_{b_{1}\cdots b_{n}}^{a_{1}\cdots a_{n}} = \delta_{b_{1}\cdots b_{n}}^{\left[  a_{1}\cdots a_{n}\right]  } = \delta_{\left[ b_{1}\cdots b_{n}\right]  }^{a_{1}\cdots a_{n}}$.

\section{Identidades de Contracci\'{o}n}

Contrayendo $n-r$ \'{\i}ndices en una delta con $n$ \'{\i}ndices se obtiene
una delta con $r$ \'{\i}ndices:
\begin{equation}
\delta_{b_{1}\cdots b_{r}a_{r+1}\cdots a_{n}}^{a_{1}\cdots a_{r}a_{r+1}\cdots
a_{n}}=\frac{\left(  d-r\right)  !}{\left(  d-n\right)  !}\delta_{b_{1}\cdots
b_{r}}^{a_{1}\cdots a_{r}}.
\end{equation}
Un caso particular de inter\'{e}s es
\begin{equation}
\delta_{a_{1}\cdots a_{n}}^{a_{1}\cdots a_{n}}=\frac{d!}{\left(  d-n\right)
!}.
\end{equation}

Sea $A^{a_{1}\cdots a_{n}}$ un tensor totalmente antisim\'{e}trico. Entonces,
\begin{align}
\delta_{b_{1}\cdots b_{n}}^{a_{1}\cdots a_{n}}A^{b_{1}\cdots b_{n}}  &
=n!A^{a_{1}\cdots a_{n}},\\
\delta_{b_{1}\cdots b_{n}}^{a_{1}\cdots a_{n}}A_{a_{1}\cdots a_{n}}  &
=n!A_{b_{1}\cdots b_{n}}.
\end{align}

El s\'{\i}mbolo de permutaciones de Levi-Civita puede ser escrito como un caso
particular de la delta de Kronecker con $n=d$:
\begin{align}
\varepsilon_{a_{1}\cdots a_{d}}  &  =\delta_{a_{1}\cdots a_{d}}^{1\cdots d},\\
\varepsilon^{a_{1}\cdots a_{d}}  &  =\delta_{1\cdots d}^{a_{1}\cdots a_{d}}.
\end{align}
El producto de dos s\'{\i}mbolos de Levi-Civita permite recuperar la delta:
\begin{equation}
\varepsilon^{a_{1}\cdots a_{d}}\varepsilon_{b_{1}\cdots b_{d}}=\delta
_{b_{1}\cdots b_{d}}^{a_{1}\cdots a_{d}}.
\end{equation}

\section{Particiones de la Delta}

Desarrollando el determinante~(\ref{deltadef}) a largo de la primera fila
encontramos
\begin{equation}
\delta_{b_{1}\cdots b_{n}}^{a_{1}\cdots a_{n}}=\sum_{p=1}^{n}\left(
-1\right)  ^{p+1}\delta_{b_{p}}^{a_{1}}\delta_{b_{1}\cdots\hat{b}_{p}\cdots
b_{n}}^{a_{2}\cdots a_{n}}, \label{dr1}%
\end{equation}
donde la notaci\'{o}n $\hat{b}_{p}$ indica que el \'{\i}ndice $b_{p}$ debe ser
excluido. Iterando una vez esta identidad podemos escribir
\begin{equation}
\delta_{b_{1}\cdots b_{n}}^{a_{1}\cdots a_{n}}=\sum_{p=1}^{n-1}\sum
_{q=p+1}^{n}\left(  -1\right)  ^{p+q+1}\delta_{b_{p}b_{q}}^{a_{1}a_{2}}%
\delta_{b_{1}\cdots\hat{b}_{p}\cdots\hat{b}_{q}\cdots b_{n}}^{a_{2}\cdots
a_{n}}. \label{dr2}%
\end{equation}

El caso general, donde se particiona una delta con $n$ \'{\i}ndices en una
delta con $r$ por una delta con $n-r$, es
\begin{align}
\delta_{b_{1}\cdots b_{n}}^{a_{1}\cdots a_{n}}  &  =\left(  -1\right)
^{r\left(  r+1\right)  /2}\sum_{p_{1}=1}^{n-r+1}\sum_{p_{2}=p_{1}+1}%
^{n-r+2}\cdots\sum_{p_{r-1}=p_{r-2}+1}^{n-1}\nonumber\\
&  \sum_{p_{r}=p_{r-1}+1}^{n}\left(  -1\right)  ^{p_{1}+\dotsb+p_{r}}%
\delta_{b_{p_{1}}\cdots b_{p_{r}}}^{a_{1}\cdots a_{r}}\delta_{b_{1}\cdots
\hat{b}_{p_{1}}\cdots\hat{b}_{p_{r}}\cdots b_{n}}^{a_{r+1}\cdots a_{n}}.
\label{dr}%
\end{align}
Por supuesto, las identidades~(\ref{dr1})--(\ref{dr}) pueden tambi\'{e}n
escribirse desarrollando el determinante por columnas, en cuyo caso obtenemos
\begin{align}
\delta_{b_{1}\cdots b_{n}}^{a_{1}\cdots a_{n}}  &  =\left(  -1\right)
^{r\left(  r+1\right)  /2}\sum_{p_{1}=1}^{n-r+1}\sum_{p_{2}=p_{1}+1}%
^{n-r+2}\cdots\sum_{p_{r-1}=p_{r-2}+1}^{n-1}\nonumber\\
&  \sum_{p_{r}=p_{r-1}+1}^{n}\left(  -1\right)  ^{p_{1}+\dotsb+p_{r}}%
\delta_{b_{1}\cdots b_{r}}^{a_{p_{1}}\cdots a_{p_{r}}}\delta_{b_{r+1}\cdots
b_{n}}^{a_{1}\cdots\hat{a}_{p_{1}}\cdots\hat{a}_{p_{r}}\cdots a_{n}}.
\end{align}

La siguiente identidad corresponde a una partici\'{o}n `impl\'{\i}cita' de la
delta:
\begin{equation}
\delta_{b_{1}\cdots b_{n}}^{a_{1}\cdots a_{n}}=\binom{n}{p}\delta_{b_{1}\cdots
b_{p}}^{[a_{1}\cdots a_{p}}\delta_{b_{p+1}\cdots b_{n}}^{a_{p+1}\cdots a_{n}%
]}=\binom{n}{p}\delta_{\lbrack b_{1}\cdots b_{p}}^{a_{1}\cdots a_{p}}%
\delta_{b_{p+1}\cdots b_{n}]}^{a_{p+1}\cdots a_{n}}.
\end{equation}

\section{La $\eta$ antisimetrizada}

Resulta muy conveniente definir
\begin{align}
\eta_{\left[  a_{1}\cdots a_{p}\right]  \left[  b_{1}\cdots b_{p}\right]  }
&  \equiv\delta_{a_{1}\cdots a_{p}}^{c_{1}\cdots c_{p}}\left(  \eta
_{b_{1}c_{1}}\cdots\eta_{b_{p}c_{p}}\right)  ,\\
\eta^{\left[  a_{1}\cdots a_{p}\right]  \left[  b_{1}\cdots b_{p}\right]  }
&  \equiv\delta_{c_{1}\cdots c_{p}}^{a_{1}\cdots a_{p}}\left(  \eta
^{b_{1}c_{1}}\cdots\eta^{b_{p}c_{p}}\right)  .
\end{align}
Estos tensores son completamente \emph{antisim\'{e}tricos} en los \'{\i}ndices
indicados ($a_{1}\cdots a_{p}$ y $b_{1}\cdots b_{p}$), siendo a la vez
\emph{sim\'{e}tricos} bajo intercambios de los dos grupos antisim\'{e}tricos
de \'{\i}ndices:
\begin{align}
\eta_{\left[  b_{1}\cdots b_{p}\right]  \left[  a_{1}\cdots a_{p}\right]  }
&  =\eta_{\left[  a_{1}\cdots a_{p}\right]  \left[  b_{1}\cdots b_{p}\right]
},\\
\eta^{\left[  b_{1}\cdots b_{p}\right]  \left[  a_{1}\cdots a_{p}\right]  }
&  =\eta^{\left[  a_{1}\cdots a_{p}\right]  \left[  b_{1}\cdots b_{p}\right]
}.
\end{align}

La $\eta$ antisimetrizada act\'{u}a como una m\'{e}trica en el espacio de los
tensores completamente antisim\'{e}tricos:
\begin{align}
\eta_{\left[  a_{1}\cdots a_{p}\right]  \left[  b_{1}\cdots b_{p}\right]
}A^{b_{1}\cdots b_{p}}  &  =p!A_{a_{1}\cdots a_{p}},\\
\eta^{\left[  a_{1}\cdots a_{p}\right]  \left[  b_{1}\cdots b_{p}\right]
}A_{b_{1}\cdots b_{p}}  &  =p!A^{a_{1}\cdots a_{p}}.
\end{align}

La contracci\'{o}n entre etas devuelve una delta:
\begin{equation}
\eta^{\left[  a_{1}\cdots a_{p}\right]  \left[  c_{1}\cdots c_{p}\right]
}\eta_{\left[  c_{1}\cdots c_{p}\right]  \left[  b_{1}\cdots b_{p}\right]
}=p!\delta_{b_{1}\cdots b_{p}}^{a_{1}\cdots a_{p}}.
\end{equation}

\chapter{\label{Ap:Cliff}El \'{A}lgebra de Clifford}

En este Ap\'{e}ndice presentamos brevemente los aspectos m\'{a}s importantes
relacionados con el \'{A}lgebra de Clifford que son ocupados a lo largo de la
Tesis, especialmente en el cap\'{\i}tulo~\ref{ch:TGFTMAlg}. Una referencia
excelente para este Ap\'{e}ndice es el art\'{\i}culo de A.~Van~Proeyen~\cite{VanPro99}.

\section{\label{Ap:DiracDef}Definiciones}

Las \emph{matrices de Dirac} en un espacio-tiempo de $d$ dimensiones son
definidas mediante el requerimiento
\begin{equation}
\left\{  \Gamma_{a},\Gamma_{b}\right\}  =2\eta_{ab}, \label{cl}%
\end{equation}
donde $\eta_{ab}=\operatorname{diag}\left(  -,+,\dotsc,+\right)  $ es la
m\'{e}trica de Minkowski. El \'{a}lgebra~(\ref{cl}) satisfecha por estas
matrices es conocida como \emph{\'{A}lgebra de Clifford}. Las matrices
$\Gamma_{a}$ son de dimensi\'{o}n $m\times m$, donde $m=2^{\left[  d/2\right]
}$ (esto puede deducirse contando cuantas matrices linealmente independientes
pueden generarse a partir de las matrices de Dirac y sus productos).

De sumo inter\'{e}s son los productos antisimetrizados de matrices de Dirac:
\begin{align}
\Gamma_{a_{1}\cdots a_{p}}  &  \equiv\Gamma_{\left[  a_{1}\right.  }%
\cdots\Gamma_{\left.  a_{p}\right]  }\\
&  =\frac{1}{p!}\delta_{a_{1}\cdots a_{p}}^{b_{1}\cdots b_{p}}\Gamma_{b_{1}%
}\cdots\Gamma_{b_{p}}.
\end{align}

Es tambi\'{e}n usual definir la matriz $\Gamma_{\ast}$ como
\begin{align}
\Gamma_{\ast}  &  \equiv\Gamma_{0}\cdots\Gamma_{d-1}\label{g51}\\
&  =\frac{1}{d!}\varepsilon^{a_{1}\cdots a_{d}}\Gamma_{a_{1}}\cdots
\Gamma_{a_{d}}\label{g52}\\
&  =\frac{1}{d!}\varepsilon^{a_{1}\cdots a_{d}}\Gamma_{a_{1}\cdots a_{d}}.
\label{g53}%
\end{align}

Esta matriz satisface las identidades
\begin{align}
\Gamma_{\ast}^{2}  &  =\left(  -1\right)  ^{\left(  d-2\right)  \left(
d+1\right)  /2},\\
\Gamma_{\ast}\Gamma_{a}  &  =\left(  -1\right)  ^{d+1}\Gamma_{a}\Gamma_{\ast
},\label{gga}\\
\Gamma_{\ast}\Gamma_{a_{1}\cdots a_{p}}  &  =\left(  -1\right)  ^{p\left(
d+1\right)  }\Gamma_{a_{1}\cdots a_{p}}\Gamma_{\ast}.
\end{align}

Como $\Gamma_{\ast}$ conmuta con todas las matrices de Dirac en dimensiones
impares, ella debe ser igual a un m\'{u}ltiplo de la matriz identidad:
$\Gamma_{\ast}=\sigma\mathbbm{1}$. M\'{a}s a\'{u}n, como en dimensi\'{o}n
$d=2n+1$ tenemos $\Gamma_{\ast}^{2}=\left(  -1\right)  ^{n+1}\mathbbm{1}$,
vemos que $\sigma=\pm i^{n+1}$.

La siguiente identidad es satisfecha:
\begin{equation}
\Gamma_{a_{1}\cdots a_{d-k}}=\frac{1}{k!}\left(  -1\right)  ^{k\left(
k-1\right)  /2}\varepsilon_{a_{1}\cdots a_{d}}\Gamma^{a_{d-k+1}\cdots a_{d}%
}\Gamma_{\ast}. \label{GeGG}%
\end{equation}
En dimensiones impares, $\Gamma_{\ast}=\pm i^{n+1}$, y~(\ref{GeGG})
proporciona una relaci\'{o}n de dualidad entre $\Gamma_{\left[  p\right]  }$ y
$\Gamma_{\left[  d-p\right]  }$ que ser\'{a} ampliamente utilizada.

El signo de $\sigma$ en $\Gamma_{\ast}=\sigma$ sirve para etiquetar las dos
representaciones inequivalentes para las matrices de Dirac que existen en
$d=2n+1$.

\section{\label{Ap:DiracSym}Simetr\'{\i}as}

Las matrices $\Gamma_{a}^{T}$ y $-\Gamma_{a}^{T}$ tambi\'{e}n satisfacen el
\'{a}lgebra de Clifford~(\ref{cl}), de modo que deben estar relacionadas con
$\Gamma_{a}$ por medio de una transformaci\'{o}n de (anti) similaridad:
\begin{equation}
\Gamma_{a}^{T}=\xi C\Gamma_{a}C^{-1},
\end{equation}
donde $C$ es conocida como la \emph{matriz de conjugaci\'{o}n de carga}. Esta
matriz $C$ puede ser sim\'{e}trica o antisim\'{e}trica:
\begin{equation}
C^{T}=\lambda C,
\end{equation}
con $\xi^{2}=\lambda^{2}=+1$.

Puede demostrarse f\'{a}cilmente que la simetr\'{\i}a de las matrices
$\Gamma_{a_{1}\cdots a_{p}}$ queda determinada a partir de la simetr\'{\i}a de
$C$ y de $\Gamma_{a}$ a trav\'{e}s de la ecuaci\'{o}n
\begin{equation}
\left(  C\Gamma_{a_{1}\cdots a_{p}}\right)  ^{T}=\left(  -1\right)  ^{p\left(
p-1\right)  /2}\lambda\xi^{p}\left(  C\Gamma_{a_{1}\cdots a_{p}}\right)  .
\label{CGT}%
\end{equation}
De~(\ref{CGT}) vemos que la matriz $\Gamma_{\left[  p\right]  }$ tiene siempre
la misma simetr\'{\i}a que la matriz $\Gamma_{\left[  p+4\right]  }$.

\begin{table}
\begin{center}
\begin{tabular}[c]{||c|c|c|c|c||}
\hline
\hline
$d \operatorname{mod} 8$ & $\lambda$ & $\xi$ & S & A\\
\hline
\hline
$0$ & $+1$ & $+1$ & $1,4$ & $2,3$\\
$0$ & $+1$ & $-1$ & $3,4$ & $1,2$\\\hline
$1$ & $+1$ & $+1$ & $1,4$ & $2,3$\\\hline
$2$ & $+1$ & $+1$ & $1,4$ & $2,3$\\
$2$ & $-1$ & $-1$ & $1,2$ & $3,4$\\\hline
$3$ & $-1$ & $-1$ & $1,2$ & $3,4$\\\hline
$4$ & $-1$ & $+1$ & $2,3$ & $1,4$\\
$4$ & $-1$ & $-1$ & $1,2$ & $3,4$\\\hline
$5$ & $-1$ & $+1$ & $2,3$ & $1,4$\\\hline
$6$ & $-1$ & $+1$ & $2,3$ & $1,4$\\
$6$ & $+1$ & $-1$ & $3,4$ & $1,2$\\\hline
$7$ & $+1$ & $-1$ & $3,4$ & $1,2$\\
\hline
\hline
\end{tabular}
\end{center}
\caption{La matriz de conjugaci\'{o}n de carga $C$ y las matrices de Dirac
$\Gamma_{a_{1} \cdots a_{p}}$ pueden ser sim\'{e}tricas o antisim\'{e}tricas,
de acuerdo a las ecs. $C^{T} = \lambda C$ y $\left(  C \Gamma_{a_{1} \cdots
a_{p}} \right)  ^{T} = \left(  -1 \right)  ^{p \left(  p-1 \right)  /2}
\lambda\xi^{p} \left(  C \Gamma_{a_{1} \cdots a_{p}} \right)  $, con
$\lambda^{2} = \xi^{2} = +1$. Esta tabla muestra los valores permitidos para
$\lambda$ y $\xi$ seg\'{u}n la dimensi\'{o}n del espacio-tiempo. Las
\'{u}ltimas dos columnas muestran cu\'{a}les matrices $\Gamma_{a_{1} \cdots
a_{p}}$ resultan ser sim\'{e}tricas o antisim\'{e}tricas con la elecci\'{o}n
de signos indicada. Conviene recordar que $\Gamma_{\left[  p \right]  }$ tiene
siempre la misma simetr\'{\i}a que $\Gamma_{\left[  p+4 \right]  }$.}
\label{tab:Dirac}
\end{table}

Contando cuantas matrices sim\'{e}tricas y antisim\'{e}tricas hay en cada
dimensi\'{o}n es posible obtener condiciones sobre $\lambda$ y $\xi$. El
resultado se muestra en el cuadro~\ref{tab:Dirac}, donde tambi\'{e}n se indica
cu\'{a}les matrices $C\Gamma_{\left[  p\right]  }$ resultan ser sim\'{e}tricas
o antisim\'{e}tricas con la combinaci\'{o}n de signos ocupada
($p\operatorname{mod}4$).

\section{\label{Ap:DiracDirac}El \'{A}lgebra de Dirac}

Las matrices de Dirac $\Gamma_{a_{1}\cdots a_{p}}$ forman una base para todas
las matrices de $m\times m$.

En dimensiones pares $d=2n$, todas las matrices $\Gamma_{\left[  p\right]  }$,
$p=0,1,\dotsc,2n$, son linealmente independientes; en efecto, sumando
obtenemos $\sum_{p=0}^{2n}\binom{2n}{p}=2^{2n}$, que es el n\'{u}mero de
matrices de $2^{n}\times2^{n}$ linealmente independientes.

En dimensiones impares $d=2n+1$, en cambio, la mitad de las matrices es
proporcional a la otra mitad, y uno debe escoger un subconjunto de ellas como
base. Una opci\'{o}n es considerar $\Gamma_{\left[  p\right]  }$,
$p=0,1,\dotsc,n$, las cuales totalizan el n\'{u}mero correcto: $\sum_{p=0}%
^{n}\binom{2n+1}{p}=2^{2n}$, que corresponde al n\'{u}mero de matrices de
$2^{n}\times2^{n}$ linealmente independientes.

La base provista por las matrices de Dirac es ortogonal en el sentido que
\begin{equation}
\operatorname*{Tr}\left(  \Gamma^{a_{1}\cdots a_{p}}\Gamma_{b_{1}\cdots b_{q}%
}\right)  =\left(  -1\right)  ^{p\left(  p-1\right)  /2}m\delta_{b_{1}\cdots
b_{q}}^{a_{1}\cdots a_{p}}, \label{braket}%
\end{equation}
donde entendemos que la delta $\delta_{b_{1}\cdots b_{q}}^{a_{1}\cdots a_{p}}$
es cero cuando $p\neq q$.

Usando~(\ref{braket}) es directo calcular los coeficientes en una
expansi\'{o}n del tipo
\begin{equation}
M=\sum_{p\geq0}\frac{1}{p!}k_{a_{1}\cdots a_{p}}\Gamma^{a_{1}\cdots a_{p}},
\label{MDirac}%
\end{equation}
los cuales resultan ser
\begin{equation}
k_{a_{1}\cdots a_{p}}=\frac{1}{m}\left(  -1\right)  ^{p\left(  p-1\right)
/2}\operatorname*{Tr}\left(  M\Gamma_{a_{1}\cdots a_{p}}\right)  .
\end{equation}
La sumatoria en~(\ref{MDirac}) incluye s\'{o}lo las matrices linealmente
independientes, i.e. $p=0,\dotsc,2n$ para $d=2n$ y $p=0,\dotsc,n$ para
$d=2n+1$.

\section{\label{Ap:DiracMult}Multiplicando matrices de Dirac}

Van~Proeyen~\cite{VanPro99} entrega la siguiente f\'{o}rmula para multiplicar
matrices de Dirac:
\begin{equation}
\Gamma_{a_{1}\cdots a_{i}}\Gamma^{b_{1}\cdots b_{j}}=\sum_{k=\left\vert
i-j\right\vert }^{i+j}\frac{i!j!}{s!t!u!}\delta_{\lbrack a_{i}}^{[b_{1}}%
\cdots\delta_{a_{t+1}}^{b_{s}}\Gamma_{a_{1}\cdots a_{t}]}%
^{\phantom{a_{1} \cdots a_{t} ]}b_{s+1}\cdots b_{j}]}, \label{VP1}%
\end{equation}
donde
\begin{align}
s  &  =\frac{1}{2}\left(  i+j-k\right)  ,\\
t  &  =\frac{1}{2}\left(  i-j+k\right)  ,\\
u  &  =\frac{1}{2}\left(  -i+j+k\right)  .
\end{align}
La sumatoria en~(\ref{VP1}) debe realizarse con $k$ variando de dos en dos, de
manera que $s,t,u$ s\'{o}lo tomen valores enteros. Es tambi\'{e}n importante
observar que los \'{\i}ndices inferiores en las $s$ deltas de Kronecker van en
orden \emph{decreciente}, i.e. $i,i-1,\dotsc,t+2,t+1$.

La gran ventaja de~(\ref{VP1}) radica en que todos los \'{\i}ndices son
libres, de modo que el resultado es transparente. Una desventaja (por lo menos
desde nuestro punto de vista) es que es necesario realizar antisimetrizaciones
complicadas en todos los \'{\i}ndices.

Efectuando expl\'{\i}citamente las antisimetrizaciones indicadas
en~(\ref{VP1}), reordenando los \'{\i}ndices y ocupando la identidad~(\ref{dr}%
) para partir y recombinar las deltas de Kronecker resultantes, llegamos a la
f\'{o}rmula equivalente
\begin{equation}
\Gamma^{a_{1}\cdots a_{i}}\Gamma_{b_{1}\cdots b_{j}}=\sum_{s=0}^{\min\left(
i,j\right)  }\frac{1}{t!u!}\left(  -1\right)  ^{s\left(  s-1\right)  /2}%
\delta_{d_{1}\cdots d_{t}b_{1}\cdots b_{j}}^{a_{1}\cdots a_{i}e_{1}\cdots
e_{u}}\Gamma_{\phantom{d_{1} \cdots d_{t}}e_{1}\cdots e_{u}}^{d_{1}\cdots
d_{t}}, \label{VP2}%
\end{equation}
donde ahora $t$ y $u$ est\'{a}n definidos como
\begin{align}
t  &  =i-s,\\
u  &  =j-s.
\end{align}
La sumatoria sobre $s$ en~(\ref{VP2}) va de uno en uno, a diferencia de lo que
ocurr\'{\i}a en~(\ref{VP1}). La ventaja de~(\ref{VP2}) sobre~(\ref{VP1})
radica en la ausencia de par\'{e}ntesis de antisimetr\'{\i}zaci\'{o}n; el
precio a pagar por ello es que ahora cada t\'{e}rmino contiene $t+u=i+j-2s$
\'{\i}ndices mudos.

F\'{o}rmulas parecidas a~(\ref{VP2}) pueden ser escritas para el conmutador y para el anticonmutador de dos matrices de Dirac:
\begin{align}
\left[  \Gamma^{a_{1}\cdots a_{i}},\Gamma_{b_{1}\cdots b_{j}}\right] = &
\sum_{s=0}^{\min\left(  i,j\right)  }\frac{1}{t!u!}\left(  -1\right)
^{s\left(  s-1\right)  /2} \times\nonumber\\
&  \times\left[  1-\left(  -1\right)  ^{ij-s^{2}}\right]  \delta_{d_{1}\cdots
d_{t}b_{1}\cdots b_{j}}^{a_{1}\cdots a_{i}e_{1}\cdots e_{u}}\Gamma
_{\phantom{d_{1} \cdots d_{t}}e_{1}\cdots e_{u}}^{d_{1}\cdots d_{t}},
\end{align}
\begin{align}
\left\{  \Gamma^{a_{1}\cdots a_{i}},\Gamma_{b_{1}\cdots b_{j}}\right\} = &
\sum_{s=0}^{\min\left(  i,j\right)  }\frac{1}{t!u!}\left(  -1\right)
^{s\left(  s-1\right)  /2} \times\nonumber\\
&  \times\left[  1+\left(  -1\right)  ^{ij-s^{2}}\right]  \delta_{d_{1}\cdots
d_{t}b_{1}\cdots b_{j}}^{a_{1}\cdots a_{i}e_{1}\cdots e_{u}}\Gamma
_{\phantom{d_{1} \cdots d_{t}}e_{1}\cdots e_{u}}^{d_{1}\cdots d_{t}}.
\end{align}

Una desventaja com\'{u}n a las f\'{o}rmulas~(\ref{VP1}) y~(\ref{VP2}) es que
las matrices de Dirac involucradas tienen los \'{\i}ndices en posiciones
diferentes. Esta observaci\'{o}n puede parecer pueril, pero este hecho
ciertamente impide la anidaci\'{o}n de la f\'{o}rmula para obtener productos
de un n\'{u}mero mayor matrices. Ocupando la identidad~(\ref{dr}) es posible
obtener una versi\'{o}n completamente expl\'{\i}cita del producto de dos
matrices de Dirac, con la ventaja adicional de tener todos los \'{\i}ndices en
la misma posici\'{o}n:
\begin{equation}
\Gamma_{a_{1}\cdots a_{i}}\Gamma_{b_{1}\cdots b_{j}}=\sum_{s=0}^{\min\left(
i,j\right)  }D_{a_{1}\cdots a_{i}b_{1}\cdots b_{j}}\left(  s\right)  ,
\end{equation}
con
\begin{align}
D_{a_{1}\cdots a_{i}b_{1}\cdots b_{j}}\left(  s\right)  =  &  \left(
-1\right)  ^{s\left(  i-s\right)  +s\left(  s-1\right)  /2}\sum_{p_{1}%
=1}^{1+j-s}\cdots\sum_{p_{s}=p_{s-1}+1}^{j}\nonumber\\
&  \sum_{q_{1}=1}^{1+i-s}\cdots\sum_{q_{s}=q_{s-1}+1}^{i}\left(  -1\right)
^{p_{1}+\dotsb+p_{s}+q_{1}+\dotsb+q_{s}}\nonumber\\
&  \eta_{\left[  a_{q_{1}}\cdots a_{q_{s}}\right]  \left[  b_{p_{1}}\cdots
b_{p_{s}}\right]  }\Gamma_{a_{1}\cdots\hat{a}_{q_{1}}\cdots\hat{a}_{q_{s}%
}\cdots a_{i}b_{1}\cdots\hat{b}_{p_{1}}\cdots\hat{b}_{p_{s}}\cdots b_{j}}.
\end{align}

Multiplicando por dos tensores antisim\'{e}tricos arbitrarios $A^{a_{1}\cdots
a_{i}}$ y $B^{b_{1}\cdots b_{j}}$ y ocupando nuevamente la identidad~(\ref{dr}%
), obtenemos la f\'{o}rmula
\begin{align}
A_{i}B_{j}=  &  \sum_{s=0}^{\min\left(  i,j\right)  }\binom{i}{s}\binom{j}%
{s}\left(  -1\right)  ^{s\left(  s-1\right)  /2}\eta_{\left[  b_{1}\cdots
b_{s}\right]  \left[  c_{1}\cdots c_{s}\right]  }\nonumber\\
&  A^{a_{1}\cdots a_{i-s}b_{1}\cdots b_{s}}B^{c_{1}\cdots c_{s}a_{i-s+1}\cdots
a_{i+j-2s}}\Gamma_{a_{1}\cdots a_{i+j-2s}}, \label{ABD}%
\end{align}
donde hemos usado las abreviaciones
\begin{align}
A_{i}  &  =A^{a_{1}\cdots a_{i}}\Gamma_{a_{1}\cdots a_{i}},\\
B_{j}  &  =B^{b_{1}\cdots b_{j}}\Gamma_{b_{1}\cdots b_{j}}.
\end{align}

La identidad~(\ref{ABD}) en particular nos ha resultado invaluable a la hora
de calcular productos simetrizados de matrices de Dirac, como los que se
requieren para definir un tensor invariante para $\mathfrak{osp}\left(
32|1\right)  $ (y, v\'{\i}a $S$-expansi\'{o}n, para el \'{a}lgebra~M).

Los resultados obtenidos a partir de la identidad~(\ref{ABD}) que han sido utilizados en la Tesis fueron tambi\'{e}n comprobados num\'{e}ricamente haciendo uso de un paquete especializado para \textsl{Mathematica} desarrollado por U.~Gran~\cite{Gra01}.

\section{\label{Ap:MajSpin}Spinores de Majorana en $\lowercase{d}=11$}

En este Ap\'{e}ndice listamos algunas propiedades importantes de los spinores de Majorana en $d=11$ utilizadas en la Tesis.

Un spinor de Lorentz en $d=11$ es una colecci\'{o}n de 32 n\'{u}meros de Grassmann $\psi^{\alpha}$, $\alpha=1,\dotsc,32$, a menudo denotados colectivamente simplemente por $\psi$. El spinor pertenece a la representaci\'{o}n del grupo de Lorentz dada por las matrices de Dirac, i.e. bajo una rotaci\'{o}n de Lorentz transforma de acuerdo a la ley
\begin{equation}
\psi^{\prime} = \exp \left( \frac{1}{4} \lambda^{ab} \Gamma_{ab} \right) \psi,
\label{SpinLorentz}
\end{equation}
donde $\lambda^{ab}=-\lambda^{ba}$ son los par\'{a}metros de la transformaci\'{o}n. La versi\'{o}n infinitesimal de~(\ref{SpinLorentz}) es
\begin{equation}
\delta \psi = \frac{1}{4} \lambda^{ab} \Gamma_{ab} \psi.
\label{SpinLorentzInf}
\end{equation}

Una variedad espacio-temporal [i.e. con signatura $\left( -,+,\ldots,+ \right)$] de 11 dimensiones admite spinores de Majorana~\cite{Freu88}. La propiedad de Majorana del spinor significa que el spinor $\bar{\psi}$ del espacio dual puede ser escrito en t\'{e}rminos de $\psi$ en la forma
\begin{equation}
\bar{\psi}_{\alpha} = \psi^{\beta} C_{\beta \alpha},
\end{equation}
donde $C_{\alpha \beta}=-C_{\beta \alpha}$ es la matriz de conjugaci\'{o}n de carga. La relaci\'{o}n inversa es
\begin{equation}
\psi^{\alpha} = \bar{\psi}_{\beta} C^{\beta \alpha},
\end{equation}
donde $C^{\alpha \beta}$ es la matriz inversa de $C_{\alpha \beta}$,
\begin{equation}
C^{\alpha \gamma} C_{\gamma \beta} = C_{\beta \gamma} C^{\gamma \alpha} = \delta_{\beta}^{\alpha}.
\end{equation}

Sean $\chi$ y $\zeta$ dos spinores de Majorana. Las siguientes identidades de contracci\'{o}n son satisfechas:
\begin{align}
\bar{\chi}   \zeta & =  \bar{\zeta}   \chi, \label{chizeta} \\
\bar{\chi} S \zeta & = -\bar{\zeta} S \chi, \label{chiSzeta} \\
\bar{\chi} A \zeta & =  \bar{\zeta} A \chi, \label{chiAzeta}
\end{align}
donde $S$ es una matriz sim\'{e}trica y $A$ una matriz antisim\'{e}trica, en el sentido que $C_{\beta \gamma} S^{\gamma}_{\phantom{\gamma} \alpha} = C_{\alpha \gamma} S^{\gamma}_{\phantom{\gamma} \beta}$, $C_{\beta \gamma} A^{\gamma}_{\phantom{\gamma} \alpha} = -C_{\alpha \gamma} A^{\gamma}_{\phantom{\gamma} \beta}$. Las identidades~(\ref{chizeta})--(\ref{chiAzeta}) son v\'{a}lidas cuando las componentes de $\chi$ y $\zeta$ son $0$-formas de Grassmann. Un factor de signo adicional $\left( -1 \right)^{pq}$ debe ser agregado para el caso de una $p$-forma con una $q$-forma.

Las propiedades de los spinores de Majorana en $d=11$ son usadas principalmente en el cap\'{\i}tulo~\ref{ch:TGFTMAlg}.

\end{document}